\documentclass[10pt]{article}

\usepackage[mathscr]{eucal} 
\usepackage{amssymb,latexsym,amsmath,bbm,stmaryrd}
\usepackage{enumitem}
\usepackage{tikz,tikz-qtree}
  \usetikzlibrary{automata,positioning}

\newtheorem{theorem}{Theorem}[section]
\newtheorem{fewtheorem}{Theorem}
\newtheorem{fact}[theorem]{Fact}
\newtheorem{proposition}[theorem]{Proposition}

\newtheorem{defi}[theorem]{Definition}

\newtheorem{conv}[theorem]{Convention}
\newtheorem{rema}[theorem]{Remark}
\newtheorem{exam}[theorem]{Example}

\newenvironment{definition}{\begin{defi}\rm}{\hfill $\lhd$\end{defi}}
\newenvironment{convention}{\begin{conv}\rm}{\end{conv}}
\newenvironment{remark}{\begin{rema}\rm}{\hfill $\lhd$\end{rema}}
\newenvironment{example}{\begin{exam}\rm}{\hfill $\lhd$\end{exam}}

\newenvironment{ourlist}{\begin{list}{}%
    {\setlength{\topsep}{1mm}\setlength{\itemsep}{0mm}\setlength{\parsep}{0mm}}
    }{\end{list}}

\newenvironment{urlist}{\begin{enumerate}[topsep=0pt,itemsep=-1ex,partopsep=1ex,parsep=1ex,%
    label={\arabic*)}]
    }{\end{enumerate}}

\newenvironment{proof}{\begin{trivlist}\item[]{\bf
Proof.}}{\hfill {\sc qed}\end{trivlist}}

\newenvironment{proofof}[1]{\begin{trivlist}\item[\hskip\labelsep{\bf
Proof~of~{#1}.\ }]}{\hspace*{\fill} {\sc qed}\end{trivlist}}

\newtheorem{claim2}{\sc Claim}
\newenvironment{claim}{\begin{claim2}\rm}{\end{claim2}\rm}
\newenvironment{claimfirst}{\setcounter{claim2}{0}
               \begin{claim2}\rm}{\end{claim2}\rm}

\newenvironment{pfclaim}{\begin{trivlist}\item[]{\sc Proof of
Claim}}{\hfill {\mbox{$\blacktriangleleft$}}\end{trivlist}}

\newtheorem{exer}{Exercise}[section]

\tikzset{shorten >=1pt,
   initial text={},
   node distance=24mm,
   every state/.style= {inner sep=0mm,thick},
   on grid, auto,
   every edge/.style={draw,->,>=stealth,thick}
   }

\newcommand{\mathstr}[1]{\mathbb{#1}}

\newcommand{\bbA}{\mathstr{A}}
\newcommand{\bbB}{\mathstr{B}}
\newcommand{\bbC}{\mathstr{C}}
\newcommand{\bbD}{\mathstr{D}}

\newcommand{\bbF}{\mathstr{F}}
\newcommand{\bbG}{\mathstr{G}}
\newcommand{\bbH}{\mathstr{H}}

\newcommand{\bbP}{\mathstr{P}}

\newcommand{\bbS}{\mathstr{S}}
\newcommand{\bbT}{\mathstr{T}}

\newcommand{\Propvar}{\ensuremath{\mathsf{Prop}}}
\newcommand{\Prop}{\ensuremath{\mathsf{Q}}}        

\newcommand{\ML}{\ensuremath{\mathtt{ML}}}

\newcommand{\TC}{\ensuremath{\mathrm{TC}}}

\newcommand{\Lit}{\mathtt{Lit}}
\newcommand{\At}{\mathtt{At}}

\newcommand{\muML}{\ensuremath{\mu\ML}}    

\newcommand{\AH}[2]{\Theta^{#1}_{#2}}

\newcommand{\lneg}[1]{\ol{#1}}

\newcommand{\bw}{\bigwedge}

\newcommand{\fopp}[1]{\ol{#1}}

\newcommand{\dia}{\Diamond}

\newcommand{\hs}{\heartsuit}

\newcommand{\sforeq}{\trianglelefteqslant}
\newcommand{\sfor}{\triangleleft}
\newcommand{\fsforeq}{\sforeq_{f}}
\newcommand{\fsfor}{\sfor_{f}}
\newcommand{\psfor}{\lhd}                                
\newcommand{\Sfor}{\ensuremath{\mathit{Sfor}}}
   \newcommand{\NSfor}{\ensuremath{\mathit{NSfor}}}
\newcommand{\Clos}{\ensuremath{\mathit{Clos}}}
\newcommand{\cla}{\rightarrow_{C}}
\newcommand{\clat}{\twoheadrightarrow_{C}}

\newcommand{\closeq}{\equiv_{C}}
\newcommand{\Cluster}{C}
\newcommand{\clpr}{\sqsubset_{C}}
\newcommand{\clpreq}{\sqsubseteq_{C}}
\newcommand{\FV}[1]{\mathit{FV}(#1)}
\newcommand{\BV}[1]{\mathit{BV}(#1)}
\newcommand{\sk}{\mathrm{sk}}
\newcommand{\gOm}{\Om_{g}}

\newcommand{\len}[1]{|#1|^{\ell}}

\newcommand{\ssz}[1]{|#1|^{s}_{0}} 
   \newcommand{\sszal}[1]{|#1|^{s}_{\al}} 
\newcommand{\csz}[1]{|#1|^{c}_{0}} 
   \newcommand{\cszal}[1]{|#1|^{c}_{\al}} 

\newcommand{\ad}{\mathit{ad}}
\newcommand{\adup}{h^{\uparrow}}
\newcommand{\cd}{\mathit{cd}}
\newcommand{\cdh}{h^{\downarrow}}
\newcommand{\fdep}[1]{\mathsf{fd}(#1)}

\renewcommand{\phi}{\varphi} 
\newcommand{\isbnf}{\;::=\;}
\newcommand{\divbnf}{\;\mid\;}

\newcommand{\foeq}{\stackrel{.}{=}}
\newcommand{\vdal}{\vdash_{\al}}
\newcommand{\eqal}{=_{\al}}
\newcommand{\eqalc}[1]{\llparenthesis #1 \rrparenthesis}
\newcommand{\pol}[1]{\mathrm{ren}_{s}(#1)}
\newcommand{\spol}[1]{\mathrm{ren}_{c}(#1)}
\newcommand{\spolform}[2]{\mathrm{ren}^{#1}_{c}(#2)}
\newcommand{\subren}[2]{\mathrm{ren}_{#2}(#1)}

\newcommand{\tr}{\mathtt{tr}}

\newcommand{\sat}{\Vdash}

\newcommand{\mng}[1]{[\![ #1 ]\!]}

\newcommand{\query}{\mathcal{Q}}

\newcommand{\funP}{\mathsf{P}}
   
\newcommand{\funPom}{\funP_{\omega}}

\newcommand{\ind}{\mathit{ind}}

\newcommand{\init}[1]{\langle#1\rangle}

\newcommand{\idx}{\mathit{ind}}
\newcommand{\Clus}{\mathit{Clus}}

\newcommand{\vtx}[3]{#1{\mid}^{#2}_{#3}}

\newcommand{\AG}{\mathcal{A}}

\newcommand{\EG}{\mathcal{E}}
   \newcommand{\EGs}{\mathcal{E}_s}
   \newcommand{\EGc}{\mathcal{E}_c}

\newcommand{\eloi}{\exists}
\newcommand{\abel}{\forall}

\newcommand{\Win}{\mathrm{Win}}

\newcommand{\PM}[1]{\mathrm{PM}_{#1}}
\newcommand{\Inf}{\mathit{Inf}}

\newcommand{\rst}[1]{\!\upharpoonright_{#1}\,}
\newcommand{\nada}{\varnothing}

\newcommand{\sse}{\subseteq}

\newcommand{\size}[1]{|#1|}

\newcommand{\last}{\mathit{last}}

\newcommand{\isdef}{\mathrel{:=}}

\newcommand{\Dom}{\mathsf{Dom}}
\newcommand{\Ran}{\mathsf{Ran}}

\newcommand{\parto}{\stackrel{\circ}{\to}}

\newcommand{\De}{\Delta}
\newcommand{\Th}{\Theta}
\newcommand{\Si}{\Sigma}
\newcommand{\Om}{\Omega}
\newcommand{\al}{\alpha}
\newcommand{\be}{\beta}
\newcommand{\de}{\delta}

\newcommand{\ka}{\kappa}
\newcommand{\la}{\lambda}
\newcommand{\si}{\sigma}
\newcommand{\om}{\omega}

\newcommand{\ol}[1]{\overline{#1}}
\newcommand{\ul}[1]{\underline{#1}}

\newcommand{\wh}[1]{\widehat{#1}}

\newcommand{\sql}{\sqsubset}
\newcommand{\sqlq}{\sqsubseteq}

\newcommand{\rng}[2]{[#1, .. ,#2]}

\newcommand{\eps}{\epsilon}

\newcommand{\ouriff}{\text{ iff }}

\title{Size matters in the modal $\mu$-calculus}

\author{%
Clemens Kupke and Johannes Marti and Yde Venema}

\date{\today}

\begin{document}

\maketitle

\begin{abstract}
We discuss and compare various complexity measures related to the modal 
$\mu$-calculus $\muML$, focusing on the notions of \emph{size} and 
\emph{alternation depth}.
As a yardstick for our measurements we take Wilke's alternating tree automata,
which we shall call \emph{parity formulas} in the text.

Building on work by Bruse, Friedmann \& Lange, we compare two size measures for
$\mu$-calculus formulas: subformula-size, that is, the number of subformulas of 
the given formula, and closure-size, viz., the size of its (Fischer-Ladner) 
closure.
These notions of size correspond to the representation of a $\muML$-formula as
a parity formula based on, respectively, its subformula dag, and its closure 
graph.
What distinguishes our approach is that we are explicit and precise about the 
role of the notion of alphabetic equivalence when studying size matters; we
motivate this by showing that naively renaming bound variables easily leads to 
an exponential blow-up in size.
In addition, we take care to match the formula's alternation depth with the 
index of the associated parity formula.

We start our discussion in a setting where alphabetical variants are not
identified.
We define subformula-size and closure-size for those formulas where this 
makes sense.
We recall the well-known fact that a $\mu$-calculus formula can be transformed 
into a parity formula of size that is linear with respect to subformula size, 
and give a construction that transforms a $\mu$-calculus formula into an 
equivalent parity formula that is linear with respect to closure-size.
In the opposite direction, there is a standard transformation which produces
a $\mu$-calculus formula of exponential subformula-size but linear closure-size 
in terms of the size of the original parity formula.
We isolate a subclass of so-called \emph{untwisted} parity formulas for which
a linear transformation exists, in terms of subformula-size.

We then discuss in detail how these notions of size interact with taking 
alphabetical variants, and we see how the subformula-size and closure-size of an 
arbitrary formula can be defined in the setting where alphabetical variants are 
considered to be identical.
We transfer the result of Bruse \emph{et alii}, showing that also in this 
setting the closure-size of a formula can be exponentially smaller than its 
subformula-size.
This example further provides formulas that cannot be renamed into alphabetical 
variants, where every bound variable corresponds to a unique subformula,
without incurring an exponential blow-up, measured in closure-size.
On the other hand we show that we may cautiously rename the bound variables 
of a formula to the effect that alphabetic equivalence boils down to syntactic
identity on the closure set of the renamed formula.

As a final topic we review the complexity of guarded transformations, that is, 
effective constructions that produce a \emph{guarded} equivalent for an 
arbitrary input formula.
Partly explicitizing existing results we give an exponential transformation 
on arbitrary parity formula, and a quadratic transformation that takes
untwisted parity formulas as input but produces possibly twisted output.
We also show that a polynomial guarded transformation of arbitrary parity
formulas would give rise to a polynomial algorithm for solving parity games.
This implies a similar result for $\mu$-calculus formulas, measured by 
closure-size.

Finally, in our concluding section we discuss the relative advantages and 
disadvantages of taking subformula-size, respectively closure-size, as the 
definition of the size of a $\mu$-calculus formula.
\end{abstract}


\newcommand{\voegin}[1]{\include{#1}}

\section{Introduction}
\label{sec:int}

\subsubsection*{The modal $\mu$-calculus}

The modal $\mu$-calculus $\muML$, introduced by 
Kozen~\cite{koze:resu83} and surveyed in for instance
\cite{arno:rudi01,grae:auto02,brad:moda06,demr:temp16}, is a 
logic for describing properties of processes that are modelled by (labelled) 
transition systems.
It extends the formalism of propositional modal logic by means of least- and 
greatest fixpoint operators; this addition significantly increases the 
expressive power of the logic, permitting the expression of various properties
of ongoing processes~\cite{stir:moda01}.
The $\mu$-calculus is generally regarded as a ``universal'' specification 
language for reactive systems, since it embeds most, if not all, other logics 
such as \textsc{pdl}, \textsc{ltl}, \textsc{ctl} or \textsc{ctl}$^{*}$.
Next to this key position in the landscape of process logics, the modal 
$\mu$-calculus has also been recognised as an important \emph{logic}, with
many interesting properties, 
such as the finite model property~\cite{koze:fini88}, 
uniform interpolation~\cite{dago:logi00},
an interesting model theory~\cite{font:mode18}, 
an intuitive finite axiomatisation~\cite{koze:resu83,walu:comp00} 
and a cut-free proof calculus~\cite{afsh:cutf17}.
From basic modal logic it inherits a very favourable balance between expressive 
power --- it is the bisimulation-invariant fragment of monadic second-order 
logic~\cite{jani:expr96} --- and good computational behaviour.

Concerning the latter: in line with the role of $\muML$ as a specification
language, various computational aspects of the formalism have been investigated 
rather intensively.
The two problems at the centre of these investigations concern 
\emph{satisfiability} (given a $\mu$-calculus formula $\xi$, is it satisfiable
in some model?) and \emph{model checking} (given a model $\bbS$, a state $s$ in
$\bbS$ and a $\mu$-calculus formula $\xi$, is $\xi$ true at $s$ in $\bbS$?)
The satisfiability problem was rather quickly shown to be 
decidable~\cite{koze:deci83}, 
while some years later Emerson \& Jutla~\cite{emer:tree91} used automata to
give an exponential time algorithm for satisfiability checking.
One of the most challenging problems about $\muML$ turned out to be the 
complexity of its model checking problem; note that this problem is often 
reformulated in game-theoretic terms, revealing close links between $\muML$ 
model checking and the problem of solving parity games.
There is an obvious model checking algorithm that runs in time $(k\cdot n)^{d}$, 
where $k,n$ and $d$ are respectively the size of the model, the size of the 
formula, and the alternation depth of the formula, i.e., the maximal length 
of a chain of alternating least- and greatest fixpoint operators in the formula.
The key question concerns the existence of an algorithm that is entirely
\emph{polynomial} in the size of the formula; fairly recently, a breakthrough 
was obtained by Calude et alii~\cite{calu:deci17} who gave a 
\emph{quasi-polynomial} algorithm for deciding parity games.

\subsubsection*{Size matters}

Given such intensive complexity-theoretic investigations, one would expect that
all basic questions concerning complexity \emph{measures} of $\mu$-calculus 
formulas would have been sorted out long ago.
More specifically, one would assume the existence of a solid and comprehensive
theoretical framework which provides a natural mathematical environment for 
well-known complexity measures such as length, size and alternation depth; 
makes a clear connection to the actual algorithms and complexity results related
to fixpoint logic, and allows transparent definitions of the important 
constructions that support the theory of the formalism, such as guarded 
transformations and (disjunctive/nondeterministic) simulations.
Last but not least, one would assume such a framework to deal explicitly with
characteristic features of the modal $\mu$-calculus, such as the fact that its
syntax crucially involves \emph{binding}.

And indeed, the literature has dealt with most (but not all) of these desiderata,
with authors generally providing well-defined complexity measures that link up
neatly with the algorithms for solving the mentioned problems of model checking
and satisfiability. 
Parts of the framework, however, still appear to be surprisingly undeveloped.
This was brought to light by Bruse, Friedmann \& Lange~\cite{brus:guar15} in a 
publication that discusses and relates the two most widely used size measures: 
subformula-size (the number of subformulas of the given formula), and 
closure-size (the number of formulas in the so-called closure set of the
formula).
While the closure-size of a formula was well-known to never exceed its 
subformula-size, and generally assumed to be roughly the same, Bruse et alii 
showed that the closure-size of a formula can in fact be \emph{exponentially 
smaller} than its subformula-size.
This result raises a number of fundamental questions concerning size matters,
and indicates that there is a gap between the world of formulas on the
one hand, and that of automata and games on the other.

The aim of this paper is to help bridging this gap by studying some key
complexity measures of $\mu$-calculus formulas in detail.
We will focus on the notion of \emph{size}, since we believe that this needs 
some further clarification.
In this sense, then, part of our work is a continuation of~\cite{brus:guar15},
in that we make a comparison between subformula-size and closure-size as size
measures for fixpoint formulas.
As a second complexity measure, throughout our narrative we take along the
notion of \emph{alternation depth}.
In order to provide a coherent and reasonably complete picture, we shall mention
and prove quite a few results that are in fact (well-)known; however, we will 
also prove some results that we believe to be original.
In particular, and as we shall see in a moment, what distinguishes our approach
is that we will be quite explicit and precise concerning the notion of 
\emph{alphabetical equivalence} of $\mu$-calculus formulas.


\subsubsection*{Parity formulas}

As the backbone of our framework we take a version of Wilke's \emph{alternating
tree automata}~\cite{wilk:alte01,grae:auto02}, which we shall call \emph{parity
formulas}.\footnote{%
   We will reserve the name `automaton' for devices that display a  more
   operational nature due to the presence of some kind of transition map.
   }
For a quick description of these devices: parity formulas are like ordinary
(modal) formulas, with the difference that (i) the underlying structure of a
parity formula is a directed graph, possibly with cycles, rather than a tree; 
and (ii) one adds a priority labelling to this syntax graph, to ensure 
a well-defined game-theoretical semantics in terms of parity games.
Parity formulas can be seen to provide a normal form for the syntax of the modal
$\mu$-calculus, and as such they are closely related to a number of other 
formalisms, such as hierarchical equation systems, $\mu$-calculus in vectorial
form, and modal automata --- later on we will provide some more details about
these connections.

The motivation for taking these parity formulas as the backbone of the framework,
is that 
(1) they have a completely straightforward connection with the \emph{parity 
games} that lie at the heart of the algorithmic/complexity-theoretic theory of
the modal $\mu$-calculus, and
(2) this connection allows for a clear and completely unambiguous definition 
of the most relevant complexity measures of parity formulas: size and index.
In particular, the \emph{size} $\size{\bbG}$ of a parity formulas is simply 
defined as the number of vertices of its underlying graph, and its 
\emph{index} corresponds to the maximal length of some suitably defined
alternating chain of states.

This algorithmic transparency of parity formulas makes them very suitable as 
\emph{yardsticks} for comparing various size measures for $\mu$-calculus 
formulas. 
Taking this approach, we can further develop the complexity-theoretical 
framework of the $\mu$-calculus by investigating the links between 
$\mu$-calculus formulas and their parity formula counterparts.
In particular, in our framework:

\begin{enumerate}
\item[(\dag)]
we consider an attribute of $\mu$-calculus formulas a \emph{size measure}
if it is induced by a construction associating a parity formula with a given 
$\muML$-formula,
\end{enumerate}
where we say that $s: \muML \to \om$ is \emph{induced} by a construction $\xi 
\mapsto \bbG_{\xi}$ (associating a parity formula $\bbG_{\xi}$ with a given 
$\muML$-formula $\xi$) if $s(\xi) = \size{\bbG_{\xi}}$.

As is (well) known, the notions of \emph{length}, \emph{subformula-size} and 
\emph{closure-size} meet the constraint (\dag), since they correspond to the 
representation
of a $\muML$-formula as a parity formula based on, respectively, its syntax 
tree, its subformula dag, and its closure graph.
We give the latter two representations explicitly, taking care to match the 
alternation depth of the formula with the index of its representation as a
parity formula.
In the opposite direction, there is a standard transformation which produces
a $\mu$-calculus formula of exponential subformula-size but linear closure-size 
in terms of the size of the original parity formula.
Where the correspondence between parity formulas and $\mu$-calculus formulas
thus generally involves an exponential blow-up in terms of subformula size,
as an original contribution we isolate a subclass of so-called \emph{untwisted} 
parity formulas for which this correspondence is linear.

\subsubsection*{Alphabetical equivalence}

There is a key feature of the modal $\mu$-calculus which is missing from the
preceding discussion, viz., the fact that the syntax of the formalism is based
on \emph{variable binding}.
Let us mention two relevant issues arising as manifestations of this feature.
First of all, some of the most basic definitions presuppose that formulas have 
a certain syntactic format, defined in terms of free and bound variables.
For instance, as we will argue below, it will only make sense to take the number
of subformulas as a size measure, if each subformula makes an unambiguous
contribution to the meaning of the formula. 
Here, we shall call a formula \emph{tidy} if its free and bound variables form 
disjoint sets,\footnote{%
   In the literature, some authors make a distinction between proposition 
   letters (which can only occur freely in a formula), and propositional 
   variables, which can be bound.
   Our tidy formulas correspond to \emph{sentences} in this approach, that is, 
   formulas without free variables.
   }
and \emph{clean} or \emph{well-named} if in addition, each of its bound 
variables is bound exactly once.\footnote{%
   Formulated like this, the definition is somewhat ambiguous, since it does
   not mention explicitly \emph{which representation} of the formula (syntax
   tree or subformula graph) has a unique binding occurrence of each bound 
   variable.
   Below we will give an unambiguous definition of the notion of 
   cleanness.
   }
As we will see,  we seem to need the cleanness and tidyness conditions to 
guarantee, respectively, a correct definition of a formula's size if this is 
to be based on the formula's set of subformulas, respectively, its closure set.

As a second example, consider the concept of $\al$-equivalence:
Recall that, roughly, two formulas are \emph{$\al$-equivalent} or 
\emph{alphabetical variants} of one another if we can obtain one from the other
by a suitable renaming of bound variables.
This concept is ubiquitous in the $\mu$-calculus, starting with the definition 
of substitution (which in its turn is crucial in the definition of the closure
of a formula). 
It is easy to see that the length of a formula is invariant under 
$\al$-equivalence, but how does $\al$-equivalence relate to other size 
measures?
In fact, there are very simple examples of $\al$-equivalent formulas 
with an exponential difference in size.

These two issues are actually closely related: virtually every definition 
of size in the literature uses alphabetical variants to define the size of 
arbitrary formulas.
Apparently, the received opinion is that $\al$-equivalent formulas can be
considered to be identical.
While we certainly don't want to argue against such an identification in
principle, we do want to draw attention to some undesirable side effects
of a naive identification of $\al$-equivalent formulas.
In particular, when defining the subformula size of an arbitrary formula one
needs to consider a clean alphabetical variant.
However, we will show that, when defining the closure size of an arbitrary 
formula, we may incur an unnecessary exponential blow-up if we turn to an
alphabetical variant which is clean.

More in general, the identification of $\al$-equivalent formulas should have 
some repercussions which, to the best of our knowledge, have not received any 
attention in the literature on the $\mu$-calculus.
Our point is that, if $\alpha$-equivalent formulas are taken to be identical
in order to define some size measure for arbitrary formulas, then this 
perspective should be reflected in the definition itself.
For instance, as the definition of the closure-size of a formula we cannot just
take the size of its closure set, or that of the closure set of an alphabetical
variant of the original formula, since these sets themselves might contain 
alphabetical variants. 
Instead, we should take the number of formulas in either set, but \emph{modulo
$\al$-equivalence}.
Furthermore, if $\al$-equivalent formulas are to be identified, then they 
certainly should have the same size --- but how to achieve this?

The main novelty of this paper is that we aim to introduce and study notions of
size that are \emph{invariant under $\al$-equivalence}, in the sense that it
meets the conjunction (\ddag) of the following two constraints:
\begin{enumerate}
\item[(\ddag a)]
$\al$-equivalent formulas have the same size;
\item[(\ddag b)]
when calculating the size of a formula, $\al$-equivalent formulas are considered 
to be identical.
\end{enumerate}
Note that it is not a priori clear whether these constraints are compatible with 
the condition (\dag) formulated earlier.

As observed already, the standard definition of \emph{length} already meets 
the constraints (\dag) and (\ddag) (with the understanding that condition 
(\ddag b) is irrelevant in this case).
In section~\ref{sec:aleq} we provide two other notions of size that are invariant
under $\al$-equivalence, for respectively subformula-size and closure-size.
In both cases our definition of the size of an arbitrary formula is based on 
first taking a suitable renaming of the original formula, and then calculating,
respectively, the number of subformulas, and the size of the closure set of this
renaming.
As a consequence, these definitions do satisfy the conditions (\dag) and (\ddag)
simultaneously.
To mention some of our results about these size measures, we will prove that they
interact well with the operation of taking substitution, and that the result of 
Bruse et alii (viz., that the closure-size of a formula can be exponentially 
smaller than its subformula size) extends to this setting.

\subsubsection*{Guarded transformation}

Finally, as an example of an important construction on $\mu$-calculus formulas,
we consider the operation of guarded transformation.
Recall that a $\mu$-calculus formula is \emph{guarded} if every occurrence of
a bound variable is in the scope of a modal operator which resides inside 
the variable's defining fixpoint formula.
Intuitively, the effect of this condition is that, when evaluating a guarded 
formula in some model, between any two iterations of the same fixpoint variable, 
one has to make a transition in the model.
Many constructions and algorithms operating on $\mu$-calculus formulas 
presuppose that the input formula is in guarded form, which explains the need
for low-cost \emph{guarded transformations}, that is, efficient procedures for 
bringing a $\mu$-calculus formula into an equivalent guarded form.

In fact, one of the main contributions of Bruse, Friedmann \& Lange
in~\cite{brus:guar15} is to discuss size issues related to guarded 
transformations, and in the process to correct some mistaken claims in the
literature.
We believe that a transfer of guarded transformation results to the setting 
of parity formulas may help to further clarify\footnote{%
   Actually, some of the formulations in~\cite{brus:guar15} are slightly 
   puzzling. 
   We will come back to these in section~\ref{sec:gua}.
   }
the situation.

In this article we will give two straightforward guarded transformation
constructions, directly on parity formulas.
First we give an exponential construction that transforms an arbitrary parity
formula into an equivalent guarded one; subsequently we define an operation
that takes an untwisted parity formula as input, and returns an equivalent 
guarded parity formula of quadratic size.
Finally, we adapt an argument from \cite{brus:guar15} to show that a polynomial
guarded transformation of arbitrary parity formulas would give rise to a 
polynomial algorithm for solving parity games.
As a corollary of these observations, what we can say about the existence of a 
\emph{polynomial} guarded transformation is the following: there is a 
\emph{quadratic} guarded  transformation on $\mu$-calculus formulas indeed,
but only if one measures the input formula of the construction as 
subformula-size, and the output formula as closure-size.
If we measure the input and output formula in the same way, then all known 
constructions are exponential, and if this measure is closure-size, then any
effective guarded transformation must be as hard as solving parity games.

\subsection*{Overview}

This article is organised as follows: In Section~\ref{sec:prel} we
discuss the necessary preliminaries concerning the modal $\mu$-calculus.
Section~\ref{sec:bas} contains basic definitions and results about the
notions of subformulas, closure and substitution that are used throughout
the paper and it introduces first measures for the size of formulas. In
Section~\ref{sec:ad} we review basic results about the alternation depth
of formulas and how it interacts with the notions introduced in the
previous section. In Section~\ref{sec:par} we introduce parity formulas
that play a crucial role in the setting of this article. We also define
the subclass of untwisted parity formulas. In Section~\ref{s:fixpar} we
show that the subformula graph and the closure graph of a formula in the
$\mu$-calculus gives rise to equivalent parity formulas. For both
constructions we show that they also preserve the alternation depth.
Section~\ref{s:parfix} contains the converse constructions showing that
for every parity formula there is a equivalent formula in the
$\mu$-calculus of the same alternation depth such that the size of its
closure is linear in the size of the parity formula. If the parity
formula is untwisted then we have an analogous constructions with a
linear bound on the number of subformulas. In Sections
\ref{sec:aleq}~and~\ref{sec:skel} we then discuss alphabetical
equivalence. In Section~\ref{sec:aleq} we define size measures that are
invariant under alphabetical equivalence and we show how to obtain for
every formula an alphabetical variant that is minimal in subformula size
among all of its variants. Section~\ref{sec:skel} contains an analogous
construction showing how for every formula we can define an alphabetical
variant of minimal closure size.

\section{The modal $\mu$-calculus}
\label{sec:prel}

In this paper we assume familiarity with the basic theory of the modal 
$\mu$-calculus, as presented in for
instance~\cite{koze:resu83,arno:rudi01,grae:auto02,brad:moda06,vene:lect18}.
In this section we fix some notation and terminology concerning the most
basic syntactic and semantic definitions.

\subsection{Structures}

\begin{convention}
Throughout the text we fix an infinite set $\Propvar$ of propositional variables,
of which we often single out a finite subset $\Prop$.
Subsets of $\Prop$ will sometimes be called \emph{colors}, and we shall regard
the size $\size{\Prop}$ of the set $\Prop$ as constant.
\end{convention}

\begin{definition}
Given a set $S$, a \emph{$\Prop$-coloring} on $S$ is a map $m: S \to \funP
\Prop$; a \emph{$\Prop$-valuation} on $S$ is a map $V: \Prop \to \funP S$.
A valuation $V : \Prop \to \funP S$ gives rise to its \emph{transpose coloring} 
$V^\dagger : S \to \funP \Prop$ defined by $V^\dagger(s) := \{p \in \Prop \mid
s \in V(p) \}$, and dually colorings give rise to valuations in the same manner. 
\end{definition}

Since colorings and valuations are interchangeable notions, we will often
switch from one perspective to the other, based on what is more convenient in 
context.

\begin{definition}
A \emph{Kripke structure} over a set $\Prop$ of proposition letters is a triple 
$\bbS = (S,R,V)$ such that $S$ is a set of objects called \emph{points}, $R 
\sse S \times S$ is a binary relation called
the \emph{accessibility} relation, and $V$ is a $\Prop$-valuation on $S$.
A \emph{pointed} Kripke structure is a pair $(\bbS,s)$ where $s$ is a point of 
$\bbS$.

Given a Kripke structure $\bbS = (S,R,V)$, a propositional variable $x$ and a 
subset $U$ of $S$, we define $V[x\mapsto U]$ as the $\Prop \cup \{x\}$-valuation 
given by
\[
V[x \mapsto U](p) \isdef \left\{
\begin{array}{ll}
   V(p) & \text{ if } p \neq x 
\\ U    & \text{otherwise},
\end{array}\right.
\]
and we let $\bbS[x\mapsto U]$ denote the structure $(S,R,V[x\mapsto U])$.
\end{definition}

\begin{definition}
The reflexive/transitive closure and the transitive closure of a binary relation 
$R$ are denoted as $R^{*}$ and $R^{+}$, respectively; elements of the sets
$R^{*}[s]$ and $R^{+}[s]$ are called \emph{descendants} and \emph{proper
descendants} of $s$, respectively.

A \emph{path} through a directed graph $(V,R)$ is a sequence 
$(s_i)_{i < \ka}$ such that $(s_i,s_{i+1}) \in R$ for all $i$ with $i+1 < \ka$; 
here $\ka \leq\om$ is the \emph{length} of the path.
A directed graph $(V,R)$ is a \emph{tree} (with \emph{root} $r$) if $S =
R^{*}[r]$ and every node $t\neq r$ has a unique predecessor.
A \emph{branch} of a tree $\bbT$ is a maximal path through $\bbT$, starting
at the root.
\end{definition}

\subsection{Syntax}

The most concise way of defining the language of the modal $\mu$-calculus is 
by means of the following grammar:
\begin{equation}
\label{eq:mu-syn1}
\phi \isbnf p 
   \divbnf \neg\phi \divbnf \phi\lor\phi 
   \divbnf \dia\phi 
   \divbnf \mu x.\phi,
\end{equation}
where $p$ and $x$ are propositional variables, and the formation of the formula 
$\mu x.\phi$ is subject to the constraint that 
all occurrences of $x$ in $\phi$ are in the scope of an even number of negations.
It will be more convenient for us, however, to assume that our formulas are in
so-called negation normal form, and that we allow $\bot$ and $\top$ as atomic
formulas.

\begin{definition}
\label{d:syntnnf}
The formulas of the modal $\mu$-calculus $\muML$ are given by the following
grammar:
\[
\phi \isbnf 
   p \divbnf \lneg{p} 
   \divbnf \bot \divbnf \top
   \divbnf (\phi\lor\phi) \divbnf (\phi\land\phi) \divbnf
   \dia\phi \divbnf \Box\phi 
   \divbnf \mu x\, \phi \divbnf \nu x\, \phi,
\]
where $p$ and $x$ are propositional variables, and the formation of the formulas
$\mu x\, \phi$ and $\nu x\, \phi$ is subject to the constraint that $\phi$ is 
\emph{positive} in $x$, i.e., there are no occurrences of $\lneg{x}$ in $\phi$.
Elements of $\muML$ will be called \emph{$\mu$-calculus formulas} or simply
\emph{formulas}.
Formulas of the form $\mu x. \phi$ or $\nu x. \phi$ will be called 
\emph{fixpoint formulas}.

We write $\BV{\xi}$ and $\FV{\xi}$ for, respectively, the set of \emph{bound}
and \emph{free variables} of a formula $\xi$, and  we let $\muML(\Prop)$ denote
the set of $\mu$-formulas of which all free variables belong to the set $\Prop$.
A variable $x$ is \emph{fresh} for a formula $\xi$ if $x \not\in \BV{\xi} \cup 
\FV{\xi}$.

We define $\Lit(\Prop) \isdef \{ p, \lneg{p} \mid p \in \Prop \}$ as the set of 
\emph{literals} over $\Prop$, and $\At(\Prop) \isdef \{ \bot, \top \} \cup 
\Lit(\Prop)$ as the set of \emph{atomic formulas} over $\Prop$.

The \emph{fixpoint depth} $\fdep{\phi}$ of a formula $\phi$ is defined as 
follows:
$\fdep{\phi} \isdef 0$ if $\phi$ is atomic, 
$\fdep{\phi_{0} \odot \phi_{1}} \isdef \max(\fdep{\phi_{0}},\fdep{\phi_{1}})$,
$\fdep{\hs\phi} \isdef \fdep{\phi}$, and 
$\fdep{\eta x. \phi} \isdef 1 + \fdep{\phi}$.
The \emph{modal depth} of a formula is defined similarly.
\end{definition}

It will be convenient for us to have some additional terminology and notation
pertaining to the fixpoint operators $\mu$ and $\nu$.
We will generally use the symbols $\eta$ and $\la$ to range over $\mu$ and $\nu$.

\begin{definition}
\label{d:munu}
We will associate $\mu$ and $\nu$ with the odd and even numbers, respectively; 
formally we will say that a number $k \in \om$ has \emph{parity} $\mu$ ($\nu$) if 
it is odd (even, respectively).
We will also associate $\mu$ and $\nu$ with the formulas $\bot$ and $\top$,  
defining $\wh{\mu} \isdef \bot$ and $\wh{\nu} \isdef \top$.
For $\eta \in \{ \mu, \nu \}$ define $\fopp{\eta}$ by putting $\fopp{\mu} \isdef
\nu$ and $\fopp{\nu} \isdef \mu$.
\end{definition}

Where the main point of this article is to define, discuss and compare various 
ways to define the size of a $\mu$-calculus formula, there is one such 
definition which is entirely unproblematic.
The \emph{length} of a formula $\xi$ is defined as the number of symbols 
occurring in $\xi$, where we disregard brackets, and count a fixpoint operator
$\eta x$ as one symbol.

\begin{definition}
The following formula induction:
\[\begin{array}{llll}
   \len{\phi} &\isdef& 1 
   & \text{if } \phi \text{ is atomic }
\\ \len{(\phi_{0}\odot\phi_{1})} &\isdef & \len{\phi_{0}} + \len{\phi_{1}}
   & \text{where } \odot \in \{ \land, \lor\}
\\ \len{\hs\phi} &\isdef & 1 + \len{\phi}
   & \text{where } \hs \in \{ \Box, \dia\}
\\ \len{\eta x\, \phi} &\isdef & 2 + \len{\phi}
   & \text{where } \eta \in \{ \mu, \nu \}
\end{array}\]
defines the \emph{length} $\len{\xi}$ of a $\mu$-calculus formula 
$\xi$.
\end{definition}

\begin{convention}
In order to increase readability by reducing the number of brackets, we adopt 
some standard scope conventions.
We let the unary (propositional and modal) connectives, $\neg,\dia$ and $\Box$,
bind stronger than the binary propositional connectives $\land$, $\lor$ and 
$\to$, and use associativity to the left for the connectives $\land$ and $\lor$.
Furthermore, we use `dot notation' to indicate that the fixpoint operators 
preceding the dot have maximal scope. 
For instance, $\mu x. \lneg{p} \lor \Box x \lor y \lor \nu y. q \land \Box (x \lor y)$
stands for $\mu x\, \Big(
\big( (\lneg{p} \lor \Box x) \lor y \big) \lor \nu y\, (q \land \Box (x \lor y))
\Big)$.
\end{convention}

In some of the literature on the modal $\mu$-calculus, a distinction is made 
between between proposition letters (corresponding to the free variables in our 
approach) and proposition variables (corresponding to the bound variables in our 
approach).
We do not make this distinction, but (as we will discuss in the next section),
it will often be convenient to confine attention to formulas for which the sets
of free and bound variables have no overlap.
These correspond to the \emph{sentences} in the alternative approach.

\begin{definition}
We call a $\mu$-formula $\xi$ \emph{tidy} if $\BV{\xi}\cap \FV{\xi} = \nada$,
and we let $\muML^{t}$ denote the set of tidy formulas.
\end{definition}

\begin{remark}
It should be obvious that every formula $\xi \in \muML$ is equivalent to a tidy
formula $\xi'$ of the same length: simply associate with every bound variable $x$
a fresh variable $x'$, and replace, for each $x \in \BV{\xi}$
(1) every occurrence of a fixpoint operators $\eta x$ in $\xi$ with $\eta x'$, 
and (2) every \emph{bound} occurrence of $x$ with $x'$.
\end{remark}

\begin{figure}[htb]
\begin{center}

\begin{tikzpicture}
\tikzset{sibling distance=4mm,
   edge from parent/.append style={->,thick},
   every node/.style= {circle,inner sep=.5mm,thick},}

\Tree [.\node (mu) [draw] {$\mu x$};
    [.$\lor$
        [.$\lor$ 
            [.$\overline{p}$ ]
            [.$\Diamond$ 
               [.\node (x1) {$x$}; ] ] 
        ]
        [.\node (nu) [draw]{$\nu y$} ;
            [.$\land$ 
                [.$q$ ]
                [.$\Box$ 
                    [.$\lor$ 
                       [.\node (x2) {$x$}; ] 
                       [.\node (y) {$y$}; ]
                    ]
                ]
            ]
        ]
    ]
]  
\end{tikzpicture}

\label{fig:x21}
\caption{A syntax tree}
\end{center}
\end{figure}
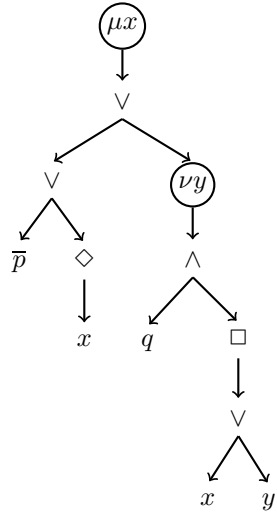

Furthermore, we assume that the reader is familiar with the concept of the 
\emph{syntax tree} or \emph{construction tree} $\bbT_{\xi}$ of a formula $\xi$.
We will not give a formal definition of this structure, but confine ourselves 
to an example: in Figure~\ref{fig:x21} we display the syntax tree of the 
$\mu$-calculus formula 
$\mu x. (\ol{p} \lor \dia x) \lor \nu y. (q \land \Box(x \lor y))$.
Note that the \emph{length} of a formula corresponds to the number of nodes of
its syntax tree, and that an \emph{occurrence} of a certain symbol in a formula
may be associated with some node in the formula's syntax tree that is labelled 
with that symbol; occurrences of literals correspond to \emph{leaves} of the
tree.

\vspace*{-10mm}

\subsection{Semantics}

The semantics of the modal $\mu$-calculus is defined as follows.

\begin{definition}
By induction on the complexity of $\mu$-calculus formulas, we define a meaning
function $\mng{\cdot}$, which assigns to a formula $\phi \in \muML(\Prop)$ 
its \emph{meaning} $\mng{\phi}^{\bbS} \sse S$ in any Kripke structure $\bbS = 
(S,R,V)$ over $\Prop$.
The clauses of this definition are standard:
\[\begin{array}{lllclll}
   \mng{p}^{\bbS} &\isdef& V(p)
 && \mng{\lneg{p}}^{\bbS} &\isdef& S \setminus V(p)
\\ \mng{\bot}^{\bbS} &\isdef& \nada
 && \mng{\top}^{\bbS} &\isdef& S 
\\ \mng{\phi\lor\psi}^{\bbS} &\isdef& \mng{\phi}^{\bbS} \cup \mng{\psi}^{\bbS} 
 && \mng{\phi\land\psi}^{\bbS} &\isdef& \mng{\phi}^{\bbS} \cap \mng{\psi}^{\bbS}
\\ \mng{\dia\phi}^{\bbS} &\isdef& 
     \{ s \in S \mid R[s] \cap \mng{\phi}^{\bbS} \neq \nada \}
 && \mng{\Box\phi}^{\bbS} &\isdef& 
     \{ s \in S \mid R[s] \sse \mng{\phi}^{\bbS} \}
\\ \mng{\mu x.\phi}^{\bbS} &\isdef& 
     \bigcap \{ U \in \funP S \mid \mng{\phi}^{\bbS[x\mapsto U]}\sse U \}
   &&  \mng{\nu x.\phi}^{\bbS} &\isdef& 
     \bigcup \{ U \in \funP S \mid \mng{\phi}^{\bbS[x\mapsto U]}\supseteq U \}.
\end{array}\]
If a point $s \in S$ belongs to the set $\mng{\phi}^{\bbS}$, we write 
$\bbS,s \sat \phi$, and say that $\phi$ is \emph{true at} $s$ or \emph{holds} 
at $s$, or that $s$ \emph{satisfies} $\phi$.
Two formulas $\phi$ and $\psi$ are \emph{equivalent}, notation: $\phi \equiv 
\psi$, if $\mng{\phi}^{\bbS} = \mng{\psi}^{\bbS}$ for any structure $\bbS$.
\end{definition}

Note that $\mng{\mu x.\phi}^{\bbS}$ and $\mng{\nu x.\phi}^{\bbS}$ are equal
to, respectively, the least and greatest fixpoint of the monotone map 
$\phi^{\bbS}_{x}: \funP S \to \funP S$ which is given by 
\begin{equation}
\label{eq:unfeq}
\phi^{\bbS}_{x}(U) \isdef \mng{\phi}^{\bbS[x \mapsto U]}.
\end{equation}

We will usually take a \emph{game-theoretic} perspective on the semantics of the
modal $\mu$-calculus, but this will be defined in the next section.

\section{Basics}
\label{sec:bas}

In this section we define some of the most basic syntactic attributes of 
$\mu$-calculus formulas, including the notions of subformula, substitution, 
and closure set.
In particular, we introduce the three size measures of length, subformula-size, 
and closure-size (for those formulas for which these concepts make sense).

\subsection*{Subformulas}

Obviously, if we want to define size in terms of number of subformulas, 
we need a closer look at the notion of subformula.
First its definition.

\begin{definition}
\label{d:sfor}
We define the set $\Sfor_{0}(\xi)$ of \emph{direct subformulas} of a formula 
$\xi \in \muML$ via the following case distinction:
\[\begin{array}{llll}
   \Sfor_{0}(\xi) & \isdef & \nada 
   & \text{ if } \xi \in \At(\Prop)
\\ \Sfor_{0}(\xi_{0} \odot \xi_{1}) & \isdef & \{ \xi_{0}, \xi_{1} \}
   & \text{ where } \odot \in \{ \land, \lor \}
\\ \Sfor_{0}(\hs\xi_{0}) & \isdef & \{ \xi_{0} \}
   & \text{ where } \hs \in \{ \dia, \Box \}
\\ \Sfor_{0}(\eta x. \xi_{0}) & \isdef & \{ \xi_{0} \}
   & \text{ where } \eta \in \{ \mu, \nu \},
\end{array}\]
and we write $\phi \psfor_{0} \xi$ if $\phi \in \Sfor_{0}(\xi)$.

For any formula $\xi \in \muML$, $\Sfor(\xi)$ is the least set of formulas which
contains $\xi$ and is closed under taking direct subformulas.
Elements of the set $\Sfor(\xi)$ are called \emph{subformulas} of $\xi$,
and we write $\phi\sforeq\xi$ ($\phi\psfor\psi$) if $\phi$ is a subformula 
(proper subformula, respectively) of $\xi$.
\end{definition}

Since we are dealing with syntax where \emph{binding} plays an important role,
it is important to understand how the notion of subformula interacts with
this.
Consider for example the formula
$\xi = \dia p \land (\mu p. q \lor \dia p) \land \nu p. \Box\dia p$, and 
its subformulas $\dia p$ and $p$.
It should be obvious that each of the various occurrences of these two 
subformulas in $\xi$ has a different contribution to the meaning of $\xi$.
If we want to avoid such a situation, that is, if we want every subformula 
of the ambient formula to have a fixed meaning, it makes sense to restrict 
attention to so-called clean formulas.
Recall that a $\mu$-formula $\xi$ is \emph{tidy} if $\BV{\xi}\cap \FV{\xi} = 
\nada$.

\begin{definition}
A tidy $\mu$-calculus formula $\xi$ is \emph{clean} if with every bound variable
$x$ of $\xi$ we may associate a \emph{unique} subformula of the form $\eta x.
\delta$ (with $\eta \in \{\mu,\nu \}$).
This unique subformula will be denoted as $\eta_{x} x.\delta_{x}$, and we call 
$x$ a $\mu$-variable if $\eta_{x}=\mu$, and a $\nu$-variable if $\eta_{x}=
\nu$.
\end{definition}

The semantics of modal $\mu$-calculus formulas is often presented in 
\emph{game-theoretic} terms (for basic definitions concerning the infinite 
(parity) games involved, the reader is referred to 
Appendix~\ref{sec:games}).

Suppose that we are interested in the truth of a formula $\xi$ in a point $s$ 
of a Kripke model $\bbS$.
In case $\xi$ is \emph{clean} formulas, we can check this by means of an 
evaluation game where the positions are simply given as pairs consisting of 
a subformula of the ambient formula $\xi$ and a point of the Kripke model 
under consideration.
The winning conditions of this game are formulated in terms of the following
priority order that can be defined on the collection of bound variables of a 
clean formula.

\begin{definition}
Let $\xi$ be a clean $\mu$--formula.
The \emph{dependency order} $<_{\xi}$ on the bound variables of $\xi$ is defined
as the least strict partial order such that $x <_{\xi} y$ if $\delta_{x}$ is a 
proper subformula of $\delta_{y}$ with $y \in \FV{\de_{x}}$.
\end{definition}

\begin{definition}
\label{d:evg}
Let $\bbS=(S,R,V)$ be a Kripke model and let $\xi$ be a clean formula in $\muML$.
We define the \emph{evaluation game} $\EGs(\xi,\bbS)$ as the parity game 
$(G,E,\Om)$ of which 
the board consists of the set $\Sfor(\xi) \times S$, and the game graph (i.e.,
the partitioning of $\Sfor(\xi) \times S$ into positions for the two players, 
together with the set $E(z)$ of admissible moves at each position), is given
in Table~\ref{tb:1s}.

\begin{table}[htb]
\begin{center}
\begin{tabular}{|ll|c|l|}
\hline
\multicolumn{2}{|l|}{Position} & Player & Admissible moves\\
\hline
   \multicolumn{2}{|l|}{$(\phi \lor \psi,s)$}   & $\eloi$   
   & $\{ (\phi,s), (\psi,s) \}$ 
\\ \multicolumn{2}{|l|}{$(\phi \land \psi,s)$} & $\abel$ 
   & $\{ (\phi,s), (\psi,s) \}$ 
\\ \multicolumn{2}{|l|}{$(\dia \phi,s)$}        & $\eloi$ 
   & $\{ (\phi,t) \mid sRt \}$ 
\\ \multicolumn{2}{|l|}{$(\Box \phi,s) $}       & $\abel$ 
   & $\{ (\phi,t) \mid sRt \}$ 
\\   $(p,s)$        & with $p\in \FV{\xi}$ and $s \in V(p)$         
   & $\abel$ & $\nada$ 
\\   $(p,s)$        & with $p\in \FV{\xi}$ and $s \notin V(p)$      
   & $\eloi$ & $\nada$ 
\\   $(\lneg{p},s)$   & with $p\in \FV{\xi}$ and $s \in V(p)$    
   & $\eloi$ & $\nada$ 
\\   $(\lneg{p},s)$  & with $p\in \FV{\xi}$ and $s \notin V(p)$ 
   & $\abel$ & $\nada$ 
\\ \multicolumn{2}{|l|}{$(\eta x . \phi,s)$}    & - 
   & $\{ (\phi,s) \}$ 
\\   $(x,s)$       & with $x\in \BV{\xi}$ 
   & - & $\{ (\delta_x,s) \}$ 
\\ \hline
\end{tabular}
\end{center}
\caption{The subformula evaluation game $\EGs(\xi,\bbS)$}
\label{tb:1s}
\end{table}

To define the priority map $\Om$ of $\EGs(\xi,\bbS)$, consider an infinite match
$\Si = (\phi_{n},s_{n})_{n\in\om}$, and let $\mathit{Inf}(\Si)$ denote the set
of (bound) variables that get unfolded infinitely often during the match.
This set contains a highest variable $x$ (with respect to the dependency order
$<_{\xi}$), and the winner of $\Si$ is $\eloi$ if $\eta_{x} = \mu$, and $\abel$ 
if $\eta_{x}=\nu$.
It is not difficult to define a priority map $\Om: (\Sfor \times S) \to \om$
that is compatible with this condition --- we will come back to the details later.
\end{definition}

Note that we do not need to assign a player to positions that admit a single 
move only, and that if we assign a position without admissible moves to a 
player this means that the player will immediately loose any match arriving
at this position.

The following fact states the \emph{adequacy} of the game semantics.

\begin{fact}
\label{f:0s}
Let $\xi$ be a clean formula of the modal $\mu$-calculus, and let 
$(\bbS,s)$ be some pointed Kripke structure.
Then
$\bbS,s \sat \xi \ouriff (\xi,s) \in \Win_{\eloi}(\EGs(\xi,\bbS))$.
\end{fact}      

\begin{remark}
\label{r:sfsm}
Observe that this definition of the evaluation game only makes sense for 
\emph{clean} formulas.
Consider for example the earlier formula
$\xi = \dia p \land (\mu p. q \lor \dia p) \land \nu p. \Box\dia p$, 
which has three occurrences of each of the subformulas
$\dia p$ and $p$.
It should be obvious that we cannot define an adequate evaluation game for this
formula based on the set $\Sfor(\xi) \times S$ as positions (where $S$ is the 
set of points in the Kripke model under consideration), since there is no 
reasonable definition of a legitimate move at a position of the form $(p,s)$.

This is not a problem per se: as a straightforward adaptation of 
Definition~\ref{d:evg}, for a dirty formula like $\xi$ one may define a variant
of the evaluation game where positions are pairs consisting of a node in the 
construction tree of the formula and a point in the Kripke model.
Alternatively, one may `clean up' $\xi$ by considering an \emph{alphabetical}
variant which is clean; in the example one might take the formula $\xi' = \dia p
\land (\mu p_{0}. q \lor \dia p_{0}) \land \nu p_{1}. \Box\dia p_{1}$, and 
then consider the game $\EGs(\xi',\bbS)$.

However, this gets more delicate when size matters.
In particular, many complexity-theoretic questions in the modal $\mu$-calculus 
are studied using exactly the kind of games that we just defined.
But if we cannot base the definition of an evaluation game on the collection of
subformulas of a dirty formula, then it does not make sense to take the number
of subformulas as an adequate measure of the size of such a formula.
\end{remark}

\subsection*{Substitution}

The syntactic operation of substitution is ubiquitous in any account of the 
modal $\mu$-calculus, first of all because it features in the basic operation 
of unfolding a fixpoint formula.
As usual in the context of a formal language featuring operators that \emph{bind}
variables, the definition of a substitution operation needs some care.

In particular, we want to protect the substitution operation from variable
capture.
To give a concrete example, suppose that we would naively define a substitution 
operation $\psi/x$ by defining $\phi[\psi/x]$ to be the formula we obtain
from the formula $\phi$ by replacing every free occurrences of $x$ with the 
formula $\psi$. 
Now consider the formula $\phi(q) = \mu p. q \lor \dia 
p$ expressing the reachability of a $q$-state in finitely many steps.
If we substitute $p$ for $q$ in $\phi$, we would expect the resulting formula
to express the reachability of a $p$-state in finitely many steps, 
but the formula we obtain is $\phi[p/q] = \mu p. p \lor \dia p$, which says 
something rather different (in fact, it happens to be equivalent to $\bot$).
Even worse, the substitution $[\ol{p}/q]$ would produce a syntactic string 
$\phi[\ol{p}/q] = \mu p. \ol{p} \lor \dia p$ which is not even a well-formed
formula.

To avoid such anomalies, for the time being we shall only consider substitutions
$\psi/x$ applied to formulas where $\psi$ is free for $x$.

\begin{definition}
\label{d:free2}
Let $\psi,\xi$ and $x$ be respectively two modal $\mu$-calculus formulas and a 
propositional variable.
We say that $\psi$ is \emph{free for $x$ in $\xi$} if $\xi$ is positive in 
$x$\footnote{%
   Strictly speaking, this condition is not needed.
   In particular, as a separate atomic case of our inductive definition,
   we could define the outcome of the substitution $\ol{p}[\psi/p]$
   to be the \emph{negation} of the formula $\psi$ (suitably defined).
   However, in this paper we will only need to look at substitutions
   $\phi[\psi/z]$ where we happen to know that $\phi$ is positive in $z$. 
   As a result, our simplified definition does not impose a real restriction.
   }
and for every variable $y \in \FV{\psi}$, every occurrence of $x$ in a 
subformula $\eta y. \chi$ of $\xi$ is in the scope of a fixpoint operator
$\lambda x$ in $\xi$, i.e., bound in $\xi$ by some occurrence of $\lambda x$.
\end{definition}

\begin{definition}
\label{d:subst}
Let $\{ \psi_{z} \mid z \in Z \}$ be a set of modal $\mu$-calculus formulas, indexed 
by a set of variables $Z$, let $\phi \in \muML$ be positive in each $z \in Z$, and assume that each $\psi_{z}$ is free for $z$ in $\phi$.
We inductively define the \emph{simultaneous substitution}
$[\psi_{z}/z \mid z \in Z]$ as the following operation on $\muML$:
\[\begin{array}{lll}
\phi [\psi_{z}/z \mid z \in Z] &\isdef & 
   \left\{ \begin{array}{ll}
      \psi_{p} & \text{ if } \phi = p \in Z
   \\ \phi     & \text{ if } \phi \text{ is atomic but } \phi \not\in Z
   \end{array}\right.
\\[3mm] (\hs\phi)[\psi_{z}/z \mid z \in Z]       & \isdef & \hs\phi[\psi_{z}/z \mid z \in Z]
\\[1mm] (\phi_{0}\odot\phi_{1})[\psi_{z}/z \mid z \in Z] & \isdef & 
   \phi_{0}[\psi_{z}/z \mid z \in Z] \odot \phi_{1}[\psi_{z}/z \mid z \in Z]
\\[1mm] (\eta x.\phi)[\psi_{z}/z \mid z \in Z]   & \isdef & 
      \eta x. \phi[\psi_{z}/z \mid z \in Z \setminus \{ x\}] 
\end{array}\]
In case $Z$ is a singleton, say $Z = \{ z \}$, we will simply write $\phi[\psi_{z}/z]$.
\end{definition}

With the restriction that we only substitute formulas that are free for the 
variables that they replace, our definition of substitution produces 
syntactically well-formed formulas, and is semantically correct, in the sense 
of the following statement.

\begin{proposition}
\label{p:subs1}
Let $\psi$ and $\xi$ be modal $\mu$-calculus formulas, and let $x$ be a 
propositional variable such that $\psi$ is free for $x$ in $\xi$.
Then for every model $\bbS = (S,R,V)$ we have
\begin{equation}
\label{eq:subs}
\mng{\xi[\psi/x]}^{\bbS} = \mng{\xi}^{\bbS[x \mapsto 
\mng{\psi}^{\bbS}]}.
\end{equation}
\end{proposition}

\begin{remark}
In case $\psi$ is not free for some $z \in Z$ in $\xi$, we can define 
a (syntactically and semantically) correct version of the substitution 
$\xi[\psi/x]$ by taking some (canonically chosen) alphabetical variant $\xi'$
of $\xi$ such that each $\psi_{z}$ is free for $z$ in $\xi'$, and setting
\[
\xi[\psi_{z}/z \mid z \in Z] \isdef \xi'[\psi_{z}/z \mid z \in Z].
\]
It is then easy to verify \eqref{eq:subs}.
We will come back to this issue in Section~\ref{sec:aleq}, when we have
properly defined the notion of alphabetical variant.
\end{remark}

The following proposition is a well known observation in areas where syntax
is used that features variable binding.
Note however that our version below is a bit subtler than usual since we do not
allow the renaming of bound variables.

\begin{proposition} 
\label{p:commutingsubst}
Let $\phi,\chi$ and $\rho$ be $\mu$-calculus formulas, and let $x$ and $y$ be 
distinct variables such that $x$ is free in $\phi$ but not in $\rho$.
Furthermore, assume that $\chi$ is free for $x$ in $\phi$ and that $\rho$ is
free for $y$ in $\phi[\chi/x]$.

Then $\rho$ is free for $y$ in both $\phi$ and $\chi$, $\chi[\rho/y]$ is free
for $x$ in $\phi[\rho/y]$, and we have
\begin{equation}
\label{eq:commsubst}
\phi[\chi/x][\rho/y] = \phi[\rho/y][(\chi[\rho/y])/x].
\end{equation}
\end{proposition}

\begin{proof}
The proposition can be proved by a straightforward but rather tedious induction
on the complexity of $\phi$. 
We omit details.
\end{proof}

\subsection*{Unfolding}

The reason that the modal $\mu$-calculus, and related formalisms, are called 
\emph{fixpoint logics} is that, for $\eta = \mu/\nu$, the meaning of the formula
$\eta x. \chi$ in a model $\bbS$ is given as the least/greatest \emph{fixpoint}
of the semantic map $\chi^{\bbS}_{x}$ expressing the dependence of 
the meaning of $\chi$ on (the meaning of) the variable $x$.
As a consequence, the following equivalence lies at the heart of semantics of 
$\muML$:
\begin{equation}
\label{eq:unf}
\eta x. \chi \equiv \chi[\eta x.\chi/x]
\end{equation}

\begin{definition}
Given a formula $\eta x.\chi \in \muML$, we call the formula 
$\chi[\eta x.\chi/x]$ its \emph{unfolding}.
\end{definition}

\begin{remark}
\label{r:unf}
Unfolding is the central operation in taking the closure of a formula that we
are about to define.
Unfortunately, the collection of clean formulas is not closed under unfolding.
Consider for instance the formula $\phi(p) = \nu q. \dia q \land p$, then we 
see that the formula $\mu p. \phi$ is clean, but its unfolding
$\phi[\mu p. \phi/p] = \nu q. \dia q \land \mu p\,\nu q. \dia q \land p$
is not.
Furthermore, our earlier observation that the naive version of substitution may 
produce `formulas' that are not well-formed applies here as well.
For instance, with $\chi$ denoting the formula
$\ol{p} \land \nu p. \Box (x \lor p)$, unfolding the formula $\mu x. \chi$ 
would produce the ungrammatical $\ol{p} \land \nu p. \Box (
\mu x. \ol{p} \land \nu p. \Box (x \lor p))$.
\end{remark}

Fortunately, the condition of \emph{tidyness} guarantees that we may calculate
unfoldings without moving to alphabetical variants, since we can prove that the
formula $\eta x. \chi$ is free for $x$ in $\chi$, whenever $\eta x. \chi$ is 
tidy. 
In addition, tidyness is preserved under taking unfoldings.

\begin{proposition}
\label{p:subst0}
Let $\eta x. \chi \in \muML$ be a tidy formula.
Then 

\begin{urlist}

\item \label{it:subst0-1}
	$\eta x. \chi$ is free for $x$ in $\chi$;

\item \label{it:subst0-2}
 $\chi[\eta x. \chi/x]$ is tidy as well.
\end{urlist}
\end{proposition}

\begin{proof}
For part \ref{it:subst0-1}, take a variable $y \in \FV{\eta x. \chi}$.
Then obviously $y$ is distinct from $x$, while $y \not\in \BV{\eta x. \chi}$
by tidyness. 
Clearly then we find $y \not\in \BV{\chi}$; in other words, $\chi$ has \emph{no}
subformula of the form $\la y. \psi$.
Hence it trivially follows that $\eta x.\chi$ is free for $x$ in $\chi$.

Part 2) is immediate by the following identities:
\[\begin{array}{lllll}
   \FV{\chi[\eta x.\chi/x]} & = 
   & (\FV{\chi} \setminus \{ x \}) \cup \FV{\eta x.\chi} & =
   & \FV{\eta x.\chi} 
\\ \BV{\chi[\eta x.\chi/x]} & = 
   & \BV{\chi} \cup \BV{\eta x.\chi} & =
   & \BV{\eta x.\chi}
\end{array}\]
which can easily be proved.
\end{proof}


\subsection*{Closure}

We are now ready to define the \emph{closure} of a (tidy) $\mu$-calculus 
formula.
In words, we define $\Clos(\psi)$ as the smallest set containing $\psi$ which
is closed under direct boolean and modal subformulas, and under unfoldings of 
fixpoint formulas.
It will be convenient to define this set in terms of so-called \emph{traces}.

\begin{definition}
\label{d:clos}
Let $\cla$ be the binary relation between tidy $\mu$-calculus formulas given 
by the following exhaustive list:

1) $(\phi_{0}\odot\phi_{1}) \cla \phi_{i}$, for any $\phi_{0},\phi_{1} 
   \in \muML^{t}$, $\odot \in \{ \land, \lor \}$ and $i \in \{0,1\}$;
   
2) $\hs\phi \cla \phi$, for any $\phi \in \muML^{t}$ and
   $\hs \in \{ \dia, \Box \}$);
   
3) $\eta x. \phi \cla \phi[\eta x. \phi /x]$, for any 
   $\eta x. \phi \in \muML^{t}$, with $\eta \in \{ \mu, \nu \}$.
 
\noindent
We define the relation $\clat$ as the reflexive and transitive closure of $\cla$,
and define
\[
\Clos(\psi) \isdef \{ \phi \mid \psi \clat \phi \};
\]
formulas in this set are said to be \emph{derived} from $\psi$.
Given a set of formulas $\Psi$, we put $\Clos(\Psi) \isdef 
\bigcup_{\psi \in \Psi} \Clos(\psi)$.

Finally, we call a $\cla$-path $\psi_{0} \cla \psi_{1} \cla \cdots \cla 
\psi_{n}$ a \emph{(finite) trace}; similarly, an \emph{infinite trace} is 
a sequence $(\psi_{i})_{i<\om}$ such that $\psi_{i} \cla \psi_{i+1}$ for all
$i<\om$.
\end{definition}

Clearly then, a formula $\chi$ belongs to the closure of a formula $\psi$ iff
there is a trace from $\psi$ to $\chi$. 
This trace perspective will be particularly useful when we need to prove 
statements about the formulas belonging to the closure of a certain formula.
We will occasionally think of Definition~\ref{d:clos} as a derivation system 
for statements of the form $\phi \in \Clos(\psi)$, and of a trace $\psi = 
\chi_{0} \cla \chi_{1} \cla \cdots \cla \chi_{n} = \phi$ as as derivation of
the statement that $\phi \in \Clos(\psi)$.

\begin{remark}
\label{r:mottidy}
The final example of Remark~\ref{r:unf} shows that the closure of a non-tidy
formula may not even be defined --- unless we work with alphabetical variants.
But then, as argued in the introduction, it makes much more sense to define a 
notion of size that is invariant under $\al$-equivalence.
\end{remark}

The following example will be instructive for understanding the concept of 
closure, and its relation with subformulas.

\begin{example}
\label{ex:clos}
Consider the following formulas:
\[\begin{array}{llr@{}c@{}l}
   \xi_{1} 
   & \isdef & \mu x_{1} \nu x_{2} \mu x_{3} 
   & . & \big((x_{1} \lor x_{2} \lor x_{3}) \land 
         \Box(x_{1} \lor x_{2} \lor x_{3})\big)
\\ \xi_{2} 
   & \isdef & \nu x_{2} \mu x_{3}
   & . & \big((\xi_{1} \lor x_{2} \lor x_{3}) \land 
         \Box(\xi_{1} \lor x_{2} \lor x_{3})\big)
\\ \xi_{3} 
   & \isdef & \mu x_{3}
   & .  & \big((\xi_{1} \lor \xi_{2} \lor x_{3}) \land 
      \Box(\xi_{1} \lor \xi_{2} \lor x_{3})\big)
\\ \xi_{4} 
   & \isdef &
   &   & \big((\xi_{1} \lor \xi_{2} \lor \xi_{3}) \land 
      \Box(\xi_{1} \lor \xi_{2} \lor \xi_{3})\big)
\\ \al 
   & \isdef &
   &   & \xi_{1} \lor \xi_{2} \lor \xi_{3}
\\ \be
   & \isdef &
   &   & \xi_{1} \lor \xi_{2},
\end{array}\]
and let $\Phi$ be the set $\Phi \isdef \{ \xi_{1}, \xi_{2}, \xi_{3}, \xi_{4},
\Box\al, \al, \be \}$.
In Figure~\ref{fig:cl3-1} we depict the \emph{closure graph} of $\xi_{1}$, i.e.,
the graph based on the set $\Clos(\xi_{1})$, that takes the trace relation for 
its edges.

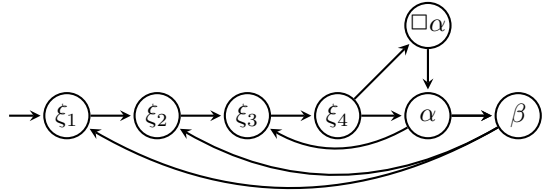
\begin{figure}[th]
\begin{center}

\begin{tikzpicture}[node distance= 12mm]
   \tikzset{every state/.append style={inner sep=-1mm,minimum size=6mm}}
   
   \node[state,initial] (x1) at (0,0) {$\xi_1$};
   \node[state, right of=x1] (x2) {$\xi_2$};
   \node[state, right of=x2] (x3) {$\xi_3$};
   \node[state, right of=x3] (x4) {$\xi_4$};
   \node[state, right of=x4] (a) {$\al$};
   \node[state, above of=a] (bxa) {$\Box\al$};
   \node[state, right of=a] (b) {$\be$};

   \path[->]
    (x1) edge     node {}      (x2)
    (x2) edge     node {}      (x3)
    (x3) edge     node {}      (x4)
    (x4) edge     node {}      (a)
    (x4) edge     node {}      (bxa)
    (a)  edge      node {}      (b)
    (bxa)  edge      node {}      (a)
    (a)  edge      node {}      (b)
    (a) edge[bend left]     node {}      (x3)
    (b) edge[bend left]     node {}      (x2)
    (b) edge[bend left]     node {}      (x1)
  ; 
\end{tikzpicture}
\caption{A closure graph}
\label{fig:cl3-1}
\end{center}
\end{figure}

For $i = 1,2,3$, the formula $\xi_{i+1}$ is the unfolding of the formula 
$\xi_{i}$.
Thus we find $\Clos(\xi_{1}) = \Phi$; in fact, we have $\Clos(\phi) = \Phi$
for every formula $\phi \in \Phi$.
Observe that the formulas $\xi_{1}, \xi_{2}, \xi_{3}$ and $\xi_{4}$ are 
equivalent to one another, and hence also to $\al$.
Note too that the formula $\xi_{1}$ is the only clean formula in $\Phi$.

Finally, it is immediate that the formula
\[
\chi \isdef \mu x_{3}. \big((\xi_{1} \lor x_{2} \lor x_{3}) \land 
         \Box(\xi_{1} \lor x_{2} \lor x_{3})\big)
\]
is a subformula of $\xi_{2} = \nu x_{2}.\chi$, while at the same time it can 
easily be verified that $\xi_{2}$ belongs to the closure of $\chi$.
\end{example}

In many respects the closure and subformula maps behave in similar ways.
In particular, we may also define an \emph{evaluation game} using the closure
set of a (tidy) formula.
This motivates in fact the choice of the number of elements in a formula's 
closure set as a suitable size measure.

The winning conditions of this alternative evaluation game can be presented 
via a priority map, making it into a parity game.
This definition is rather tricky however --- it will take a large part of 
section~\ref{s:fixpar} to get the details right.
For the time being, we can define the winning condition using the observation
that for any infinite trace $(\xi_{n})_{n<\om}$ of tidy formulas there is a 
unique fixpoint formula $\xi = \eta x.\chi$ which occurs infinitely often on
the trace and is a subformula of $\xi_{n}$ for cofinitely many $n$.

\begin{definition}
\label{d:evgc}
Let $\bbS=(S,R,V)$ be a Kripke model and let $\xi$ be a tidy formula in $\muML$.
We define the \emph{evaluation game} $\EGc(\xi,\bbS)$ as the game $(G,E,\Om)$ 
of which the board consists of the set $\Clos(\xi) \times S$, and the game graph
(i.e., the partitioning of $\Clos(\xi) \times S$ into positions for the two 
players, together with the set $E(z)$ of admissible moves at each position), 
is given in Table~\ref{tb:1c}.

\begin{table}[htb]
\begin{center}
\begin{tabular}{|ll|c|l|}
\hline
\multicolumn{2}{|l|}{Position} & Player & Admissible moves\\
\hline
   \multicolumn{2}{|l|}{$(\phi \lor \psi,s)$}   & $\eloi$   
   & $\{ (\phi,s), (\psi,s) \}$ 
\\ \multicolumn{2}{|l|}{$(\phi \land \psi,s)$} & $\abel$ 
   & $\{ (\phi,s), (\psi,s) \}$ 
\\ \multicolumn{2}{|l|}{$(\dia \phi,s)$}        & $\eloi$ 
   & $\{ (\phi,t) \mid sRt \}$ 
\\ \multicolumn{2}{|l|}{$(\Box \phi,s) $}       & $\abel$ 
   & $\{ (\phi,t) \mid sRt \}$ 
\\   $(p,s)$        & with $p\in \FV{\xi}$ and $s \in V(p)$         
   & $\abel$ & $\nada$ 
\\   $(p,s)$        & with $p\in \FV{\xi}$ and $s \notin V(p)$      
   & $\eloi$ & $\nada$ 
\\   $(\lneg{p},s)$   & with $p\in \FV{\xi}$ and $s \in V(p)$    
   & $\eloi$ & $\nada$ 
\\   $(\lneg{p},s)$  & with $p\in \FV{\xi}$ and $s \notin V(p)$ 
   & $\abel$ & $\nada$ 
\\ \multicolumn{2}{|l|}{$(\eta x . \phi,s)$}    & - 
   & $\{ (\phi[\eta x\, \phi/x],s) \}$ 
\\ \hline
\end{tabular}
\end{center}
\caption{The closure evaluation game $\EGc(\xi,\bbS)$}
\label{tb:1c}
\end{table}

To define the winner of an infinite match $\Si = (\xi_{n},s_{n})_{n\in\om}$, 
let $\xi = \eta x.\chi$ be the fixpoint formula that occurs infinitely often 
on the trace $(\xi_{n})_{n\in\om}$ and is a subformula of $\xi_{n}$ for 
cofinitely many $n$.
Then we declare $\eloi$ ($\abel$) as the winner of $\Si$ if $\eta = \nu$ (if
$\eta = \mu$, respectively).
\end{definition}

The following fact, which goes back to the work of Emerson \& 
Jutla~\cite{emer:tree91}, states the \emph{adequacy} of this game.

\begin{fact}
\label{f:0c}
Let $\xi$ be a tidy formula of the modal $\mu$-calculus, and let $(\bbS,s)$ 
be some pointed Kripke structure.
Then
$\bbS,s \sat \xi \ouriff (\xi,s) \in \Win_{\eloi}(\EGc(\xi,\bbS))$.
\end{fact}

Despite the similarities, there are notable differences between the operations 
of taking subformulas and closure.
For instance, the subformula relation is noetherian (conversely well-founded),
which makes formula induction into a useful tool.
The closure relation may have cycles, as witnessed in Example~\ref{ex:clos}.
The same example also shows that, while $\Sfor(\xi) = \Sfor(\chi)$ implies 
$\xi = \chi$, it may well be the case that $\Clos(\xi) = \Clos(\chi)$ while 
$\xi \neq \chi$, or even $\xi \not\equiv \chi$.

We gather some basic observations on the closure map.
To start with, while we saw that the closure of a clean formula will generally 
not consist of clean formulas only, tidyness is preserved, however.
Furthermore, a formula that is free for a certain variable in some tidy formula 
$\xi$ will be free for that variable in any formula in the closure of $\xi$.

\begin{proposition}
\label{p:clos5}
Let $\xi \in \muML$ be a tidy formula, and let $\phi$ be a formula in 
$\Clos(\xi)$.
Then
\begin{urlist}

\item \label{i:clos5-1}
$\BV{\phi} \sse \BV{\xi}$ and $\FV{\phi} \sse \FV{\xi}$;

\item \label{i:clos5-2}
$\phi$ is tidy;

\item \label{i:clos5-3}
if $\psi$ is free for $x$ in $\xi$ and $\xi[\psi/x]$ is tidy, then
   $\psi$ is free for $x$ in $\phi$, and $\phi[\psi/x]$ is tidy;

\item \label{i:clos5-4}
if $\xi$ is of the form $\xi = \eta x.\chi$, then $\xi$ is free for $x$ in 
  every formula in $\Clos(\xi)$.
\end{urlist}
\end{proposition}

\begin{proof}
The proofs of the items \ref{i:clos5-1} and \ref{i:clos5-2} proceed by a 
straightforward induction on the trace $\xi \clat \phi$.
For instance, for the preservation of tidyness it suffices to prove 
that $\chi$ is tidy if $\hs\chi$ is so (where $\hs \in \{ \dia, \Box \}$),
that $\chi_{0}$ and $\chi_{1}$ are tidy if $\chi_{0}\odot\chi_{1}$ is so (where
$\odot \in \{ \land, \lor \}$), and that the unfolding of a tidy formula is tidy
again.
The proofs of the first two claims are easy, and the third claim was stated in
Proposition~\ref{p:subst0}.

For item \ref{i:clos5-3} assume that $\psi$ is free for $x$ in $\xi$, and that $\xi[\psi/x]$ 
is tidy.
We will prove that, for any formula $\phi \in \Clos(\xi)$, $\psi$ is free for
$x$ in $\phi$ and the formula $\phi[\psi/x]$ is tidy. 
We argue by induction on the length of the trace from $\xi$ to $\phi$, and the
key case is where $\phi$ is of the form $\phi = \chi[\eta y\, \chi/y]$.
Inductively we assume that $\psi$ is free for $x$ in $\eta y\, \chi \in 
\Clos(\xi)$ and that $(\eta y\, \chi)[\psi/x]$ is tidy.
Here we make a case distinction. 
If $x = y$ then we have $x \not\in \FV{\chi[\eta y\, \chi/y]}$, and so $\psi$ is
free for $x$ in $\chi[\eta y\, \chi/y]$ in a trivial way.

Now assume that $x$ and $y$ are distinct variables, and suppose for 
contradiction that some free variable $z \in \FV{\psi}$ is captured by the
substitution $[\psi/x]$ applied to the formula $\chi[\eta y\, \chi /y]$.
This means that some free occurrence of $x$ in $\chi[\eta y\, \chi /y]$ lies in
the scope of a binder $\la z$; for concreteness, let $t$ and $u$ be the nodes in 
the syntax tree $\bbT$ of the formula $\chi[\eta y\, \chi/y]$, associated with 
$x$ and $\la x$ respectively, that witness this configuration.

It is not hard to see that $t$ cannot be part of the $\chi$-part of the 
construction tree, since it would imply that $u$ is also situated in this part,
and this would contradict the assumption that $\psi$ is free for $x$ in 
$\eta y\, \chi$.
It follows that $t$ is a leaf of the $\eta y\, \chi$ part of $\bbT$.
This means, however, that the node $u$ must lie in the $\chi$ part of $\bbT$;
otherwise, we could again get a contradiction with the assumption that $\psi$
is free for $x$ in $\eta y\, \chi$.
Now consider the occurrence of $x$ in the $\chi$ part of $\bbT$ that 
corresponds to $t$ --- recall that after all, the syntax tree of $\chi$ is a 
subtree of that of $\eta x\, \chi$.
Using the same reasoning as before, we may derive that this occurrence cannot
lie in the scope of a binder $\la z$. 
However, this means that the formula $(\eta y\, \chi)[\psi/x]$ would have both
free and bound occurrences of $z$, contradicting its tidyness.

It is left to prove that the formula $\chi[\eta y\, \chi/y][\psi/x]$ is tidy, 
but this is rather straightforward since this formula has the same free and 
bound variables as the formula $(\eta y\, \chi)[\psi/x]$ that is tidy per 
induction hypothesis.

Finally, item~\ref{i:clos5-4} of the Proposition is immediate by 
Proposition~\ref{p:subst0}(\ref{it:subst0-1} and part~\ref{i:clos5-3}.
\end{proof}

Second, $\Clos$ is indeed a closure operation (on the set of tidy formulas):

\begin{proposition}
\label{p:clos1}
$\Clos$ is a closure operation on the collection of tidy formulas:

\begin{urlist}
\item $\Phi \sse \Clos(\Phi)$;
\item $\Clos$ is monotone: $\Phi \sse \Psi$ implies $\Clos(\Phi) \sse \Clos (\Psi)$;
\item $\Clos(\Clos(\Phi)) \sse \Clos(\Phi)$.
\end{urlist}
\end{proposition}

In the proposition below we see how the closure map interacts with various 
connectives and formula constructors of the $\mu$-calculus.

\begin{proposition}
\label{p:clos3}
Let $\chi$ and $\xi$ be tidy formulas.
Then the following hold:

\begin{urlist}
\item \label{it:clos3-1}
if $\chi \sforeq \xi$ is a literal then $\chi \in \Clos(\xi)$;

\item \label{it:clos3-2}
if $\xi = \hs \chi$, then $\Clos(\xi) = \{ \hs\chi \} \cup \Clos(\chi)$,
   where $\hs \in \{ \dia, \Box \}$;
    
\item \label{it:clos3-3}
if $\xi = \chi_{0}\odot\chi_{1}$ then $\Clos(\xi) =
\{ \chi_0\odot\chi_1 \} \cup \Clos(\chi_0) \cup \Clos(\chi_{1})$, 
   where $\odot \in \{ \land, \lor \}$;

\item \label{it:clos3-4}
if $\xi = \chi[\psi/x]$ then $\Clos(\xi) =
   \{ \phi[\psi/x] \mid \phi \in \Clos(\chi) \} \cup \Clos(\psi)$,
      provided $x \in FV(\chi)$ and $\psi$ is free for $x$ in $\chi$;

\item \label{it:clos3-5}
if $\xi = \eta x.\chi$ then $\Clos(\xi) = \{ \eta x.\chi \} \cup \{ \phi[\eta x.\chi/x] \mid 
   \phi \in \Clos(\chi) \}$, where $\eta \in \{ \mu, \nu \}$.
\end{urlist}
\end{proposition}

\begin{proof}
Leaving the relatively easy proofs of the second and third claim to the reader,
we first prove the fourth and fifth item, 
The first statement is an instance of Proposition~\ref{p:11-3}, which we will 
prove later.

For the proof of 4), assume that $x \in FV(\chi)$ and that $\psi$ is free for 
$x$ in $\chi$.
By Proposition~\ref{p:clos5}(\ref{i:clos5-3}, the formula $\psi$ is free for 
$x$ in every $\phi \in \Clos(\chi)$.
To prove the inclusion $\sse$ it suffices to show that the set $\{ \phi[\psi/x]
\mid \phi \in \Clos(\chi) \} \cup \Clos(\psi)$ has the required closure
properties. 
This is easily verified, and so we omit the details.

For the opposite inclusion, we first show that
\begin{equation}
\label{eq:cl11}
\phi[\psi/x] \in \Clos(\chi[\psi/x]), \text{ for all } \phi \in \Clos(\chi),
\end{equation}
and we prove this  by induction on the trace from $\xi$ to $\chi$.
It is immediate by the definitions that $\chi[\psi/x] \in \Clos(\chi[\psi/x])$,
which takes care of the base case of this induction.

In the inductive step we distinguish three cases. 
First, assume that $\phi \in \Clos(\chi)$ because the formula $\hs\phi \in 
\Clos(\chi)$, with $\hs \in \{ \dia, \Box \}$.
Then by the inductive hypothesis we find $\hs\phi[\psi/x] = (\hs\phi)[\psi/x] 
\in \Clos(\chi[\psi/x])$; but then we may immediately conclude that $\phi[\psi/x]
\in \Clos(\chi[\psi/x])$ as well.
The second case, where we assume that $\phi \in \Clos(\chi)$ because there is 
some formula $\phi\odot\phi'$ or $\phi'\odot\phi$ in $\Clos(\chi)$ (with $\odot 
\in \{ \land, \lor \}$), is dealt with in a similar way.

In the third case, we assume that $\phi \in \Clos(\chi)$ is of the form
$\phi = \rho[\la y.\rho/y]$, with $\la \in \{ \mu,\nu \}$ and $\la y.\rho \in 
\Clos(\chi)$.
Then inductively we may assume that $(\la y.\rho)[\psi/x] \in 
\Clos(\chi[\psi/x])$.
Now we make a case distinction: if $x = y$ we find that $(\la y.\rho)[\psi/x]
= \la y.\rho$, while
at the same time we have $\phi[\psi/x] = \rho[\la y.\rho/y][\psi/x] = 
\rho[\la y.\rho/y]$, so that it follows by the closure properties that 
$\phi[\psi/x] \in \Clos(\chi)$ indeed.
If, on the other hand, $x$ and $y$ are distinct variables, then we find 
$(\la y.\rho)[\psi/x] = \la y. \rho[\psi/x]$, and so it follows by the closure 
properties that the formula $(\rho[\psi/x])\big[\la y.\rho[\psi/x]/y\big]$ 
belongs to $\Clos(\chi[\psi/x])$.
But since $\psi$ is free for $x$ in $\chi$, the variable $y$ is not free in
$\psi$, and so a straightforward calculation shows that
$(\rho[\psi/x])\big[\la y.\rho[\psi/x]/y\big] =
\rho [\la y.\rho/y][\psi/x] = \phi[\psi/x]$, and so we find that $\phi[\psi/x]
\in \Clos(\chi[\psi/x])$ in this case as well.
\smallskip

Now we turn to claim 5) of the proposition.
First observe that by Proposition~\ref{p:subst0}(\ref{it:subst0-1},
the formula $\eta x.\chi$ is free for $x$ in $\chi$, so that we may apply
part 4) without any problem.
For the proof of the inclusion `$\sse$' it suffices to show that the set 
$\{ \eta x.\chi \} \cup \{ \phi[\eta x.\chi/x] \mid \phi \in \Clos(\chi) \}$ 
has the right closure properties, which is easy.
For the opposite inclusion `$\supseteq$', it is immediate by the definitions
that $\Clos(\eta x.\chi) = \{ \eta x.\chi \} \cup \Clos(\chi[\eta x.\chi/x])$.
But we saw in Proposition~\ref{p:subst0} that the formula $\eta x.\chi$ is 
free for $x$ in $\chi$.
It then follows by 4) that 
$\Clos(\eta x.\chi) = \{ \eta x.\chi \} \cup 
\{ \phi[\eta x.\chi/x] \mid \phi \in \Clos(\chi) \} \cup \Clos(\eta x.\chi)$,
whence the inclusion `$\supseteq$' is immediate.
\end{proof}

\subsection*{Subformulas \& derived formulas}

We now have a look at the relation between the sets $\Sfor(\xi)$ and 
$\Clos(\xi)$.

Our first observation concerns the question which subformulas of a formula 
$\psi$ also belong to its closure.
For this purpose we introduce the notion of a \emph{free} subformula.

\begin{definition}
\label{d:11-1}
Let $\phi$ and $\psi$ be $\mu$-calculus formulas. 
We say that $\phi$ is a \emph{free} subformula of $\psi$, notation: $\phi
\fsforeq \psi$, if $\psi = \psi'[\phi/x]$ for some formula $\psi'$ such that 
$x \in \FV{\psi'}$ and $\phi$ is free for $x$ in $\psi'$.
\end{definition}

Note that in particular all literals occurring in $\psi$ are free subformulas
of $\psi$.
The following characterisation is handy.

\begin{proposition}
\label{p:11-3}
Let $\phi$ and $\psi$ be $\mu$-calculus formulas.
If $\psi$ is tidy, then the following are equivalent:

\begin{urlist}

\item \label{it:11-3-1}
$\phi \fsforeq\psi$;

\item \label{it:11-3-2}
$\phi \sforeq \psi$ and $\FV{\phi} \cap \BV{\psi} = \nada$;

\item \label{it:11-3-3}
$\phi \sforeq \psi$ and $\psi \clat \phi$.
\end{urlist}
\end{proposition}

\begin{proof}
We will prove the equivalence of the statements \ref{it:11-3-1} 
- \ref{it:11-3-3} to a fourth
statement, viz.:

4) there is a $\sfor_{0}$-chain 
$\phi = \chi_{0} \sfor_{0} \chi_{1} \sfor_{0} \cdots \sfor_{0} \chi_{n}
= \psi$,
such that no $\chi_{i}$ has the form $\chi_{i} = \eta y. \rho_{i}$ with
$y \in \FV{\phi}$.
\medskip

For the implication \ref{it:11-3-1} $\Rightarrow$ 4), assume that $\phi \fsforeq \psi$, then 
by definition $\psi$ is of the form $\psi'[\phi/x]$ where $x \in \FV{\psi'}$ 
and $\phi$ is free for $x$ in $\psi'$.
But if $x \in \FV{\psi}$, then it is easy
to see that there is a $\sfor_{0}$-chain
$x = \chi'_{0} \sfor_{0} \chi'_{1} \sfor_{0} \cdots \sfor_{0} \chi'_{n}
= \psi'$ such that no $\chi'_{i}$ is of the form $\chi'_{i} = \la x. 
\rho'$.
Assume for contradiction that one of the formulas $\chi'_{i}$ is of the form
$\chi_{i} = \eta y. \rho_{i}$ where $y \in \FV{\phi}$.
Since $\phi$ is free for $x$ in $\psi'$ this would mean that there is a formula 
of the form $\la x. \chi$ with $\eta y. \rho_{i} \sforeq \la x. \chi \sforeq
\psi'$.
However, the only candidates for this would be the formulas $\chi'_{j}$ with 
$j > i$, and we just saw that these are not of the shape $\la x. \rho'$.
This provides the desired contradiction.
\medskip

For the opposite implication  4) $\Rightarrow$ \ref{it:11-3-1}, assume that 
there is a $\sfor_{0}$-chain 
$\phi = \chi_{0} \sfor_{0} \chi_{1} \sfor_{0} \cdots \sfor_{0} \chi_{n}
= \psi$ as in the formulation of 4).
One may then show by a straightforward induction that $\phi \fsforeq \chi_{i}$,
for all $i \geq 0$.
\medskip

For the implication \ref{it:11-3-2} $\Rightarrow$ 4), assume that $\phi 
\sforeq \psi$ and $\FV{\phi} \cap \BV{\psi} = \nada$.
It follows from $\phi \sforeq \psi$ that there is a $\sfor_{0}$-chain
$\phi = \chi_{0} \sfor_{0} \chi_{1} \sfor_{0} \cdots \sfor_{0} \chi_{n}
= \psi$.
Now suppose for contradiction that one of the formulas $\chi_{i}$ would be of 
the form $\chi_{i} = \eta y. \rho_{i}$ with $y \in \FV{\phi}$.
Then we would find $y \in \FV{\phi} \cap \BV{\psi}$, contradicting the 
assumption that $\FV{\phi} \cap \BV{\psi} = \nada$.
\medskip

In order to prove the implication 4) $\Rightarrow$ \ref{it:11-3-3}, it suffices
to show, for any $n$, 
that if $(\chi_{i})_{0\leq i \leq n}$ is an $\sfor_{0}$-chain of length $n+1$
such that no $\chi_{i}$ has the form $\chi_{i} = \eta y. \rho_{i}$ with $y \in
\FV{\chi_{0}}$, then $\chi_{n} \clat \chi_{0}$.
We will prove this claim by induction on $n$.
Clearly the case where $n = 0$ is trivial.

For the inductive step we consider a chain
\[
\chi_{0} \sfor_{0} \chi_{1} \sfor_{0} \cdots \sfor_{0} \chi_{n} \sfor_{0}
\chi_{n+1}
\]
such that no $\chi_{i}$ has the form $\chi_{i} = \eta y. \rho_{i}$ with
$y \in \FV{\chi_{0}}$, and we make a case distinction as to the nature of 
$\chi_{n+1}$.
Clearly $\chi_{n+1}$ cannot be an atomic formula.

If $\chi_{n+1}$ is of the form $\rho_{0} \odot \rho_{1}$ with $\odot \in \{
\land, \lor\}$, then since $\chi_{n} \sfor_{0} \chi_{n+1}$, the first formula
must be of the form $\chi_{n} = \rho_{i}$ with $i \in \{ 0,1 \}$.
But since it follows by the induction hypothesis that $\chi_{n} \clat \chi_{0}$, 
we obtain from $\chi_{n+1} \cla \chi_{n}$ that $\chi_{n+1} \clat \chi_{0}$ as
required.
The case where $\chi_{n+1}$ is of the form $\hs\rho$ with $\hs \in \{ 
\dia, \Box \}$ is handled similarly.

This leaves the case where $\chi_{n+1} = \lambda y. \rho$ is a fixpoint formula.
Then since $\chi_{n} \sfor_{0} \chi_{n+1}$ it must be the case that $\chi_{n} =
\rho$.
Furthermore, it follows from the assumption in 4) that $y \not\in \FV{\chi_{0}}$.
From this it is not so hard to see that
\[
\chi_{0} \sfor_{0} \chi_{1}[\chi_{n+1}/y] \sfor_{0} \cdots \sfor_{0} 
\chi_{n}[\chi_{n+1}/y]
\]
is a $\sfor_{0}$-chain to which the induction hypothesis applies.
It follows that $\chi_{n}[\chi_{n+1}/y] \clat \chi_{0}$.
From this and the observation that $\chi_{n+1} \cla \chi_{n}[\chi_{n+1}/y]$ we 
find that $\chi_{n+1} \clat \chi_{0}$ indeed.
This finishes the proof of the implication 4) $\Rightarrow$ \ref{it:11-3-3}.
\medskip

Finally, it follows from Proposition~\ref{p:clos5}(\ref{i:clos5-1} that $\psi 
\clat \phi$ implies $\FV{\phi} \cap \BV{\psi} \sse \FV{\psi} \cap \FV{\psi} = 
\nada$. 
From this the implication \ref{it:11-3-3} $\Rightarrow$ \ref{it:11-3-2} is
immediate.
\end{proof}

The following rather technical proposition is stated here for future 
reference.

\begin{proposition} \label{p:always comparable}
If $\xi$ is tidy and $\xi \cla \psi$ then it holds for every
$\phi \fsforeq \psi$ that either $\phi \fsforeq \xi$ or 
$\xi \fsforeq \phi$, and in the latter case $\xi$ is a fixpoint formula.
\end{proposition}

\begin{proof}
Consider formulas $\xi$, $\psi$ and $\phi$ with $\xi \cla \psi$ and 
$\phi \fsforeq \psi$.
Distinguish cases depending on whether $\xi$ is a fixpoint formula. 
If it is not, then it is immediate from the definition of $\cla$ that 
$\psi \sforeq \xi$ and thus $\phi \sforeq \xi$.
Since we also have $\xi \cla \psi \clat \phi$ we obtain $\phi \fsforeq 
\xi$ as required.

On the other hand, if $\xi = \eta x. \chi$ is a fixpoint formula then $\psi
= \chi[\xi/x]$. 
We can distinguish three further cases for the free subformula $\phi$ of 
$\psi$.
If $\phi$ is a free subformula of one of the $\xi$s that has been
substituted for an $x$ in $\chi$ then we are done. 
If $\phi$ is a free subformula of $\chi[\xi/x]$ that contains a part of 
the outer $\chi$ plus a substituted instance of $\xi$ that has been 
substituted for one of the free occurrences of $x$ in $\chi$, then 
it is easy to see that $\xi \fsforeq \phi$.
Finally, if $\phi$ contains a part of the outer $\chi$ into which no
copy of $\xi$ has been substituted then the substitution had no
effect on this subformula and hence $\phi$ is still a free subformula of
$\xi$.
\end{proof}

The following proposition states that under some mild conditions, the 
substitution operation $[\xi/x]$ is in fact injective.

\begin{proposition}
\label{p:subst1}
Let $\phi_{0}, \phi_{1}$ and $\xi$ be formulas such that $\xi$ is free for 
$x$ in both $\phi_{0}$ and $\phi_{1}$, and not a free subformula of either 
$\phi_{i}$.
Then 
\begin{equation}
\label{eq:subst1}
\phi_{0}[\xi/x] = \phi_{1}[\xi/x] \text{ implies } \phi_{0} = \phi_{1}.
\end{equation}
\end{proposition}

\begin{proof}
We first observe that, with $\phi_{0}, \phi_{1}, \xi$ and $x$ as in the 
statement of the proposition, it holds that
\begin{equation}
\label{eq:subst1a}
\phi_{0}[\xi/x] = \phi_{1}[\xi/x] \text{ implies that }
x \in \FV{\phi_{0}} \text{ iff } x \in \FV{\phi_{1}}.
\end{equation}
This is in fact easy to see: if $x \in \FV{\phi_{0}} \setminus \FV{\phi_{1}}$,
then we would obtain a contradiction from $\xi \fsforeq \phi_{0}[\xi/x] = 
\phi_{1}[\xi/x] = \phi_{1}$.

We now turn to \eqref{eq:subst1}, which we will prove by a straightforward 
induction on the complexity of $\phi_{0}$.
Note that by \eqref{eq:subst1a} we only need to worry if $x \in \FV{\phi_{0}}
\cap \FV{\phi_{1}}$; if this is not the case then \eqref{eq:subst1} holds
trivially. 

In particular, this means that the base step of the inductive proof of
\eqref{eq:subst1} is reduced to the case where $\phi_{0} = \phi_{1} = x$, so 
that \eqref{eq:subst1} holds trivially.

For the inductive step we first consider the case where $\phi_{0}$ is of the
form $\psi_{0} \land \chi_{0}$.
Then we obtain $\phi_{0}[\xi/x] = \psi_{0}[\xi/x] \land \chi_{0}[\xi/x]$.
But if $\phi_{1}[\xi/x] = \psi_{0}[\xi/x] \land \chi_{0}[\xi/x]$ and $\xi
\not\fsforeq \phi_{1}$, it must be the case that $\phi_{1}$ is of the form
$\phi_{1} = \psi_{1} \land \chi_{1}$, with $\psi_{1}[\xi/x] = \psi_{0}[\xi/x]$ 
and $\chi_{1}[\xi/x] = \chi_{0}[\xi/x]$.
By the induction hypothesis we obtain $\psi_{0} = \psi_{1}$ and $\chi_{0} = 
\chi_{1}$, so that $\phi_{0} = \phi_{1}$ indeed.

The cases for disjunction and the modal operators are very similar to this, so
we omit the details.

This leaves the case where $\phi_{0}$ is of the form $\phi_{0} = \eta y. 
\psi_{0}$.
Restricting to the case where $x \in \FV{\phi_{0}} \cap \FV{\phi_{1}}$
we may assume that $x$ and $y$ are distinct variables, so we have 
$\phi_{0}[\xi/x] = \eta y. \psi_{0}[\xi/x]$.
But it follows from $\eta y. \psi_{0}[\xi/x] = \phi_{1}[\xi/x]$ and $\xi 
\not\fsforeq \phi_{1}$, that $\phi_{1}$ must be of the form $\phi_{1} = \eta y.
\psi_{1}$ for some formula $\psi_{1}$ with $\psi_{0}[\xi/x] = \psi_{1}[\xi/x]$.
Thus the inductive hypothesis yields that $\psi_{0} = \psi_{1}$, which 
immediately implies that $\phi_{0} = \phi_{1}$ as required.
\end{proof}

Finally, an important observation, going back to Kozen~\cite{koze:resu83},
concerns the existence of a surjective map from $\Sfor(\xi)$ to $\Clos(\xi)$, at 
least for a clean formula $\xi$.
Recall that, given a clean formula $\xi$, we define the \emph{dependency 
order} $<_{\xi}$ on the bound variables of $\xi$ as the least strict partial 
order such that $x <_{\xi} y$ if $\de_{x} \psfor \de_{y}$ and $y \sforeq 
\delta_{x}$.

\begin{definition}
\label{d:kozmap}
Let $\xi$ be a clean formula.
Writing $\mathit{BV}(\xi) = \{ x_1, \dots, x_n\}$, where we may assume that
$i<j$ if $x_{i} <_{\xi} x_{j}$, we define the \emph{expansion} $\exp_{\xi}(\phi)$
of a subformula $\phi$ of $\xi$ as:
\[
\exp_{\xi}(\phi) \isdef \phi[\eta_{x_{1}} x_{1}.\delta_{x_1} / x_1] \dots 
   [\eta_{x_{n}} x_{n}. \delta_{x_n} / x_n].
\]
That is, we substitute first $x_1$ by $\eta_{x_{1}} x_{1}.\delta_{x_1}$ in 
$\phi$; in the resulting formula, we substitute $x_2$ by $\eta_{x_{2}} x_{2}.
\delta_{x_2}$, etc.
If no confusion is likely we write $\exp(\phi)$ instead of $\exp_{\xi}(\phi)$.
\end{definition}

Without proof we mention the following result.

\begin{proposition}
Let $\xi \in \muML$ be a clean formula and $\bbS$ a pointed Kripke structure.
Then for all subformulas $\phi \sforeq \xi$ and all states $s$ in $\bbS$ we have
\begin{equation*}
(\phi,s) \in \Win_{\eloi}(\EGs(\xi,\bbS)) \ouriff \bbS,s \sat \exp_{\xi}(\phi).
\end{equation*}
\end{proposition}

\begin{proposition}
\label{p:clos2}
Let $\xi$ be a clean $\muML$-formula.
Then
\begin{equation}
\label{eq:exp}
\Clos(\xi) = \{ \exp_{\xi}(\phi) \mid \phi \sforeq \xi \}.
\end{equation}
\end{proposition}

\begin{proof}
This statement was proved by Kozen~\cite{koze:resu83}, so here we confine 
ourselves to a brief sketch.
For the inclusion $\sse$ it suffices to show that the set $\{ 
\exp_{\xi}(\phi)  \mid \phi \sforeq \xi \}$ has the relevant closure properties.
This is a fairly routine proof.
For the opposite inclusion it suffices to prove that $\exp_{\xi}(\phi) \in 
\Clos(\xi)$, for every $\phi \in \Sfor(\xi)$, which can be done by a 
straightforward induction.
\end{proof}

As an immediate corollary of Proposition~\ref{p:clos2} we find that the closure
set of a clean $\mu$-calculus formula is always smaller than its set of 
subformulas, and thus, in particular, finite.

\subsection*{Size measure: first definitions}

We are now ready to define two notions of size for $\mu$-calculus formulas:
subformula-size and closure-size.
There are some issues concerning both notions.
First of all, recall from Remark~\ref{r:sfsm} that it only makes sense to define
size as the number of subformulas in case of a \emph{clean} formula.
Similarly, it follows from the observations in Remark~\ref{r:unf} that the 
closure of an untidy formula may not even be well defined, so that it is not
a priori clear how to define the closure-size of an untidy formulas.
Of course, one could then choose to define the subformula-size of a dirty 
formula $\xi$, or the closure-size of an untidy formula $\xi$, by moving to 
some canonically chosen alphabetical variant $\xi'$ of $\xi$ that is clean
in the case of subformula size and tidy in the case of closure-size.
But bringing alphabetical variants into the picture naturally raises the 
question whether we should not consider the sets $\Sfor(\xi')$ and $\Clos(\xi')$
\emph{modulo $\alpha$-equivalence}.
These questions are obviously of interest, and will be discussed in detail
in the Sections~\ref{sec:aleq} and~\ref{sec:skel}.
For now, we restrict the definition of subformula-size to clean formulas,
and that of closure-size to tidy ones.

\begin{definition}
We define the following two size measures for a formula $\xi \in \muML$:

- if $\xi$ is clean, its \emph{subformula-size} $\ssz{\xi}$ is simply given
as its number of subformulas, $\ssz{\xi}\isdef \size{\Sfor(\xi)}$;

- if $\xi$ is tidy, its \emph{closure-size} $\csz{\xi} \isdef \size{\Clos(\xi)}$ is 
given as the size of its closure set.
\end{definition}

In the following theorem, which basically formulates known results, we compare
these different size measures.  
In short, we see that the subformula size is always smaller than or equal to the 
length of a formula, and can in fact be exponentially more succinct; and 
similarly, that closure size is always smaller than or equal to the subformula 
size, and that it can be exponentially smaller than the number of subformulas.

\begin{proposition}
\label{p:szbas}
\begin{urlist}
\item \label{it:szbas-1}
Every clean formula $\xi \in \muML$ satisfies $\csz{\xi} \leq \ssz{\xi} \leq 
\len{\xi}$.

\item \label{it:szbas-2}
There is a family of clean formulas $(\xi_{n})_{n\in\om}$ with 
$\ssz{\xi_{n}} \leq n+1$ while $\len{\xi_{n}} \geq 2^{n}$, for each $n$.

\item \label{it:szbas-3}
There is a family of tidy formulas $(\xi_{n})_{n\in\om}$ with $\csz{\xi_{n}} 
\leq 2\cdot n$ while $\size{\Sfor(\xi_{n})} \geq 2^{n}$, for each $n$.
\end{urlist}
\end{proposition}

\begin{proof}
For part~\ref{it:szbas-1}, it is immediate by a result of Kozen (here formulated
as Proposition~\ref{p:clos2}), that $\csz{\xi} \leq \ssz{\xi}$, while the fact 
that $\ssz{\xi} \leq \len{\xi}$ follows by a straightforward formula induction.
(To the best of our knowledge, the latter result is folklore.)

For part~\ref{it:szbas-2} consider the formulas that are inductively given by
$\xi_{0} \isdef p$, $\xi_{n+1} \isdef (\xi_{n} \land \xi_{n})$.

Part~\ref{it:szbas-3} of the Theorem was shown by Bruse, Friedmann \& 
Lange~\cite[Theorem~3.1]{brus:guar15}; see also our Example~\ref{ex:bfl}.
\end{proof}

For reference later on, we mention the following result which links the two
size measures to the syntactic operation of substitution.

\begin{proposition}
Let $\xi$ and $\psi$ be $\mu$-calculus formulas such that $\psi$ is free for $x$ 
in $\xi$.
Then we have

1) $\ssz{\xi[\psi/x]} \leq \ssz{\xi} + \ssz{\psi}$ if $\xi[\psi/x]$ is clean; 

2) $\csz{\xi[\psi/x]} \leq \csz{\xi} + \csz{\psi}$ if $\xi[\psi/x]$ is tidy. 
\end{proposition}

\begin{proof}
By a routine verification one may check that 
\begin{equation}
\Sfor(\xi[\psi/x]) \sse \{ \phi[\psi/x] \mid \phi \sforeq \xi \} \cup \Sfor(\psi)
\end{equation}
and from this, part 1) is immediate.
Similarly, part 2) follows from Proposition~\ref{p:clos3}(\ref{it:clos3-4}.
\end{proof}

\section{Alternation depth}
\label{sec:ad}

Next to its size, the most important complexity measure of a $\mu$-calculus 
formula is its \emph{alternation depth}, that is, the maximal number of 
alternations between least and greatest fixpoint operators.
There are various ways to make this notion precise; here we shall work with the 
most widely used definition, which originates with
Niwi\'{n}ski~\cite{niwi:fixp86}.
Recall our notation $\fopp{\mu} = \nu$, $\fopp{\nu} = \mu$.

\begin{definition}
\label{d:ad}
By natural induction we define classes $\AH{\mu}{n},\AH{\nu}{n}$ of 
$\mu$-calculus formulas.
With $\eta, \la \in \{ \mu, \nu \}$ arbitrary, we set:
\begin{enumerate}
\setlength{\itemsep}{0mm}
\setlength{\topsep}{0mm}
\item \label{adr:1}
all atomic formulas belong to $\AH{\eta}{0}$;
\item \label{adr:2}
if $\phi_{0},\phi_{1} \in \AH{\eta}{n}$, then
$\phi_{0} \lor \phi_{1}, \phi_{0} \land \phi_{1}, \dia\phi_{0}, \Box\phi_{0}
\in \AH{\eta}{n}$;
\item \label{adr:3}
if $\phi \in \AH{\eta}{n}$ then $\fopp{\eta} x. \phi \in \AH{\eta}{n}$;
\item \label{adr:4}
if $\phi(x), \psi \in \AH{\eta}{n}$, then $\phi[\psi/x] \in \AH{\eta}{n}$,
   provided that $\psi$ is free for $x$ in $\phi$;
\item \label{adr:5}
all formulas in $\AH{\la}{n}$ belong to $\AH{\eta}{n+1}$.
\end{enumerate}
The \emph{alternation depth} $\ad(\xi)$ of a formula $\xi$ is defined as the 
least $n$ such that $\xi \in \AH{\mu}{n} \cap \AH{\nu}{n}$.
\end{definition}

Intuitively, the class $\AH{\eta}{n}$ consists of those $\mu$-calculus formulas
where $n$ bounds the length of any alternating nesting of fixpoint operators 
of which the most significant formula is an $\eta$-formula.
The alternation depth is then the maximal length of an alternating nesting of 
fixpoint operators.
We will make this intuition more precise further on.

\begin{example}
Observe that the basic modal (i.e., fixpoint-free) formulas are exactly the ones
with alternation depth zero. 
Formulas that use $\mu$-operators or $\nu$-operators, but not both, have 
alternation depth $1$.
For example, observe that $\mu x. p \lor x$ belongs to $\AH{\nu}{0}$ but 
not to $\AH{\mu}{0}$: none of the clauses in Definition~\ref{d:ad} is applicable.
On the other hand, using clause \eqref{adr:5} it is easy to see that $\mu x. p
\lor x \in \AH{\nu}{1} \cap \AH{\mu}{1}$, from which it is immediate that 
$\ad(\mu x. p \lor x) = 1$.

Consider the formula $\xi_{1} = \mu x. (\nu y. p \land \Box y) \land \dia x$.
Taking a fresh variable $q$, we find $\mu x. q \land \dia x \in \AH{\nu}{0} \sse
\AH{\nu}{1}$ and $\nu y. p \land \Box y \in \AH{\mu}{0} \sse \AH{\nu}{1}$, so 
that by the substitution rule we have 
$\xi_{1} = (\mu x. q \land \dia x)[\nu y. p \land \Box y/q] \in \AH{\nu}{1}$.
Similarly we may show that $\xi_{1} \in \AH{\mu}{1}$, so that $\xi_{1}$ has 
alternation depth $1$.

The formula $\xi_{2} = \nu x. \mu y. (p \land \dia x) \lor \dia y$ is of higher 
complexity.
It is clear that the formula $\mu y. (p \land \dia x) \lor \dia y$ belongs to 
$\AH{\nu}{0}$ but not to $\AH{\mu}{0}$.
From this it follows that $\xi_{2}$ belongs to $\AH{\mu}{1}$ but there is no way
to place it in $\AH{\nu}{1}$.
Hence we find that $\ad(\xi_{2}) = 2$.

As a third example, consider the formula 
\[
\xi_{3} = \mu x. \nu y. (\Box y \land  \mu z. (\dia x \lor z)).
\]
This formula looks like a $\mu/\nu/\mu$-formula, in the sense that it contains
a nested fixpoint chain $\mu x /\nu y / \mu z$.
However, the variable $y$ does not occur in the subformula $\mu z. (\dia x \lor
z)$, and so we may in fact consider $\xi_{3}$ as a $\mu/\nu$-formula.
Formally, we observe that 
$\mu z. \dia x \lor z \in \AH{\nu}{0} \sse \AH{\nu}{1}$ and 
$\nu z. \Box y \land p \in \AH{\mu}{0} \sse \AH{\nu}{1}$;
from this it follows by the substitution rule that the formula
$\nu y. (\Box y \land  \mu z. (\dia x \lor z))$ belongs to the set 
$\AH{\nu}{1}$ as well; from this it easily follows that $\xi_{3} \in 
\AH{\nu}{1}$.
It is not hard to show that $\xi_{3} \not\in \AH{\mu}{1}$, so that we find
$\ad(\xi_{3}) = 2$.
\end{example}

\begin{remark}
In the literature one usually sees the alternation hierarchy defined in terms of
classes $\Si_{n}$ and $\Pi_{n}$, with the notation taken from the arithmetical 
hierarchy.
Our notation, which uses $\mu$ and $\nu$ as superscripts, allows to exploit the
symmetry between $\mu$ and $\nu$ more directly, and may thus make the definition
of alternation depth a bit easier.
\end{remark}

In the propositions below we make some observations on the sets $\AH{\eta}{n}$
and on the notion of alternation depth.
First we show that each class $\AH{\mu}{n}$ is closed under subformulas and 
derived formulas.

\begin{proposition}
\label{p:ad1}
Let Let $\xi$ and $\phi$ be $\mu$-calculus formulas. 
\begin{enumerate}[topsep=0pt,itemsep=-1ex,partopsep=1ex,parsep=1ex,%
    label={\arabic*)}]
\item \label{it:ad1-1}
If $\phi \sforeq \xi$ and $\xi \in \AH{\eta}{n}$ then $\phi \in \AH{\eta}{n}$.
\item \label{it:ad1-2}
If $\xi \clat \phi$ and $\xi \in \AH{\eta}{n}$ then $\phi \in \AH{\eta}{n}$.
\end{enumerate}
\end{proposition}

\begin{proof}
We prove the statement in part~\ref{it:ad1-1} by induction on the derivation of 
$\xi \in \AH{\eta}{n}$.
In the base case of this induction we have that $n = 0$ and $\xi$ is an atomic
formula.
But then obviously all subformulas of $\xi$ are atomic as well and thus belong
to $\AH{\eta}{n}$.

In the induction step of the proof it holds that $n>0$; we make a case 
distinction as to the applicable clause of Definition~\ref{d:ad}.

In case $\xi \in \AH{\eta}{n}$ because of clause \eqref{adr:2} in 
Definition~\ref{d:ad},
we make a further case distinction as to the syntactic shape of $\xi$.
First assume that $\xi$ is a conjunction, say, $\xi = \xi_{0} \land \xi_{1}$,
with $\xi_{0},\xi_{1} \in \AH{\eta}{n}$. 
Now consider an arbitrary subformula $\phi$ of $\xi$; it is not hard to see 
that either $\phi = \xi$ or $\phi \sforeq \xi_{i}$ for some $i \in \{0,1\}$.
In the first case we are done, by assumption that $\xi \in \AH{\eta}{n}$; in the
second case, we find $\phi \in \AH{\eta}{n}$ as an immediate 
consequence of the induction hypothesis.
The cases where $\xi$ is a disjunction, or a formula of the form $\Box\psi$ or 
$\dia\psi$ are treated in a similar way.

If $\xi \in \AH{\eta}{n}$ because of clause \eqref{adr:3} of the definition, 
then $\xi$ must be of the form $\xi = \eta x. \chi$, with $\chi \in \AH{\eta}{n}$.
We proceed in a way similar to the previous case: any subformula $\phi \sforeq
\xi$ is either equal to $\xi$ (in which case we are done by assumption), or 
a subformula of $\chi$, in which we are done by one application of the induction
hypothesis.

In the case of clause \eqref{adr:4}, assume that $\xi$ is of the form
$\chi[\psi/x]$, where $\psi$ is free for $x$ in $\chi$, and $\chi$ and $\psi$ 
are in $\AH{\eta}{n}$.
Then by the induction hypothesis all subformulas of $\chi$ and $\psi$ belong to
$\AH{\eta}{n}$ as well.
Now consider an arbitrary subformula $\phi$ of $\xi$; it is easy to see that 
either $\phi \sforeq \chi$, $\phi \sforeq \psi$ or else $\phi$ is of the form 
$\phi = \phi'[\psi/x]$ where $\phi' \sforeq \chi$.
In either case it is straightforward to prove that $\phi \in \AH{\eta}{n}$, as
required.

Finally, in case $\xi$ is in $\AH{\eta}{n}$ because of clause \eqref{adr:5},
it belongs to $\AH{\la}{n-1}$ for some $\la \in \{ \mu, \nu\}$.
Then by induction hypothesis all subformulas of $\xi$ belong to $\AH{\la}{n-1}$.
We may then apply the same clause \eqref{adr:5} to see that any such $\phi$ also
belongs to the set $\AH{\eta}{n}$.
\medskip

To prove part~\ref{it:ad1-2}, it suffices to show that the class $\AH{\eta}{n}$
is closed under unfoldings, since by part~\ref{it:ad1-1} we already know it to
be closed under subformulas.
So assume that $\la x . \chi \in \AH{\eta}{n}$ for some $n$ and $\la \in \{\mu,
\nu\}$. 
Because $\chi \sforeq \eta x . \chi$ it follows from part~\ref{it:ad1-1} that 
$\chi \in \AH{\la}{n}$. 
But then we may apply clause \eqref{adr:4} from Definition~\ref{d:ad} and
conclude that $\chi[\eta. \chi / x] \in \AH{\la}{n}$.
\end{proof}

As an immediate corollary of Proposition~\ref{p:ad1} we find the following.

\begin{proposition}
\label{p:ad2}
Let $\xi$ and $\chi$ be $\mu$-calculus formulas. 
Then
\begin{enumerate}[topsep=0pt,itemsep=-1ex,partopsep=1ex,parsep=1ex,%
    label={\arabic*)}]
\item \label{it:ad2-1}
if $\chi \in \Sfor(\xi)$ then $\ad(\chi) \leq \ad(\xi)$;
\item \label{it:ad2-2}
if $\chi \in \Clos(\xi)$ then $\ad(\chi) \leq \ad(\xi)$.
\end{enumerate}
\end{proposition}

Clause~\eqref{adr:4} of Definition~\ref{d:ad} states that the classes 
$\AH{\eta}{n}$ are closed under substitution.
The following proposition states a kind of converse to this. 

\begin{proposition}
\label{p:ahsubst}
Let $\xi$ and $\chi$ be $\mu$-calculus formulas such that $\xi$ is free for $x$ 
in $\chi$.
If $\chi[\xi/x] \in \AH{\eta}{k}$ then $\chi \in \AH{\eta}{k}$.
Furthermore, if $x \in \FV{\chi}$ then we also have $\xi \in \AH{\eta}{k}$.
\end{proposition}

\begin{proof}
The result about $\xi$ is immediate by Proposition~\ref{p:ad1}.
We prove the statement about $\chi$ by induction on the number $\#_{x}(\chi)$ of
free occurrences of the variable $x$ in $\chi$.
In the base of this induction we assume that  $\#_{x}(\chi) = 1$, and we 
proceed via a subinduction on the length of the derivation that the formula 
$\chi[\xi/x]$ belongs to the set $\AH{\eta}{k}$.
We make a case distinction as to the last applied clause of 
Definition~\ref{d:ad} in this derivation.
The cases where we applied clause \eqref{adr:2} or \eqref{adr:5} are relatively 
easy and therefore left as an exercise.

In case we applied clause \eqref{adr:1}, the formula $\chi[\xi/x]$ is atomic.
But then certainly the formula $\chi$ is atomic as well, ensuring that $\chi \in
\AH{\eta}{0}$ as required.

If, by clause \eqref{adr:3}, we have $\chi[\xi/x] = \fopp{\eta}y. \phi$, for 
some formula $\phi\in \AH{\eta}{k}$, we make a further case distinction.
First, if $\chi$ is of the form $\chi = x$ (and hence $\xi = \fopp{\eta}y. \phi$)
then clearly we have $\chi \in \AH{\eta}{0} \sse \AH{\eta}{n}$.
Second, the case where $\chi \neq x$ but $x = y$ cannot occur since we would
obtain that $\chi[\xi/x] = \fopp{\eta}x. \phi$ which must mean that $\chi$
itself is of the form $\chi = \fopp{\eta}x.\psi$  contradicting the assumption
that $x \in \FV{\chi}$.
This leaves the case where $\chi \neq x$ and $x \neq y$.
Now we find $\chi = \fopp{\eta}y.\phi'$, where $\phi = \phi'[\xi/x]$.
Then by the inductive hypothesis we obtain that $\phi' \in \AH{\eta}{k}$,
so that one application of clause \eqref{adr:3} yields that $\chi \in 
\AH{\eta}{k}$.

Finally, assume that $\chi[\xi/x] \in \AH{\eta}{k}$ because of clause 
\eqref{adr:4}, i.e., $\chi[\xi/x] = \phi[\psi/y]$ for some formulas $\phi,\psi 
\in \AH{\eta}{k}$.
Call a node $t$ of the syntax tree $T$ of $\chi[\xi/x]$ \emph{critical} if $t$
also exists as a node in the syntax tree of $\phi$, where it is labelled 
with a free occurrence of the variable $y$.
Furthermore, let $r_{\xi}$ be the unique node of $T$ which, as a node of the
syntax tree $T_{\chi}$ of $\chi$, is labelled with $x$. 
Observe that the subtree of $T$ generated by $r_{\xi}$ is isomorphic to the 
syntax tree of $\xi$ and that, similarly, any subtree of $T$ generated by a 
critical node is isomorphic to the syntax tree of $\psi$.
We now distinguish three case, as to the relative position of $r_{\xi}$ with
respect to critical nodes of $T$.

\textit{Case 1:} The node $r_{\xi}$ has a critical descendant in $T$.
This can only happen if $\xi = \xi'[\psi/y]$ for some formula $\xi'$.
Let $\rho$ be the (uniquely determined) formula such that $\phi = \rho[\xi'/x]$
and $\chi = \rho[\psi/y]$.
It follows by the inner induction hypothesis that $\rho \in \AH{\eta}{k}$,
and so we immediately find that $\chi = \rho[\psi/y]$ belongs to $\AH{\eta}{k}$
as well.

\textit{Case 2:} The node $r_{\xi}$ has a critical ancestor in $T$.
In this case we must have $\psi = \psi'[\xi/x]$ for some formula $\psi'$.
We may assume that $\psi'$ has a \emph{unique} free occurrence of $x$ (if not,
then $\chi$ would also have multiple free occurrences of $x$).
This means that we may apply the inner induction hypothesis to $\psi = 
\psi'[\xi/x]$ and obtain $\psi' \in \AH{\eta}{k}$.
Furthermore, we may take a variation of the formula $\phi$ by relabelling the
unique critical ancestor of $r_{\xi}$ by a fresh variable $y'$.
This gives a formula $\phi'$ such that $\phi = \phi'[y/y']$, while $\chi = 
\phi'[\psi'/y',\psi/y]$.
Then another application of the induction hypothesis yields that $\phi' \in 
\AH{\eta}{k}$, and so we find $\chi = \phi'[\psi'/y'][\psi/y] \in \AH{\eta}{k}$
by two applications of clause \eqref{adr:5}.

\textit{Case 3:} The node $r_{\xi}$ has neither critical descendants, nor a 
critical ancestor in $T$.
This means that there is a formula $\chi'$ such that $\chi = \chi'[\psi/y]$
and $\phi = \chi'[\xi/x]$.
It then follows by the inner induction hypothesis that $\chi' \in \AH{\eta}{k}$,
so that $\chi \in \AH{\eta}{k}$ by an application of clause \eqref{adr:5}.
\medskip

Finally, in the induction step of the outer induction we are dealing with the
situation that $\#_{x}(\chi) > 1$, that is, $\chi$ has multiple free occurrences
of the variable $x$.
Let $\chi^{-}$ be a variation of $\chi$ where we replace some (but not all) 
of these occurrences with a fresh variable $x'$.
Clearly then we obtain that $\xi$ is free for $x'$ in $\chi^{-}$ and $\chi = 
\chi^{-}[x/x']$, so that $\chi[\xi/x] = \chi^{-}[\xi/x][\xi/x']$.
On the other hand we have 
$\#_{x'}(\chi^{-}[\xi/x]) < \#_{x}(\chi)$ since $x'$ is fresh for $\xi$,
and $\#_{x}(\chi^{-}[\xi/x]) < \#_{x}(\chi)$.
But then two applications of the induction hypothesis yield that, respectively, 
the formulas $\chi^{-}[\xi/x]$ and $\chi^{-}$ belong to $\AH{\eta}{k}$.
From this it is immediate that the formula $\chi = \chi^{-}[x/x']$ also
belongs to $\AH{\eta}{k}$.
\end{proof}

In the case of a \emph{clean} formula there is a simple characterisation of
alternation depth, making precise the intuition about alternating chains,
in terms of the formula's dependency order on the bound variables.
Although this characterisation seems to be a well-known result, we could
not find a proper reference for it.

\begin{definition}
Let $\xi \in \muML$ be a clean formula.
A \emph{dependency chain} in $\xi$ of \emph{length $d$} is a sequence $\ol{x} =
x_{1}\cdots x_{d}$ such that $x_{1} <_{\xi} x_{2} \cdots <_{\xi} x_{d}$; such a
chain is \emph{alternating} if $x_{i}$ and $x_{i+1}$ have different parity, for 
every $i < d$.
For $\eta \in \{ \mu, \nu\}$, we call an alternating dependency chain $x_{1}\cdots x_{d}$
an \emph{$\eta$-chain} if $x_{d}$ is an $\eta$-variable, and we let 
$d_{\eta}(\xi)$ denote the length of the length of the longest $\eta$-chain in
$\xi$; we write $d_{\eta}(\xi) = 0$ if $\xi$ has not such chains. 
\end{definition}

\begin{proposition}
\label{p:adcf}
Let $\xi$ be a clean formula. 
Then for any $k \in \om$ and $\eta \in \{ \mu, \nu\}$ we have
\begin{equation}
\label{eq:adcf11}
\xi \in \AH{\eta}{k} \text{ iff } d_{\eta}(\xi) \leq k,
\end{equation}
As a corollary, the alternation depth of $\xi$ is equal to the length of its
longest alternating dependency chain.
\end{proposition}

One of the key insights in the proof of this Proposition is that, 
with $\psi$ free for $x$ in $\phi$, any dependency chain in $\phi[\psi/x]$ 
originates entirely from either $\phi$ or $\psi$.
Recall from Definition~\ref{d:munu} that we write $\fopp{\mu} = \nu$
and $\fopp{\nu} = \mu$.

\begin{proof}
We prove the implication from left to right in \eqref{eq:adcf11} by induction on 
the derivation that $\xi\in \AH{\eta}{k}$.
In the base step of this induction (corresponding to clause \eqref{adr:1} in 
the definition of alternation depth) $\xi$ is atomic, so that we immediately
find $d_{\eta}(\xi) = 0$ as required.

In the induction step of the proof, we make a case distinction as to the last
applied clause in the derivation of $\xi\in \AH{\eta}{k}$, and we leave the 
(easy) cases, where this clause was either \eqref{adr:2} or \eqref{adr:3}, for
the reader.

Suppose then that $\xi\in \AH{\eta}{k}$ on the basis of clause \eqref{adr:4}.
In this case we find that $\xi = \xi'[\psi/z]$ for some formulas $\xi',\psi$
such that $\psi$ is free for $z$ in $\xi'$ and $\xi', \psi \in \AH{\eta}{k}$.
By the `key insight' mentioned right after the formulation of the Proposition, 
any $\eta$-chain in the formula $\xi$ is a $\eta$-chain in either
$\xi'$ or $\psi$.
But then by the induction hypothesis it follows that the length of any such 
chain must be bounded by $k$. 

Finally, consider the case where $\xi\in \AH{\eta}{k}$ on the basis of clause 
\eqref{adr:5}.
We make a further case distinction.
If $\xi \in \AH{\eta}{k-1}$, then by the induction hypothesis we may conclude 
that $d_{\eta}(\xi) \leq k-1$, and from this it is immediate that 
$d_{\eta}(\xi) \leq k$.
If, on the other hand, $\xi \in \AH{\fopp{\eta}}{k-1}$ then the induction
hypothesis yields $d_{\fopp{\eta}}(\xi) \leq k-1$.
But since $d_{\eta}(\xi) \leq d_{\fopp{\eta}}(\xi) + 1$ we obtain 
$d_{\eta}(\xi) \leq k$ indeed.
\medskip

The opposite,  right-to-left, implication in \eqref{eq:adcf11} is proved by 
induction on  $k$.
In the base step of this induction we have $d_{\eta}(\xi) = 0$, which means that
$\xi$ has no $\eta$-variables; from this it is easy to derive that $\xi \in 
\AH{\eta}{0}$.

For the induction step, we assume as our induction hypothesis that
\eqref{eq:adcf11} holds for $k \in \om$, and we set out to prove the same 
statement for $k+1$ and an arbitrary $\eta \in \{ \mu, \nu\}$:
\begin{equation}
\label{eq:adcf12}
\text{if } d_{\eta}(\xi) \leq k+1 \text{ then }  \xi \in \AH{\eta}{k+1}.
\end{equation}
We will prove \eqref{eq:adcf12} by an `inner' induction on the length of $\xi$.
The base step of this inner induction is easy to deal with: if $\len{\xi} = 1$
then $\xi$ must be atomic so that certainly $\xi \in \AH{\eta}{k+1}$.

In the induction step we are considering a formula $\xi$ with $\len{\xi} > 1$.
Assume that $d_{\eta}(\xi) \leq k+1$.
We make a case distinction as to the shape of $\xi$. 
The only case of interest is where $\xi$ is a fixpoint formula, say, $\xi =
\eta x. \chi$ or $\xi = \fopp{\eta} x. \chi$.
If $\xi = \fopp{\eta} x. \chi$, then obviously we have $d_{\eta}(\xi) = 
\de_{\eta}(\chi)$, so by the inner induction hypothesis we find $\chi \in 
\AH{\eta}{k+1}$.
From this we immediately derive that $\xi = \fopp{\eta} x. \chi \in 
\AH{\eta}{k+1}$ as well.

Alternatively, if $\xi = \eta x. \chi$, we split further into cases:
If $\chi$ has an $\fopp{\eta}$-chain $y_{1}\cdots y_{k+1}$ of length $k+1$, then 
obviously we have $x \not\in \FV{\de_{k+1}}$ (where we write $\de_{k+1}$ instead
of $\de_{y_{k+1}}$), for otherwise we would get $x >_{\xi} y_{k+1}$, so that 
we could add $x$ to the $\fopp{\eta}$-chain $y_{1}\cdots y_{k+1}$ and obtain an
$\eta$-chain $y_{1}\cdots y_{k+1}x$ of length $k+2$.
But if $x \not\in \FV{\de_{k+1}}$ we may take some fresh variable $z$ and write 
$\xi = \xi'[\fopp{\eta} y_{k+1}. \de_{k+1}/z]$ for some formula $\xi'$ where
the formula $\fopp{\eta} y_{k+1}. \de_{k+1}$ is free for $z$.
By our inner induction hypothesis we find that both $\xi'$ and $\eta y_{k+1}. 
\de_{k+1}$ belong to $\AH{\eta}{k+1}$.
But then by clause \eqref{adr:4} of Definition~\ref{d:ad} the formula $\xi$ 
also belongs to the set $\AH{\eta}{k+1}$.

If, on the other hand, $\chi$ has \emph{no} $\fopp{\eta}$-chain of length $k+1$,
then we clearly have $d_{\fopp{\eta}}(\chi) \leq k$.
Using the outer induction hypothesis we infer $\chi \in \AH{\fopp{\eta}}{k}$, 
and so by clause \eqref{adr:3} of Definition~\ref{d:ad} we also find $\xi =
\eta x. \chi \in \AH{\fopp{\eta}}{k}$.
Finally then, clause \eqref{adr:5} gives $\xi \in \AH{\eta}{k+1}$.
\end{proof}

One may prove a similar characterisation in the wider setting of tidy formulas,
cf.~Proposition~\ref{p:ahandchains} below.
Both the formulation and the proof of this result require some preparation, 
which is why we postpone this characterisation to 
section~\ref{s:fixpar}.

\section{Parity formulas}
\label{sec:par}

In this section we introduce the syntactic structures that serve as the 
yardsticks in our framework, viz., parity formulas.
Intuitively, these are like ordinary (modal) formulas, with the difference that
(i) the underlying structure of a parity formula is a directed graph, possibly 
with cycles, rather than a tree; and (ii) we add a priority labelling to this 
syntax graph, to ensure that, in the game semantics of parity formulas, all full
matches in the two-player evaluation games can be assigned a winner.

Parity formulas are basically the same as Wilke's \emph{alternating tree
automata}~\cite{wilk:alte01,grae:auto02};
they are also very similar to so-called \emph{hierarchical equation systems}
(see for instance~\cite{arno:rudi01,demr:temp16}, and references therein).
At the end of this section we discuss these connections in some more detail.

\subsection*{Syntax}
We start with the basic definition of a parity formula.
Recall that, given a set $\Prop$ of proposition letters, $\At(\Prop) = \{ \bot, 
\top \} \cup \{q, \ol{q} \mid q \in \Prop \}$ denotes the set of \emph{atomic}
formulas over $\Prop$.

\begin{definition}
\label{d:pf}
Let $\Prop$ be a finite set of proposition letters.
A \emph{parity formula over $\Prop$} is a quintuple $\bbG = (V,E,L,\Om,v_{I})$, 
where

a) $(V,E)$ is a finite, directed graph, with $\size{E[v]} \leq 2$ for 
every vertex $v$;

b) $L: V \to \At(\Prop) 
      \cup \{ \land, \lor, \dia, \Box, \epsilon \}$ is a labelling function;

c) $\Om: V \parto \om$ is a partial map, the \emph{priority} map of $\bbG$; and 
 
d) $v_{I}$ is a vertex in $V$, referred to as the \emph{initial} node of $\bbG$;

\noindent
such that 

1) $\size{E[v]} = 0$ if $L(v) \in \At(\Prop)$, and 
   $\size{E[v]} = 1$ if $L(v) \in \{ \dia, \Box\} \cup \{ \epsilon \}$;
   
2) every cycle of $(V,E)$ contains at least one node in $\Dom(\Om)$.

\noindent
A node $v \in V$ is called \emph{silent} if $L(v) = \epsilon$,
\emph{constant} if $L(v) \in \{ \top,\bot\}$,
\emph{literal} if $L(v) \in \Lit(\Prop)$, 
\emph{atomic} if it is either constant or literal,
\emph{boolean} if $L(v) \in \{ \land, \lor \}$,
and \emph{modal} if $L(v) \in \{ \dia, \Box \}$.
The elements of $\Dom(\Om)$ will be called \emph{states}.
We say that a proposition letter $q$ \emph{occurs} in $\bbG$ if $L(v) 
\in \{ q, \ol{q} \}$ for some $v \in V$.
A parity formula $\bbG = (V,E,L,\Om,v_{I})$ is \emph{$\epsilon$-free} if 
$L^{-1}(\epsilon) = \nada$.
\end{definition}

\begin{figure}[th]
\begin{center}

\begin{tikzpicture}
\tikzset{sibling distance=4mm,
   edge from parent/.append style={->,thick},
   every node/.style= {circle,inner sep=0mm,thick},}

\Tree [.\node (mu) [draw] {$\epsilon{\mid} 1$};
    [.$\lor$
        [.$\lor$ 
            [.$\overline{p}$ ]
            [.\node (a) {$\Diamond$};] 
        ]
        [.\node (nu) [draw]{$\epsilon{\mid} 0$} ;
            [.$\land$ 
                [.$q$ ]
                [.$\Box$ 
                    [.$\lor$ 
                       [.\node (x) {$x$}; ] 
                       [.\node (y) {$y$}; ]
                    ]
                ]
            ]
        ]
    ]
]  
\path[->]
    (y) edge     [out=0,in=0]                 node {}      (nu)
    (a) edge     [out=270,in=135]        node {}      (x)
    (x) edge     [out=180,in=180]         node {}      (mu)
;
\end{tikzpicture}
\quad\quad
\begin{tikzpicture}

   \node[state,initial] (v0) at (0,4) {$\lor{\mid}0$};
   \node[state] (v1) at (6,4) {$\lor{\mid}1$};
   \node[state] (v2) at (2,2) {$\lor{\mid}2$};
   \node[state] (v3) at (4,2) {$\lor{\mid}3$};
   \node[state] (v4) at (0,0) {$\lor{\mid}4$};
   \node[state] (v5) at (6,0) {$\lor{\mid}5$};


   \path[->]
    (v0) edge     node {}      (v1)
    (v0) edge     node {}      (v2)
    (v1) edge     node {}      (v3)
    (v1) edge     node {}      (v5)
    (v2) edge[bend left]     node {}      (v3)
    (v2) edge[bend left]     node {}      (v1)
    (v3) edge[bend left]     node {}      (v2)
    (v3) edge[bend left]     node {}      (v4)
    (v4) edge     node {}      (v0)
    (v4) edge     node {}      (v2)
    (v5) edge     node {}      (v4)
    (v5) edge     node {}      (v3)
 ; 
\end{tikzpicture}
\caption{Two parity formulas}
\label{fig:x53}
\end{center}
\end{figure}
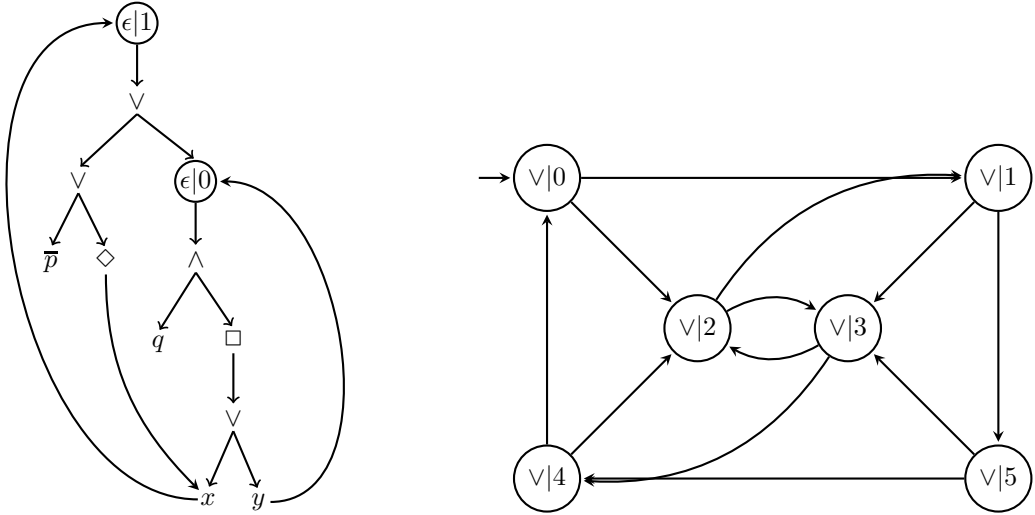

\begin{example}
In Figure~\ref{fig:x53} we give two examples of parity formulas.
The picture on the left displays a parity formula that is directly based
on a $\mu$-calculus formula, viz., the formula 
$\xi = \mu x. (\ol{p} \lor \dia x) \lor \nu y. (q \land \Box(x \lor y))$, by
adding \emph{back edges} to the subformula dag of $\xi$.
Recall that the syntax tree of this formula was displayed in 
Figure~\ref{fig:x21}.
The picture on the right displays a parity formula that is based on a rather
more entangled graph.
\end{example}

The definition of parity formulas needs little explanation.
Condition 2) says that every cycle must pass through at least one state; as we
will see below, this is needed to provide a winner for infinite matches of the
evaluation games that we use to define the semantics of parity formulas.
In the remark below, which can be skipped upon a first reading,  we discuss
some design choices.

\begin{remark}
1) The reason for allowing $\epsilon$-nodes is for technical convenience.
As will be immediately clear once we have defined the semantics of parity 
formulas, the $\epsilon$-free parity formulas have the same expressive power as
the full class.
To see this, we may easily transform a parity formula $\bbG = (V,E,L,\Om,v_{I})$
with silent steps into an equivalent $\epsilon$-free $\bbG'\isdef (V,E,L',\Om,
v_{I}\}$, by replacing the silent steps with `dummy' disjunctions:
\[
L'(v) \isdef 
\left\{ \begin{array}{ll}
   L(v) & \text{ if } L(v) \neq \epsilon
\\ \lor & \text{ if } L(v) = \epsilon.
\end{array}\right.
\]
Observe that instead of disjunctions, we could have taken dummy conjunctions 
just as well.

2) Similarly, we could have omitted constant nodes from the definition, in favour
of boolean nodes with no successors that are labelled with, respectively, $\land$
(for $\top$) and $\lor$ (for $\bot$).

3) Conversely, we could have fixed the number of successors of boolean nodes
to two, replacing boolean nodes with zero and one successors with constant 
and silent nodes, respectively.
In addition, and in order to stay as close as possible to the syntax of regular
formulas, we could have chosen to assign an explicit left- and right successor
to each boolean node, instead of an unranked set.
We have chosen our set-up to keep the structure of parity formulas as simple
as possible, and to stay close to the definition of Wilke's alternating tree 
automata.
In cases where the symmetry between distinct successors of a Boolean node needs
to be broken, we can employ some standard trick, see Convention~\ref{conv:fpv}.

4) Finally, we could have allowed boolean nodes to have an \emph{arbitrary}
number of successors.
However, in this approach the size of the edge relation would no longer be 
linear but quadratic in the number of vertices of the graph, with the
disadvantage that it would be less clear how to define the size of a parity 
formula.
\end{remark}

\subsection*{Semantics}

The semantics of parity formulas is given in terms of an \emph{evaluation 
game}, which is defined as the following two-player infinite parity 
game.\footnote{%
   Basic notions concerning infinite games~\cite{grae:auto02} are given in the
   appendix~\ref{sec:games}.
   }
The rules (admissible moves) in this evaluation game are completely obvious.

\begin{definition}
\label{d:pfgam}
Let $\bbS=(S,R,V)$ be a Kripke model for a set $\Prop$ of proposition letters, 
and let $\bbG = (V,E,L,\Om,v_{I})$ be a parity $\Prop$-formula.
We define the \emph{evaluation game} $\EG(\bbG,\bbS)$ as the parity game 
$(G,E,\Om')$ of which 
the board consists of the set $V \times S$, 
the priority map $\Om': V \times S \to \om$ is given by
\[
\Om'(v,s) \isdef \left\{ \begin{array}{ll}
      \Om(v) & \text{if } v \in \Dom(\Om)
   \\ 0      & \text{otherwise},
\end{array}\right.
\]
and the game graph is given in Table~\ref{tb:2}.
Note that we do not need to assign a player to positions that admit a single 
move only.
\end{definition}

\begin{table}[t]
\begin{center}
\begin{tabular}{|ll|c|c|}
\hline
\multicolumn{2}{|l|}{Position} & Player  & Admissible moves 
\\\hline
     $(v,s)$ & with $L(v) = p$ and $s \in V(p)$         
   & $\abel$ & $\nada$ 
\\   $(v,s)$ & with $L(v) = p$ and $s \notin V(p)$      
   & $\eloi$ & $\nada$ 
\\   $(v,s)$ & with $L(v) = \lneg{p}$ and $s \in V(p)$    
   & $\eloi$ & $\nada$ 
\\   $(v,s)$ & with $L(v) = \lneg{p}$ and $s \notin V(p)$ 
   & $\abel$ & $\nada$ 
\\   $(v,s)$ & with $L(v) = \bot$ 
   & $\eloi$ & $\nada$ 
\\   $(v,s)$ & with $L(v) = \top$ 
   & $\abel$ & $\nada$ 
\\   $(v,s)$ & with $L(v) = \epsilon$ 
   & - & $E[v] \times \{ s \}$ 
\\   $(v,s)$ & with $L(v) = \lor$ 
   & $\eloi$ & $E[v] \times \{ s \}$ 
\\   $(v,s)$ & with $L(v) = \land$ 
   & $\abel$ & $E[v] \times \{ s \}$ 
\\   $(v,s)$ & with $L(v) = \dia$ 
   & $\eloi$ & $E[v] \times R[s]$ 
\\   $(v,s)$ & with $L(v) = \Box$ 
   & $\abel$ & $E[v] \times R[s]$ 
\\ \hline
\end{tabular}
\end{center}
\caption{The evaluation game $\EG(\bbG,\bbS)$}
\label{tb:2}
\end{table}

\begin{remark}
Perhaps a more natural definition of a priority map for the evaluation game
$\EG(\bbG,\bbS)$ than the function $\Om'$ of Definition~\ref{d:pfgam} would
have been to simply put $\Om''(v,s) \isdef \Om(v)$.
However, this would have required a (minor) adaptation of the definition of 
a parity game, allowing its priority map to be \emph{partial}, but requiring,
similar to condition (2) in Definition~\ref{d:pf}, that every cycle in a 
parity game graph hits a position for which the priority is defined.
\end{remark}

\begin{definition}
We say that a parity formula $\bbG = (V,E,L,\Om,v_{I})$ \emph{holds at} or 
\emph{is satisfied by} a pointed Kripke model $(\bbS,s)$, notation: $\bbS,s \sat
\bbG$, if the pair $(v_{I},s)$ is a winning position for $\eloi$ in 
$\EG(\bbG,\bbS)$.
We let $\query(\bbG)$ denote the \emph{query} of $\bbG$, that is, the class 
of pointed Kripke models where $\bbG$ holds, and we call two parity formulas 
$\bbG$ and $\bbG'$ \emph{equivalent} if they determine the same query, notation:
$\bbG \equiv \bbG'$.
We will use the same terminology and notation to compare parity formulas with
standard formulas.
\end{definition}

\subsection*{Basic notions}

As we mentioned in the abstract/introduction, the two key complexity measures 
of a parity formula, viz., size and index, both have perspicuous definitions.
We will introduce these measures here, together some other useful notions 
pertaining to parity formulas.

\begin{definition}
The \emph{size} of a parity formula  $\bbG = (V,E,L,\Om,v_{I})$ is defined as
its number of nodes: $\size{\bbG} \isdef \size{V}$.
\end{definition}

Next to size, as the second fundamental complexity measure for a parity formula
we need is its \emph{index}, which corresponds to the alternation depth of 
regular formulas.
It concerns the degree of alternation between odd and even positions in an
infinite match of the evaluation game, and it is thus directly related to the 
range of the priority map of the formula.
The most straightforward approach would be to define the index of parity formula
as the size of this range; a slightly more sophisticated approach would be a
\emph{clusterwise} version of this.

\begin{definition}
Let $\bbG = (V,E,L,\Om,v_{I})$ be a parity formula, and let $u$ and $v$ be 
vertices in $V$.
We say that $v$ is \emph{active} in $u$ if $E^{+}uv$, and we let ${\bowtie_{E}}
\sse V \times V$ hold between $u$ and $v$ is $u$ is active in $v$ and vice 
versa, i.e., ${\bowtie_{E}} \isdef E^{+} \cap (E^{-1})^{+}$.
We let $\equiv_{E}$ be the equivalence relation generated by $\bowtie_{E}$; 
the equivalence classes of $\bowtie_{E}$ will be called \emph{clusters}.
A cluster $C$ is called \emph{degenerate} if it is a singleton $\{ v\}$ such 
that $v$ is not active in itself, and \emph{nondegenerate} otherwise.

The collection of clusters of a parity formula $\bbG$ is denoted as 
$\Clus(\bbG)$, and we say that a cluster $C$ is \emph{higher} than another
cluster $C'$ if for every $u \in C$ there is a $u \in C'$ such that $E^{+}uu'$.
\end{definition}

\begin{remark}
Recall that in graph theory, a subset $S$ of a directed graph $(V,E)$ is called 
a \emph{strongly connected component} if there is a path, in each direction, 
between any two vertices in $S$.
Intuitively, the strongly connected components of a parity formula consist of 
those sets of vertices that can feature as the ones that occur infinitely often
in a match of an evaluation game associated with the formula.
The clusters of a parity formula correspond to those strongly connected 
components of its underlying graph that are \emph{maximal} with respect to this
strong connectedness condition.

Note that in a nondegenerate cluster there is a nontrivial path between any pair
of vertices.
Finally, observe that the `higher than' relation between clusters is a strict 
partial order.
\end{remark}

The index of a parity formula is sometimes defined as the maximum value of the
sizes of the set $\Ran(\Om\rst{C})$, where $C$ ranges over all clusters of the
formula.
As it turns out, however, it will be more convenient for us to define the index
in terms of the maximal length of so-called alternating $\Om$-chains.

\begin{definition}
Let $\bbG = (V,E,L,\Om,v_{I})$ be a parity formula. 
An \emph{alternating $\Om$-chain} of \emph{length} $k$ in $\bbG$ is a finite
sequence $v_{1}\cdots v_{k}$ of states that all belong to the same cluster, and 
satisfy, for all $i < k$, that $\Om(v_{i}) < \Om(v_{i+1})$ while $v_{i}$ and
$v_{i+1}$ have different parity.
Such a chain is called an $\eta$-chain if $\Om(v_{k})$ has parity $\eta$ (where
we recall that we associate even numbers with $\nu$ and odd numbers with $\mu$).
\end{definition}
Note that a parity formula $\bbG$ has alternating chains iff it has states,
i.e., $\Dom(\Om) \neq \nada$.

\begin{definition}
\label{d:ind}
Let $\bbG = (V,E,L,\Om,v_{I})$ be a parity formula, and let $C$ be a cluster of
$\bbG$.
For $\eta \in \{ \mu,\nu\}$ we define $\idx_{\eta}(C)$, the $\eta$-index of $C$,
as the maximal length of an alternating $\eta$-chain in $C$, and the \emph{index}
of $C$ as $\idx_{\bbG}(C) \isdef \max\big(\idx_{\mu}(C),\idx_{\nu}(C)\big)$.
Finally, we define 
\[
\idx(\bbG) \isdef \max \{ \idx_{\bbG}(C) \mid C \in \Clus(\bbG)\},
\]
as the maximal length of an alternating $\Om$-chain in $\bbG$.
\end{definition}

Observe that if $\bbG$ is cycle-free then we can assume that the range of $\Om$ 
is empty.
Thus, every-cycle free parity is equivalent to one with index zero.
\medskip

A useful consequence of this definition is that parity formulas that are
\emph{parity variant} will have the same index.

\begin{definition}
\label{d:parvar}
A \emph{parity variant} of a parity formula $\bbG = (V,E,L,\Om,v_{I})$ is a 
parity formula $\bbG = (V,E,L,\Om',v_{I})$ such that (i) $\Om(v) \equiv_{2}
\Om'(v)$, for all $v$, and (ii) $\Om(u) < \Om(v)$ iff $\Om'(u) < \Om'(v)$, 
for all $u$ and $v$ that belong to the same cluster but have different parity.
\end{definition}

It is easy to see that parity variants are semantically equivalent, and have
the same index.
From this it follows that there are certain normal forms for parity formulas.

\begin{definition}
\label{d:intitv}
A parity formula $\bbG = (V,E,L,\Om,v_{I})$ is called \emph{linear} if $\Om$ is
injective, and \emph{tight} if for any cluster $C$, the range of $\Om$ on $C$
is connected, that is, of the form $\Ran(\Om\rst{C}) = \rng{k}{n}$ for some
natural numbers $k,n$ with $k \leq n$.
Here we define $\rng{k}{n} \isdef \{ i \in \om \mid k \leq i \leq n \}$.
\end{definition}

It is not hard to see that every parity formula can be effectively transformed
into either a linear or a tight parity variant; for the tight case, see 
Proposition~\ref{p:tight} below.
Furthermore, it is rather obvious that for a tight parity formula $\bbG = 
(V,E,L,\Om,v_{I})$, we have
\begin{equation}
\idx_{\bbG}(C) = \size{\Ran(\Om\rst{C})},
\end{equation}
so that for these devices our definition of index matches the one we mentioned
earlier, viz., in terms of the clusterwise size of the range of the priority 
map.

\subsection*{Priority maps and parity preorders}

Quite often the priority function of a parity formula is induced by some kind of
preorder.
It will be convenient to introduce some terminology.
Recall that a \emph{preorder} is a structure $(P,\sqlq)$ such that $\sqlq$ is a 
reflexive and transitive relation on $P$; given such a relation we will write 
$\sql$ for the irreflexive version of $\sqlq$ (given by $u \sql v$ iff $u \sqlq
v$ but not $v \sqlq u$) and $\equiv$ for the equivalence relation induced by 
$\sqlq$ (given by $u \equiv v$ iff $u \sqlq v$ and $v \sqlq u$).
A preorder is directed if for any two points $u$ and $v$ there is a $w$ such
that $u \sqlq w$ and $v \sqlq w$.

\begin{definition}
\label{d:ppo}
A \emph{parity preorder} is a structure $\bbP = (P,\sqlq,p)$, where $(P,\sqlq)$
is a directed preorder and $p: P \to \{ 0,1 \}$ is a map such that $u \equiv v$ 
implies $p(u) = p(v)$.
\end{definition}

Observe that by directedness, every parity preorder has an $\equiv$-cell of
\emph{$\sqlq$-maximal} elements, and that these points all have the same parity.

\begin{definition}
\label{d:Omppo}
Fix a parity preorder $\bbP = (P,\sqlq,p)$.
An \emph{alternating chain} in $\bbP$ of \emph{length} $k$ in $\bbP$ is a finite
sequence $v_{1}\cdots v_{k}$ of states such that, for all $i < k$, $v_{i} \sqlq
v_{i+1}$ while $v_{i}$ and $v_{i+1}$ have different parity.
Given a point $p \in P$ we define $\adup(p)$ as the maximal length of an 
alternating chain starting at $v$, and we let $\ad(\bbP)$ denote the 
\emph{alternation depth} of $\bbP$, i.e., the maximal length of an alternating 
chain in $\bbP$.

We define the following map $\Om_{\bbP}: P \to \om$:
\begin{equation}
\label{eq:Omppo1}
\Om_{\bbP}(v) \isdef \left\{\begin{array}{ll}
    \ad(\bbP) - \adup(v)     &
       \text{if $\ad(\bbP) - \adup(v) \equiv_{2} p(v)$}
\\  \ad(\bbP) - \adup(v) + 1     &
       \text{if $\ad(\bbP) - \adup(v) \not\equiv_{2} p(v)$},
\end{array} \right.
\end{equation}
and we will call this map the \emph{priority map induced by} $\bbP$.
\end{definition}

Intuitively, we define $\Om_{\bbP}(v)$ to be $\ad(\bbP) - \adup(v)$, possibly 
with a corrective `+ 1' to ensure the right parity.
As a fairly direct consequence of this definition, it follows that $u \sql v$ 
implies $\Om(u) \leq \Om(v)$, with an inequality holding if $u$ and $v$ have 
different parity.
In particular, all $\sqlq$-maximal points obtain the same priority which is the 
maximal $\Om$-value reached.
More information about the construction is provided by the next proposition.

\begin{proposition}
\label{p:Omppo}
Let $\bbP = (P,\sqlq,p)$ be a parity preorder, and let $\Om$ be its induced
parity map.
Then for every $u,v \in P$, it holds that $\Om(v) \equiv_{2} p(v)$, 
that $u \sql v$ implies $\Om(u) \leq \Om(v)$,
and that $u \sql v$ and $p(u) \neq p(v)$ implies $\Om(u) < \Om(v)$, 
Furthermore, $\Ran(\Om)$ is connected, and  
\begin{equation}
\label{eq:pOmppo2}
\size{\Ran(\Om)} = \ad(\bbP).
\end{equation}
\end{proposition}

\begin{proof}
We leave it for the reader to verify that, for every $u,v \in P$, we have
that $\Om(v) \equiv_{2} p(v)$, that $u \sqlq v$ implies $\Om(u) \leq \Om(v)$,
that $u \sql v$ implies $\Om(u) < \Om(v)$, and that $\Ran(\Om)$ is connected,
We focus on the proof of \eqref{eq:pOmppo2}.

Abbreviate $H \isdef \ad(\bbP)$.
For the inequality $\leq$, first make the rather obvious observation that
$\adup(v) \in \rng{1}{H}$, and that from this it easily follows that $\Ran(\Om) 
\sse \rng{0}{H}$.
It then suffices to prove that we cannot have both $0$ and $H$ in $\Ran(\Om)$.

To see why this is the case, assume that $0 \in \Ran(\Om)$; by definition of 
$\Om$ this can only be the case if there is a point $v \in P$ such that $\adup(v)
= H$ and $p(v) = 0$.
Now consider an alternating path starting from $v$ and of maximal length, say
$\pi: u_{0} \sql u_{1} \cdots \sql u_{n}$ with $v = u_{0}$ and $n = \adup(v) 
- 1$.
Without loss of generality we may assume that $u_{n}$ is a \emph{maximal} 
element of $\bbP$, so that $\adup(u_{n}) = 1$ and $\Om$ reaches its maximal
value at $u_{n}$.
By the assumed maximal length of $\pi$ we find that $n = H - 1$, while it 
follows from the alternating nature of $\pi$, and the fact that $p(u_{0}) = 0$,
that either $n$ is even and $p(u_{n}) = 0$ or $n$ is odd and $p(u_{n}) = 1$.
In other words, we have $n \equiv_{2} p(u_{n})$, and since $n = H - 1$ this 
means that for the definition of $\Om(u_{n})$ we are in the first case of 
\eqref{eq:Omppo1} so that we obtain $\Om(u_{n}) = H - \adup(u_{n}) = 
\ad(\bbP) - 1$.
This means that $\Ran(\Om) = \rng{0}{\ad(\bbP)-1}$ as required.

For the opposite inequality $\geq$ of \eqref{eq:pOmppo2}, let 
$u_{1} \sql u_{2} \cdots \sql u_{n}$ be an alternating $\bbP$-chain of 
maximal length.
We leave it for the reader to verify that $n = \ad(\bbP)$ and that 
\[
\Om(v_{i}) = \left\{ \begin{array}{ll}
   i   & \text{if } \ad(\bbP) \equiv_{2} i
\\ i+1 & \text{if } \ad(\bbP) \not\equiv_{2} i
\end{array}\right.\]
for all $i \in \rng{1}{n}$.
From this it easily follows that one of the sets $\rng{1}{\ad(\bbP)}$ or 
$\rng{0}{\ad(\bbP)-1}$ is contained in $\Ran(\Om)$. 
In either case we find that $\size{\Ran(\Om) } \geq \ad(\bbP)$.

As an alternative proof of \eqref{eq:pOmppo2} the reader is invited to verify
directly that either $\Ran(\Om)$ equals either $\rng{0}{H-1}$ or $\rng{1}{H}$, 
depending on the parity of $H$ and the parity value $p(m)$ for any 
$\sqlq$-maximal point $m$:
\[\begin{array}{l|l|l}
                          & p(m) \text{ even } & p(m) \text{ odd }
\\[1mm] \hline H \text{ even } & \rng{1}{H}         & \rng{0}{H{-}1}
\\[1mm] H \text{ odd }  & \rng{0}{H{-}1}         & \rng{1}{H}
\end{array}\]
\end{proof}

As a first application of this Proposition, we show that every parity formula
is equivalent to a tight one.

\begin{proposition}
\label{p:tight}
For every parity formula $\bbG$ there is a tight parity formula $\bbG'$
such that $\bbG' \equiv \bbG$ and $\idx(\bbG') \leq \idx(\bbG)$.
\end{proposition}

\begin{proof}
Let $\bbG = (V,E,L,\Om,v_{I})$ be some parity formula.
For each cluster $C$ of $\bbG$, consider the parity preorder $\sqlq^{C}_{\Om}$
on the set $C \cap \Dom(\Om)$ of states in $C$, given by $u \sqlq^{C}_{\Om} v$
iff $\Om(u) \leq \Om(v)$.
The parity map $p$ we consider is given by $p(v) \isdef \Om(v)\mod 2$.
Let $\Om'_{C}: C \cap \Dom(\Om) \to \om$ be the partial map on $C$ that is 
induced by this order, and let $\Om': V \to \om$ be the union of these maps.
It then follows by Proposition~\ref{p:Omppo} that $\Ran(\Om)$ is connected, 
and that $\idx(\bbG') \leq \idx(\bbG)$ and that $\bbG'$ is a parity variant of
$\bbG$.
Furthermore, it is obvious that $\Dom(\Om') = \Dom(\Om)$, so that we may
unambiguously talk about \emph{states}.

To prove that $\bbG' \equiv \bbG$ we first show that for all $u,v \in \Dom(\Om)$
we have that
\begin{equation}
\label{eq:omrefl}
\text{ if } u \equiv_{E} v \text{ then }
\Om'(u) < \Om'(v) \mbox{ implies } \Om(u) < \Om(v).
\end{equation}
To see this, assume that $\Om(u) \not< \Om(v)$, so that $\Om(v) \leq \Om(u)$.
From this it follows that $v \sqlq^{C}_{\Om} u$, where $C$ is the cluster of $u$
and $v$.
But then we obtain $\Om'(v) \leq \Om'(u)$, contradicting our assumption. 

It remains to show that on every infinite trace $\tau = (v_{n})_{n\in\om}$ in
$(V,E)$ the parity of the maximal priority occurring infinitely often is the 
same under $\Om'$ as under $\Om$. 
Let $v$ be one of the states such that $\Om(v)$ is maximal among the 
$\Om$-priorities of all the states that occur infinitely often on $\tau$. 
We argue that $\Om'(v)$ is also maximal among the $\Om'$-priorities of all the 
states that occur infinitely often on $\tau$. 
Assume for a contradiction that this is not the case, that is, there is some
$u$ that occurs infinitely often on $\tau$ such that $\Om'(u) > \Om'(v)$. 
Clearly $u$ and $v$ must belong to the same cluster.
With \eqref{eq:omrefl} it follows that $\Om(u) > \Om(v)$, which contradicts the
assumption on $v$.
\end{proof}

\subsection*{Operations on parity formulas}

Parity formulas are interesting logical objects in their own right, and so 
one might want to develop their theory.
To start with, it is fairly easy to define various operations on parity 
formulas, such as modal and boolean operations (including negation), least and 
fixpoint operations, and substitution.
Furthermore, it would be of interest to study various \emph{structural} notions
of equivalence between parity formulas; as a starting point, we introduce a
notion of \emph{morphism} between parity formulas.
Given the coalgebraic flavour of parity formula, it should come as no surprise
that this definition takes elements of the notion of a bounded morphism between 
Kripke models.

\begin{definition}
\label{d:mf}
Let $\bbG = (V,E,L,\Om,v_{I})$ and $\bbG' = (V',E',L',\Om',v'_{I})$ be two 
parity formulas.
A \emph{morphism} from $\bbG$ to $\bbG'$ is a map $f: V \to V'$ satisfying the
following conditions, for all $u,v \in V$:

\begin{urlist}
\item \label{it:mf-1}
$L(u) = L'(f(u))$
\item \label{it:mf-2}
if $Euv$ then $E'f(u)f(v)$
\item \label{it:mf-3}
if $E'f(u)v'$ then $Euv$ for some $v$ with $f(v) = v'$
\item \label{it:mf-4}
$\Om(u) = \Om'(f(u))$
\item \label{it:mf-5}
$f(v_{I}) = v'_{I}$.
\end{urlist}
We write $f: \bbG \to \bbG'$ to denote that $f$ is a morphism from $\bbG$ to
$\bbG'$.
\end{definition}

As an example, here is the parity formula version of a subformula.

\begin{definition}
Let $\bbG = (V,E,L,\Om,v_{I})$ be a parity formula, and let $v$ be a vertex in
$V$.
We let $\bbG\init{v} \isdef (V,E,L,\Om,v)$ denote the variant of $\bbG$ that
takes $v$ as its initial node;
we define $V_{v}$ to be the smallest subset of $V$ which contains $v$ and is 
closed under taking successors, and we we call $\bbG_{v} \isdef 
(V_{v}, E\rst{V_{v}},L\rst{V_{v}}, \Om\rst{V_{v}},v)$ the \emph{subformula} of
$\bbG$ that is \emph{generated} from $v$.
\end{definition}

The following proposition will be needed further on; we omit its proof, since it
is straightforward.

\begin{proposition}
\label{p:2}
Let $f:\bbG \to \bbG'$ be a morphism of parity formulas.
Then for any node $v$ in $\bbG$ it holds that 
\[
\bbG\init{v} \equiv \bbG'\init{f(v)}.
\]
In particular, for every node $v \in V$ we have
\[
\bbG\init{v} \equiv \bbG_{v}.
\]
\end{proposition}

\begin{definition}
\label{d:parbis}
Let $\bbG = (V,E,L,\Om,v_{I})$ be a parity formula.
An equivalence relation ${\sim} \sse V \times V$ is called  a \emph{congruence}
on $\bbG$ if it satisfies the following conditions, for every pair of 
nodes $u,u'$ such that $u \sim u'$:

\begin{urlist}
\item \label{it:pb-1}
$L(u) = L(u')$
\item \label{it:pb-2}
if $Euv$ then there is a $v'$ such that $Eu'v'$ and $v \sim v'$
\item \label{it:pb-4}
$\Om(u) = \Om(u')$
\end{urlist}

\noindent
Given such a congruence $\sim$,
we can correctly define the following maps on the collection $\ol{V}$ of 
$\sim$-cells:
\[\begin{array}{lll}
   \ol{L}(\ol{v})   & \isdef & L(v)
\\ \ol{E}(\ol{v})   & \isdef & \{ \ol{u} \mid Evu \}
\\ \ol{\Om}(\ol{v}) & \isdef & \Om(v).
\end{array}\]
We shall refer to the resulting structure $\bbG/{\sim} \isdef 
(\ol{V},\ol{E},\ol{L},\ol{\Om},\ol{v}_{I})$ as the \emph{quotient} of $\bbG$
under $\sim$.
\end{definition}

We leave it for the reader to verify that a quotient of a parity formula
it itself indeed a parity formula. 
\medskip

As is to be expected, there are many connections between morphisms and
quotients.
Without proof we mention the following two propositions; recall that the kernel
of a function $f$ consists of those pairs that are identified by $f$.

\begin{proposition}
\label{p:523}
Let $f:\bbG \to \bbG'$ be a morphism of parity formulas.
Then its \emph{kernel} is a congruence on $\bbG$.
\end{proposition}

\begin{proposition}
\label{p:524}
Let $\sim$ be a congruence on the parity formula $\bbG$. 
Then the quotient map sending nodes of $\bbG$ to their $\sim$-cells is a 
morphism of parity formulas.
\end{proposition}

Combined with the previous proposition, the next observation can be used for 
a \emph{minimisation construction} on parity formulas.

\begin{proposition}
\label{p:525}
Let $\bbG$ be a parity formula.
The collection of congruences on $\bbG$ is closed under unions and thus forms
a bounded lattice.
In particular, $\bbG$ has a largest congruence.
\end{proposition}

We hasten to remark, however, that the quotient of a parity formula $\bbG$ under
its largest quotient is not necessarily the smallest parity formula equivalent
to $\bbG$.

\begin{remark}
Definition~\ref{d:mf} and~\ref{d:parbis} provide what are probably the most
straightforward versions of, respectively, a morphism between parity formulas, 
and a congruence on a parity formula.
There are alternatives, however.
Note that given the non-local nature of the winning conditions in the evaluation
games for parity formulas, the respective conditions on the priority map could
be relaxed quite a bit.
For instance, as an alternative to clause \ref{it:mf-4} in Definition~\ref{d:mf}
we could take the following two clauses:

4a) $\Om(u) \equiv_{2} \Om'(f(u))$

4b) if $u \equiv_{E} v$ and $u \not\equiv_{2} v$ then $\Om(u) < \Om(v)$ 
iff $\Om'(f(u)) < \Om'(f(v))$.

\noindent 
Similarly, clause \ref{it:pb-4} of Definition~\ref{d:parbis} could be replaced 
with

3a) $\Om(u) \equiv_{2} \Om(u')$

3b) for every pair of nodes $v,v'$ such that $v\sim v'$, $u \equiv_{E} v$, 
$u' \equiv_{E} v'$ and $u \not\equiv_{2} v$, we have 
$\Om(u) < \Om(v)$ iff $\Om(u') < \Om(v')$.

With these definitions we can easily prove analogues of the Proposition~\ref{p:524}
and~\ref{p:524}.
It is less straightforward, however, to prove, for an arbitrary parity formula,
the existence of a maximal congruence of this type, but we plan to address this 
issue in future work.
\end{remark}

\subsection*{Formula dags with back edges and untwisted formulas}

In the next section we will associate, with an arbitrary $\mu$-calculus formula
$\xi$, a parity formula $\bbH_{\xi}$ which is based on the subformula dag of 
$\xi$.
These parity formulas are rather special, and below we isolate some of their 
properties; as we shall see in section~\ref{s:parfix}, these guarantee the 
subformula size.
In the definition below, we call a vertex $v$ a \emph{root} of a directed 
acyclic graph if every vertex of the graph is reachable from $v$.

\begin{definition}
\label{d:dagcyc}
A parity formula $\bbG = (V,E,L,\Om,v_{I})$ is \emph{based on a dag with back 
edges} if there are \emph{witnessing} binary relations $D,B \sse V \times V$
such that $E = D \uplus B$,  and 

1) $(V,D)$ is a finite dag (directed acyclic graph) with root $v_{I}$;

2) $B$ is a functional relation (i.e., $\size{B(v)} \leq 1$, for all $v \in V$);

3) $B \sse (D^{-1})^{+}$;

4) if $D^{*}vv'$ and $v \equiv_{E} v'$ then $\Om(v) \geq \Om(v')$ (if defined).

5) $\Ran(B) = \Dom(\Om)$.

Pairs in $D$ and $B$ may be called \emph{downward} and \emph{back} edges, 
respectively, and the relation $B$ is called the \emph{back edge} relation.
If $u \in B(v)$ we call $u$ the \emph{companion} of $v$.
A sequence $(v_{i})_{i\leq k}$ is called a \emph{downward path} if $Dv_{i}
v_{i+1}$ for all $i < k$.

A dag-based parity formula is called \emph{untwisted} if it satisfies the 
following condition:

6) if $Bvu$ then all downward paths from $v_{I}$ to $v$ pass through $u$.
\end{definition}

\begin{example}
The parity formula on the left in Figure~\ref{fig:x53} is in fact untwisted,
the one on the right is not.
\end{example}

The conditions 1) -- 5) more or less speak for themselves.
Another way of formulating condition 6) is that all companion nodes are 
so-called articulation points of the dag $(V,D)$, that is: removal of any such
node results in a disconnected graph.
Note that this condition is superfluous in case the dag underlying the parity 
formula is actually a tree.

\begin{remark}
(1)
Note that if $\bbG = (V,E,L,\Om,v_{I})$ is an untwisted parity formula, the
relations $D$ and $B$ witnessing the conditions 1) - 4) and 6) of 
Definition~\ref{d:dagcyc} are unique.

To see this, suppose for contradiction that there would be two distinct pairs 
of relations, $(D_{0},B_{0})$ and $(D_{1},B_{1})$, both witnessing the untwisted 
nature of $\bbG$.
Then there is a node $v$ such that $(D_{0}[v],B_{0}[v]) \neq (D_{1}[v]
,B_{1}[v])$.
Without loss of generality we may assume the existence of a downward path
$v_{0}Dv_{1}D \cdots v_{n}$ such that $v_{0} = v_{I}$, $v_{n} = v$, and 
$(D_{0}[v_i],B_{0}[v_i]) = (D_{1}[v_i],B_{1}[v_i])$ for all $i<n$.
(These identities also mean that we may use the term `downward path' without 
ambiguity, i.e. $v_{0}\cdots v_{n}$ is both a $D_{0}$ and a $D_{1}$-path).

Now it is not hard to prove that in fact $D_{0}[v] \sse D_{1}[v]$.
To see this, assume for contradiction that $u \in D_{0}[v] \setminus D_{1}[v]$.
Using the fact that $E = D_{i} \uplus B_{i}$ for $i = 0,1$, we easily find that 
$u$ is the (unique) $B_{1}$-successor of $u$, so that by untwistedness, $u$ must
be one of the vertices $v_{0},\ldots,v_{n-1}$, say, $u = v_{k}$.
But then the path $u = v_{k}v_{k+1}\cdots v_{n} u$ is a $D_{0}$-cycle,
contradicting condition (1).

Similarly, we can prove that $D_{1}[v] \sse D_{0}[v]$, $B_{0}[v] \sse B_{1}[v]$
and $B_{1}[v] \sse B_{0}[v]$.
But this contradicts the assumption on $v$.

(2) The priority function $\Om$ of an untwisted parity formula is basically
determined by the relations $D$ and $B$ (and thus, by its graph structure), 
together with the parity of the priority map.
More precisely, we can prove that if $\bbG = (V,E,L,\Om,v_{I})$ and $\bbG' = 
(V,E,L,\Om',v_{I})$ are untwisted parity formulas such that, for every vertex
$v$,  $\Om(v)$ and $\Om'(v)$ have the same parity, then $\bbG$ and $\bbG'$ are 
priority variants, cf.~Definition~\ref{d:parvar}.
\end{remark}

The following proposition makes a useful observation on the paths going through
an untwisted formula.

\begin{proposition}
\label{p:dagcyc}
Let $\bbG = (V,E,L,\Om,v_{I})$ be an untwisted parity formula, based on the dag 
$(V,D)$ with back edges $B$.

1) Let $\rho = (v_{i})_{i\leq k}$ be a downward path through $\bbG$ and let 
  $(v_{k},u)$ be a back edge.
  Then either $D^{*}uv_{0}$ or $u = v_{i}$ for some $i \in \rng{0}{k}$.

2) Let $\rho = (v_{i})_{i\leq k}$ be any path through $\bbG$. 
  Then there is an $n \in \rng{0}{k}$ such that $D^{*}v_{n}v_{i}$ for 
  all $i \in \rng{0}{k}$.
\end{proposition}

\begin{proof}
1) Consider any downward path $\si$ from $v_{I}$ to $v_{0}$, and compose this
path with $\rho$.
By untwistedness, this combined path must pass through $u$, so $u$ either lies
on $\si$, implying $D^{*}uv_{0}$, or on $\rho$, meaning that $u = v_{i}$ for 
some $i \in \rng{0}{k}$.

2) This statement is proved by a straightforward induction.
Omitting the trivial base case, consider a path $\rho' = \rho\cdot v_{k+1}$, and 
inductively assume that $n = n_{\rho} \in \rng{0}{k}$ is such that $D^{*}v_{n}v_{i}$,
for all $i \in \rng{0}{k}$.

Now distinguish cases. 
If $Dv_{k}v_{k+1}$ then by $D^{*}v_{n}v_{k}$ we have $D^{+}v_{n}v_{k+1}$, and 
so we may take $n_{\rho'} \isdef n$.
If, on the other hand, $(v_{k},v_{k+1})$ is a back edge, then by part 1) we have
either $D^{*}v_{n}v_{k+1}$, in which case we take $n_{\rho'} \isdef n$,
or else $D^{*}v_{k+1}v_{n}$, in which case we take $n_{\rho'} \isdef k+1$.
\end{proof}

\subsection*{Formulas, automata and equation systems}

As mentioned earlier, our parity formulas are in fact tightly related to the 
alternating tree automata introduced by Wilke~\cite{wilk:alte01}, and also to
hierarchical equation systems (see for instance~\cite{arno:rudi01,demr:temp16},
and references therein).
We now provide some details.

An \emph{alternating tree automaton} or \textsc{ata} over a set $\Prop$ of 
proposition letters is a quadruple $\bbA = (A,\Th,\Om,a_{I})$,
where $A$ is a finite set of objects called \emph{states}, $a_{I} \in A$ is an
element of $A$ called the \emph{initial state}, $\Om: A \to \om$ is a (total)
priority function, and $\Th: A \to \TC(\Prop,A)$ is the \emph{transition map}
of the automaton.
Here $\TC(\Prop,A)$ is a logical language of usually rather simple formulas that
are called \emph{transition conditions} over ($\Prop$ and) $A$.
The exact definition of this language comes with some variations; here we 
mention the version of~\cite{wilk:alte01,grae:auto02}.
That is, we define the set $\TC(\Prop,A)$ via the following grammar:
\begin{equation}
\label{eq:TC0a}
\al \isbnf \bot \divbnf \top 
   \divbnf p \divbnf \ol{p}
   \divbnf a \divbnf \dia a \divbnf \Box a
   \divbnf a \land b \divbnf a \lor b,
\end{equation}
where $p \in \Prop$ and $a,b \in A$.

The (operational) semantics of alternating tree automata is formulated in terms
of an acceptance game $\AG(\bbA,\bbS)$, for an \textsc{ata} $\bbA$ and a 
Kripke model $\bbS = (S,R,V)$.
The positions of this game are given as the set 
\[
\{ (\al,s) \in A \times S \mid 
   a \in A \text{ or } \al = \Th(a), \text{ for some } a \in A \},
\]
while its rules and winning conditions are listed in Table~\ref{tb:3}.

\begin{table}[htb]
\begin{center}
\begin{tabular}{|ll|c|l|}
\hline
\multicolumn{2}{|l|}{Position} & Player & Admissible moves\\
\hline
   $(a,s)$        & with $\Th(a) = \bot$                
   & $\eloi$ & $\nada$ 
\\ $(a,s)$        & with $\Th(a) = \top$                
   & $\abel$ & $\nada$ 
\\ $(a,s)$        & with $\Th(a) = p$ and $s \in V(p)$                
   & $\abel$ & $\nada$ 
\\ $(a,s)$        & with $\Th(a) = p$ and $s \not\in V(p)$                
   & $\eloi$ & $\nada$ 
\\ $(a,s)$        & with $\Th(a) = \lneg p$ and $s \in V(p)$                
   & $\eloi$ & $\nada$ 
\\ $(a,s)$        & with $\Th(a) = \lneg p$ and $s \not\in V(p)$                
   & $\abel$ & $\nada$ 
\\ $(a,s)$        & with $\Th(a) = b$   
   & -       & $\{ (b,s) \}$ 
\\  $(a,s)$        & with $\Th(a) = (b_{0} \lor b_{1})$ 
   & $\eloi$ & $\{ (b_{0},s), (b_{1},s) \}$ 
\\  $(a,s)$        & with $\Th(a) = (b_{0} \land b_{1})$ 
   & $\abel$ & $\{ (b_{0},s), (b_{1},s) \}$ 
\\  $(a,s)$        & with $\Th(a) = \dia b$ 
   & $\eloi$ & $\{ (a,t) \mid sRt \}$ 
\\  $(a,s)$        & with $\Th(a) = \Box b$ 
   & $\abel$ & $\{ (a,t) \mid sRt \}$ 
\\ \hline
\end{tabular}
\end{center}
\caption{The acceptance game $\AG(\bbA,\bbS)$}
\label{tb:3}
\end{table}

From this definition it is fairly obvious that we may identify parity formulas 
and \textsc{ata}s: with an alternating tree automaton $\bbA$ we may associate
the parity formula $\bbG_{\bbA} = (A,E,L,\Om,a_{I})$ given by
\[
E(a) \isdef 
\left\{\begin{array}{ll}
  \nada     
\\ \{ b \} 
\\ \{ b \}  
\\ \{ a_{0},a_{1}\} 
\end{array}\right.
\quad\text{ and } \quad
L(a) \isdef 
\left\{\begin{array}{lll}
  \Th(a) \hspace*{10mm}    & \text{if } \Th(a) \in \At(\Prop)
\\ \epsilon & \text{if } \Th(a) = b, \text{ for some } b \in A
\\ \hs      & \text{if } \Th(a) = \hs b, \text{ for some } b \in A
\\ \odot    & \text{if } \Th(a) = a_{0}\odot a_{1}, \text{ for some } a_{0},a_{1} \in A
\end{array}\right.
\]
Conversely, it is equally simple to turn a parity formula into an alternating 
tree automaton which is based on the set of vertices of the formula.
Note that if we simply take the size of an automaton to be its number of states, 
the constructions that we just outlined respect size in either direction.
\medskip

Another perspective on alternating tree automata is that of \emph{hierarchical 
equation systems} or \textsc{hes}s.
We need not go into detail here, but to get the basic idea, fix an \textsc{ata}
$\bbA = (A,\Th,\Om,a_{I})$.
Think of the states in $A$ as \emph{variables}, then with each state $a \in A$ 
we may associate an \emph{equation} of the form $a = \Th(a)$, and make this 
equation inherit the priority $\Om(a)$.
Then we may organise this collection of equations by grouping equations of the 
same priority together in so-called \emph{blocks}, and order these blocks 
according to their priorities.
The result of this operation is known as a hierarchical equation system.
It should be clear from this description that there is a 1-1 correspondence
between alternating tree automata and these hierarchical equations systems.
\medskip

Finally, the definition \textsc{hes}'s and \textsc{ata}s can be 
modified/generalised in two directions.
First of all, one may vary the set of admissible transition conditions.
For instance, Demri, Goranko \& Lange~\cite{demr:temp16} work with the concept 
of a so-called \emph{modal equation system} or MES.
An MES corresponds to an alternating tree automaton $\bbA = (A,\Th,\Om,a_{I})$
where the transition map $\Th$ maps every state $a \in A$ to an arbitrary 
modal formula as given by the following grammar
\begin{equation}
\label{eq:TC0b}
\al,\be \isbnf \bot \divbnf \top 
   \divbnf p \divbnf \ol{p}
   \divbnf a \divbnf \dia \al \divbnf \Box \al
   \divbnf \al \land \be \divbnf \al \lor \be,
\end{equation}
where $p \in \Prop$ and $a\in A$.
The acceptance game associated with these automata are obtained via a minor 
modification of the one given in Table~\ref{tb:3}.
Note, however, that in this case one cannot simply take the size of the
\textsc{ata} to be its number of states; one has to take the size of the
formulas in the range of the transition map into account as well.

Second, one may change the nature of the transition map so that it shifts the 
propositions from its output (transition conditions) to the input.
That is, given a finite set $\Prop$ of proposition letters, think of its power
set $\funP \Prop$ as an alphabet of \emph{colours}.
The transition map of an \textsc{ata} can then be given as a map $\Th: A \times
\funP\Prop \to \TC(A)$ mapping pairs of states and colours to transition 
conditions that now may only involve the states of the automata as propositional 
variables.
More details about these devices, which look a bit more like classical automata,
can be found in Venema~\cite{vene:lect18}.

\section{From regular formulas to parity formulas}
\label{s:fixpar}

Now that we have a yardstick in place in the form of the parity formulas 
introduced in the previous section, it is time to start using these for
measuring regular $\mu$-calculus formulas.
Given a formula $\xi \in \muML$, the challenge is to find a graph structure
on which to construct an equivalent parity formula; additionally, we want the
index of this parity formula to be as close as possible to the alternation depth
of the input formula.
Roughly, there are three candidates for such a graph: next to the syntax tree,
these are the \emph{subformula dag} and the \emph{closure graph} of the formula
$\xi$.
These structures correspond, respectively, to taking the length, the 
subformula-size, and the closure-size of $\xi$ as it basic size measure.
Since we will not be interested much in working with length as a size measure,
this means that in this section we will focus on the latter two graph 
structures.

\begin{definition}
The \emph{subformula dag} of a clean formula $\xi$ is the pointed graph 
$\bbD_{\xi} \isdef (\Sfor(\xi),\rhd_{0},\xi)$, where $\rhd_{0}$ is the converse 
of the direct subformula relation $\psfor_{0}$.
The \emph{closure graph} of a tidy formula $\xi \in \muML$ is the structure
$\bbC_{\xi} = (\Clos(\xi), E^{C}_{\xi},\xi)$, where for a given formula $\phi 
\in \Clos(\xi)$, the relation $E^{C}_{\xi}$ is the trace relation $\cla$, 
restricted to the set $\Clos(\xi)$.

Both structures can be expanded with a natural labelling, which is the
appropriate restriction of the labelling $L_{n}: \muML(\Prop) \to \At(\Prop)
\cup \{ \land, \lor,\dia,\Box, \epsilon\}$ which we define as follows:
\[
L_{n}(\phi) \isdef
\left\{\begin{array}{lll}
   \phi     & \text{if } \phi \text{ is atomic}
\\ \hs      & \text{if } \phi = \hs\psi
\\ \odot    & \text{if } \phi = \psi_{0}\odot\psi_{1}
\\ \epsilon & \text{if } \phi = \eta x.\psi
\end{array}\right.
\]
With $\Om: \Clos(\xi) \to \om$ a priority map, we let $(\bbC_{\xi},\Om)$ denote
the parity formula $(\Clos(\xi), E^{C}_{\xi}, L_{C},\Om,\xi)$, and we will
use similar notation for parity formulas based on the subformula dag of $\xi$.
\end{definition}

Since the only labellings that we will consider here will be (restrictions of)
the natural labelling $L_{n}$, we will permit ourselves to be sloppy regarding 
the distinction between the labelled and unlabelled versions of the subformula
dag and the closure graph.
\medskip

The following theorem, which is essentially a reformulation of
Wilke~\cite[Theorem~1]{wilk:alte01}, shows that for a clean formula, we can
obtain an equivalent parity formula which is based on its \emph{subformula dag}.
Our (minor) contribution here is the observation that the associated parity
formula is \emph{untwisted}.

\begin{fewtheorem}
\label{t:1}
There is an algorithm that constructs, for a clean formula $\xi \in \muML(\Prop)$,
an equivalent untwisted parity formula $\bbH_{\xi}$ over $\Prop$ which is based
on the subformula dag of $\xi$, and such that $\size{\bbH_{\xi}} = \ssz{\xi}$;
in addition we have $\idx(\bbH_{\xi}) = \ad(\xi)$.
\end{fewtheorem}

\begin{proof}
We will obtain $\bbH_{\xi} = (V_{\xi},E_{\xi},L_{\xi},\Om_{\xi},v_{\xi})$ by
adding back edges to the subformula dag $\bbD_{\xi}$ of $\xi$.
More specifically, we set $V_{\xi} \isdef \Sfor(\xi)$, $v_{\xi} \isdef \xi$
and $E_{\xi} \isdef D_{\xi} \cup B_{\xi}$, where $D_{\xi} \isdef {\rhd_{0}}$
(that is, the converse of the direct
subformula relation $\psfor_{0}$), and $B_{\xi} \isdef \{ (x, \de_{x}) \mid
x \in \BV{\xi}\}$.
Furthermore, the labelling map $L_{\xi}$ is defined via the following case
distinction:
\[
L_{\xi}(\phi) \isdef
\left\{ \begin{array}{lll}
   \odot    & \text{ if } \phi \text{ is of the form } \phi_{0} \odot \phi_{1}
                \text{ with } \odot \in \{ \land,\lor \}
\\ \hs      & \text{ if } \phi \text{ is of the form } \hs\psi
                \text{ with } \hs \in \{ \dia, \Box \}
\\ \epsilon & \text{ if } \phi \text{ is of the form } \eta_{x} x. \de_{x}
                \text{ with } \eta \in \{ \mu, \nu \}
\\ \phi     & \text{ if } \phi \in \{ \top, \bot \} \cup
                 \{ \pm p \mid p \in \FV{\xi} \}
\\ \epsilon & \text{ if } \phi \in \BV{\xi}
\end{array}\right.
\]
Finally, the priority map $\Om_{\xi}$ will only be defined on nodes of the form
$\de_{x}$ (for some bound variable $x$ of $\xi$), that is, $\Dom(\xi) \isdef
\{ \de_{x} \mid x \in \BV{\xi} \}$.
For the definition of the value of $\Om_{\xi}(\de_{x})$, where $x \in \BV{\xi}$,
let $d$ be the maximal length of an alternating fixpoint chain starting at
$x$; then we set
\[
\Om_{\xi}(\de_{x}) \isdef
   \left\{ \begin{array}{ll}
      d   & \text{ if $d$ has parity $\eta_{x}$}
   \\ d+1 & \text{ otherwise},
   \end{array}\right.
\]
where we recall that $\mu$ and $\nu$ have, respectively, odd and even parity.

It is easy to see that the parity formula $\bbH_{\xi}$ satisfies the conditions
1) to 5) of Definition~\ref{d:dagcyc} --- we leave it for the reader to check
the details.
To show that $\bbH_{\xi}$ also satisfies the untwistedness condition 6), observe
that any back edge of $\bbH_{\xi}$ must be of the form $(x,\de_{x})$ with $x \in
\BV{\xi}$.
By definition of the relation $D_{\xi}$ it therefore suffices to prove that any
subformula $\phi \sforeq \xi$ satisfies the condition (*) that $x \sforeq \phi$
implies that either $\phi \sforeq \de_{x}$ or $\eta_{x} x.\de_{x} \sforeq \phi$.

But since $\xi$ is \emph{clean}, we can in fact show, by a straightforward
induction on the complexity of $\phi\sforeq \xi$, that $x \sforeq \phi$ implies
that either $x \in \FV{\phi}$ and $\phi \sforeq \de_{x}$, or else
$x \in \BV{\phi}$ and $\eta_{x} x.\de_{x} \sforeq \phi$.
For the key observation in the proof of this statement, consider the inductive
case where $\phi = \eta y. \delta$, and distinguish cases.
If $x = y$ then $\phi = \eta_{x} x. \de_{x}$; in this case we find $x \in
\BV{\phi}$ and $\eta_{x}x. \de_{x} = \phi \sforeq \phi$.
If, on the other hand, $x$ and $y$ are distinct variables, it follows from
$x \sforeq \phi$ that $x \sforeq \de_{y}$.
Then by the induction hypothesis we have either (i) $x \in \FV{\de_{y}}$ and
$\de_{y} \leq \de_{x}$, implying $x \in \FV{\phi}$ and $\phi \sforeq \de_{x}$,
or (ii) $x \in \BV{\de_{y}} \sse \BV{\phi}$ and $\eta_{x}x. \de_{x} \sforeq
\eta y. \de_{y} \sforeq \phi$.
In both cases we are done.

The proof of the equivalence of $\xi$ and $\bbH_{\xi}$ is essentially the
same as in~\cite{wilk:alte01}, so we omit the details.
It is immediate from the definitions that $\size{\bbH_{\xi}} =
\size{\Sfor(\xi)} = \ssz{\xi}$.
Finally, to show that $\idx(\bbH_{\xi}) = \ad(\xi)$ we use
Proposition~\ref{p:adcf} together with the observation that, for any two bound
variables $x$ and $y$ of $\xi$, we have that $x \leq_{\xi} y$ iff $\de_{x}$ and
$\de_{y}$ belong to the same cluster of $\bbH_{\xi}$ and satisfy
$\Om_{\xi}(\de_{x}) \leq \Om_{\xi}(\de_{y})$, with $<$ holding if $x$ and $y$
are of different type (i.e., $\eta_{x} \neq \eta_{y}$).
\end{proof}

The next theorem states that for an arbitrary tidy formula, we can find an
equivalent parity formula that is based on the formula's \emph{closure graph},
and has an index which is bounded by the alternation depth of the formula.
This is in fact the main result that bridges the gap, mentioned in the
introduction, between the world of formulas and that of automata and parity 
games.
For instance, as a corollary of Theorem~\ref{t:clur} and the quasi-polynomial
complexity results on the problem of solving parity games, we obtain a
quasi-polynomial upper bound on the complexity of the model checking problem 
for the modal $\mu$-calculus.

\begin{fewtheorem}
\label{t:clur}
There is a construction transforming an arbitrary tidy formula $\xi \in \muML$
into an equivalent parity formula $\bbG_{\xi}$ which is based on the closure
graph of $\xi$, so that $\size{\bbG} = \csz{\xi}$; in addition we have
$\idx(\bbG_{\xi}) \leq \ad(\xi)$.
\end{fewtheorem}

In the absence of a cleanness assumption on the input formula, and under the
constraint of an exact match of index with alternation depth, the proof of this 
correspondence result turned out to be quite hard, as we will see now.
In fact, we believe that ours is the \emph{first} proof of this result, in the 
sense that the construction below (1) makes an exact match of alternation 
depth and index, while at the same time (2) we do \emph{not} assume that our
input formula is clean.
When it comes to complexity results on the model checking of the modal
$\mu$-calculus, approaches that involve games or automata usually involve an,
often rather implicit, pre-processing step where the input formula is replaced
with an alphabetical variant that is `well-named', i.e., clean. 
As we will see further on in Proposition~\ref{p:closexp}, such a pre-processing 
step generally incurs an unnecessary exponential blow-up of the formula's 
closure-size.

Note that the priority map $\Om_{C}$ that we will define on the closure graph 
of a tidy formula is in fact \emph{global} in the sense that it can be defined
uniformly for all (tidy) formulas, independently of any ambient formula.
For the definition of this uniform priority map, we need some auxiliary notation 
and definitions.
Recall that we write $\phi \cla \psi$ if there is a trace arrow from $\phi$
to $\psi$, and $\phi \clat \psi$ if $\psi\in \Clos(\phi)$, or equivalently, if
there is a trace (possibly of length zero) from $\phi$ to $\psi$.

\begin{definition}
\label{d:clstuff}
We let $\closeq$ denote the equivalence relation generated by the relation
$\cla$, in the sense that: $\phi \closeq \psi$ if $\phi \clat \psi$ and $\psi
\clat \phi$.
We will refer to the equivalence classes of $\closeq$ as \emph{(closure) 
clusters}, and denote the cluster of a formula $\phi$ as $\Cluster(\phi)$.

We define the closure priority relation $\clpreq$ on fixpoint formulas by putting
$\phi \clpreq \psi$ precisely
if $\psi \clat^{\psi} \phi$, where $\clat^{\psi}$ is the relation given by
$\rho \clat^{\psi} \si$ if there is a trace $\rho = \chi_{0} \cla
\chi_{1} \cla \cdots \cla \chi_{n} = \si$ such that $\psi \fsforeq
\chi_{i}$, for every $i \in \rng{0}{n}$. 
We write $\phi \clpr \psi$ if $\phi \clpreq \psi$ and 
$\psi \not\clpreq \phi$.
\end{definition}

As we will see further on, intuitively $\phi \clpr \psi$ means that $\psi$ is
more significant than $\phi$.
We first make some basic observations on the relations $\closeq$ and $\clpreq$
in Proposition~\ref{p:2001}; to avoid confusion we already mention here that
$\closeq$ is not necessarily the equivalence relation induced by $\clpr$:
For starters, $\clpr$ is only defined on fixpoint formulas.
For our observations we need the following proposition.

\begin{proposition} 
\label{p:fsfordirected}
For any trace $\rho_n \cla \dots \cla \rho_1$ of tidy formulas there exists 
a unique $\rho \in \{\rho_{1},\dots,\rho_{n}\}$ such that $\rho_n \clat^{\rho}
\rho_1$. 
Moreover, if $\rho_1$ is a fixpoint formula then so is $\rho$.
\end{proposition}

\begin{proof}
We prove the proposition by induction over $n$, and note that we only need to
worry about existence: if there would be a $\rho$ and a $\rho'$ meeting the 
constraints, we would find $\rho \fsforeq \rho'$ and $\rho' \fsforeq \rho$, 
implying $\rho = \rho'$.

The base case, where $\rho_n = \rho_1$, is trivial.
For the induction step consider a trace $\rho_{n + 1} \cla \rho_n \cla
\dots \cla \rho_1$ and assume that the induction hypothesis holds for
$\rho_n \cla \dots \cla \rho_1$. 
Thus there is a $\rho_{i}$ among $\rho_1,\dots,\rho_n$ such that 
$\rho_n \clat^{\rho_{i}} \rho_1$. 
We want to find a $j$ such that $\rho_{n+1} \clat^{\rho_j} \rho_j \clat^{\rho_j} 
\rho_1$.

Because $\rho_{n + 1} \cla \rho_n$ we can use Proposition~\ref{p:always
comparable} to deduce that for every free subformula $\psi'$ of $\rho_n$
either $\psi' \fsforeq \rho_{n + 1}$ or $\rho_{n + 1} \fsforeq \psi'$.
We have $\rho_i \fsforeq \rho_n$ since $\rho_n \clat^{\rho_i} \rho_i$.
Hence, we get either $\rho_i \fsforeq \rho_{n + 1}$, in which case we
can set $j \isdef i$, or we get $\rho_{n + 1} \fsforeq \rho_i$, in
which case we can set $j \isdef n + 1$, because then $\rho_n
\clat^{\rho_i} \rho_i \clat^{\rho_i} \rho_1$ implies $\rho_n
\clat^{\rho_j} \rho_i \clat^{\rho_j} \rho_1$.

The `moreover'-part is trivial in the base case. For the inductive step
observe that we only reassign the $\rho_j$ to $\rho_{n + 1}$ in the second
case of the case distinction. But then Proposition~\ref{p:always comparable}
gives us that $\rho_{n + 1}$ must be a fixpoint formula.
\end{proof}

\begin{proposition}
\label{p:2001}
\begin{enumerate}[topsep=0pt,itemsep=-1ex,partopsep=1ex,parsep=1ex,%
    label={\arabic*)}]

\item \label{i:porder}
The closure order $\clpreq$ is a partial order; hence
$\phi \clpr \psi$ iff $\phi \clpreq \psi$ and $\phi\neq \psi$.

\item  \label{i:clpreqincluster}
The closure order is included in the closure equivalence
relation: $\phi \clpreq \psi$ implies $\phi \closeq \psi$.

\item  \label{i:clpreqincvfsf}
The closure order is included in the converse free subformula relation: 
$\phi \clpreq \psi$ implies $\psi \fsforeq \phi$.

\item 
Every cell of $\closeq$ contains a unique fixpoint formula $\xi = \eta x. 
\chi$ such that $\xi \not \in \Clos(\chi)$. 
This formula is the $\clpreq$-maximum element of its cluster.
\label{i:max}
\end{enumerate}
\end{proposition}

\begin{proof}
For item~\ref{i:porder} we need to show that $\clpreq$ is reflexive,
transitive and antisymmetric. 
Reflexivity is obvious, and antisymmetry follows from 
\ref{i:clpreqincvfsf}.
For transitivity assume that $\phi \clpreq
\psi$ and $\psi \clpreq \chi$ hold. By definition this means that $\psi
\clat^\psi \phi$ and $\chi \clat^\chi \psi$. The latter entails that $\chi
\fsforeq \psi$ and the former means that there is some $\cla$-trace from
$\psi$ to $\phi$ such that $\psi$ is a free subformula of every formula
along this trace. Because $\chi \fsforeq \psi$ and $\fsforeq$ is
transitive it then also holds that $\chi$ is a free subformula of every
formula on the trace from $\psi$ to $\phi$. 
Composing this trace with the one from $\chi$ to $\psi$ we obtain a trace 
from $\chi$ to $\phi$ such that $\chi$ is a free subformula of all formulas
along this trace. Hence $\chi \clat^\chi \phi$ and so $\phi \clpreq \chi$.

For item~\ref{i:clpreqincluster} we assume that $\phi \clpreq \psi$ and
need to show that $\phi \clat \psi$ and $\psi \clat \phi$. The
assumption $\phi \clpreq \psi$ means that $\psi \clat^\psi \phi$ which
clearly entails $\psi \clat \phi$. But, as already observed above, $\psi
\clat^\psi \phi$ also entails that $\psi \fsforeq \phi$, from which
$\phi \clat \psi$ follows by Proposition~\ref{p:11-3}.

Item \ref{i:clpreqincvfsf} is immediate by the definition of $\clpreq$.

Finally, consider item~\ref{i:max}. We first argue that every cluster of
$\cla$ contains a fixpoint formula $\xi = \eta x . \chi$ such that $\xi
\not \in \Clos(\chi)$. 
To this aim we claim that if $\xi = \eta x.\chi$ and $\xi \in \Clos(\chi)$ 
then there is some fixpoint formula $\psi \in \Cluster(\xi)$ with $\xi \clpr
\psi$. 
If a cluster would only contain fixpoint formulas $\xi = \eta x . \chi$ with 
$\xi \in \Clos(\chi)$ then this would allow us to construct an infinite 
$\clpr$-chain in the cluster, which is impossible as all clusters are finite.

Thus, consider some $\xi = \eta x . \chi$ with $\xi \in \Clos(\chi)$.
Then we have a trace $\chi \clat \xi$ to which we can apply
Proposition~\ref{p:fsfordirected} to obtain a fixpoint formula $\psi$
with $\chi \clat^\psi \psi$ and $\psi \clat^\psi \xi$. From the latter
we have that $\xi \clpreq \psi$ and from the former we get $\psi
\fsforeq \chi$ which entails $\psi \neq \xi$ because $\xi = \eta x .
\chi$.

Uniqueness of $\xi$ is immediate by the following claim:
\begin{equation}
\label{eq:2002}
\text{if $\xi = \eta x . \chi$ and $\xi \notin \Clos(\chi)$ then $\xi
\fsforeq \rho$, for all $\rho \in \Cluster(\xi)$}.
\end{equation}
For the proof of \eqref{eq:2002}, take an arbitrary $\rho \in \Clos(\xi)$.
It follows by Proposition~\ref{p:clos3}~5) that either $\rho = \xi$, in which 
case we are done, or $\rho = \gamma[\xi/x]$ for some $\gamma \in \Clos(\chi)$.
Now $x \in \FV{\gamma}$ because otherwise $\chi \clat \gamma =
\gamma[\xi/x] = \rho \clat \xi$ contradicting $\xi \notin \Clos(\chi)$.
But if $\rho = \gamma[\xi/x]$ and $x \in \FV{\gamma}$ then by definition
we have $\xi \fsforeq \rho$.
This finishes the proof of \eqref{eq:2002}.

For the $\clpreq$-maximality of $\xi$ consider an arbitrary fixpoint formula 
$\psi \in \Cluster(\xi)$.
Then we have $\xi \clat \psi$, and since every formula $\rho$ on this trace 
$\xi \clat \psi$ is in $\Cluster(\xi)$ it follows from \eqref{eq:2002} that 
$\xi \fsforeq \rho$ and thus $\xi \clat^\xi \psi$. 
By definition this means $\psi \clpreq \xi$.
\end{proof}

We are now ready to define the priority map of the parity formula corresponding
to a fixpoint formula.
As mentioned, we can in fact define a global map $\gOm$, which uniformly assigns
a priority to any tidy formula.
In fact, this map will be (clusterwise) induced by the closure order $\clpr$,
as in Definition~\ref{d:Omppo}.

\begin{definition}
\label{d:gOm}
An \emph{alternating $\clpr$-chain} of length $n$ is a sequence 
$(\eta_{i} x_{i}. \chi_{i})_{i \in \rng{1}{n}}$ of tidy fixpoint formulas
such that $\eta_{i}x_{i}. \chi_{i} \clpr \eta_{i+1}x_{i+1}.\chi_{i+1}$ and 
$\eta_{i+1} = \fopp{\eta_i}$ for all $i \in \rng{0}{n-1}$.
We say that such a chain \emph{starts at} $\eta_{1}x_{1}.\chi_{1}$ and 
\emph{leads up to} $\eta_{n}x_{n}.\chi_{n}$.

Given a tidy fixpoint formula $\xi$, we let $\adup(\xi)$ and $\cdh(\xi)$ denote
the maximal length of any alternating $\clpr$-chain starting at, respectively 
leading up to, $\xi$. 
Given a closure cluster $C$, we let $\cd(C)$ denote the \emph{closure depth} of
$C$, i.e., the maximal length of any alternating $\clpr$-chain in $C$.

The \emph{global priority function} $\gOm: \muML^{t} \to \omega$ is defined
cluster-wise, as follows.
Take an arbitrary tidy fixpoint formula $\psi = \eta y. \phi$, and define
\begin{equation}
\label{eq:gom}
\gOm(\psi) \isdef \left\{\begin{array}{ll}
   \cd(C(\psi)) - \adup(\psi)    &
       \text{if $\cd(C(\psi)) - \adup(\psi)$ has parity $\eta$}
\\    \cd(C(\psi)) - \adup(\psi) + 1    &
       \text{if $\cd(C(\psi)) - \adup(\psi)$ has parity $\fopp{\eta}$}. 
\end{array} \right.
\end{equation}
Here we recall that we associate $\mu$ and $\nu$ with odd and even parity, 
respectively.

If $\psi$ is not of the form $\eta y. \phi$, we leave $\gOm(\psi)$ undefined. 

Finally, given a tidy formula $\xi$, we define $\bbG_{\xi} \isdef (\bbC_{\xi},
\gOm\rst{\Clos(\xi)})$.
\end{definition}

The next Proposition gathers some facts about $\gOm$, all of which are immediate
consequences of Proposition~\ref{p:Omppo}.
Recall that the \emph{index} of a cluster in a parity formula is defined as the 
maximal length of an alternating chain in $C$, where alternation is expressed in 
terms of the priority map.
With our definition of the global priority map $\gOm$, the index of any cluster
corresponds to the size of the range of $\gOm$, restricted to the cluster.

\begin{proposition}
\label{p:gOm1}

\begin{enumerate}[topsep=0pt,itemsep=-1ex,partopsep=1ex,parsep=1ex,%
    label={\arabic*)}]

\item  \label{i:gOm1-1}
Let $\xi = \eta x. \chi$ be a tidy fixpoint formula.
Then $\gOm(\xi)$ has parity $\eta$.

\item \label{i:gOm1-2}
Let $\phi$ and $\psi$ be tidy fixpoint formulas such that $\phi \clpreq \psi$.
Then $\gOm(\phi) \leq \gOm(\psi)$, and $\gOm (\phi) < \gOm(\psi)$ if $\phi$ and 
$\psi$ have different parity.

\item \label{i:gOm1-3}
For any closure cluster $C$ it holds that $\cd(C) = \idx(C) = 
\size{\Ran(\gOm\rst{C})}$.
\end{enumerate}
\end{proposition}

The following proposition shows that the global priority map indeed captures 
the right winner of infinite matches of the evaluation game.

\begin{proposition}
\label{p:inftr}
Let $\tau = (\xi_{n})_{n\in\om}$ be an infinite trace of tidy formulas.
Then 

\begin{enumerate}[topsep=0pt,itemsep=-1ex,partopsep=1ex,parsep=1ex,%
    label={\arabic*)}]

\item \label{i:inftr-1}
there is a unique fixpoint formula $\xi = \eta x. \chi$ which occurs
infinitely often on this trace and satisfies $\xi_{n} \clpreq \xi$ for 
cofinitely many $n$.

\item \label{i:inftr-2}
$\max\Big(\{ \Om(\phi) \mid \phi \text{ occurs infinitely often on } \tau \} 
\Big)$ is even iff $\eta = \nu$.

\end{enumerate}
\end{proposition}

\begin{proof}
Part~\ref{i:inftr-1} is more or less immediate by 
Proposition~\ref{p:fsfordirected} and the definitions.
From this it follows by Proposition~\ref{p:Omppo} that $\gOm(\phi) \leq 
\gOm(\xi)$ for all $\phi$ that occur infinitely often on $\tau$.
Finally, that $\gOm(\xi)$ has the right parity was stated in 
Proposition~\ref{p:gOm1}.
This proves Part~\ref{i:inftr-2}.
\end{proof}

\begin{remark}
The definition of the priority map $\gOm$ and of the priority order $\clpr$ on
which it is based, may look overly complicated.
In fact, simpler definitions would suffice if we are only after the equivalence
of a tidy formula with an associated parity formula that is based on its closure 
graph, i.e., if we do not need an exact match of index and alternation depth.

In particular, we could have introduced an alternative priority order $\clpr'$ 
by putting $\phi \clpr' \psi$ if $\phi \closeq \psi$ and $\psi \fsfor \phi$.
If we would base a priority map $\gOm'$ on this priority order instead of on 
$\clpr$, then we could prove the equivalence of any tidy formula $\xi$ with the 
associated parity formula $\bbG'_{\xi} \isdef (\bbC_{\xi},
\gOm'\rst{\Clos(\xi)})$.
However, we would not be able to prove that the index of $\bbG'_{\xi}$ is
bounded by the alternation depth of $\xi$.

To see this, consider the following formula:
\[
\al_{x} \isdef 
\nu x. \big( (\mu y. x \land y) \lor \nu z. ( z \land \mu y. x \land y ) \big).
\]
We leave it for the reader to verify that this formula has alternation depth 
two, and that its closure graph looks as follows:
\medskip

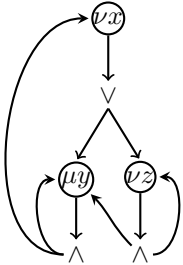
\begin{tikzpicture}
\tikzset{sibling distance=4mm,
   edge from parent/.append style={->,thick},
   every node/.style= {circle,inner sep=0mm,thick},}

\Tree [.\node (alx) [draw] {$\nu x$};
    [.$\lor$
        [.\node (aly) [draw] {$\mu y$} ; 
               [.\node (y1) {$\land$}; ] 
        ]
        [.\node (z) [draw]{$\nu z$} ;
            [.\node (z1) {$\land$};            
            ]
        ]
    ]
]  
\path[->]
    (y1) edge   [out=180,in=180]               node {}      (aly)
    (y1) edge   [out=180,in=180]               node {}      (alx)
    (z1) edge    [out=0,in=0]               node {}      (z)
    (z1) edge    [out=135,in=-45]               node {}      (aly)
;
\end{tikzpicture}
\medskip

Let $\al_{y}$ and $\al_{z}$ be the other two fixpoint formulas in the cluster
of $\al_{x}$, that is, let $\al_{y} \isdef \mu y. \al_{x} \land y$ and 
$\al_{z} \isdef \nu z. z \land \al_{y}$.
These formulas correspond to the nodes in the graph that are labelled $\mu y$
and $\nu z$, respectively.
Now observe that we have $\al_{x} \fsfor \al_{y} \fsfor \al_{z}$, so that 
this cluster has an alternating $\clpr'$-chain of length \emph{three}: 
$\al_{z} \clpr' \al_{y} \clpr' \al_{x}$.
Note however, that any trace from $\al_{y}$ to $\al_{z}$ must pass through
$\al_{x}$, the $\clpr$-maximal element of the cluster.
In particular, we do \emph{not} have $\al_{z} \clpr \al_{y}$, so that there is
\emph{no} $\clpr$-chain of length three in the cluster.

A different kind of simplification of the definition of the global priority map 
would be to define
\begin{equation}
\label{eq:gom0}
\gOm''(\psi) \isdef \left\{\begin{array}{ll}
     \cdh(\psi)     &
        \text{if $\cdh(\psi)$ has parity $\eta$}
\\ \cdh(\psi) - 1 &
        \text{if $\cdh(\psi)$ has parity $\fopp{\eta}$}. 
\end{array} \right.
\end{equation}
Using this definition for a priority map $\gOm''$, we would again obtain the
equivalence of $\xi$ and the resulting parity formula $\bbG''_{\xi} \isdef 
(\bbC_{\xi},\gOm''\rst{\Clos(\xi)})$.
In addition, we would achieve that the index of the parity formula 
$\bbG''_{\xi}$ satisfies $\ind(\bbG''_{\xi}) \leq \ad(\xi) + 1$.
However, the above formula $\al_{x}$ would be an example of a formula $\xi$ 
where $\ind(\bbG''_{\xi})$ exceeds $\ad(\xi)$:
We leave it for the reader to verify that we would get $\gOm''(\al_{z}) = 0$,
$\gOm''(\al_{y}) = 1$ and $\gOm''(\al_{x}) = 2$, implying that $\ind(\bbG''_{\xi})
= 3$.

With our definition of the priority map $\gOm$, we find the same values for 
$\al_{y}$ and $\al_{x}$ as with $\gOm''$, but we obtain $\gOm(\al_{z}) = 2$, 
implying that $\ind(\bbG_{x}) = 2 = \ad(\xi)$ as required.
\end{remark}


Our first goal will be to prove the equivalence of any formula $\xi$ to its
associated parity formula $\bbG_{\xi}$, but for this purpose we need some 
auxiliary results.
Our main lemma will be Proposition~\ref{p:clx1} below, which concerns the 
relation between the structures $\bbG_{\eta x.\chi}$ and $\bbG_{\chi}$.
In order to prove this Proposition, we need some preliminary observations
concerning the interaction of the notion of substitution with the 
operations of taking free subformulas and closure, respectively.
We first consider the free subformula relation.

\begin{proposition} 
\label{p:subst preserves sfor}
Let $\phi$, $\psi$ and $\xi$ be formulas in $\muML$ such that $x \in \FV{\phi}$,
and $\xi$ is free for $x$ in both $\phi$ and $\psi$.
Then 
\begin{enumerate}[topsep=0pt,itemsep=-1ex,partopsep=1ex,parsep=1ex,%
    label={\arabic*)}]

\item \label{it-610-1}
$\phi \fsforeq \psi$ implies $\phi[\xi/x] \fsforeq \psi[\xi/x]$;

\item \label{it-610-2}
 $\phi[\xi/x] \fsforeq \psi[\xi/x]$ implies $\phi \fsforeq \psi$, provided
that $\xi \not\fsforeq \phi, \psi$.
\end{enumerate}
\end{proposition}

\begin{proof}
For part~\ref{it-610-1}, assume that $\phi \fsforeq \psi$, then $\psi = 
\psi'[\phi/y]$ for some formula $\psi'$ such that $y \in \FV{\psi'}$ and 
$\phi$ is free for $y$ in $\psi'$.
Without loss of generality we may assume that $y$ does not occur in $\psi$.
Then by Proposition~\ref{p:commutingsubst} we obtain that $\psi[\xi/x] = 
\psi'[\xi/x][\phi[\xi/x]/y]$, while $\phi[\xi/x]$ is free for $y$ in 
$\psi'[\xi/x]$, again by Proposition~\ref{p:commutingsubst}.
This means that $\phi[\xi/x] \fsforeq \psi[\xi/x]$ indeed.
\medskip

For part~\ref{it-610-2} we first observe that for any pair of tidy formulas
$\al$ and $\be$, we have $\al \fsforeq \be$ iff for some fresh variable $y$,
$\be$ is of the form $\be = \rho[\al/y]$ for some formula $\rho$ that has a 
\emph{single} occurrence of $y$.
Let $\muML^{t}(y)$ denote the collection of such formulas, then by an obvious 
induction we may define, for any formula $\rho \in \muML^{t}(y)$, the 
\emph{depth} $d_{y}(\rho)$ of $y$ in $\rho$ .
It now clearly suffices to prove the following statement, for any formula 
$\rho \in \muML^{t}(y)$:
\begin{equation}
\label{eq:610}
\text{ if $\psi[\xi/x] = \rho[\phi[\xi/x]/y]$, then $\psi = \rho'[\phi/y]$
for some $\rho' \in \muML^{t}(y)$ with $d_{y}(\rho') = d_{y}(\rho)$.
}
\end{equation}
We will prove \eqref{eq:610} by induction on $d_{y}(\rho)$.

In the base step of this induction the only case to consider is where
$\rho = y$.
This means that $\rho[\phi[\xi/x]/y] = \phi[\xi/x]$, so that by assumption we
have $\psi[\xi/x] = \phi[\xi/x]$.
But since we also assumed that $\xi \not\fsforeq \phi, \psi$, it follows from 
Proposition~\ref{p:subst1} that $\phi = \psi$, so that we may take $\rho' \isdef
\rho$ and obtain $\rho[\phi/y] = \psi$, as required.

For the induction step we first consider the case where $\rho$ is of the form 
$\rho = \rho_{0} \land \rho_{1}$.
Without loss of generality assume that the single occurrence of $y$ in $\rho$ 
is in $\rho_{0}$.
It then follows from the assumption $\psi[\xi/x] = \rho[\phi[\xi/x]/y]$ that 
$\psi[\xi/x] = \rho_{0}[\phi[\xi/x]/y] \land \rho_{1}$.
Note that, since $y \in \FV{\rho_{0}}$ and $x \in \FV{\phi}$, $\xi$ is a proper
free subformula of the formula $\rho_{0}[\phi[\xi/x]/y] \land \rho_{1}$, and 
so we cannot have that $\psi = x$, since this would imply that $\xi$ 
is a proper subformula of itself.
This means that $\psi$ must be of the form $ \psi = \psi_{0} \land \psi_{1}$, 
with $\psi_{0}[\xi/x] = \rho_{0}[\phi[\xi/x]/y]$ and $\psi_{1}[\rho/x] =
\rho_{1}$.
From the first observation it follows by the induction hypothesis that there is 
some formula $\rho_{0}'\in \muML^{t}(y)$ with $\psi_{0} = \rho_{0}'[\phi/y]$ 
and $d_{y}(\rho_{0}') = d_{y}(\rho_{0})$.
Now define $\rho' \isdef \rho_{0}' \land \psi_{1}$, then obviously we have 
$d_{y}(\rho') = d_{y}(\rho_{0}') + 1 = d_{y}(\rho_{0}) + 1 = d_{y}(\rho)$, 
while $\rho'[\phi/y] = \rho_{0}'[\phi/y] \land \psi_{1}[\phi/y] = 
\psi_{0} \land \psi_{1} = \psi$.

We omit the inductive cases for disjunction and the modalities since these are
analogous to the case for conjunction.

This leaves the case where $\rho$ is a fixpoint formula, say, $\rho = \la z.
\si$.
It now follows from the assumption $\psi[\xi/x] = \la z. \rho_{0}[\phi[\xi/x]/y]$
that $\psi$ must be of the form $\psi = \la z. \psi_{0}$ with $\psi_{0}[\xi/x] 
= \rho_{0}[\phi[\xi/x]/y]$.
Then by the induction hypothesis we obtain $\psi_{0} = \rho_{0}'[\phi/y]$ for 
some formula $\rho_{0}'$ such that $d_{y}(\rho_{0}') = d_{y}(\rho_{0})$.
Now define $\rho' \isdef \la z. \rho_{0}$, then we find
$d_{y}(\rho') = d_{y}(\rho_{0}') + 1 = d_{y}(\rho_{0}) + 1 = d_{y}(\rho)$, 
while $\rho'[\phi/y] = \la z.\rho_{0}[\phi/y] = \la z. \psi_{0} = \psi$ as
required.

This finishes the proof of \eqref{eq:610} and hence, that of part~\ref{it-610-2}
of the proposition.
\end{proof}

\noindent
In the next proposition we consider the closure relation.

\begin{proposition} \label{p:substbm}
Let $\xi$ and $\chi$ be tidy $\mu$-calculus formulas such that
$\BV{\chi} \cap \FV{\xi} \neq \nada$
and $\chi[\xi/x]$ is tidy.
Then the substitution operation ${\xi/x} : \Clos(\chi) \to \muML$ satisfies the
following back- and forth condition, for every $\phi \in \Clos(\chi) \setminus
\{ x\}$:
\begin{equation}
\label{eq:1002}
\{\chi \mid \phi[\xi/x] \cla \chi\} = \{\psi[\xi/x] \mid \phi \cla \psi\}.
\end{equation}
\end{proposition}

\begin{proof}
We distinguish cases depending on the shape of $\phi$.

If $\phi \neq x$ is atomic then there are no $\psi$ with $\phi \cla \psi$.
Because there is also no $\cla$-successor of $\phi[\xi/x] = y [\xi/x] = \phi$
the claim holds trivially.

The cases where $\phi = \phi_0 \odot \phi_1$ with $\odot \in \{ \land, \lor
\}$, or $\phi = \hs\psi$ with $\hs \in \{ \dia, \Box \}$ are straightforward.

If $\phi = \eta y . \rho$, then the unique $\cla$-successor of $\phi$ is
the formula $\rho[\phi/y]$.
We distinguish further cases depending on whether $y = x$.
If $y=x$ then we have that $\phi[\xi/x] = \phi$, which has again $\rho[\phi/x]$
as its only $\cla$-successor.
Because $x \notin \FV{\phi}$ we also have $\rho[\phi/x][\xi/x] = \rho[\phi/x]$
and thus the claim holds trivially.

If $y \neq x$ then we find $\phi[\xi/x] = \eta y . \rho[\xi/x]$, and the unique
$\cla$-successor of this formula is its unfolding
$\big(\rho[\xi/x]\big)[\phi[\xi/x]/y]$.
For the right hand side of \eqref{eq:1002}, obviously the unique $\cla$-successor
of $\phi$ is its unfolding $\rho[\phi/y]$.
It is thus left to show that
\begin{equation}
\label{eq:1003}
\big(\rho[\xi/x]\big)[\phi[\xi/x]/y] = \big(\rho[\phi/y]\big)[\xi/x].
\end{equation}
But since $\BV{\phi} \sse \BV{\chi}$ by Proposition~\ref{p:clos5}, it follows
from $y \in \BV{\phi}$ and the assumptions that $y \not\in \FV{\xi}$.
From this and the fact that $y \neq x$, \eqref{eq:1003} follows by
Proposition~\ref{p:commutingsubst}.
\end{proof}

Proposition~\ref{p:adqc} below states the equivalence of any tidy formula $\xi$
to its associated parity formula $\bbG_{\xi}$.
The proof of the main statement in this proposition proceeds by induction on
the complexity of $\xi$, and the next proposition is the main technical 
ingredient in the key inductive step of this proof, where $\xi$ is of the 
form $\eta x.\chi$.
Roughly, Proposition~\ref{p:clx1} states that the substitution $\xi/x$ is 
`almost an isomorphism' between  $\bbG_{\chi}$ and $\bbG_{\xi}$; note, however, 
that actually, rather than $\chi$ we consider its variant $\chi' \isdef 
\chi[x'/x]$ --- this guarantees tidyness.
Recall that the alternation height $\cdh(\xi)$ of a formula $\xi$ was 
introduced in Definition~\ref{d:gOm}.

\begin{proposition}
\label{p:clx1}
Let $\xi = \eta x. \chi$ be a tidy fixpoint formula such that $x \in \FV{\chi}$
and $\xi \notin \Clos (\chi)$.
Furthermore, let $\chi' \isdef \chi[x'/x]$ for some fresh variable $x'$.
Then $\chi'$ is tidy and the following hold.
\begin{enumerate}[topsep=0pt,itemsep=-1ex,partopsep=1ex,parsep=1ex,%
    label={\arabic*)}]

\item \label{i:substbij}
the substitution $\xi/x'$ is a bijection between $\Clos(\chi')$ and $\Clos(\xi)$.
\end{enumerate}

\noindent
Let $\phi,\psi \in \Clos(\chi')$. Then we have
\begin{enumerate}[topsep=0pt,itemsep=-1ex,partopsep=1ex,parsep=1ex,%
    label={\arabic*)}]
\addtocounter{enumi}{1}

\item  \label{i:substiso}
if $\phi \neq x'$, then 
$\phi \cla \psi \text{ iff } \phi[\xi/x'] \cla \psi[\xi/x']$
and $L_{C}(\phi) = L_{C}(\phi[\xi/x'])$;

\item \label{i:samesfor}
if  $x' \in \FV{\phi}$ then 
$\phi \fsforeq \psi \text{ iff } \phi[\xi/x'] \fsforeq \psi[\xi/x']$;

\item \label{i:sameclpr}
if $\phi$ and $\psi$ are fixpoint formulas then 
$\psi \clpreq \phi$ iff $\psi[\xi/x'] \clpreq \phi[\xi/x']$;

\item \label{i:sameh}
if $\phi$ is a fixpoint formula then $\cdh(\phi) = \cdh(\phi[\xi/x'])$;

\item \label{i:samewin}
if $(\phi_{n})_{n\in\om}$ is an infinite trace through $\Clos(\chi')$, then 
$(\phi_{n})_{n\in\om}$ has the same winner as 
$(\phi_{n}[\xi/x'])_{n\in\om}$.
\end{enumerate}
\end{proposition}

\begin{proof}
Let $\xi = \eta x. \chi$ be a tidy fixpoint formula such that $x \in \FV{\chi}$
and $\xi \notin \Clos (\chi)$, and let $\chi' \isdef \chi[x'/x]$ for some fresh
variable $x'$.
We leave it for the reader to verify that $\chi'$ is tidy, and first make the
following technical observation:

\begin{equation}
\label{eq:xinincl}
\text{if } \phi \in \Clos(\chi') \text{ then }
\xi \notin \Clos(\phi) \text{ and } \xi \not\fsforeq \phi.
\end{equation}
To see this, take an arbitrary $\phi \in \Clos(\chi')$, and first assume for
contradiction that $\xi \in \Clos(\phi)$.
Combining this with the assumption that $\phi \in \Clos(\chi')$ we get that
$\xi \in \Clos(\chi')$.
By item~4) of Proposition~\ref{p:clos3} it holds that $\Clos(\chi[x'/x]) =
\{\rho[x'/x] \mid \rho \in \Clos(\chi)\} \cup \Clos(x')$.
Thus, $\xi = \rho[x'/x]$ for some $\rho \in \Clos(\chi)$ and because $x' \notin
\FV{\xi}$ it follows that $\xi = \rho \in \Clos(\chi)$.
But this contradicts the assumption that $\xi \notin\Clos(\chi)$.
In other words, we have proved that $\xi \not\in \Clos(\phi)$. 
To see that also $\xi \not\fsforeq \phi$ note that by Proposition~\ref{p:11-3}
$\xi \fsforeq \phi$ would entail $\xi \in \Clos(\phi)$.
\medskip

\noindent
We now turn to proving the respective items of the Proposition.
\medskip

\textit{Item~\ref{i:substbij}:}
We leave it for the reader to verify that the substitution $\xi/x'$ is 
well-defined, i.e., that 
\begin{equation}
\label{eq:2003}
\xi \text{ is free for } x' \text{ in every } \phi \in \Clos(\chi'),
\end{equation}
and that $\phi[\xi/x'] \in \Clos(\xi)$, for all $\phi \in \Clos(\chi')$.

For injectivity of the substitution, suppose that $\phi_{0}[\xi/x'] = 
\phi_{1}[\xi/x']$, where $\phi_{0},\phi_{1} \in \Clos(\chi')$.
It follows by \eqref{eq:xinincl} that $\xi$ is not a free subformula of either
$\phi_{0}$ or $\phi_{1}$.
But then it is immediate by Proposition~\ref{p:subst1} that $\phi_{0} =
\phi_{1}$.

For surjectivity, it suffices to show that $\xi$ belongs to the set $\Phi \isdef
\{ \phi[\xi'/x] \mid \chi' \clat \phi \}$, and that the set $\Phi$ is closed,
i.e., $\Phi \sse \Clos(\Phi)$.
But since we have $x' \in \FV{\chi'}$, we obtain $\chi' \clat x'$ by
Proposition~\ref{p:clos3}(1), and so we have $\xi = x'[\xi/x'] \in \Phi$.
The proof that $\Phi$ is closed is routine, and left as an exercise.
\medskip

\textit{Item~\ref{i:substiso}:}
This follows immediately from Proposition~\ref{p:substbm} and
item~\ref{i:substbij}.
Note that the condition of Proposition~\ref{p:substbm} (viz., that $\BV{\chi'}
\cap \FV{\xi} = \nada$) follows because $\xi$ is tidy and $\BV{\chi'} =
\BV{\chi} \subseteq \BV{\xi}$, where the latter inclusion is item~1) of
Proposition~\ref{p:clos5}.

The claim that $L_{C}(\phi) = L_{C}(\phi[\xi/x'])$ is rather trivial.
\medskip

\textit{Item~\ref{i:samesfor}}
This is Proposition~\ref{p:subst preserves sfor}. The assumption $\xi
\not \fsforeq \psi$ and $\xi \not \fsforeq \phi$ follows from
\eqref{eq:xinincl}.
\medskip

\textit{Item~\ref{i:sameclpr}:}
For the left-to-right direction assume that $\psi \clpreq \phi$. 
By definition there is some trace $\phi = \rho_0 \cla \rho_1 \cla \dots \cla
\rho_n = \psi$ such that $\phi \fsforeq \rho_i$ for all $i \in \rng{0}{n}$. 
It is clear that none of the $\rho_i$ is equal to $x$ because $x$ has no 
outgoing $\cla$-edges and $\psi \neq x$. 
Thus we can use item~\ref{i:substiso} to obtain a trace $\phi[\xi/x'] = 
\rho_0[\xi/x'] \cla \rho_1[\xi/x'] \cla \dots \cla \rho_n[\xi/x'] = 
\psi[\xi/x']$.
By Proposition \ref{p:subst preserves sfor}~
it follows
from $\phi \fsforeq \rho_i$ that $\phi[\xi/x'] \fsforeq \rho_i[\xi/x']$, for all
$i \in \rng{0}{n}$. 
That is, we have shown that $\phi[\xi/x'] \clat^{\phi[\xi/x']} \psi[\xi/x']$.

Before we turn to the opposite direction we show that, for all $\rho,\sigma
\in \Clos(\chi')$, we have
\begin{equation} \label{eq:xfree}
\mbox{ if } \rho[\xi/x'] \clat \sigma[\xi/x'] \mbox{ and } x' \in
\FV{\sigma} \mbox{ then } x' \in \FV{\rho}.
\end{equation}
This claim holds because, since $\xi$ is free for $x'$ in $\si$ by 
\eqref{eq:2003}, by 
definition of $\fsforeq$ it follows from $x' \in \FV{\sigma}$ that $\xi \fsforeq
\si[\xi/x']$, and thus we find $\si[\xi/x'] \clat \xi$ by
Proposition~\ref{p:11-3}.
If it were the case that $x' \notin \FV{\rho}$ then we would have that $\rho =
\rho[\xi/x'] \clat \si[\xi/x'] \clat \xi$, contradicting \eqref{eq:xinincl}. 

Turning to the right-to-left direction of item~\ref{i:sameclpr}, assume that
$\psi[\xi/x'] \clpreq \phi[\xi/x']$. 
This means that there is a trace $\phi[\xi/x'] = \rho'_0 \cla \dots \cla \rho'_m 
= \psi[\xi/x']$ with $\phi[\xi/x'] \fsforeq \rho'_i$ for all $i \in \rng{0}{m}$. 
By Proposition~\ref{p:2001} we have $\psi[\xi/x'] \closeq \phi[\xi/x']$. 
It follows from \eqref{eq:xfree} and $\psi[\xi/x'] \closeq \phi[\xi/x']$ that 
$x'$ is either free in both $\phi$ and $\psi$, or free in neither of the two 
formulas. 
In the second case we obtain $\phi = \phi[\xi/x']$ and $\psi = 
\psi[\xi/x']$, so that the statement of this item holds trivially.

We now focus on the case where $x' \in \FV{\phi} \cap \FV{\psi}$.
Our first claim is that $\rho'_i \neq \xi$ for all $i \in \rng{0}{m}$.
This follows from the fact that $\phi[\xi/x'] \fsforeq \rho'_{i}$, which holds 
by assumption, and the observation that $\xi$ is a proper free subformula of 
$\phi[\xi/x']$, which holds since $\phi$ is a fixpoint formula and hence, 
distinct from $x'$.
But if $\rho'_i \neq \xi$ for all $i \in \rng{0}{m}$, we may use the items 
\ref{i:substbij}~and~\ref{i:substiso} to obtain a trace $\phi = \rho_0 \cla 
\dots \cla \rho_m = \psi$ such that $\rho_i[\xi/x'] = \rho'_i$ for all $i \in
\rng{0}{m}$. 
Furthermore, by Proposition~\ref{p:clos5} it follows from $x' \in \FV{\psi}$
that $x' \in \FV{\rho_{i}}$, and so we may use item~\ref{i:samesfor} to obtain
$\phi \fsforeq \rho_i$, for all $i \in \rng{0}{m}$. 
This suffices to show that $\psi \clpreq \phi$.
\medskip

\textit{Item~\ref{i:sameh}:} 
First observe that we may focus on the case where $x' \in \FV{\phi}$:
If $x' \notin \FV{\phi}$ then $\phi[\xi/x'] = \phi$ and the claim that 
$\cdh(\phi) = \cdh(\phi[\xi/x'])$ holds trivially.

To show that $\cdh(\phi) \leq \cdh(\phi[\xi/x'])$ consider an alternating chain
$\psi_1 \clpr \psi_1 \clpr \dots \clpr \psi_n$ with $\psi_n = \phi$ of maximal
length $n = \cdh(\phi)$. 
Because all the $\psi_i$ for $i \in \rng{1}{n}$ are fixpoint formulas they must 
all be different from $x'$. 
Moreover, they are all in $\Clos(\chi')$ because $\clpr$ only relates formulas
that are in the same $\cla$-cluster and $\phi \in \Clos(\chi')$. 
Hence, we can apply item~\ref{i:sameclpr} to obtain the chain $\psi_1[\xi/x'] 
\clpr \psi_1[\xi/x'] \clpr \dots \clpr \psi_n[\xi/x']$ which leads up to 
$\phi[\xi/x']$. 
Because the substitution preserves the parity of fixpoint formulas, this is also
an alternating chain, showing that $n \leq \cdh(\phi[\xi/x'])$.

For the opposite inequality $\cdh(\phi) \geq \cdh(\phi[\xi/x'])$, consider an
alternating chain $\psi'_1 \clpr \psi'_1 \clpr \dots \clpr \psi'_n$ with 
$\psi'_n = \phi[\xi/x']$ of maximal length $n = \cdh(\phi[\xi/x'])$.

Because ${\clpr} \sse {\closeq}$ and $\psi'_n \in \Clos(\xi)$ it follows that 
$\psi'_i \in \Clos(\xi)$ for all $i$.
Because of item~\ref{i:substbij} we can then find for all $i \in \rng{1}{n}$ 
some $\psi_i \in \Clos(\chi')$ such that $\psi'_i = \psi_i[\xi/x']$. 
We now argue that $\psi_0 \clpr \psi_1 \clpr \dots \clpr \psi_n$ is an 
alternating chain of length $n$ leading up to $\psi_n$. 
Because the only formulas that the substitution $\xi/x'$ maps to fixpoint 
formulas are fixpoint formulas, and the map $\xi/x$ preserving the parity, it
suffices to show that $\psi_i \clpr \psi_{i + 1}$ for all $i \in \rng{1}{n-1}$.
This follows from item~\ref{i:sameclpr}, provided we can argue that 
$\psi_i \neq x'$ for all $i \in \rng{1}{n - 1}$.
For every $i \in \rng{1}{n - 1}$ we have that $\psi_i[\xi/x'] = \psi'_i \clpr
\psi'_n = \phi[\xi/x']$ and hence $\phi[\xi/x'] \fsfor \psi_i[\xi/x']$ by the 
definition of $\clpr$. 
But then $\psi_i = x'$ would imply $\phi[\xi/x'] \sfor \xi$, contradicting the
assumption that  $x' \in \FV{\phi}$.
\medskip

\textit{Item \ref{i:samewin}:}
This observation is immediate by item~\ref{i:sameclpr} and
Proposition~\ref{p:inftr}.
\end{proof}

\begin{proposition}
\label{p:adqc}
Let $\xi$ be a tidy $\mu$-calculus formula. Then $\xi \equiv \bbG_{\xi}$.
\end{proposition}

\begin{proof}
It will be convenient for us to consider the \emph{global} formula graph $\bbG
\isdef (\muML^{t},\cla,L_{C},\gOm)$, where $\muML^{t}$ is the set of all tidy
formulas using a fixed infinite set of variables, and $L_{C}$ is the obviously
defined global labelling function.
We may assign a semantics to this global graph using an equally obvious
definition of an acceptance game, where the only non-standard aspect is that the
carrier set of this `formula' is infinite.
For each tidy formula $\phi$ we may then consider the structure $\bbG\init{\phi}
\isdef (\muML^{t},\cla,L_{C},\gOm,\phi)$ as an initialised (generalised) parity
formula.
Note that all structures of this form have the same (infinite) set of vertices,
but that the only vertices that are accessible in $\bbG\init{\phi}$ are the
formulas in the (finite) set $\Clos(\phi)$.
It is then easy to see that $\bbG\init{\phi} \equiv \bbG_{\xi}\init{\phi}$, for
any pair of tidy formulas $\phi,\xi$ such that $\phi \in \Clos(\xi)$.

In order to prove the Proposition, it therefore suffices to show that every tidy
formula $\xi$ satisfies the following:
\begin{equation}
\label{eq:adqc}
\bbG\init{\phi} \equiv \phi, \text{ for all } \phi \in \Clos(\xi).
\end{equation}
We will prove \eqref{eq:adqc} by induction on the length of $\xi$.
In the base step of this induction we have $\len{\xi} = 1$, which means that
$\xi$ is an atomic formula.
In this case it is easy to see that \eqref{eq:adqc} holds.

In the induction step of the proof we assume that $\len{\xi} > 1$, and we make
a case distinction.
The cases where $\xi$ is of the form $\xi = \xi_{0} \odot \xi_{1}$ with $\odot
\in \{ \land, \lor \}$ or $\xi = \hs \xi_{0}$ with $\hs \in \{ \dia, \Box \}$,
are easy and left as exercises for the reader.

In the case where $\xi$ is of the form $\xi = \eta x. \chi$ with $\eta \in \{
\mu, \nu \}$ we make a further case distinction.
If $\xi$ belongs to the closure set of $\chi$, then we have $\Clos(\xi) \sse
\Clos(\chi)$, so that \eqref{eq:adqc} immediately follows from the induction
hypothesis, applied to the formula $\chi$.

This leaves the case where $\xi$ is of the form $\eta x. \chi$, while
$\xi \not\in \Clos(\chi)$. Let $x'$ be some fresh variable, then
obviously we may apply the induction hypothesis to the (tidy) formula
$\chi' \isdef \chi[x'/x]$. The statement that $\xi \equiv
\bbG\init{\xi}$ now follows by a routine argument, based on the
observations in Proposition~\ref{p:clx1}.
\end{proof}


It is left to show that the index of $\bbG_{\xi}$ does not exceed the
alternation depth of the formula $\xi$.
For this purpose it suffices to prove Proposition~\ref{p:ahandchains} below,
which links the alternation hierarchy to the maximal length of alternating 
$\clpr$-chains.
We need quite a bit of preparation to get there.
\medskip

Our first auxiliary proposition states that, when analysing the alternation
depth of a tidy formula of the form $\chi[\xi/x]$, we may without loss of 
generality assume that $\xi$ is not a free subformula of $\chi$.
Recall that $\ad_{\eta}(\xi)$ denotes the least $k$ such that $\xi \in 
\AH{\eta}{k}$.

\begin{proposition}
\label{p:cancelsfor1}
Let $\xi$ and $\chi$ be $\mu$-calculus formulas such that $\xi$ is free for $x$ 
in $\chi$, $x \in \FV{\chi}$, $\len{\xi} > 1$, and $\chi[\xi/x]$ is tidy.
Then there is a tidy formula $\chi'$ such that $\xi$ is free for $x'$ in $\chi'$, 
$\chi'[\xi/x'] = \chi[\xi/x]$, $\len{\chi'} \leq \len{\chi}$, $\ad_{\eta}(\chi') 
\leq \ad_{\eta}(\chi)$ for $\eta \in \{ \mu, \nu \}$, and $\xi \not\fsforeq  \chi'$.
\end{proposition}

\begin{proof}
The proof proceeds by induction on the length of $\chi$.
In case $\xi \not\fsforeq \chi$ we are done immediately, since then we can
simply take $\chi' \isdef \chi$. We will
therefore assume that $\xi \fsforeq \chi$. Then by definition we have
$\chi = \phi[\xi/y]$ for some tidy formula $\phi$ such that $y \in
\FV{\chi}$ and $\xi$ is free for $y$ in $\phi$. Since $\len{\xi} > 1$ we
find that $\len{\phi} < \len{\chi}$, and so we may apply the induction
hypothesis to $\phi$. That means that we can find a tidy formula $\phi'$
such that $\chi = \phi'[\xi/y']$ for some variable $y \in \FV{\phi'}$,
$\xi$ is free for $y'$ in $\phi'$, $\len{\phi'} \leq \len{\phi}$,
$\ad_{\eta}(\phi') \leq \ad_{\eta}(\phi)$ and $\xi \not\fsforeq  \phi'$.

Now define $\chi' \isdef \phi'[x'/x,x'/y']$, for some fresh variable
$x'$. Then clearly we have that $\len{\chi'} = \len{\phi'} \leq
\len{\phi} < \len{\chi}$ and $\ad_{\eta}(\chi') = \ad_{\eta}(\phi') \leq
\ad_{\eta}(\phi) \leq \ad_{\eta}(\chi)$, where the last inequality follows
from Proposition~\ref{p:ahsubst} and the fact that $\chi = \phi[\xi/y]$.
We leave it for the reader to convince themselves that $\xi \not
\fsforeq \phi'$ entails $\xi \not \fsforeq \chi'$.
Finally, it is not hard to verify that $\chi'[\xi/x'] =
\phi'[x'/x,x'/y'][\xi/x'] = \phi'[\xi/y'][\xi/x] = \chi[\xi/x]$ as
required.
\end{proof}

Our main auxiliary proposition concerns the relation between parity formulas
of the form $\bbG_{\chi}$ and $\bbG_{\chi[\xi/x]}$, respectively.
Roughly, it states that the substitution $\xi/x$ is a `local isomorphism'
between these two structures, i.e., it is an isomorphism at the level of 
certain clusters.
Recall that $\Cluster(\psi)$ denotes the $\closeq$-cluster of a formula
$\psi$, cf.~Definition~\ref{d:clstuff}.

\begin{proposition} \label{p:clustertocluster}
Let $\xi$ and $\chi$ be formulas such that $\xi$ is free for $x$ in 
$\chi$, $\xi \not \fsforeq \chi$, and $x \not\in \FV{\xi}$.
Furthermore, let $\psi \in \Clos(\chi)$ be such that $\psi[\xi/x] \notin
\Clos(\chi) \cup \Clos(\xi)$. 
Then the following hold:
\begin{enumerate}[topsep=0pt,itemsep=-1ex,partopsep=1ex,parsep=1ex,%
	label={\arabic*)}]

\item the substitution $\xi/x : \Cluster(\psi) \to \Cluster(\psi[\xi/x])$ is
a bijection between $\Cluster(\psi)$ and $\Cluster(\psi[\xi/x])$.
\label{i:substclustbij}
\end{enumerate}

\noindent
Let $\phi_{0},\phi_{1} \in \Clos(\chi')$. Then we have
\begin{enumerate}[topsep=0pt,itemsep=-1ex,partopsep=1ex,parsep=1ex,%
    label={\arabic*)}]
\addtocounter{enumi}{1}

\item \label{i:substclustiso}
$\phi_0 \cla \phi_1$ iff $\phi_0[\xi/x] \cla
\phi_1[\xi/x]$ and $L_C(\phi_0) = L_C(\phi_0[\xi/x])$;

\item \label{i:substclustsfor}
$\phi_0 \fsforeq \phi_1$ iff $\phi_0[\xi/x] \fsforeq
\phi_1[\xi/x]$;

\item \label{i:substclustcdh}
$\cdh(\phi_{0}) = \cdh(\phi_{0}[\xi/x])$, if $\phi_{0}$ is a fixpoint formula. 
\end{enumerate}
\end{proposition}

\begin{proof}
Fix $\xi$, $\chi$ and $\psi$ as in the formulation of the Proposition.
We start with a technical observation that will be of use throughout the proof:
\begin{equation}
\label{eq:xinotsfor}
\text{if $\psi \clat \phi$ and $\phi[\xi/x] \clat \psi[\xi/x]$ then
$x \in \FV{\phi}$ and $\xi \not \fsforeq \phi$, for every formula $\phi$.}
\end{equation}
For a proof of this claim, we first argue that $x \in \FV{\phi}$. If
this was not the case then $\phi[\xi/x] = \phi$ and thus $\phi \clat
\psi[\xi/x]$. Because $\psi \clat \phi$ and $\chi \clat \psi$ it follows
that $\chi \clat \psi[\xi/x]$, contradicting our assumption that
$\psi[\xi/x] \notin \Clos(\chi)$.

To prove $\xi \not \fsforeq \phi$ we show that
\begin{equation} \label{eq:ihtrace}
 \xi \fsforeq \rho \mbox{ implies } \xi \fsforeq \chi \mbox{ for all }
\rho \in \Clos(\chi) \mbox{ with } x \in \FV{\rho}.
\end{equation}
From this it follows that $\xi \not \fsforeq \phi$ because $\xi \not
\fsforeq \chi$ by assumption and $\phi \in \Clos(\chi)$ because $\phi
\in \Clos(\psi)$ and $\psi \in \Clos(\chi)$. 

We show \eqref{eq:ihtrace} with an induction on the length of the trace $\chi 
\clat \rho$.
The base case, where $\rho = \chi$, is trivial.
In the inductive step we have that $\chi \clat \rho' \cla \rho$ for some
$\rho'$. Because of item~1) in Proposition~\ref{p:clos5} it follows from
$\rho' \cla \rho$ and $x \in \FV{\rho}$ that $x \in \FV{\rho'}$. Thus we
can apply the induction hypothesis to $\rho'$.
Now assume that $\xi \fsforeq \rho$. Because of $\rho' \cla \rho$ we can
apply Proposition~\ref{p:always comparable} to get that either $\xi
\fsforeq \rho'$ or $\rho' \fsforeq \xi$. The latter is not possible
because $x \in \FV{\rho'} \setminus \FV{\xi}$. Thus, we find that $\xi
\fsforeq \rho'$ and with the induction hypothesis we obtain $\xi
\fsforeq \chi$.

This shows \eqref{eq:ihtrace} and hence it finishes the proof 
of \eqref{eq:xinotsfor}.
\medskip

\noindent
We now turn to the various items of the Proposition.

\textit{Item~\ref{i:substclustbij}:}
We first show that the substitution is well-defined as a map
from the cluster of $\psi$ to the cluster of $\psi[\xi/x]$:
\begin{equation}
\text{if $\phi \in \Cluster(\psi)$ then $\psi[\xi/x] \in 
\Cluster(\psi[\xi/x])$}.
\end{equation}
To see this, assume that $\psi \clat \phi$ and $\phi \clat \psi$.
Note that none of the traces $\psi \clat \phi$ and $\phi \clat \psi$ can 
pass over the variable $x$, because $x$ has no outgoing $\cla$-edges. 
Thus we can use Proposition~\ref{p:substbm} to push these traces down 
along the substitution, yielding $\psi[\xi/x] \clat \phi[\xi/x]$ and 
$\phi[\xi/x] \clat \psi[\xi/x]$, as required.

Injectivity of the substitution $\xi/x : \Cluster(\psi) \to 
\Cluster(\psi[\xi/x])$ is immediate by the claim below.
We need the rather peculiar formulation in the proof of surjectivity
below.
\begin{equation} 
\label{eq:injective}
\text{if
$\psi \clat \phi_i$ and $\phi_i[\xi/x] \clat \psi[\xi/x]$ for 
$i \in \{0,1\}$ then $\phi_0[\xi/x] = \phi_1[\xi/x]$ entails
$\phi_0 = \phi_1$}.
\end{equation}
For a proof, observe that for both $i \in \{0,1\}$ we have $\xi \not 
\fsforeq \phi_i$, by \eqref{eq:xinotsfor}.
The injectivity claim of \eqref{eq:injective} is then immediate by 
Proposition~\ref{p:subst1}.

It is left to prove surjectivity:
\begin{equation}
\label{eq:surjectivity}
\text{for every $\phi' \in \Cluster(\psi[\xi/x])$ there is a $\phi
\in \Cluster(\psi)$ such that $\phi[\xi/x] = \phi'$}.
\end{equation}
To show this, take an arbitrary formula $\phi'$ such that $\psi[\xi/x]
\clat \phi'$ and $\phi' \clat \psi[\xi/x]$.
First observe that the trace $\psi[\xi/x] \clat \phi'$ can not pass
over $\xi$ because otherwise we would have $\xi \clat \phi' \clat \psi[\xi/x]$,
contradicting the assumption that $\psi[\xi/x] \notin \Clos(\xi)$.
Thus we can use the back- and forth property \eqref{eq:1002} from 
Proposition~\ref{p:substbm} to lift the trace, $\psi[\xi/x] \clat \phi'$,
up along the substitution $\xi/x$ until we obtain a $\phi$ with 
$\psi \clat \phi$ and $\phi[\xi/x] = \phi'$.

Similarly, the trace $\phi[\xi/x] = \phi' \clat \psi[\xi/x]$ can
not pass over $\xi$ because otherwise $\xi \clat \psi[\xi/x]$,
contradicting $\psi[\xi/x] \notin \Clos(\xi)$. Thus, we can use
Proposition~\ref{p:substbm} again to lift this trace and get a $\rho$
with $\rho[\xi/x] = \psi[\xi/x]$ and $\phi \clat \rho$.

It remains to be shown that $\rho = \psi$. To this aim note that both
$\rho$ and $\psi$ satisfy the conditions in \eqref{eq:injective}.
For $\psi$ this is trivially the case and for $\rho$ observe that $\psi
\clat \phi \clat \rho$ and that $\rho[\xi/x] \clat \psi[\xi/x]$, as
$\rho[\xi/x] = \psi[\xi/x]$. 
Thus \eqref{eq:injective} gives $\rho = \psi$.
This finishes the proof of item~\ref{i:substclustbij}.
\medskip

 \textit{Item~\ref{i:substclustiso}:}
It is clear that the substitution $\xi/x$, preserving the main connective
of every formula (with the exception of $x$ itself), preserves the labels
in the closure graph. 
Note that $x$ does not belong to the cluster of $\psi$, because $x$ has no
outgoing $\cla$-edges. 
That the substitution $\xi/x$ is an
isomorphism for $\cla$-edges that stay inside the cluster follows from 
Proposition~\ref{p:substbm} and item~\ref{i:substclustbij}.

 \textit{Item~\ref{i:substclustsfor}:}
The preservation of the free subformula relation is just 
Proposition~\ref{p:subst preserves sfor}(1). 
Thus it is left to argue that $\phi_0[\xi/x] \sforeq \phi_1[\xi/x]$
implies $\phi_0 \sforeq \phi_1$ for all $\phi_0, \phi_1 \in
\Cluster(\psi)$. 
To this aim, observe that by \eqref{eq:xinotsfor} we have $x \in 
\FV{\phi_0}$, $\xi \not \fsforeq \phi_0$ and $\xi \not
\fsforeq \phi_1$.
We may now apply the second part of Proposition~\ref{p:subst preserves
sfor} to obtain the desired property.

 \textit{Item~\ref{i:substclustcdh}:}
Observe that $\cdh$ depends just on the alternating $\clpr$-chains in the 
cluster of its argument and that the relation $\clpr$ between such formulas
is defined in terms of the $\cla$-relation and the $\fsforeq$-relations in 
this cluster. 
Thus the claim follows by the items \ref{i:substclustiso}~and%
~\ref{i:substclustsfor}.
\end{proof}

The following Proposition is the key observation linking the alternation depth
of a formula to the index of its associated automaton, and thus to the maximal
length of alternating $\clpr$-chains in the closure graph of the formula.
It is thus the result, announced at the end of section~\ref{sec:ad}, that
corresponds to Proposition~\ref{p:adcf} but applies to the wider class of tidy
formulas.

To formulate and prove this observation, we need to refine some of our earlier 
definitions.

\begin{definition}
Let $C$ be a closure cluster.
For $\eta \in \{ \mu,\nu\}$, define $\cd_{\eta}(C)$ as the maximal length 
of an alternating $\clpr$-chain in $C$ leading up to an $\eta$-formula. 
Given a formula $\xi$, let $\cd_{\eta}(\xi)$ and $\cd(\xi)$ be defined as the
maximum value of $\cd_{\eta}(C)$ and $\cd(C)$, respectively, where $C$ ranges 
over all clusters of $\Clos(\xi)$.
\end{definition}

Clearly then we have $\cd(C) = \max(\cd_{\mu}(C),\cd_{\nu}(C))$, and, similarly,
$\cd(\xi) = \max(\cd_{\mu}(\xi),\cd_{\nu}(\xi))$.

\begin{proposition}
\label{p:ahandchains}
For any tidy formula $\xi$ and $\eta \in \{ \mu, \nu \}$, we have
\begin{equation}
\label{eq:cdad}
\cd_{\eta}(\xi) \leq n \text{ iff } \xi \in \AH{\eta}{n}.
\end{equation}
As a corollary, the alternation depth of $\xi$ is equal to the length of its
longest alternating $\clpr$-chain.
\end{proposition}

\begin{proof}
For the proof of the left-to-right direction of \eqref{eq:cdad}, we proceed by
an outer induction on $n$, and an inner induction on the length $\len{\xi}$ of
the formula $\xi$.
We focus on the outer inductive case, leaving the base case, where $n = 0$, to
the reader.

First of all, it is easy to see that every fixpoint formula $\xi'$ in the 
cluster of $\xi$ satisfies $\cd_{\eta}(\xi') = \cd_{\eta}(\xi)$, while it 
follows from Proposition~\ref{p:ad1} that $\xi' \in \AH{\eta}{n}$ iff $\xi \in
\AH{\eta}{n}$.
For this reason we may, without loss of generality, confine our attention to 
the case where $\xi$ is the $\clpr$-maximal element of its cluster.
Now distinguish cases, as to the parity of $\xi$.

First we consider the case where $\xi$ is of the form $\xi = 
\fopp{\eta} x. \chi$.
Let 
\[
\eta_{1}x_{1}. \psi_{1} \clpr \eta_{2} x_{2}.\psi_{2} \clpr \cdots \clpr
\eta_{k}x_{k}. \psi_{k}
\]
be a maximal alternating $\eta$-chain in $\Clos(\chi)$.
Then
\[
\eta_{1}x_{1}. \psi_{1}[\xi/x] \clpr \eta_{2} x_{2}.\psi_{2} \clpr \cdots \clpr
\eta_{k}x_{k}. \psi_{k}[\xi/x]
\]
is an alternating $\eta$-chain in $\Clos(\xi)$, and so we have $k \leq n$.
It then follows by the inner induction hypothesis that $\chi \in \AH{\eta}{n}$, 
and so by definition of the latter set we find $\xi = \fopp{\eta} x. \chi \in
\AH{\eta}{n}$, as required.

The other case to be discussed is where $\xi$ is of the form $\xi = \eta x.
\chi$.
Now let 
\[
\eta_{1}x_{1}. \psi_{1} \clpr \eta_{2} x_{2}.\psi_{2} \clpr \cdots \clpr
\eta_{k}x_{k}. \psi_{k}
\]
be a maximal alternating $\fopp{\eta}$-chain in $\Clos(\chi)$.

We now make a further case distinction.
If $x$ is a free variable of some formula in this chain, it is in fact a free 
variable of every formula in the chain; from this it follows that 
\[
\eta_{1}x_{1}. \psi_{1}[\xi/x] \clpr \eta_{2} x_{2}.\psi_{2} \clpr \cdots \clpr
\eta_{k}x_{k}. \psi_{k}[\xi/x] \clpr \xi
\]
is an alternating $\eta$-chain in $\Clos(\xi)$.
Since this chain has length $k+1$, it follows by our assumption on $\xi$ that
$k+1 \leq n$, and so $k \leq n-1$.
Alternatively, if $x$ is not a free variable of any formula in this chain, 
then the chain is itself an alternating $\fopp{\eta}$-chain in $\Clos(\xi)$, 
and from this and the assumption that $\cd_{\eta}(\xi) \leq n$ it readily 
follows that $k \leq n-1$.

In both cases we find that $k \leq n-1$, which means that 
$\cd_{\fopp{\eta}}(\chi) \leq n-1$.
By the outer induction hypothesis we thus find that 
$\chi \in \AH{\fopp{\eta}}{n-1}$. 
From this it is then easy to derive that $\xi = \eta x. \chi \in \AH{\eta}{n}$.
\medskip

For a proof of the opposite, right-to-left direction `$\Leftarrow$' of
\eqref{eq:cdad}, the argument proceeds by induction on the length of $\phi$.
In the base case $\phi$ is atomic and hence the claim is trivially true.

In the inductive step we make a case distinction depending on the clause of
Definition~\ref{d:ad} that was applied in the last step of the derivation of 
$\phi \in \AH{\eta}{k}$. 
We leave the easy cases, for the clauses \ref{adr:1}~and~\ref{adr:2}, to the
reader.
\smallskip

If clause~\ref{adr:3} is used to derive $\phi \in \AH{\eta}{n}$ then
$\phi = \fopp{\eta} x . \chi$ for some $\chi \in \AH{\eta}{n}$. First
define $\chi' = \chi[x'/x]$ for an $x'$ that is fresh for $\chi$ and
$\phi$. Note that the length of $\chi'$ is equal to the length of
$\chi$, which is shorter than the length of $\phi$. By
Proposition~\ref{p:ahsubst} we also have that $\chi' \in \AH{\eta}{n}$.
Moreover, $\chi'$ is tidy because $\phi$ is tidy, $\BV{\chi'} =
\BV{\chi} \subseteq \BV{\phi}$, $\FV{\chi'} = (\FV{\chi} \setminus
\{x\}) \cup \{x'\} \subseteq \FV{\phi} \cup \{x'\}$, and $x'$ is
fresh for $\phi$. This means that we can apply the inductive
hypothesis to $\chi'$, obtaining that $\cd_{\eta}(\chi') \leq n$

We then distinguish cases depending on whether $\phi \in \Clos(\chi)$ or not.

If $\phi \in \Clos(\chi)$ then it is not hard to prove that 
$\phi \in \Clos(\chi')$ as well.
It is then easy to see that every alternating chain in $\bbG_\phi$ also exists
in $\bbG_{\chi'}$, and thus it follows that
$\cd_{\eta}(\phi) \leq n$.

If $\phi \notin \Clos(\chi)$ we distinguish further cases depending on whether
$x \in \FV{\chi}$. 
If this is not the case then $\chi' = \chi$ and $\bbG_\phi$ is just like 
$\bbG_{\chi}$  with an additional vertex for $\phi$ that forms a degenerate 
cluster on its own and is connected just with an outgoing $\cla$-edge to the 
vertex of $\chi'$ in $\bbG_{\chi'}$. 
Thus, every alternating chain in a cluster of
$\bbG_\phi$ also exists in $\bbG_{\chi'}$ and thus
$\cd_{\eta}(\phi) \leq n$ follows from $\cd_{\eta}(\chi') \leq n$.

The last case is where $\phi \notin \Clos(\chi)$ and $x \in \FV{\chi}$.
To prove $\cd_{\eta}(\phi) \leq n$ consider an alternating $\clpr$-chain 
$\eta_1 x_1 . \rho_1 \clpr \cdots \clpr \eta_m x_m .\rho_m$, of length $m$ 
and with $\eta_m = \eta$ in some cluster of $\bbG_{\phi}$. 
We now argue that $m \leq n$. Because $\eta_i x_i .
\rho_i \in \Clos(\phi)$ for all $i \in \rng{1}{m}$ it follows by
Proposition~\ref{p:2001}(\ref{i:max}) that the only possibility for $\phi$ 
to be among the $\eta_i x_i . \rho_i$ in this chain is if $\phi = \eta_m x_m .
\rho_m$.
This would lead to a contradiction however, because $\eta_m = \eta$ while
we assumed that $\phi = \fopp{\eta} x . \chi$.
We may therefore conclude that $\phi$ is not among the $\eta_i x_i . \rho_i$ 
for $i \in \rng{1}{m}$. 
By the items \ref{i:substbij}, \ref{i:substiso} and \ref{i:sameclpr} of
Proposition~\ref{p:clx1} it follows that there is an alternating $\clpr$-chain
$\eta_1 x_1 . \sigma_1 \clpr \cdots \clpr \eta_m x_m. \sigma_m$ in $\Clos(\chi')$
such that $(\eta_i x_i . \sigma_i) [\xi/x'] = \eta_i x_i . \rho_i$ for all $i 
\in \rng{1}{m}$. 
Because $\cd_{\eta}(\chi') \leq n$ it follows that $m \leq n$.
\smallskip

If clause~\ref{adr:4} is used to derive $\phi \in \AH{\eta}{n}$ then $\phi$ is
of the form $\phi = \chi[\xi/x]$ such that $\chi,\xi \in \AH{\eta}{n}$. 
First observe that we may assume that $x \in \FV{\chi}$ and $\len{\xi} > 1$,
otherwise the claim trivialises.
Furthermore, because of Proposition~\ref{p:cancelsfor1} we may without loss 
of generality assume that in addition $\chi$ is tidy as well, that $x$ is
fresh for $\xi$, and that $\xi \not \fsforeq \chi$.
Finally, since $\len{\xi} > 1$ we find that the length of $\chi$ is smaller
than that of $\phi = \chi[\xi/x]$, so that we may apply the inductive hypothesis,
which gives that $\cd_{\eta}(\chi) \leq n$ and $\cd_{\eta}(\xi) \leq n$.

To show that $\cd_{\eta}(\chi[\xi/x]) \leq n$ consider then a fixpoint formula 
$\eta x . \rho \in \Clos(\chi[\xi/x])$ that is at the top of a maximal 
alternating $\clpr$-chain in $\bbG_{\chi[\xi/x]}$. 
In order to show that $\cdh(\eta y . \rho) \leq n$, we claim that 
\begin{equation}
\label{eq:liftr}
\cdh(\lambda y . \rho) = \cdh(\lambda y . \rho') \text{ for some }
   \lambda y . \rho' \in \Clos(\chi) \cup \Clos(\xi).
\end{equation}
To see this, first note that we may assume that $\lambda y . \rho \notin 
\Clos(\chi) \cup \Clos(\xi)$ because otherwise we can just set $\rho' \isdef
\rho$.
By Proposition~\ref{p:clos3} we obtain that
\[
\Clos(\chi[\xi/x]) = \{\psi[\xi/x] \mid \psi \in \Clos(\chi)\} \cup
\Clos(\xi).
\]
Therefore, since $\lambda y . \rho \in \Clos(\chi[\xi/x])$, and we assume that
$\lambda y. \rho \notin \Clos(\xi)$, it follows that $\lambda y . \rho = 
\psi[\xi/x]$ for some $\psi \in \Clos(\chi)$. 
We are thus in a position to apply Proposition~\ref{p:clustertocluster}, which
describes how the $\cla$-cluster of $\psi$ relates under the substitution 
$\xi/x$ to the $\cla$-cluster of $\lambda y . \rho = \psi[\xi/x]$.
Note that $\psi \neq x$ because otherwise we would have $\lambda y . \rho = \xi$,
contradicting the assumption that $\lambda x . \rho \notin \Clos(\xi)$. 
This means that $\psi = \lambda y. \rho'$ for some formula $\rho'$, since by
item~\ref{i:substclustiso} of Proposition~\ref{p:clustertocluster} the 
substitution $\xi/x$ preserves the main connective of formulas other than $x$. 
Finally, it follows from item~\ref{i:substclustcdh} of
Proposition~\ref{p:clustertocluster} that $\cdh(\lambda y . \rho) =
\cdh(\lambda y . \rho')$.

As an immediate consequence of \eqref{eq:liftr} we obtain that 
$\cdh(\eta y . \rho') \leq n$ because $\eta y . \rho'$ is either in $\bbG_\chi$
or in $\bbG_\xi$, where the inductive hypothesis applies.
This finishes the proof for the case of clause~\ref{adr:4}.
\smallskip

We leave the last case, where clause~\ref{adr:5} is used to derive that
$\phi \in \AH{\eta}{n}$, to the reader.
\end{proof}

Now that we have proved the main technical lemma, our desired result about 
the index of the parity formula $\bbG_{\xi}$ is almost immediate.

\begin{proposition}
For every tidy formula $\xi$ it holds that $\idx(\bbG_\xi) \leq \ad(\xi)$.
\end{proposition}

\begin{proof}
Take an arbitrary fixpoint formula $\xi$, and assume that $\ad(\xi) \leq n$.
Clearly it suffices to show that $\idx(\bbG_{\xi}) \leq n$.

For this purpose, first observe that by $\ad(\xi) \leq n$ we find that
$\xi \in \AH{\mu}{n} \cap \AH{\nu}{n}$.
Then it follows by
Proposition~\ref{p:ahandchains} that $\cd(\xi) \leq n$, so that by 
Proposition~\ref{p:gOm1} we obtain $\idx(\bbG_\xi) \leq n$.
\end{proof}

\section{From parity formulas to regular formulas}
\label{s:parfix}

In the previous section we saw constructions that, for a given regular formula, 
produce equivalent parity formulas based on, respectively, the subformula dag 
and the closure graph of the original formula.
We will now move in the opposite direction.
We first discuss in Theorem~\ref{t:cyc-fix} a well-known construction that turns
an arbitrary parity formula $\bbG$ into an equivalent regular formula 
$\xi_{\bbG} \in \muML$; basically this construction takes parity formulas as
systems of equations, and solves these equations by a Gaussian elimination of
variables.
What is of interest for us is that here we encounter a significant difference 
between our two size measures: whereas the closure-size of the resulting formula
$\xi_{\bbG}$ is \emph{linear} in the size of $\bbG$, its subformula-size is
only guaranteed to be exponential.
And in fact, Example~\ref{ex:bfl} (essentially due to Bruse, Friedmann \& 
Lange~\cite{brus:guar15}) shows that there is a family of parity formulas
for which the translation actually reaches this exponential subformula-size.
Finally, as an original result, in Theorem~\ref{t:cyc-fix-utw} we show that if
we restrict to untwisted formulas, we can translate parity formulas to 
equivalent regular formulas of linear subformula-size.

But as mentioned, we first consider a translation that works for arbitrary 
parity formulas.

\begin{fewtheorem}
\label{t:cyc-fix}
There is an effective procedure providing for any parity formula $\bbG= (V,E,L,
\Om,v_{I})$ over some set $\Prop$ of proposition letters, a map $\tr_{\bbG}: V
\to \muML(\Prop)$ such that 

1) $\bbG\init{v} \equiv \tr_{\bbG}(v)$, for every $v \in V$;

2) $\csz{\tr_{\bbG}(v)} \leq 2 \cdot \size{\bbG}$;

3) $\size{\Sfor(\tr_{\bbG}(v))}$ is at most exponential in $\size{\bbG}$;

4) $\ad(\tr_{\bbG}(v_{I})) \leq \idx(\bbG)$.
\end{fewtheorem}

Clearly, the algorithm mentioned in the Theorem will produce, given a parity 
formula $\bbG = (V,E,L,\Om,v_{I})$, an equivalent $\mu$-calculus formula 
\[
\xi_{\bbG} \isdef \tr_{\bbG}(v_{I})
\]
of linear closure-size with respect to $\bbG$, and with alternation depth bounded by the
index of $\bbG$.
Note that, although the definition of the translation map $\tr_{\bbG}$ involves
may substitution operations, it does \emph{not} involve any renaming of 
variables.

\begin{remark}
\label{r:cyc-fix}
Note that in item 3) of Theorem~\ref{t:cyc-fix} we cannot state that the 
\emph{subformula-size} of $\tr_{\bbG}$ is at most exponential in the size of
$\bbG$ since the formula $\tr_{\bbG}$ will generally not be clean, and so its
subformula-size may not be defined.
For this reason we compare the number of subformulas of $\tr_{\bbG}$ to the size
of $\bbG$.
We will come back to this issue in the next section, see Proposition~\ref{p:cfal}.
\end{remark}

It will be convenient for us to make a certain (harmless) assumption on the 
nature of the vertices of the parity formulas under consideration.

\begin{convention}
\label{conv:fpv}
In the sequel we shall assume, without loss of generality, that the nodes of 
parity formulas are taken from some (fixed) countably infinite set of objects
that we shall call \emph{potential vertices}, and that we have fixed a certain
enumeration of this set.
This enumeration then induces a natural enumeration of any set of vertices of 
a given parity formula.
\end{convention}

Both the definition of the collection of translation maps $\tr_{\bbG}$ and the
proofs of their most important properties will proceed by an induction on a 
certain complexity measure of parity formulas that we shall call its 
\emph{weight}.

\begin{definition}
We define the \emph{weight} of a parity formula $\bbG = (V,E,L,\Om,v_{I})$ as 
the pair $(\size{\Dom(\Om)},\size{\bbG})$ consisting of, respectively, the 
number of states and the size of $\bbG$.
Pairs of this form will be ordered lexicographically.
\end{definition}

\begin{definition} 
\label{d:cyc-tr}
The goal of this definition is to provide a map 
\[
\tr_{\bbG}: V \to \muML(\Prop)
\]
for every parity formula $\bbG = (V,E,L,\Om,v_{I})$.
The family of these maps $\tr_{\bbG}$ is defined by induction on the weight of 
$\bbG$.

Consider a parity formula $\bbG = (V,E,L,\Om,v_{I})$, and let $T$ be the top 
cluster of $\bbG$, that is, the cluster of the initial state $v_{I}$.
We make the following case distinction.
\medskip

\noindent
\textit{Case 1: $T$ is degenerate.}
In this case we must have $T = \{ v_{I} \}$, with $v_{I} \not\in \Ran(E)$,
and for every $u \neq v_{I}$ we may apply the induction hypothesis to the 
parity formula $G^{u}$ generated by $u$, since $\bbG$ has at least as many
states as $\bbG_{u}$, and more vertices.
We define
\[
\tr_{\bbG}(u) \isdef \tr_{\bbG^{u}}(u)
\]
for $u \neq v_{I}$, while for $v_{I}$ we set\footnote{%
   Note that the formulation of the boolean clause of our definition (i.e., the 
   case where $L(v) = \odot \in \{ \land, \lor \}$) is a bit sloppy, since our 
   language only has binary conjunctions and disjunctions, no conjunctions or 
   disjunctions over finite sets. 
   A more precise definition can be given as follows; assume that $\odot = 
   \land$, the case where $\odot = \lor$ is treated analogously. 
   We put $\tr_{\bbG}(v) = \top$ if $E(v) = \nada$, 
   $\tr_{\bbG}(v) = \tr_{\bbG}(u)$ if $E(v) = \{ u \}$, and
   $\tr_{\bbG}(v) = \tr_{\bbG}(u_{0}) \land \tr_{\bbG}(u_{1})$ if  $E(v) =
   \{ u_{0}, u_{1} \}$ and $u_{0}$ is preceding $u_{1}$ in the enumeration of 
   potential vertices, cf.~Convention~\ref{conv:fpv}.
}
\[
\tr_{\bbG}(v_{I}) \isdef 
\left\{\begin{array}{ll}
L(v_{I}) & \text{if } L(v_{I}) \in \At(\Prop)
\\ \hs\tr_{\bbG^{u}}(u)
   & \text{if } L(v) = \hs \in \{ \dia, \Box \} \text{ and } E(v) = \{ u \}
\\ \bigodot \{ \tr_{\bbG^{u}}(u) \mid u \in E(v) \}
   & \text{if } L(v) = \odot \in \{ \land, \lor \}
\\ \tr_{\bbG^{u}}(u)
   & \text{if } L(v) = \epsilon \text{ and } E(v) = \{ u \}
\end{array}\right.
\]
\smallskip

\noindent
\textit{Case 2: $T$ is nondegenerate.}
In this case we have $T \cap \Dom(\Om) \neq \nada$; let $m_{T} \isdef 
\max(\Ran(\Om\rst{T}))$ be the maximum priority reached in $T$, and
define $M \isdef T \cap \Om^{-1}(m_{T})$ as the set of states in $T$ reaching
this priority.

Inductively we will consider the elements of $M$ as propositional variables, 
and define $\bbG^{-} = (V^{-},E^{-},L^{-},\Om^{-},v_{I})$ as the parity formula
over $\Prop \cup M$, given by 
\[\begin{array}{lll}
   V^{-} & \isdef & V \cup \{ z^{*} \mid z \in M \}
\\ E^{-} & \isdef & 
  \{ (v,x) \mid (v,x) \in E, x \not\in M\}
      \cup \{ (v,z^{*}) \mid (v,z) \in E, z \in M \}
\\ \Om^{-} & \isdef & \Om\rst{V\setminus M},
\end{array}\]
while its labelling $L^{-}$ is defined by putting
\[
L^{-}(v) \isdef
  \left\{\begin{array}{ll}
     L(v) & \text{if } v \in V
  \\ z    & \text{if } v = z^{*} \text{ for some } z \in M.
  \end{array}\right.
\]
Since $\size{\Dom(\Om^{-})} < \size{\Dom(\Om)}$, inductively we may assume a map 
$\tr_{\bbG^{-}}: V^{-} \to \muML(\Prop \cup M)$.
For the definition of $\tr_{\bbG}$ let $z_{1},\ldots,z_{k}$ be the enumeration
of $M$ that is induced by the enumeration of the collection of all potential 
vertices (cf.~Convention~\ref{conv:fpv}), and let $\eta$ be the parity of (each 
state in) $M$.
We now use an inner induction on $i$ to define, for each $i \in \rng{0}{k}$,
a translation $\tr^{i}: V \cup \{ z_{i+1},\ldots,z_{k} \} \to \muML(\Prop \cup
\{ z_{i+1},\ldots,z_{k} \})$: \footnote{%
   Since the definition of these maps involve substitutions, and up until now 
   we have only defined substitution as a partial operation, we need to argue 
   that our translations are well defined.
   But a straightforward inductive proof shows, that for all $i \in \rng{0}{k}$,
   and for all $v \in V \cup \{ z_{1+1}, \ldots, z_{k} \}$, we have that 
   $\FV{\tr^{i}(v)} \sse \Prop \cup \{ z_{i+1}, \ldots, z_{k} \}$, while
   $\BV{\tr^{i}(v)} \sse (V \setminus M) \cup \{ z_{0}, \ldots, z_{i} 
   \}$.
   From this it immediately follows that the free variables of
   $\tr^{i+1}(z_{i+1})$ are disjoint from the bound variables of $\tr^{i}(v)$,
   for any node $v \in V \cup \{ z_{i+2}, \ldots, z_{k} \}$, so that 
   $\tr^{i+1}(z_{i+1})$ is free for $z_{i+1}$ in $\tr^{i}(v)$.
   This ensures that $\tr^{i}$ is well-defined, for all $i$.
   }
\[\begin{array}{llll}
   \tr^{0}(v)   & \isdef & \tr_{G^{-}}(v) 
   & \text{ for all } v
\\ \tr^{i+1}(z_{i+1})  & \isdef & \eta z_{i+1}. \tr^{i}(z_{i+1})
\\ \tr^{i+1}(v) & \isdef & \tr^{i}(v)[\tr^{i+1}(z_{i+1})/z_{i+1}]
   & \text{ for any node } v \in V \cup \{ z_{i+2}, \ldots, z_{k} \}
\end{array}\]
Finally, we set $\tr_{\bbG} \isdef \tr^{k}$.
\end{definition}

\begin{remark}
Readers who are familiar with the \emph{vectorial} $\mu$-calculus will have 
realised that the inductive step of the above definition, in the case of a 
nondegenerate top cluster of the automaton $\bbG$, is basically an 
application of the \emph{Beki\v{c} principle}.
That is, in the notation of the definition, the formulas $\tr_{\bbG}(z)$ that 
we are defining, one by one, for all states $z \in M$, are extremal solutions 
of the family of `equations'
\[
\{ z \equiv \tr_{\bbG^{-}}(z) \mid a \in M \},
\]
where we are after the least fixpoints in case $\eta = \mu$, and after the 
greatest fixpoints if $\eta = \nu$.

Note that we could have simplified the definition by focusing on parity 
formulas that are \emph{linear}, i.e., have an injective priority map.
In this case the set $M$ of states of maximal priority would have been a
singleton, so that the `internal' induction would trivialise.
\end{remark}

The following propositions state that this map is truth-preserving, of linear
size with respect to the closure size, and of exponential size with respect to
the number of subformulas of the formula.

\begin{proposition}
\label{p:cyclin1}
Let $\bbG$ be a parity formula. 
Then 
\begin{equation}
    \bbG\init{v} \equiv \tr_{\bbG}(v)
\end{equation}
for all vertices $v$ of $\bbG$.
\end{proposition}

\begin{proof}
The proof of this proposition is routine.
\end{proof}

\begin{proposition}
\label{p:cyclin2}
Let $\bbG$ be a parity formula. 
Then 
\begin{equation}
    \csz{\tr_{\bbG}(v)} \leq 2\cdot \size{\bbG}.
\end{equation}
for all vertices $v$ of $\bbG$.
\end{proposition}

\begin{proof}
We will show that every parity formula $\bbG = (V,E,L,\Om,v_{I})$ satisfies
\begin{equation}
\label{eq:cycfix101}
\size{\Clos(\bbG)} \leq \size{\bbG} + \size{\Dom(\Om)},
\end{equation}
where we define
\[
\Clos(\bbG) \isdef \bigcup \big\{ \Clos(\tr_{\bbG}(v)) \mid v \in V \big\}.
\]
We will prove \eqref{eq:cycfix101} by induction on the weight of $\bbG$.
As in Definition~\ref{d:cyc-tr}, we let $T$ be the top cluster of 
$\bbG$, and make a case distinction.

Leaving the case where $T$ is degenerate as an exercise, we focus on the case
where $T$ is nondegenerate. 
Let $M$ and $z_{1},\ldots,z_{k}$ be as in Definition~\ref{d:cyc-tr}, and write 
$\tau_{i}$ for the substitution $\tr_{\bbG}(z_{i})/z_{i}$.
Furthermore, we let $\bbG^{-}$ be as in Definition~\ref{d:cyc-tr}, so that 
we may apply the induction hypothesis to $\bbG^{-}$.

Our main claim is now that 
\begin{equation}
\label{eq:cycfix102}
\size{\Clos(\bbG)} \leq  \size{\Clos(\bbG^{-})},
\end{equation}
and we will prove \eqref{eq:cycfix102} by first showing that, for each $i 
\in \rng{0}{k}$ we have 
\begin{equation}
\label{eq:cycfix102i}
\Clos^{i+1} \sse \big\{ \phi[\tau_{i+1}] \mid \phi \in \Clos^{i} \big\},
\end{equation}
where, for each $j \in \rng{0}{k}$, we define 
\[
\Clos^{j} \isdef \bigcup \Big\{ \Clos(\tr^{j}(v) \mid v \in V \cup 
\{ z_{j+1}^{*},\ldots,z_{k}^{*} \} \Big\}.
\]
To prove \eqref{eq:cycfix102i}, we take a node $v \in V \cup \{ z_{i+2}^{*},
\ldots,z_{k}^{*} \}$, and make a case distinction.
If $v = z_{i+1}$ then we find
\begin{align*}
\Clos(\tr^{i+1}(z_{i+1})) 
   & = \{ \tr^{i+1}(z_{i+1}) \} \cup 
       \{ \phi[\tau_{i+1}] \mid \phi \in \Clos(\tr^{i}(z_{i+1})) \}
   & \text{(Proposition~\ref{p:clos3}(5))}
\\ & = \{ z_{i+1}[\tau_{i+1}] \}  \cup 
       \{ \phi[\tau_{i+1}] \mid \phi \in \Clos(\tr^{i}(z_{i+1})) \}
\\ & \sse \big\{ \phi[\tau_{i+1}] \mid \phi \in \Clos^{i} \big\}
   & \text{(since $z_{i+1} \in \Clos^{i}$)}
\end{align*}
For $v \neq z_{i+1}$ we obtain by Proposition~\ref{p:clos3}(4) and 
the definition of $\tr^{i+1}(v)$:
\[
\Clos(\tr^{i+1}(v)) = 
    \{ \phi[\tau_{i+1}] \mid \phi \in \Clos(\tr^{i}(v)) \}
    \cup \Clos(\tr^{i+1}(z_{i+1})).
\]
But by the earlier result for $z_{i+1}$ we already know that 
$\Clos(\tr^{i+1}(z_{i+1})) \sse \big\{ \phi[\tau_{i+1}] \mid \phi \in 
\Clos^{i} \big\}$, so that also for 
$v \neq z_{i+1}$ we now obtain
\[
\Clos(\tr^{i+1}(v)) \sse \{ \phi[\tau_{i+1}] \mid \phi \in \Clos^{i}\}.
\]
But then we are done proving \eqref{eq:cycfix102i}, and from this it easily
follows that $\size{\Clos^{i+1}} \leq \size{\Clos^{i}}$, for all $i < k$.
From this, \eqref{eq:cycfix102} is immediate: 
$\size{\Clos(\bbG)} = \size{\Clos^{k}} \leq \cdots \leq 
\size{\Clos^{0}} = \size{\Clos(\bbG^{-})}$.

Similarly, from \eqref{eq:cycfix102i}, the proof of \eqref{eq:cycfix101}
is straightforward:
\begin{align*}
\size{\Clos(\bbG)} 
   & \leq \size{\Clos(\bbG^{-})}
   & \text{(by \eqref{eq:cycfix102})}
\\ & \leq \size{\bbG^{-}} + \size{\Dom(\Om^{-})}
   & \text{(induction hypothesis)}
\\ & = (\size{\bbG} + \size{M}) + (\size{\Dom(\Om)} - \size{M})
   & \text{(definition of $\bbG^{-}$)}
\\ & = \size{\bbG} + \size{\Dom(\Om)}
\end{align*}
Finally, the Proposition itself is an immediate consequence of \eqref{eq:cycfix101}.
\end{proof}

\begin{proposition}
\label{p:cyclin3}
Let $\bbG = (V,E,L,\Om,v_{I})$ be a parity formula. 
Then 
\begin{equation}
\label{eq:szcf0}
\size{\Sfor(\tr_{\bbG}(v))} \leq (\size{\bbG}+2)^{\size{\Dom(\Om)}+2}
\end{equation}
for all vertices $v$ of $\bbG$.
\end{proposition}

\begin{proof}
In this proof it will be convenient to work with the collection $\NSfor(\phi)$ 
of \emph{non-atomic} subformulas of a formula $\phi$.
We leave it for the reader to prove that this notion interacts with that of
taking substitutions as follows:
\begin{equation}
\label{eq:sbsc}
\NSfor(\phi[\psi/x]) \subseteq \{ \phi'[\psi/x] \mid \phi' \in \NSfor(\phi) \}
\cup \NSfor(\psi),
\end{equation}
whenever $\psi$ is free for $x$ in $\phi$ and $x \not\in \BV{\phi}$.
    
The main statement that we shall prove is that for every parity formula $\bbG =
(V,E,L,\Om,v_{I})$ we have
\begin{equation}
\label{eq:szcf1}
\size{\NSfor(\bbG)} \leq (m_{\bbG}+2)^{\ell_{\bbG}+1}.
\end{equation}
Here we let $\NSfor(\bbG)$ denote the set $\NSfor(\bbG) \isdef \bigcup 
\{ \NSfor(\tr_{\bbG}(v)) \mid v \in V \}$, and we write $m_{\bbG}$ and
$\ell_{\bbG}$ for, respectively, the number of \emph{non-atomic} vertices and 
the number of states in $\bbG$ (that is, $\ell_{\bbG} = \size{\Dom(\Om)}$). 
We will prove \eqref{eq:szcf1} by induction on the weight $(\size{\Dom(\Om)},
\size{\bbG})$ of $\bbG$.
Let, again as in Definition~\ref{d:cyc-tr}, $T$ be the top cluster of $\bbG$,
and distinguish cases.
\smallskip

\noindent
\textit{Case 1}
If $T$ is degenerate, we let $z \isdef v_{I}$ be the unique member of $T$, and 
we make a further case distinction.
If in fact $v_{I}$ is the only vertex of $\bbG$, then it is easy to see that 
\eqref{eq:szcf1} holds because $\NSfor(\bbG) = \nada$.

In case $\bbG$ has other vertices besides $v_{I}$, let $w_{I}$ be the first one 
of these in the fixed enumeration of potential nodes 
(cf.~Convention~\ref{conv:fpv}), and let $\bbG^{-}$ denote the parity formula
generated by $\bbG$ on the set $V \setminus \{ v_{I} \}$, taking $w_{I}$ as its
initial vertex.

We make a further case distinction as to whether $v_{I}$ is atomic or not.
In the first case, it is easy to see that $\NSfor(\bbG) = \NSfor(\bbG^{-})$,
$m_{\bbG} = m_{\bbG^{-}}$, and $\ell_{\bbG} = \ell_{\bbG^{-}}$.
From this \eqref{eq:szcf1} follows readily.

Assume then that $v_{I}$ is non-atomic, and note that this implies, for $v_{I}$
itself, and for an arbitrary $v \neq v_{I}$, respectively:
\[\begin{array}{lll}
\NSfor(\tr_{\bbG}(v_{I}))
   & = & \{ \tr_{\bbG}(v_{I}) \} \cup 
      \bigcup \{ \NSfor(\tr_{\bbG^{-}}(u)) \mid u \in E(v) \}
\\[1mm] \NSfor(\tr_{\bbG}(v))
   & = & \NSfor(\tr_{\bbG^{-}}(v)).
\end{array}\]
From this it is easy to see that 
$\size{\NSfor(\bbG)} \leq 1 + \size{\NSfor(\bbG^{-})}$.
Using the facts that $m_{\bbG} = m_{\bbG^{-}}+1$ and $\ell_{\bbG} = 
\ell_{\bbG^{-}}$, we may then derive the following 
\begin{align*}
\size{\NSfor(\bbG)}
   & \leq 1 + \size{\NSfor(\bbG^{-})}
\\ & \leq 1 + ({m_{\bbG^{-}}}+2)^{\ell_{\bbG^{-}}+1}
   & \text{(induction hypothesis})
\\ & \leq 1 + ({m_{\bbG}}+1)^{\ell_{\bbG}+1}
\\ & \leq (m_{\bbG}+2)^{\ell_{\bbG}+1}.
\end{align*}

\noindent
\textit{Case 2}
If $T$ is nondegenerate, let $M, z_{1},\ldots,z_{k}$ and $\bbG^{-}$ be as in 
Definition~\ref{d:cyc-tr}, and abbreviate $\ell \isdef \ell_{\bbG}$ and $m \isdef
m_{\bbG}$.
Clearly we may apply the induction hypothesis to the parity formula $\bbG^{-}$.
We now claim that, for all $i \in \rng{0}{k}$, and all $v \in V \cup 
\{ z_{i+1}^{*},\ldots, z_{k}^{*} \}$, we have
\begin{equation}
\label{eq:fcssz}
\size{\NSfor^{i}(\bbG)} \leq (m+2)^{\ell-k+i+1}
\end{equation}
where we define $\NSfor^{i}(\bbG) \isdef \bigcup \big\{ \NSfor(\tr^{i}(v)) \mid 
v \in V \cup \{ z_{i+1}, \ldots, z_{k} \} \big\}$.
Clearly this suffices to take care of the (outer) induction step, since we 
defined $\tr_{\bbG}$ as the map $\tr^{k}$.

We will prove \eqref{eq:fcssz} by induction on $i$.
To take care of the base case (where $i=0$) we observe that $\tr^{0} = 
\tr_{\bbG^{-}}$, and that $\ell_{\bbG^{-}} = \ell-k$ and $m_{\bbG^{-}} = m$.
Hence, we find
\begin{align*}
  \size{\NSfor^{0}(\bbG))}
   & = \size{\NSfor(\bbG^{-})}
   & \text{(definition $\tr^{0}(v)$)}
\\ & \leq (m+2)^{\ell-k+1}
   & \text{(induction hypothesis on $\bbG^{-}$)}
\end{align*}
as required.

For the induction step we first make the following observations, based on
\eqref{eq:sbsc} and the definition of the map $\tr^{i+1}$, for the 
state $z_{i+1}$ and for an arbitrary vertex $v \neq z_{i+1}$, respectively:
\[\begin{array}{lll}
\NSfor(\tr^{i+1}(z_{i+1}))
   & = & \{ \tr^{i+1}(z_{i+1}) \} \cup 
       \NSfor(\tr^{i}(z_{i+1}))
\\[1mm] \NSfor(\tr^{i+1}(v))
   & = & \{ \phi[\tau_{i+1}] \mid \phi \in \NSfor(\tr^{i}(v)) \} 
      \cup \NSfor(\tr^{i+1}(z_{i+1}))
\end{array}\]
where $\tau_{i+1}$ is the substitution $\tr^{i+1}(z_{i+1})/z_{i+1}$.
From this it follows that
\[
\NSfor^{i+1}(\bbG)
   \; = \; \{ \tr^{i+1}(z_{i+1}) \} \cup 
       \NSfor(\tr^{i}(z_{i+1}) \cup
       \{ \phi[\tau_{i+1}] \mid \phi \in \NSfor(\tr^{i}(v)) \},
\]
and so we find
\begin{align*}
  \size{\NSfor^{i+1}(\bbG)}
   & \leq 1 + (m+2)^{\ell-k+i+1} + (m+2)^{\ell-k+i+1}
   & \text{(induction hypothesis on $\tr^{i}$)}
\\ & \leq (m+2)^{\ell-k+i+2}
\end{align*}
This finishes the proof of \eqref{eq:fcssz}, and therefore, that of 
\eqref{eq:szcf1}.

Finally, to see why the Proposition follows from \eqref{eq:szcf1}, let $V_{a}$
denote the set of atomic vertices in $\bbG$, observe that
\begin{align*}
  \size{\Sfor(\bbG)}
   & \leq \size{\NSfor(\bbG)} + \size{V_{a}} 
   & \text{(obvious)}
\\ & \leq (m_{\bbG}+2)^{\ell_{\bbG}+1} + \size{\bbG}
   & \text{(statement \eqref{eq:szcf1})}
\\ & \leq (\size{\bbG}+2)^{\ell_{\bbG}+2}
\end{align*}
and recall that $\ell_{\bbG} = \size{\Dom(\Om)}$.
\end{proof}

We now turn to the proof of the final item of Theorem~\ref{t:cyc-fix}.

\begin{proposition} 
For any parity formula $\bbG$ and for any vertex $v$ in $\bbG$ we have 
$\ad(\tr_{\bbG}(v)) \leq \idx(\bbG)$.
\end{proposition}

\begin{proof}
Let $C$ be a cluster of $\bbG$, and let $\eta$ be either $\mu$ or $\nu$.
Recall from Definition~\ref{d:ind} that $\idx_{\eta}(C)$ denotes the maximal 
length of an alternating $\eta$-chain in $C$, with $\idx_{\eta}(C) \isdef 0$ if
$C$ has no such chains.

\begin{claimfirst}
\label{cl:adcf}
Let $d$ and $\eta$ be such that $\idx_{\eta}(C) \leq d$ for every cluster $C$ of 
$\bbG$.
Then $\tr_{\bbG}(v) \in \AH{\eta}{d}$.
\end{claimfirst}

\begin{pfclaim}
We prove the claim by induction on the pair $(\size{\Dom(\Om)},\size{\bbG})$.
Let $T$ be the top cluster of $\bbG$, and make a case distinction.
We leave the case where $T$ is degenerate as an exercise, and focus on the case
where $T$ is nondegenerate.
Let $m_{T}$ and $M$ be as in Definition~\ref{d:cyc-tr}.

Let $\bbG_{T} = (V, E_{T},L_{T},\Om_{T},v_{I})$ be the parity formula given by
$E_{T} \isdef E \cap (T \times V)$, $L_{T} \isdef {L\rst{T}} \cup \{ (u,u) \mid 
u \in V\setminus T\}$ and $\Om_{T} \isdef \Om\rst{T}$.
In words, $\bbG_{T}$ is the parity formula we obtain from $\bbG$ by focusing on
the top cluster $T$, replacing, for every vertex $u \not\in T$, the generated 
subgraph $G^{u}$ with the `atomic' parity formula representing the atom $u$.  
It is not hard to see that, for all $u \in V\setminus T$ we have
\begin{equation}
\label{eq:adcf1}
\tr_{\bbG}(u) = \tr_{\bbG^{u}}(u),
\end{equation}
while the point of the construction is that for every $t \in T$ we get:
\begin{equation}
\label{eq:adcf2}
\tr_{\bbG}(t) = \tr_{\bbG_{T}}(t)[\tr_{\bbG^{u}}(u)/u \mid u \in V\setminus T].
\end{equation}
Now suppose that we can prove, for all $t \in T$, that
\begin{equation}
\label{eq:adcf3}
\tr_{\bbG_{T}}(t) \in \AH{\eta}{d}.
\end{equation}
Note that by the induction hypothesis, applied to the parity formulas 
$\bbG^{u}$ with $u \in V \setminus T$, we have $\tr_{\bbG^{u}}(u) \in 
\AH{\eta}{d}$.
Then we may use clause \eqref{adr:4} of Definition~\ref{d:ad} to derive from
\eqref{eq:adcf2} and  \eqref{eq:adcf3} that $\tr_{\bbG}(v) \in 
\AH{\eta}{d}$ as required.

It is thus left to prove \eqref{eq:adcf3}, and for this purpose we shall apply
the induction hypothesis to the parity formula $\bbG_{T}^{-}$.
Let $\la_{M}$ be the parity of the states in $M$.
The key observation on the relation between $\tr_{\bbG}$ and $\tr_{\bbG_{T}}$
is the following claim, which is proved by a straightforward induction on the 
definition of $\tr_{\bbG_{T}}$:
\begin{equation}
\label{eq:adcf4}
\text{if } \Ran(\tr_{\bbG^{-}}) \sse \AH{\fopp{\la_{M}}}{e} 
\text{ then } \Ran(\tr_{\bbG}) \sse \AH{\fopp{\la_{M}}}{e}.
\end{equation}

Turning to the proof of \eqref{eq:adcf3}, we make a case distinction, as to 
the nature of the parity $\la_{M}$ of the states in $M$.
Our reasoning will be slightly different in either case.

First consider the case where $\la_{M} = \eta$.
This implies that every cluster $D$ of $\bbG_{T}^{-}$ satisfies 
$\idx_{\fopp{\eta}}(D) \leq d-1$. 
Then by the induction hypothesis we find that $\tr_{\bbG_{T}^{-}}(v) \in
\AH{\fopp{\eta}}{d-1}$, for all $v \in T$.
From this it follows by \eqref{eq:adcf4} that $\Ran(\tr_{\bbG}) \sse 
\AH{\fopp{\la_{M}}}{d-1} = \AH{\fopp{\eta}}{d-1}$, which means that we are done
since $\AH{\fopp{\eta}}{d-1} \sse \AH{\eta}{d}$.

If, on the other hand, we have $\la_{M} = \fopp{\eta}$, then we reason as 
follows.
Clearly, every cluster $D$ of $\bbG_{T}^{-}$ satisfies $\idx_{\eta}(D) \leq d$. 
It follows by the induction hypothesis that $\tr_{\bbG_{T}^{-}}(v) \in
\AH{\eta}{d}$, for all $v \in T$.
But now we can use \eqref{eq:adcf4} to show that every formula of the form 
$\tr_{\bbG_{T}}(t)$ belongs to $\AH{\eta}{d}$ as required.
\end{pfclaim}

Finally, it is not hard to derive the Proposition from Claim~\ref{cl:adcf}.
With $d \isdef \idx(\bbG)$ and $C$ a cluster of $\bbG$, one easily derives from
the definitions that
$\idx_{\mu}(C), \idx_{\nu}(C) \leq d$.
From this it is immediate by the Claim that, for any $v$ in $\bbG$, we have
$\tr_{\bbG}(v) \in \AH{\mu}{d} \cap \AH{\nu}{d}$, so that 
$\ad(\tr_{\bbG}(v)) \leq d$ as required.
\end{proof}


The next example, which basically stems from~\cite[Theorem~3.1]{brus:guar15}, 
shows that the translation given in Definition~\ref{d:cyc-tr} may actually 
produce formulas with exponentially many subformulas, relative to the size of 
the parity formula.
The `culprit' here is the application of the substitution operation in the
inductive step of the definition, since this may double the number of 
subformulas each time it is applied.

\begin{example}
\label{ex:bfl}
For some arbitrary but fixed number $n$, consider the parity formula $\bbF = 
(V,E,L,\Om,v_{I})$ given by
\[\begin{array}{lll}
   V & \isdef &  \{ s_{i}, v_{i} \mid 0 \leq i \leq n \} 
\\ E & \isdef & 
    \{ (s_{i},v_{i}) \mid 0 \leq i \leq n \}
    \cup  \{ (s_{i+1},s_{i}) \mid 0 \leq i \leq n - 1\}
\\ & &  \cup\;  \{ (v_{0},s_{n})\}
    \cup  \{ (v_{i},s_{i}) \mid 0 < i \leq n \}
\\ L & \isdef & \{ (s_{i},\land), (v_{i},\dia) \mid 0 \leq i \leq n \}
\\ \Om   & \isdef & \{ (v_{i},i) \mid 0 \leq i \leq n \}
\\ v_{I} & \isdef & v_{0}.
\end{array}\]
In Figure~\ref{fig:x1} we display a picture of the automaton $\bbF$, 
for $n = 4$.
\begin{figure}[htb]
\begin{center}

\begin{tikzpicture}

   \node[state] (s_0)  {$\vtx{s_0}{\land}{}$};
   \node[state] (s_1) [right=of s_0] {$\vtx{s_1}{\land}{}$};
   \node[state] (s_2) [right=of s_1] {$\vtx{s_2}{\land}{}$};
   \node[state] (s_3) [right=of s_2] {$\vtx{s_3}{\land}{}$};
   \node[state] (s_4) [right=of s_3] {$\vtx{s_4}{\land}{}$};

   \node[state,initial] (v_0) [below=of s_0] {$\vtx{v_0}{\Diamond}{0}$};
   \node[state] (v_1) [right=of v_0] {$\vtx{v_1}{\Diamond}{1}$};
   \node[state] (v_2) [right=of v_1] {$\vtx{v_2}{\Diamond}{2}$};
   \node[state] (v_3) [right=of v_2] {$\vtx{v_3}{\Diamond}{3}$};
   \node[state] (v_4) [right=of v_3] {$\vtx{v_4}{\Diamond}{4}$};


   \path[->]
    (s_0) edge                     node {}      (v_0)
    (s_1) edge [bend left]    node {}      (v_1)
    (s_2) edge [bend left]    node {}      (v_2)
    (s_3) edge [bend left]    node {}      (v_3)
    (s_4) edge [bend left]    node {}      (v_4)

    (s_1) edge      node {}      (s_0)
    (s_2) edge      node {}      (s_1)
    (s_3) edge      node {}      (s_2)
    (s_4) edge      node {}      (s_3)

    (v_0) edge  [out=45,in=225]  node {}      (s_4)
    (v_1) edge  [bend left]     node {}      (s_1)
    (v_2) edge  [bend left]     node {}      (s_2)
    (v_3) edge  [bend left]     node {}      (s_3)
    (v_4) edge  [bend left]     node {}      (s_4)
  
 ; 
\end{tikzpicture}

\caption{the parity formula $\bbF$}
\label{fig:x1}
\end{center}
\end{figure}

Our claim is that 
\begin{equation}
\label{eq:exexp}
\size{\Sfor(\tr_{\bbF}(v_{0}))} \geq 2^{n},
\end{equation}
and in order to prove \eqref{eq:exexp}, we will use the notion of \emph{fixpoint 
depth} of a formula.
Recall that we define $\fdep{\phi} \isdef 0$ if $\phi$ is atomic, 
$\fdep{\phi_{0} \odot \phi_{1}} \isdef \max(\fdep{\phi_{0}},\fdep{\phi_{1}})$,
$\fdep{\hs\phi} \isdef \fdep{\phi}$, and 
$\fdep{\eta x. \phi} \isdef 1 + \fdep{\phi}$.
It is an easy exercise to verify that any $\mu$-calculus formula $\xi$ satisfies
$\size{\Sfor(\xi)} \geq \fdep{\xi}$, so that, in order to prove \eqref{eq:exexp},
it suffices to show that 
\begin{equation}
\label{eq:exexp1}
\fdep{\tr_{\bbF}(v_{0})} \geq 2^{n}.
\end{equation}

To calculate $\tr_{\bbF}(v_{0})$ it will be useful to introduce some auxiliary 
structures.
For $k \in \rng{0}{n}$, we let $\bbF_{k}$ denote the formula $(V_{k},E_{k},L_{k},
\Om_{k},s)$ given by
\[\begin{array}{lll}
   V_{k} & \isdef & V \cup \{ u_{i} \mid k \leq i \leq n \}
\\ E_{k} & \isdef & 
    \{ (s_{i},v_{i}) \mid 0 \leq i < k \}
    \cup \{ (s_{j},u_{j}) \mid k \leq j \leq n \}
    \cup  \{ (s_{i+1},s_{i}) \mid 0 \leq i \leq n - 1\}
\\ & &  \cup\;  \{ (v_{0},s_{n})\}
    \cup  \{ (v_{i},s_{i}) \mid 0 < i \leq n \}
\\ L_{k} & \isdef & L \cup \{ (u_{j},v_{j}) \mid k \leq j \leq n \}
\\ \Om_{k}   & \isdef & \{ (v_{i},i) \mid 0 \leq i \leq k \}
\end{array}\]

For an example, see Figure~\ref{fig:x2}, which contains a picture of the 
automaton $\bbF_{2}$ in the case where $n = 4$.
\begin{figure}[htb]

\begin{center}

\begin{tikzpicture}

   \node[state] (s_0)  {$\vtx{s_0}{\land}{}$};
   \node[state] (s_1) [right=of s_0] {$\vtx{s_1}{\land}{}$};
   \node[state] (s_2) [right=of s_1] {$\vtx{s_2}{\land}{}$};
   \node[state] (s_3) [right=of s_2] {$\vtx{s_3}{\land}{}$};
   \node[state] (s_4) [right=of s_3] {$\vtx{s_4}{\land}{}$};

   \node[state,initial] (v_0) [below=of s_0] {$\vtx{v_0}{\Diamond}{0}$};
   \node[state, inner sep=0mm] (v_1) [right=of v_0] {$\vtx{v_1}{\Diamond}{1}$};
   \node[state] (v_2) [right=19mm of v_1] {$\vtx{v_2}{\Diamond}{2}$};
   \node[state] (v_3) [right=of v_2] {$\vtx{v_3}{\Diamond}{3}$};
   \node[state] (v_4) [right=of v_3] {$\vtx{v_4}{\Diamond}{4}$};

   \node[state] (u_2) [right=29mm of v_1] {$\vtx{u_2}{v_2}{}$};
   \node[state] (u_3) [right=of u_2] {$\vtx{u_3}{v_3}{}$};
   \node[state] (u_4) [right=of u_3] {$\vtx{u_4}{v_4}{}$};


   \path[->]
    (s_0) edge                     node {}      (v_0)
    (s_1) edge [bend left]    node {}      (v_1)
    (s_2) edge                     node {}      (u_2)
    (s_3) edge                     node {}      (u_3)
    (s_4) edge                     node {}      (u_4)

    (s_1) edge      node {}      (s_0)
    (s_2) edge      node {}      (s_1)
    (s_3) edge      node {}      (s_2)
    (s_4) edge      node {}      (s_3)

    (v_0) edge  [out=45,in=225]  node {}      (s_4)
    (v_1) edge  [bend left]     node {}      (s_1)
    (v_2) edge                      node {}      (s_2)
    (v_3) edge                      node {}      (s_3)
    (v_4) edge                      node {}      (s_4)
 
 ; 
\end{tikzpicture}

\caption{the parity formula $\bbF_{2}$}
\label{fig:x2}
\end{center}
\end{figure}
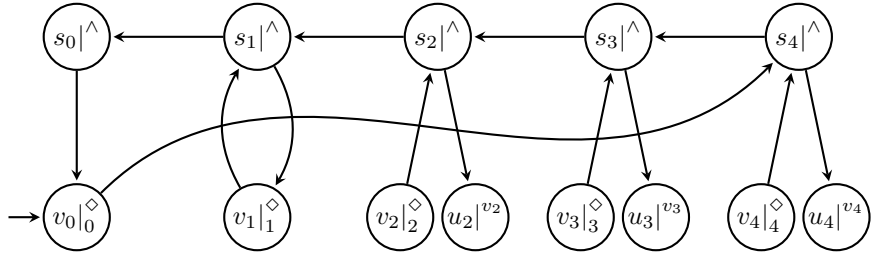
Using the notation of Definition~\ref{d:cyc-tr} (but writing $u_{i}$ for 
$v_{i}^{*}$), we have $\bbF = \bbF_{n}$ and $\bbF_{k+1}^{-} = \bbF_{k}$,
for all $k \in \rng{0}{n-1}$.

We now turn to the translation maps associated with these parity formulas.
Observe that it follows from the definitions that $M_{\bbF_{k}} = \{ v_{k} \}$,
so that we obtain the following definitions (where to avoid clutter we write
$\tr_{k}$ rather than $\tr_{\bbF_{k}}$, and omit brackets in conjunctions):
\[\begin{array}{llll}
\tr_{0}(v_{0}) &\isdef& \dia \bw_{0\leq i\leq n} v_{i}
\\ \tr_{0}(v_{m)} &\isdef& \dia \bw_{0\leq i\leq m} v_{i}
   & \text{ for all $m \in \rng{1}{n}$}
\\ \tr_{k+1}(v_{k}) &\isdef& \eta_{k}v_{k}. \tr_{k}(v_{k})
   & \text{ for all $k \in \rng{0}{n}$}
\\ \tr_{k+1}(v_{\ell}) &\isdef& \tr_{k}(v_{\ell})[\tr_{k+1}(v_{k})/v_{k}]
   & \text{ for all $k \in \rng{0}{n}$ and all $\ell\neq k$)}
\end{array}\]

In order to prove \eqref{eq:exexp}, we need an auxiliary notion of (relative) 
fixpoint depth.
Given a formula $\phi$ and variable $x$, we let $\fdep{x,\phi}$, the 
\emph{fixpoint depth of $x$ in $\phi$}, denote the maximum number of fixpoint 
operators that one may meet on a path from the root of the syntax tree of $\phi$
to a free occurrence of $x$ in $\phi$, with $\fdep{x,\phi} = -\infty$ if no such 
occurrence exists.
Formally, we set
\[\begin{array}{llll}
\fdep{x,\phi} &\isdef&
   \left\{ \begin{array}{ll}
      0       & \text{if $\phi = x$}
   \\ -\infty & \text{if $\phi$ is atomic, but $\phi \neq x$}
   \end{array}\right.
\\[2mm] \fdep{x,\phi_{0}\odot\phi_{1}} &\isdef& 
   \max \big( \fdep{x,\phi_{0}}, \fdep{x,\phi_{1}}
   \big)
   & \text{where $\odot \in \{ \land, \lor \}$}
\\[1mm] \fdep{x,\hs\phi} &\isdef& \fdep{x,\phi}
   & \text{where $\hs \in \{ \dia, \Box \}$}
\\[1mm] \fdep{x,\eta y. \phi} &\isdef&    
   \left\{ \begin{array}{ll}
      -\infty           & \text{if } x = y
   \\ 1 + \fdep{x,\phi} & \text{if } x \neq y 
   \end{array}\right.
   & \text{where $\eta \in \{ \mu, \nu \}$}
\end{array}
\]
Without proof we mention that, provided $x \neq y$ and $y$ is free for $y$ in 
$\phi$: 
\[
\fdep{x,\phi[\psi/y]} = 
\max\big(\fdep{x,\phi}, \fdep{y,\phi} + \fdep{x,\psi} \big).
\]
From this we immediately infer that 
\begin{equation}
\label{eq:mdep1}
\fdep{x,\phi[\psi/y]} \geq
\fdep{y,\phi} + \fdep{x,\psi},
\end{equation}
which is in fact the crucial observation in the proof: here we see that
the translation doubles the fixpoint depth of the formulas in every step.


\begin{claimfirst}
For all $k \in \rng{1}{n}$, and all $\ell,m \geq k$ we have that 
$\fdep{v_{\ell},\tr_{k}(v_{m})} \geq 2^{k} -1$.
\end{claimfirst}

\begin{pfclaim}
We prove the claim by induction on $k$.
For the base step of the induction, where $k = 1$, it suffices to observe that 
$\fdep{v_{\ell},\tr_{1}(v_{m})} = 1$, for all $\ell,m \geq 1$.
But this is obvious by the observation that for all $m \geq 1$ we may calculate
$\tr_{1}(v_{m}) = \dia \big( \nu v_{0}. \dia \bw_{0\leq i \leq n} 
v_{i}\big) \land \bw_{1\leq i \leq \ell}v_{i}$.

For the induction step, we consider the case for $k+1$.
Taking arbitrary numbers $\ell,m \geq k+1$, we reason as follows:
\begin{align*}
   \fdep{v_{\ell},\tr_{k+1}(v_{m})} 
   & = \fdep{v_{\ell},\tr_{k}(v_{m})[\tr_{k+1}(v_{k})/v_{k}]} 
   & \text{(definition $\tr_{k+1}(v_{m})$)}
\\ & \geq \fdep{v_{k},\tr_{k}(v_{m})} + \fdep{v_{\ell},\tr_{k+1}(v_{k})}
   & \text{(equation \eqref{eq:mdep1})}
\\ & = \fdep{v_{k},\tr_{k}(v_{m})} + \fdep{v_{\ell},\eta_{k}v_{k}.\tr_{k}(v_{k}))}
   & \text{(definition $\tr_{k+1}(v_{k})$)}
\\ & \geq \fdep{v_{k},\tr_{k}(v_{m})} + 1 + \fdep{v_{\ell},\tr_{k}(v_{k}))}
   & \text{(definition $\fdep{\cdot}$, $\ell > k$)}
\\ & \geq (2^{k}-1) + 1 + (2^{k}-1)
   & \text{(induction hypothesis, twice)}
\\ & = 2^{k+1} -1.
\end{align*}
Clearly this finishes the proof of the claim.
\end{pfclaim}

\noindent
Finally, it is easy to see how \eqref{eq:exexp1} follows from the Claim.
\end{example}

The third and final result in this section shows that for \emph{untwisted} 
formulas, we can actually give a translation into the modal $\mu$-calculus
which is \emph{linear} in terms of subformula-size.

\begin{fewtheorem}
\label{t:cyc-fix-utw}
There is an effective procedure providing for any untwisted parity formula 
$\bbG$ over some set $\Prop$ of proposition letters, a clean formula 
$\xi_{\bbG} \in \muML(\Prop)$ such that 

1) $\bbG \equiv \xi_{\bbG}$;

2) $\ssz{\xi_{\bbG}} \leq 2 \cdot \size{\bbG}$;

3) $\ad(\xi_{\bbG}) \leq \idx(\bbG)$.
\end{fewtheorem}

\begin{proof}
Let $\bbG= (V,E,L,\Om,v_{I})$ be an untwisted parity formula witnessed by the 
dag with back edges $(V,D,B)$.
The idea is to define a modal $\mu$-calculus formula $\xi$, satisfying the
conditions of the theorem, and such that $\BV{\xi} = \Dom(\Om)$.

It will be convenient to impose an additional condition on the shape of $\bbG$,
to the effect that elements of $\Dom(B)$ and of $\Dom(\Om)$ are silent nodes
(i.e., labelled with $\epsilon$); this also means that such vertices have exactly one
$E$-successor.
These assumptions can be made without loss of generality: any untwisted formula
$\bbG$ violating one or both of these conditions can easily be transformed into
an equivalent untwisted formula $\bbG'$ which does meet the conditions, and is
of size at most $2 \cdot \size{\bbG}$.

In the sequel we will consider two versions of the labelling function on $V$: 
$L$ itself and the map $L'$ given by
\[
L'(v) \isdef \left\{ \begin{array}{ll}
   L(v) & \text{if } v \not\in \Dom(B)
\\ b(v) & \text{if } v \in \Dom(B),
\end{array}\right.
\]
where we let $b(u)$ denote the unique element of $B[u]$ (it it exists).
It is helpful to think of $L'$ as the version of $L$ which takes the vertices 
in $\Dom(\Om) = \Ran(B)$ as free variables.

Since $(V,D)$ is a finite dag, we may base inductive definitions and proofs on
the relation $D$.
In particular, we may inductively define the map $\xi: V \to \muML(\Prop\cup 
\Ran(B))$ below.
For the base step of this induction, note that if $v$ has no $D$-successors,
then we have either $v \in \Dom(B)$ meaning that $L'(v) = b(v)$ or else 
$v \not\in \Dom(B)$ implying $L'(v) = L(v)$ is a literal.
\[
\xi(v) \isdef \left\{ \begin{array}{ll}
   L'(v)    
   & \text{ if } D[v] = \nada
\\ \hs\xi(u)
   & \text{ if } L(v) = \hs \in \{ \dia,\Box \} \text{ and } 
      D[v] = \{ u \}
\\ \bigodot\{\xi(u) \mid u \in D[v]\} 
   & \text{ if } L(v) = \odot \in \{ \land,\lor \} 
\\ \xi(u)
   & \text{ if } L(v) = \epsilon, D[v] = \{ u \} \text{ and } 
      v \not\in \Dom(\Om)  
\\ \mu v. \xi(u)
   & \text{ if } L(v) = \epsilon, D[v] = \{ u \},
      v \in \Dom(\Om) \text{ and } \Om(v) \text{ is odd}
\\ \nu v. \xi(u)
   & \text{ if } L(v) = \epsilon, D[v] = \{ u \},
      v \in \Dom(\Om) \text{ and } \Om(v) \text{ is even}.
\end{array}\right.
\]
Based on this definition we put
\[
\xi_{\bbG} \isdef \xi(v_{I}).
\]

As an immediate consequence of the definition we find that the formula $\xi(v)$
is clean, for every $v \in V$; this applies then in particular to $\xi_{\bbG}$.
To show that $\xi_{\bbG}$ has the right size, we first prove, by a 
straightforward induction on $D$, that 
\begin{equation}
\label{eq:sf1}
\Sfor(\xi(v)) \sse \xi[D^{*}[v]]
\end{equation}
for every $v \in V$.
It then follows that $\Sfor(\xi_{\bbG}) \sse \Ran(\xi)$ by taking $v = v_{I}$ 
in \eqref{eq:sf1}, and from this, part 2) of the theorem is immediate.
\medskip

Proving the equivalence of $\xi_{\bbG}$ and $\bbG$ is a relatively routine
exercise, and so we confine ourselves to a sketch.
Define, for $u \in V$, the parity formula $\bbH^{u} = (W^{u}, E^{u},
L^{u},\Om^{u},u)$, where $W^{u} \isdef D^{*}[u]$, $E^{u}$ and $\Om^{u}$ are
simply given as $E$ and $\Om$, respectively, restricted to $W^{u}$, and 
$L^{u}$ is given as
\[
L^{u}(v) \isdef \left\{ \begin{array}{ll}
   L(v) & \text{if } v \not\in \Dom(B)
\\ L(v) & \text{if } v \in \Dom(B) \text{ and } b(v) \in D^{*}[u]
\\ L'(v) & \text{if } v \in \Dom(B) \text{ and } u \in D^{*}[b(v)]
\end{array}\right.
\]
To understand this definition, the key observation is that by
untwistedness, $v \in D^{*}[u] \cap \Dom(B)$ implies that either $b(v)
\in D^{*}[u]$ (meaning that $b(v)$ is a bound variable of $\xi(u)$), or
$u \in D^{*}[b[v]]$ (meaning that $b(v)$ is a free variable of
$\xi(u)$). By induction on the dag relation $D$ one may now prove that 
\begin{equation}
\label{eq:sf2}
\bbH^{u} \equiv \xi(u), \text{ for all } u \in V.
\end{equation}
Part 1) of the proposition follows from this by observing that $\bbH^{v_{I}} =
\bbG$.
\medskip

Finally, to prove the constraint on alternation depth in part 3), first observe
that the bound variables of $\xi_{\bbG}$ correspond to the elements of $\Dom(B)$,
and that each such variable $u$, as a vertex $u \in \Dom(B)$, has a unique 
$B$-successor $b(u)$ which satisfies $D^{*}b(u)u$ and thus belongs to the same 
cluster as $u$.
Now if $u$ and $v$ are variables in $\xi_{\bbG}$ such that $u <_{\xi_{\bbG}} v$, 
then by definition we have $v \sforeq \de_{u} \sforeq \de_{v}$, where $\de_{u}$ 
and $\de_{v}$ are, respectively, the formulas associated with the unique 
$D$-successors of $b(u)$ and $b(v)$.
From this and the definition of $\xi$ it is easy to infer that $D^{*}b(u)v$ and
$D^{*}b(v)b(u)$.
This leads us to the key observation that for any two variables/vertices $u,v 
\in \BV{\xi_{\bbG}}$ we have that $u <_{\xi_{\bbG}} v$ implies $b(u) \equiv_{E}
b(v)$ and $\Om(b(u)) \leq \Om(b(v))$.
From this it easily follows that any alternating dependency chain in $\xi_{\bbG}$ 
originates with a single-cluster $\Om$-alternating chain in $\bbG$.
Then by Proposition~\ref{p:adcf} we may conclude that $\ad(\xi_{\bbG}) \leq 
\idx(\bbG)$.
\end{proof}

\section{Alphabetical equivalence}
\label{sec:aleq}

In formalisms that feature some kind of variable binding, the meaning of a 
syntactic expression usually does not depend on the exact choice of its bound
variables.
In such a setting $\al$-equivalent formulas, i.e., formulas that can be obtained 
from one another by a suitable renaming of bound variables, are often taken to
be identical.
As we already mentioned in the introduction, however, the consequences of such 
an identification for notions such as formula size have, to the best of our
knowledge, not received any attention in the literature on the modal 
$\mu$-calculus.
It is precisely this matter that we address in this section (and the next).
As one of our key results we show that there is a high price to pay for working
with clean formulas: in some cases, any operation of taking a clean alphabetical
variant of a tidy formula may will result in an exponentially bigger formula.
Our goal in this section, however, will be to introduce two size measures that 
are invariant under alphabetical equivalence, as expressed by the conditions
(\ddag a) and (\ddag b) in the introduction.

First, however, we give a definition of the notion of \emph{alphabetical 
variant} itself.

\begin{definition}
An equivalence relation $\sim$ on the set $\muML$ of formulas will be called 
a \emph{(syntactic) congruence} if it satisfies the following two conditions:

1) $\phi_{0}\sim\psi_{0}$ and $\phi_{1}\sim\psi_{1}$ imply
   $\phi_{0}\odot\phi_{1} \sim \psi_{0}\odot\psi_{1}$, for $\odot \in 
   \{ \lor, \land \}$;

2) $\phi\sim\psi$ implies $\hs\phi \sim \hs\psi$, for $\hs \in \{ \dia, 
   \Box \}$.
   
\noindent
We define the relation $\eqal $ as the smallest congruence $\sim$ on $\muML$
which is closed under the rule:

3) if $\phi_{0}[z/x_{0}] \sim \phi_{1}[z/x_{1}]$, where $z$ is \emph{fresh}
   for $\phi_{0}$ and $\phi_{1}$, then $\eta x_{0}. \phi_{0} \sim
   \eta x_{1}.\phi_{1}$, for $\eta \in \{ \mu, \nu \}$.

\noindent
If $\phi \eqal \psi$ we call $\phi$ and $\psi$ \emph{$\alpha$-equivalent}, or
\emph{alphabetical variants} of one another.
The $\al$-equivalence class of a formula $\phi$ is denoted as 
$\eqalc{\phi}$.
\end{definition}

It will be convenient to have a formal system in place by which we can
\emph{derive} the $\al$-equivalence of two formulas --- this will enable us to
prove statements about $\eqal$ using induction on the complexity of derivations.

\begin{definition}
With $\foeq$ denoting a formal identity symbol, an \emph{equation} is an
expression of the form $\phi \foeq \psi$ with $\phi,\psi \in \muML$.
We define $\vdal$ as the derivation system on such equations, which consists of 
the axiom $\phi \foeq \phi$ and the obvious rules corresponding to the 
conditions 1) -- 3) above.
In case an equation $\phi \foeq \psi$ is derivable in this system we write 
$\vdal \phi \foeq \psi$.
\end{definition}

Note that the absence of rules for symmetry or transitivity in $\vdal$ makes
the system a very useful proof tool.
This absence is justified by the following proposition.

\begin{proposition}
\label{p:alder}
The derivation system $\vdal$ for $\eqal$ is sound and complete for 
$\al$-equivalence, that is, for any pair of $\muML$-formulas $\phi, \psi$
we have
\[
\phi \eqal \psi \text{ iff } \vdal \phi \foeq \psi.
\]
\end{proposition}
\begin{proof}
Soundness, i.e., the implication from right to left, is obvious.
For the opposite implication, one shows by induction on $\phi$ that
$\vdal \phi \foeq \psi$ and $\vdal \psi \foeq \xi$ imply $\vdal \phi \foeq \xi$,
which obviously implies that the relation generated by $\vdal$-deductions is 
transitive.
Similarly, one can show that the relation of $\vdal$-derivable equivalence is 
symmetric.
From this it is immediate that $\phi \eqal \psi$  implies  $\vdal \phi \foeq
\psi$ as required. 
\end{proof}

In the sequel we will use the above proposition without warning; we will also be
a bit sloppy concerning notation and terminology, for instance allowing
ourselves to write that `$\phi \eqal \psi$ is derivable' if we mean that
$\vdal \phi \foeq \psi$.

\subsection*{Basic observations}

We first provide some key information about $\al$-equivalence.
The first proposition states that many basic concepts of $\mu$-calculus formulas
are invariant under $\al$-equivalence (recall that $\fdep{\phi}$ 
denotes the fixpoint depth of $\phi$.

\begin{proposition}
\label{p:aleq0}
The following hold, for any pair $\phi_{0},\phi_{1}$ of $\mu$-calculus
formulas:

\begin{urlist}

\item \label{it:aleq0-1}
if $\phi_{0} \eqal \phi_{1}$ then $\phi_{0} \equiv \phi_{1}$;

\item \label{it:aleq0-2}
if $\phi_{0} \eqal \phi_{1}$ then $\len{\phi_{0}} = \len{\phi_{1}}$;

\item \label{it:aleq0-3}
if $\phi_{0} \eqal \phi_{1}$ then $\FV{\phi_{0}} = \FV{\phi_{1}}$;

\item \label{it:aleq0-4}
if $\phi_{0} \eqal \phi_{1}$ then $\fdep{\phi_{0}} = \fdep{\phi_{1}}$.

\item \label{it:aleq0-5}
if $\phi_{0} \eqal \phi_{1}$ then $\ad(\phi_{0}) = \ad(\phi_{1})$.

\end{urlist}
\end{proposition}

We postpone the proof of Proposition~\ref{p:aleq0} until we have proved the
next proposition, which gathers some rather technical observations.

\begin{proposition}
\label{p:aleq1}
Let $\phi,\phi_{0},\phi_{1},\psi,\psi_{0},\psi_{1}$ and $\chi$ be $\mu$-calculus
formulas, and let $\eta, \eta_{0}, \eta_{1} \in \{ \mu, \nu \}$.
Then the following hold:

\begin{enumerate}[topsep=0pt,itemsep=-1ex,partopsep=1ex,parsep=1ex,%
    label={\arabic*)}]
 \item\label{aleq:fresh}  if $\phi \eqal \psi$ then
    $\phi[z/x] \eqal \psi[z/x]$ for any $z$ that is fresh for $\phi$ and $\psi$; 
 \item\label{aleq:1} if $\eta_{0} x_{0}. \phi_{0} \eqal \psi_{1}$ then $\psi_{1}$ is of the form 
   $\psi_{1} = \eta_{1} y. \phi_1$, where $\eta_{0} = \eta_{1}$;
 \item\label{aleq:2}  if $\eta x_{0}. \phi_{0} \eqal \eta x_{1}. \phi_{1}$ then
   $\phi_{0}[z/x_{0}] \eqal \phi_{1}[z/x_{1}]$, for any fresh variable $z$;
 \item\label{aleq:3} if $\eta x. \phi_{0} \eqal \eta x. \phi_{1}$ then
   $\phi_{0} \eqal \phi_{1}$;
\item\label{aleq:4a}  if $\eta x. \phi_{0} \odot \phi_{1} \eqal \eta y. \psi_{0} \odot \psi_{1}$
   then $\eta x. \phi_{i} \eqal \eta y. \psi_{i}$, 
   for $i \in \{0,1\}$ and $\odot \in \{ \land, \lor \}$;
\item\label{aleq:4b}  if $\eta x. \hs \phi \eqal \eta y. \hs\psi$
   then $\eta x. \phi \eqal \eta y. \psi$ for $\hs \in \{ \dia, \Box \}$;
\item\label{aleq:4c} if $\eta x. \la z. \phi \eqal \eta y. \la z. \psi$
   then $\eta x. \phi \eqal \eta y. \psi$ for $\la \in \{ \mu, \nu \}$;
\item\label{aleq:5}  
   if $\phi \eqal \psi$, $y \not\in \FV{\phi}$ and $y$ is free for $x$ 
   in $\psi$, then $\eta x.\phi \eqal \eta y.\psi[y/x]$; 
\item\label{aleq:6}  
   if $\phi_{0} \eqal \phi_{1}$, $\psi_{0} \eqal \psi_{1}$ and 
   $\psi_{i}$ is free for $x$ in $\phi_{i}$,
   then $\phi_{0}[\psi_{0}/x] \eqal \phi_{1}[\psi_{1}/x]$;
\item\label{aleq:7}  if $\eta x_{0}.\phi_{0} \eqal \eta x_{1}. \phi_{1}$ for tidy formulas
   $\eta x_{i}.\phi_{i}$ then 
   $\phi_{0}[\eta x_{0}.\phi_{0}/x_{0}] \eqal \phi_{1}[\eta x_{1}\phi_{1}/x_{1}]$;
\item\label{aleq:8}  
  if $\phi_{0} \eqal \phi_{1}$ then $\eta x. \phi_{0} \eqal \eta x. \phi_{1}$.
\item\label{aleq:12}
  if $\phi \eqal \psi[\chi/x]$, then $\phi = \psi'[\chi'/x']$ for some formulas
  $\psi', \chi'$ and a fresh variable $x'$ such that $\psi \eqal \psi'[x/x']$
  and $\chi \eqal \chi'$.
\end{enumerate}
\end{proposition}

\begin{proof}
Item~\ref{aleq:fresh} follows by proving that a  derivation
of $\phi \eqal \psi$ can be easily transformed into a derivation
of $\phi[z/x] \eqal \psi[z/x]$ by induction on the length of the derivation.

Item~\ref{aleq:1} is obvious by Proposition~\ref{p:alder}. 
For item~\ref{aleq:2} let $z,z'$ be fresh variables. Now observe that any
derivation of $\phi_0[z/x_0] \eqal \phi_1[z/x_1]$ can be transformed into
a derivation of $\phi_0[z'/x_0] \eqal \phi_1[z'/x_1]$. 
\smallskip

For item~\ref{aleq:3}, suppose that $\eta x. \phi_0 \eqal \eta x .\phi_1$. 
Then by the second item we have $\phi_0[z/x] \eqal \phi_1[z/x]$ for any fresh
variable $z$.
To prove the claim we show that this implies $\phi_0 \eqal \phi_1$. 
This is proved by a straightforward induction on the complexity of $\phi_0$. 

For the atomic case, make a case distinction. 
If $\phi_{0} = x$, then by $\phi_0[z/x] \eqal \phi_1[z/x]$ we find that $\phi_{1}
= x$ as well, so that we obtain $\phi_{0} \eqal \phi_{1}$ because $\phi_{0} =
\phi_{1}$.
On the other hand, if $\phi_{0} = y$ with $y \neq x$, then we have 
$\phi_{0}[z/x] = y$, so that we also find $\phi_{1}[z/x] = y$, which can only
mean that $\phi_{1} = y$ since $z$ is fresh.
So in this case we also obtain $\phi_{0} \eqal \phi_{1}$ because $\phi_{0} =
\phi_{1}$.

For the induction step, the cases where $\phi= \phi_{0} \odot \phi_{1}$ with
$\odot \in \{\land,\lor\}$ and where $\phi = \hs \phi'$ with $\hs$ a modal
operator follows easily by induction.
This leaves the case where $\phi_0 = \eta y_0. \phi'_0$. 
Then  $\phi_1 = \eta y_1. \phi'_1$ and by assumption $(\eta y_0. \phi'_0) [z/x]
\eqal (\eta y_1. \phi'_1) [z/x]$.

{\it Case} $x \not\in \FV{\eta y_0. \phi'_0}$. Then by item~\ref{p:aleq0}(3)
we also have $x \not\in\FV{ \eta y_1. \phi'_1}$ and thus
$\eta y_0. \phi'_0 \eqal  \eta y_1. \phi'_1$. 

{\it Case} $x \in \FV{\eta y_0. \phi'_0}$ and thus  by item~\ref{p:aleq0}(3) also
 $x \in\FV{ \eta y_1. \phi'_1}$. Then $y_0 \not= x$ and $y_1 \not= x$ and therefore
 we get $\eta y_0. (\phi'_0 [z/x])
\eqal \eta y_1. (\phi'_1 [z/x])$.
By definition of $\eqal$ this implies $\phi'_0 [z/x][u/y_0] \eqal 
\phi'_1 [z/x][u/y_1]$ for some fresh variable $u$.
As $u$ and $z$ are fresh this implies 
$\phi'_0 [u/y_0][z/x] \eqal \phi'_1 [u/y_1][z/x]$ and thus by the
induction hypothesis that
$\phi'_0 [u/y_0] \eqal \phi'_1 [u/y_1]$.
Applying the definition of $\eqal$ we conclude that
$\eta y_0. \phi'_0 \eqal \eta y_1. \phi'_1$ as required. 
This finishes the proof of item~\ref{aleq:3}.
\smallskip

For part~\ref{aleq:4a} assume that $\eta x. \phi_0 \odot \phi_1 \eqal \eta y. \psi_0 \odot
\psi_1$.
We leave the case where both formulas are syntactically equal as an exercise, 
and assume otherwise.
By the definition of $\eqal$ this implies that 
$(\phi_0 \odot \phi_1)[z/x] \eqal (\psi_0 \odot \psi_1) [z/y]$ for a fresh 
variable $z$.
This in turn yields $\phi_i[z/x] \eqal \psi_i[z/x]$ and thus
$\eta x. \phi_i \eqal \eta y. \psi_i$ for $i \in \{ 0,1\}$. 

Part~\ref{aleq:4b} can be proven in the same way as \ref{aleq:4a}.
For part~\ref{aleq:4c}, take a fresh variable $v$, then by item~\ref{aleq:2} we find 
$\la z.\phi[v/x] \eqal \la z. \psi[v/y]$.
By item~\ref{aleq:3} this implies $\phi[v/x] \eqal \psi[v/y]$, so that by the definition
of $\eqal$ we may conclude that $\eta x.\phi \eqal \eta y. \phi$.
\smallskip

For part~\ref{aleq:5} by assumption and item~\ref{aleq:fresh} we have $\phi[z/x] \eqal \psi[z/x]$ for
any variable $z$ that is fresh for $\phi$ and $\psi$. We choose such a $z$ that satisfies in addition that
$z \not = y$.
Furthermore, as $y \not\in \FV{\psi}$ and $y$ free for $x$ in $\psi$, we have $\psi[z/x] \eqal 
\psi [y/x][z/y]$. The latter can be easily shown by induction on $\psi$.  
Therefore we have $\phi[z/x] \eqal \psi [y/x][z/y]$ and thus we obtain 
$\eta x. \phi \eqal \eta y.\psi[y/x]$ by definition of $\eqal$. 
\smallskip

Regarding part~\ref{aleq:6}, we show that $\phi_0 \eqal \phi_1$ and $\psi_0 
\eqal \psi_1$ with $\psi_i$ free for $x$ in $\phi_i$ for $i \in \{0,1\}$ implies 
$\phi_0[\psi_0/x] \eqal \phi_1[\psi_1/x]$. 
We prove the claim by induction on the length of $\phi_0$, and omit the rather 
obvious base step, and the boolean and modal cases of the induction step.

This leaves the inductive case where $\phi_0 = \eta z_0. \phi'_0$ and, 
consequently, $\phi_1$ is of the form $\phi_1 = \eta z_1. \phi'_1$.
We only consider the situation where $x$ actually occurs freely in $\phi_{0}$.
Then $(\eta z_i. \phi'_i) [\psi_i/x] = \eta z_i. (\phi'_i [\psi_i/x])$ 
by definition of substitution.
Furthermore, by definition of $\eqal$ we have  $\phi'_0 [y/z_0] \eqal 
\phi'_1 [y/z_1]$ for some fresh variable $y$.
This implies 
\begin{align*}
\phi'_0 [\psi_0/x] [y/z_0] 
   & = \phi'_0 [y/z_0] [\psi_0/x] & \text{($y$ fresh, $\psi_0$ free for $x$ in $\phi_0'$)}
\\ & \eqal \phi'_1 [y/z_1] [\psi_1/x] 
   & \text{(induction hypothesis)}
\\ & = \phi'_{1} [\psi_{1}/x] [y/z_{1}] & \text{($y$ fresh, $\psi_1$ free for $x$ in $\phi_1'$)} 
\end{align*}
From this we obtain $\phi_{0}[\psi_{0}/x] \eqal \phi_{1}[\psi_{1}/x]$ 
by definition of $\eqal$.
This finishes the proof of~\ref{aleq:6}.
\smallskip

For part~\ref{aleq:7}, note that tidiness of the formulas implies that $\eta x_i \phi_i$ 
is free for $x_i$ in $\phi_i$. 
The claim then follows from item~\ref{aleq:6}.
\smallskip

Part~\ref{aleq:8} follows by the definition of $\eqal$ as the assumption implies
that $\phi_0[z/x] \eqal \phi_1[z/x]$ for some fresh variable $z$ by item~\ref{aleq:fresh}, and thus
$\eta x. \phi_0 \eqal \eta x. \phi_1$ by definition.
\smallskip

Finally, part~\ref{aleq:12} is proved by induction on the length of $\psi$.
In the base step of the induction, we distinguish cases. 
If $\psi = x$, then we have $\psi[\chi/x] = \chi$, so that we may take 
$\psi' \isdef x'$ (for some arbitrary fresh variable $x'$) and $\chi' \isdef 
\chi$.
If $\psi = y$ for some variable distinct from $x$, then we have $\psi[\chi/x]
= \psi$, so that we can take $\psi' \isdef y$ and $\chi' \isdef \chi$, for some
fresh $x'$.

We leave the inductive cases where $\psi$ is of the form $\psi = \psi_{0} \odot 
\psi_{1}$ with $\odot \in \{ \land, \lor\}$, of the form $\psi = \hs\psi_{0}$ 
with $\hs \in \{ \Box, \dia \}$, or of the form $\psi = \eta x. \rho$, as 
exercises. 

The key case of the inductive step is where $\psi$ is of the form $\psi = 
\eta y. \rho$ with $y \neq x$.
If $x$ is not free in $\psi$, we have $\psi[\chi/x] = \psi$, so that we may take
some arbitrary fresh variable $x'$, and define $\psi' \isdef \psi$ and $\chi' 
\isdef \chi$; then obviously we have $\phi = \psi'[\chi'/x']$ (since $x'$ is 
not free in $\psi'$), and furthermore $\psi[x/x'] = \psi = \phi \eqal 
\psi[\chi/x] = \si$, as required.

Hence, from now on we assume that $x$ does occur freely in $\psi$.
Note that since $y \neq x$ we have $\psi[\chi/x] = \eta y. \rho[\chi/x]$.
It follows from part~\ref{aleq:1} and~\ref{aleq:2} of this proposition
that $\phi$ must be of the form $\phi = \eta y'. \si$, where $\si[z/y'] \eqal 
\rho[\chi/x][z/y]$ for any fresh variable $z$.
Now observe that $y$ cannot be a free variable of $\chi$ since this 
would contradict the fact that $\chi$ is free for $x$ in $\psi$.
Because of this we find that $\rho[\chi/x][z/y] = \rho[z/y][\chi/x]$, so that 
we may apply the induction hypothesis to the $\al$-equivalence $\si[z/y'] \eqal
\rho[z/y][\chi/x]$.

This yields formulas $\rho',\chi'$ and a fresh variable $x'$ such that 
$\si[z/y'] = \rho'[\chi'/x']$, $\rho[z/y] \eqal \rho'[x/x']$ and $\chi' \eqal
\chi$.
Now observe that 
\begin{align*}
\si
   & = (\si[z/y'])[y'/z]
   & \text{($z$ fresh for $\si$)}
\\ & = (\rho'[\chi'/x'])[y'/z]
   & \text{(induction hypothesis)}
\\ & = (\rho'[y'/z])[\chi'/x']
   & (z \not\in \FV{\chi'}) 
\end{align*}
where we have $z \not\in \FV{\chi'}$ since $z$ is fresh for $\chi$ and $\chi 
\eqal \chi'$ (this is in fact Proposition~\ref{p:aleq0}(\ref{it:aleq0-3} which
can be proved by a straightforward induction).

The identity $\si = (\rho'[y'/z])[\chi'/x']$ above suggests that we define 
$\psi' \isdef \eta y'. \rho'[y'/z]$, so that we obtain $\phi' = \eta y'. \si = 
\psi'[\chi'/x']$ without effort.
It is left to prove that $\psi \eqal \psi'[x/x']$.

For this purpose, first observe that since $\si[z/y'] = \rho'[\chi'/x']$ and 
$z$ is fresh for $\si$, it is not hard to see that $y'$ is free for $z$ in 
$\rho'$ and that $z$ is fresh for $\rho'[y'/z]$.
From this it easily follows that $y'$ is free for $z$ in $\rho'[x/x'][y'/z]$,
and that $z$ is fresh for this formula as well..
Further reasoning about the syntax tree of $\rho'$ will then reveal that 
if we resubstitute $z$ for $y'$ in $\rho'[x/x'][y'/z]$, we obtain the formula 
$\rho'[x/x']$ back, that is: $\rho'[x/x'] = \rho'[x/x'][y'/z][z/y']$. 
Clearly then we have $\rho'[x/x'] = \rho'[y'/z][x/x'][z/y']$.
It then follows from $\rho[z/y] \eqal \rho'[x/x']$ (which is part of the
inductive hypothesis) that $\rho[z/y] \eqal \rho'[y'/z][x/x'][z/y']$.
Furthermore, we already saw that $z$ is fresh for $\rho'[y'/z][x/x']$, and since
$z$ is also fresh for $\rho$, it follows by clause 3) of the definition of 
$\al$-equivalence that $\eta y. \rho \eqal \eta y'. 
\rho'[y'/z][x/x']$.
But this just means that $\psi \eqal \psi'[x/x']$ as required.
\end{proof}

\begin{proofof}{Proposition~\ref{p:aleq0}}
The first four items have rather straightforward proofs, which we leave as an
exercise to the reader.

For the fifth item, it suffices to show that, for all formulas 
$\phi,\phi'$, $\eta \in \{ \mu, \nu\}$ and natural numbers $n$, it holds that
\begin{equation}
\label{eq:adaleq}
\phi \eqal \phi' \text{ and } \phi \in \AH{\eta}{n}
\text{ imply that } \phi' \in \AH{\eta}{n}.
\end{equation}
We prove this statement by induction on the derivation that $\phi \in 
\AH{\eta}{n}$.
Leaving the other cases as (relatively easy) exercises for the reader, we focus
on the case where $\phi \in \AH{\eta}{n}$ because of the substitution rule 
(i.e., clause~\eqref{adr:4} of Definition~\ref{d:ad}).
That is, $\phi$ is of the form $\phi = \psi[\chi/x]$ for some formulas 
$\psi,\chi$ in $\AH{\eta}{n}$.
Then by Proposition~\ref{p:aleq1}(\ref{aleq:12} we find formulas $\psi', \chi'$
and a variable $x'$ such that $\phi' = \psi'[\chi'/x']$, $\psi \eqal 
\psi'[x/x']$ and $\chi \eqal \chi'$.
But then it follows by the induction hypothesis that both $\psi'[x/x']$ and
$\chi'$ belong to $\AH{\eta}{n}$.
From this we easily obtain that (first $\psi'$ and then) $\phi' = 
\psi'[\chi'/x']$ belongs to $\AH{\eta}{n}$.
\end{proofof}

The following proposition shows that, although $\al$-equivalent formulas have
the same \emph{length}, their \emph{sizes} may differ exponentially.
This shows in particular that both size measures, $\ssz{\cdot}$ and $\csz{\cdot}$,
fail to meet the requirement (\ddag a) of invariance under $\al$-equivalence.

\begin{proposition}
There are sequences $(\xi_{n})_{n\in\om}$ and $(\chi_{n})_{n\in\om}$ 
of clean $\mu$-calculus formulas such that
$\xi_{n} \eqal \chi_{n}$, but 
we have $\ssz{\xi_{n}} = n+3$ while $\ssz{\chi_{n}} = 2^{n+2}-1$, and 
we have $\csz{\xi_{n}} = n+2$ while $\csz{\chi_{n}} = 2^{n+2}-2$,
for each $n$.
\end{proposition}

\begin{proof}
We define the $\chi$-formulas inductively by putting $\chi_{0} \isdef \nu x. 
\Box x$, and $\chi_{n+1} \isdef (\chi_{n} \land \chi_{n})$.
For $\xi_{n}$ then we take an alphabetical variant of $\chi_{n}$ in which
every occurrence of a subformula $\nu x\, \Box x$ is replaced with a different 
formula $\nu y\, \Box y$.
Formally, we may define, for $n \in \om$, $\si \in \{0,1\}^{*}$ and a family 
$\{ x_{\si} \mid \si \in \{0,1\}^{*} \}$ of distinct variables:
$\phi_{\si,0} \isdef \nu x_{\si}\, \Box x_{\si}$ and 
$\phi_{\si,n+1} \isdef (\phi_{\si0,n} \land \phi_{\si1,n})$, and then set
$\xi_{n} \isdef \phi_{\epsilon,n}$ (where $\epsilon$ denotes the empty string).
With these definitions, it is straightforward to verify that the sequences 
$(\xi_{n})_{n\in\om}$ and $(\chi_{n})_{n\in\om}$ meet the requirements.
\end{proof}

A less obvious but much more significant observation concerns the cost of 
renaming. 
In the literature on the modal $\mu$-calculus one often works with clean 
formulas; this assumption is not a big issue if one is mainly interested in 
logical aspects of the formalism, but it may become problematic if the main 
concern is complexity theoretic.
The point is that an algorithm which works on clean formulas, is only
guaranteed to work on arbitrary formulas after a preprocessing step in which 
the input formula is renamed into some clean alphabetical variant.
As the following proposition shows, for some formulas such a renaming incurs
an unnecessary exponential blow-up.
In short:
if one defines the size of a formula in terms of its closure set, then 
renaming tidy formulas into clean alphabetical variants comes at a rather high
cost.

\begin{proposition}
\label{p:closexp}
There is a family $(\xi_{n})_{n\in\om}$ of tidy formulas such that $\csz{\xi_{n}}
\leq 2\cdot n$, while for any sequence of clean formulas $\chi_{n}$ such that 
$\xi_{n} \eqal \chi_{n}$ for all $n$, we have $\csz{\chi_{n}} \geq 2^{n}$.
\end{proposition}

\begin{proof}
Let $\xi_{n}$ be the formula $\tr_{\bbF}(v_{0})$, where $\bbF$ is the parity
formula of Example~\ref{ex:bfl} (note that in this Example the parameter $n$ is 
left implicit for the sake of simple notation).
Recall that we showed that the fixpoint depth of $\xi_{n}$ is at least $2^{n}$.
Now let $\chi_{n}$ be a sequence of formulas that are one by one $\al$-equivalent
to the formulas $\xi_{n}$. 
It is readily seen that $\fdep{\chi_{n}} = \fdep{\xi_{n}} \geq 2^{n}$.

However, we also claim that 
\begin{equation}
\label{eq:clx1}
\text{every clean $\mu$-calculus formula $\chi$ satisfies $\csz{\chi} \geq 
\fdep{\chi}$}.
\end{equation}
For a proof of this statement,
it is not hard to show that for any subformula $\eta x. \phi \sforeq \chi$, the 
closure of $\chi$ contains a formula of the form $\eta x. \phi'$.
From this it follows $\csz{\chi} = \size{\Clos(\chi)} \geq \size{\BV{\chi}}$.
But if $\chi$ is a formula of fixpoint depth $k$, then there is a chain of 
subformulas $\eta_{1} x_{1}. \phi_{1} \sforeq \eta_{2} x_{2}. \phi_{2} \sforeq
\cdots \sforeq \eta_{k} x_{k}. \phi_{k}$, and if $\chi$ is \emph{clean}, then all
these variables must be distinct.
This shows that $\size{\BV{\chi}} \geq \fdep{\chi}$.
Combining these observations, we see that $\csz{\chi} \geq \fdep{\chi}$ indeed.

Obviously then, the proposition follows by \eqref{eq:clx1} and the earlier
observation that $\fdep{\chi_{n}} \geq 2^{n}$.
\end{proof}

\subsection*{Subformulas}

In case we define size in terms of number of subformulas, and we want to count
alphabetical variants only once, it makes sense to focus on formulas for which
$\alpha$-equivalence is the identity relation on its collection of 
subformulas.

\begin{definition}
We call a set of formulas $\Phi$ \emph{polished} if $\phi_{0} \eqal \phi_{1}$
implies $\phi_{0} = \phi_{1}$, for any pair $\phi_{0},\phi_{1}$ 
in $\bigcup_{\phi\in\Phi}\Sfor(\phi)$.
We call a formula $\xi$ \emph{polished} if the singleton set $\{ \xi \}$ is
polished.
\end{definition}

Intuitively, if $\chi$ is a (clean and) polished alphabetical variant of some
clean formula $\xi$, then $\chi$ should not have more subformulas than $\xi$.
This intuition is confirmed by the following proposition.

\begin{proposition}
\label{p:alsz}
Let $\xi$ and $\chi$ be $\al$-equivalent, clean $\muML$-formulas.
If $\chi$ is polished then $\size{\Sfor(\chi)} \leq \size{\Sfor(\xi)}$.
\end{proposition}

\begin{proof}
We confine ourselves to a sketch.
By induction on the derivation that $\phi \in \Sfor(\xi)$ (that is, using the 
definition of subformulas in terms of direct subformulas), we will (1) define a 
formula $r(\phi) \in \Sfor(\chi)$, (2) define a renaming $\rho_{\phi}: 
(\FV{\phi} \cap \BV{\xi}) \to \BV{\chi}$ and show that (3) $r(\phi) \eqal 
\phi[\rho_{\phi}]$.
Since it will be clear from our construction that this defines a surjection 
$r: \Sfor(\xi) \to \Sfor(\chi)$, this suffices to prove the proposition.

For the base step we define $r(\xi) \isdef \chi$, and let $\rho_{\xi}$ be the
empty map. 
Then (3) holds by assumption.

In the inductive step we are dealing with a proper subformula $\phi$ of $\xi$,
and make a case distinction as to the reason why $\phi$ is a subformula of
$\xi$.

\textit{(Boolean case)} There is a subformula $\psi = \psi_{0} \odot \psi_{1}$
of $\xi$ such that $\odot \in \{ \lor, \land \}$ and $\phi \in \{\psi_{0},
\psi_{1}\}$; without loss of generality assume that $\phi = \psi_{0}$.
By the induction hypothesis we have that $r(\psi) \eqal \psi[\rho_{\psi}]$.
From this it is easy to see that $r(\psi) \in \Sfor(\chi)$ is of the form 
$r(\psi) = \chi_{0} \odot \chi_{1}$.
Now put $r(\phi) \isdef \chi_{0}$, and define $\rho_{\phi}$ as the restriction 
of $\rho_{\psi}$ to the set $\FV{\phi} \cap \BV{\xi}$.
It is then straightforward to verify (3).

\textit{(Modal case)} There is a subformula $\psi = \hs\phi$ of $\xi$, with 
$\hs \in \{ \dia, \Box\}$.
This case can be dealt with similarly as the previous one.

\textit{(Fixpoint case)}
There is a subformula of the form $\eta x. \phi$ of $\xi$, for which we 
inductively have defined $r(\eta x.\phi) \sforeq \chi$ and $\rho = 
\rho_{\eta x.\phi}$ such that $r(\eta x. \phi)  \eqal (\eta x.\phi)[\rho]$.
It follows that $r(\eta x. \phi)$ must be of the form $\eta z. \psi$, 
with $\psi \sforeq \chi$.
Now define $r(\phi) \isdef \psi$, and $\rho_{\phi} \isdef \rho \cup \{ (x,z)\}$,
which is a well-defined function by cleanness.
Furthermore we can prove that $r(\phi) = \psi \eqal \phi[\rho][z/x] = 
\phi[\rho_{\phi}]$, as required.
\end{proof}

As we will see now, every formula $\xi$ can be transformed into an 
$\al$-equivalent clean, polished formula $\pol{\xi}$.
For this purpose it will be convenient to introduce a set $Z$ of fresh variables
from which we will draw the bound variables of the formulas $\pol{\xi}$.

\begin{definition}
\label{d:XZ}
Let $X$ and $Z$ be two (disjoint) sets of variables.
We let $\muML_{X}$ denote the set of $\mu$-calculus formulas taking their
variables (free or bound) from $X$, and we let $\muML_{X,Z}$ denote the set of 
formulas $\xi$ in $\muML_{X\cup Z}$ such that $\BV{\xi} \sse X$.
\end{definition}

In the sequel we will want to associate a variable $z_{E}$ with each 
$\al$-equivalence class $E$.
We will need that $z_{E}$ is not free in any formula belonging to $E$.

\begin{proposition}
\label{p:z}
Let $X$ and $Z$ be countable disjoint sets of variables.
Then there is a map $z: \muML_{X,Z}/{\eqal} \to Z$ such that for all $\phi \in
\muML_{X,Z}$ we have that $z_{\eqalc{\phi}} \not\in \FV{\phi}$.
\end{proposition}

\begin{proof}
Partition $Z$ into a countable collection of countable sets as $Z = 
\bigcup_{i\in\om}Z_{i}$, and consider a bijection $f: \funPom(X \cup Z) \to 
\om$ such that $Q \cap Z_{f(Q)} = \nada$, for any $Q \in \funPom(X \cup Z)$.
Then consider, for any $Q \in \funPom(X \cup Z)$, a bijection $z^{Q}$ between the
countable sets $\{ \phi \in \muML_{X,Z} \mid \FV{\phi} = Q \}/{\eqal}$ and 
$Z_{f(Q)}$.
Finally, given $\phi \in \muML_{X,Z}$, define
\[
z_{\eqalc{\phi}} \isdef z^{\FV{\phi}}(\eqalc{\phi}).
\]
This map is well-defined by Proposition~\ref{p:aleq0}(3), and it is straightforward 
to verify that it has the required properties.
\end{proof}

\begin{definition}\label{def:polishing}
Fix a map $z: \muML_{X,Z}/{\eqal} \to Z$ satisfying the conditions of 
Proposition~\ref{p:z}.
We define the \emph{polishing map} $\pol{-}: \muML_{X,Z} \to \muML_{Z,X}$ 
by the following induction: 
\[\begin{array}{llll}
   \pol{\phi}        & \isdef & \phi
  & \text{if $\phi$ is atomic} 
\\ \pol{\hs\phi}     & \isdef & \hs \pol{\phi} 
  & \text{where } \hs \in \{ \dia, \Box \}
\\ \pol{\phi_{0}\odot\phi_{1}} & \isdef & \pol{\phi_{0}}\odot\pol{\phi_{1}} 
  & \text{where } \odot \in \{ \lor, \land \}
\\ \pol{\eta x.\phi} & \isdef & \eta z_{E}. \pol{\phi[z_{E}/x]}
  & \text{where } \eta \in \{ \mu, \nu \} \text{ and } E = \eqalc{\eta x.\phi}.
\end{array}\]
Furthermore, for a set of formulas $\Phi \sse \muML_{X,Z}$ we define $\pol{\Phi} 
\isdef \{ \pol{\phi} \mid \phi \in \Phi \}$.
\end{definition}

We first make some straightforward observations about this definition.

\begin{proposition}
\label{p:rens1}
Let $\phi,\xi$ be formula in $\muML_{X,Z}$. Then

1) $\len{\xi} = \len{\pol{\xi}}$;

2) $z_{\eqalc{\xi}}$ is fresh for $\pol{\phi}$ whenever $\len{\phi} < \len{\xi}$;

3) if $z$ is fresh for $\xi$ and $\pol{\xi}$, then $\pol{\xi[z/y]} \eqal
   \pol{\xi}[z/y]$.
\end{proposition}

\begin{proof}
Part 1) of the proposition follows directly from the definition of the polishing
map.

For part 2), if $\phi$ is shorter than $\xi$, then no subformula of $\phi$ can
be $\al$-equivalent to $\xi$.
From this it is easy to see that $z_{\eqalc{\xi}}$ does not appear in 
$\pol{\phi}$.

Part 3) is proved by induction on the length of $\xi$.
The only case of interest is where $\xi$ is of the form $\eta x. \chi$ for some 
$\eta \in \{ \mu, \nu \}$ and $x \neq y$.
Consider the following calculation, where $E = \eqalc{\xi}$ and $E' = 
\eqalc{\xi[z/y]}$, and where part 2) of the proposition is used throughout:
\begin{align*}
\pol{\eta x. \chi[z/y]}
   & = \eta z_{E'}. \pol{\chi[z/y][z_{E'}/x]}
   & \text{(definition $\pol{-}$)}
\\ & \eqal \eta z_{E'}. \pol{\chi[z/y]}[z_{E'}/x]
   & \text{(IH, Prop.~\ref{p:aleq1}(\ref{aleq:8}))}
\\ & \eqal \eta z_{E}. \pol{\chi[z/y]}[z_{E}/x]
   & \text{(definition ${\eqal}$, 
       Prop.~\ref{p:aleq1}(\ref{aleq:6},\ref{aleq:8}))}
\\ & \eqal \eta z_{E}. \pol{\chi[z/y][z_{E}/x]}
   & \text{(IH, Prop.~\ref{p:aleq1}(\ref{aleq:8}))}
\\ & = \eta z_{E}. \pol{\chi[z_{E}/x][z/y]}
   & \text{($x \neq y$, freshness $z,z_{E}$)}
\\ & \eqal \eta z_{E}. \pol{\chi[z_{E}/x]}[z/y]
   & \text{(IH, Prop.~\ref{p:aleq1}(\ref{aleq:8}))}
\\ & = \big(\eta z_{E}. \pol{\chi[z_{E}/x]}\big)[z/y]
   & \text{(definition substitution)}
\\ & = \big(\pol{\eta x. \chi}\big)[z/y]
   & \text{(definition $\pol{-}$)}
\end{align*}
Clearly this takes care of the inductive case where $\xi = \eta x. \chi$.
\end{proof}

The following proposition can be read as stating that $\pol{\xi}$ is a canonical
representative of the $\al$-equivalence class of $\xi$.

\begin{proposition}
\label{p:rens2}
Let $\xi, \xi_{0}$ and $\xi_{1}$ be formulas in $\muML_{X,Z}$.
Then

1) $\xi \eqal \pol{\xi}$;

2) $\xi_{0} \eqal \xi_{1}$ iff $\pol{\xi_{0}} = \pol{\xi_{1}}$.
\end{proposition}

\begin{proof}
For part 1) we proceed by induction on the complexity of the formula $\xi$;
for the key inductive case, where $\xi$ is of the form $\xi = \eta x. \chi$, we
reason as follows, where we let $E \isdef \eqalc{\xi}$.
By Proposition~\ref{p:rens1}(2) the variable $z_{E}$ is fresh for 
$\pol{\chi}$, so that we may use part 3) of the same Proposition to find that 
\[
\pol{\xi} = \eta z_{E}. \pol{\chi[z_{E}/x]} \eqal \eta z_{E}. \pol{\chi}[z_{E}/x].
\]
By induction hypothesis we have $\chi \eqal \pol{\chi}$, so that by
Proposition~\ref{p:aleq1}(\ref{aleq:6},\ref{aleq:8}) we may conclude that 
\[
\eta z_{E}. \pol{\chi}[z_{E}/x] \eqal \eta z_{E}. \chi[z_{E}/x].
\]
Finally, by Proposition~\ref{p:aleq1}(\ref{aleq:5}) we see that 
\[
\eta z_{E}. \chi[z_{E}/x] \eqal \eta x. \chi = \xi.
\]
By the three equations displayed above we then find that $\xi \eqal \pol{\xi}$,
as required.
\medskip

Turning to the second part of the Proposition, it is obvious that the implication 
`$\Leftarrow$' follows from part 1).
For the opposite direction, we proceed by induction 
on the derivation of $\xi_{0} \eqal \xi_{1}$.
The crucial case is where $\xi_{i} = \eta x_{i}. \chi_{i}$ for $i = 0,1$, and 
$\xi_{0} \eqal \xi_{1}$ because $\chi_{0}[z/x_{0}] \eqal \chi_{1}[z/x_{1}]$
for some fresh variable $z$.
Let $E$ denote the equivalence class $E = \eqalc{\xi_{0}} = \eqalc{\xi_{1}}$, 
then by Proposition~\ref{p:aleq1}(\ref{aleq:6}) we may infer that $\chi_{0}[z_{E}/x_{0}]
= \chi_{0}[z/x_{0}][z_{E}/z] \eqal \chi_{1}[z/x_{1}][z_{E}/z]
= \chi_{1}[z_{E}/x_{1}]$.
From this it follows by the induction hypothesis that 
$\pol{\chi_{0}[z_{E}/x_{0}]} \eqal \pol{\chi_{1}[z_{E}/x_{1}]}$. 
But then by Proposition~\ref{p:aleq1}(\ref{aleq:8}) we obtain that $\pol{\xi_{0}} = 
\eta z_{E}. \pol{\chi_{0}[z_{E}/x_{0}]} \eqal 
\eta z_{E}. \pol{\chi_{1}[z_{E}/x_{1}]} = \pol{\xi_{1}}$.
\end{proof}

\begin{proposition}
\label{p:polcl}
Every formula of the form $\pol{\xi}$ is clean.
\end{proposition}

\begin{proof}
It suffices to show, for every formula $\xi$, that
\begin{equation}
\label{eq:polsf}
\eta x.\phi \sforeq \pol{\xi} \text{ implies } x = z_{\eqalc{\eta x.
\phi}}.
\end{equation}

We will prove \eqref{eq:polsf} by induction on the length of $\xi$, and only
consider the inductive case where $\xi$ is of the form $\xi = \la y. \chi$.
Writing $E = \eqalc{\xi}$, we have $\pol{\xi} = \lambda z_{E}. 
\pol{\chi[z_{E}/y]}$.
We make a case distinction here.
In case $\eta x. \phi$ is actually the formula $\pol{\xi}$ itself, we find
in particular that $\eta = \la$ and $x = z_{E}$.
Furthermore, it follows from $E = \eqalc{\xi} = \eqalc{\eta z. \phi}$ that 
$z_{E} = z_{\eqalc{\eta x. \phi}}$, so that we conclude that $x =
z_{\eqalc{\eta x. \phi}}$, as required.
Otherwise, if $\eta x.\phi$ is a \emph{proper} subformula of $\pol{\xi}$ it must 
be the case that $\eta x. \phi \sforeq \pol{\chi[z_{E}/y]}$, so that we find 
$z = z_{\eqalc{\eta x. \phi}}$ by applying the induction hypothesis to the 
formula $\chi[z_{E}/x]$.
\end{proof}

\begin{proposition}
\label{p:polpol}
Every formula of the form $\pol{\xi}$ is polished.
\end{proposition}

\begin{proof}
In fact, we will obtain the proposition as a special case of the following 
slightly more general statement:
\begin{equation}
\label{eq:pol}
\pol{\Phi} \text{ is polished, for any set of formulas } \Phi \sse \muML_{X}.
\end{equation}
We prove \eqref{eq:pol} by induction on $c(\Phi) \isdef \sum \big\{ \len{\phi}
\mid \phi\in\Phi,\; \phi \text{ non-atomic }\}$.
In the base step of this induction, where $c(\Phi) = 0$, all formulas in $\Phi$
are atomic, and so the statement is immediate.
In the inductive step we assume that $c(\Phi) > 0$, and we make a case 
distinction.
\medskip

\noindent
\textit{Case 1} $\Phi$ contains two distinct non-atomic formulas $\phi$ and 
$\psi$ such that $\pol{\phi} \sforeq \pol{\psi}$.

This case is easily dealt with by applying the inductive hypothesis to the 
set $\Psi \isdef \Phi \setminus \{ \phi \}$, since the case assumption implies 
$\bigcup_{\chi\in\Phi} \Sfor(\pol{\chi}) = \bigcup_{\chi\in\Psi} 
\Sfor(\pol{\chi})$.
\medskip

\noindent
\textit{Case 2} $\Phi$ contains a formula of the form $\hs\phi$ or 
$\phi_{0}\odot\phi_{1}$, with $\hs \in \{ \dia, \Box \}$, $\odot \in 
\{ \lor, \land \}$, while Case 1 does not apply.

Here we only consider the subcase where $\Phi$ contains a formula of the form 
$\phi = \phi_{0}\odot\phi_{1}$, the subcase involving modalities being very 
similar.
Define $\Psi \isdef \Phi \setminus \{ \phi \}$, then clearly we find
\[
\Sfor(\pol{\Phi}) = 
\Sfor\big( \pol{\Psi \cup \{ \phi_{0},\phi_{1} \}} \big)
\cup \{ \pol{\phi} \},
\]
where the set $\Sfor\big( \pol{\Psi \cup \{ \phi_{0},\phi_{1} \}} \big)$ is
polished by the induction hypothesis.
Hence, in order to check that the set $\Sfor(\pol{\Phi})$ is polished, we only
need to consider alphabetical variants of the formula $\pol{\phi} = 
\pol{\phi_{0}} \odot \pol{\phi_{1}}$ in the set 
$\Sfor\big( \pol{\Psi \cup \{ \phi_{0},\phi_{1} \}} \big)$.
It is easy to see that such alphabetical variants must be of the form $\chi =
\chi_{0}\odot\chi_{1}$, with $\chi \in \Sfor(\pol{\Psi})$, and $\chi_{i} \eqal
\pol{\phi_{i}}$ for $i = 0,1$.
Applying the induction hypothesis to the set $\Psi \cup \{ \phi_{0}, \phi_{1}\}$
we find that $\chi_{i} = \pol{\phi_{i}}$ for $i = 0,1$.
This means that $\chi = \pol{\phi}$, showing that $\Sfor(\pol{\Phi})$ is polished
as required.
\medskip

\noindent
\textit{Case 3} All non-atomic formulas in $\Phi$ are fixpoint formulas, while 
Case 1 does not apply.

Suppose for contradiction that the set $\Sfor(\pol{\Phi})$ is not polished.
Let $\eta x. \phi$ be a fixpoint formula in $\Phi$ of maximal length, let $E$ 
be its $\al$-equivalence class, and define $\Psi \isdef \Phi \setminus \{ \eta 
x. \phi \}$.
First observe that since Case 1 does not apply, $\eta x. \phi$ is the only 
formula in $\Phi$ whose polished version is of the form $\eta z_{E}. \chi$.
It then follows from Proposition~\ref{p:rens1}(2) that $z_{E}$ is fresh for 
every formula $\pol{\psi}$ with $\psi \in \Psi$.
Now compute
\begin{align*}
\Sfor(\pol{\Phi}) & = 
   \Sfor(\pol{\Psi}) \cup \Sfor(\eta z_{E}. \pol{\phi[z_{E}/x]})
\\ & = \{ \eta z_{E}. \pol{\phi[z_{E}/x]} \} \cup
    \Sfor\big(\pol{\Psi \cup \{ \phi[z_{E}/x] \}}\big)
\end{align*}
By the inductive hypothesis, the set $\pol{\Psi \cup 
\{ \phi[z_{E}/x] \}}$ is polished.
But then by our assumption that $\Sfor(\pol{\Phi})$ is not polished, there must 
be a formula $\chi \in \Sfor\big(\pol{\Psi \cup \{ \phi[z_{E}/x] \}}\big)$ such 
that $\chi$ and $\eta z_{E}. \pol{\phi[z_{E}/x]}$ are distinct but 
$\al$-equivalent.
Clearly then $\chi \eqal \eta x. \phi$, so that $\len{\chi} = \len{\eta x. \phi}$
by Proposition~\ref{p:aleq0}(2).
But by our choice of $\eta x.\phi$ as a formula of maximal length in $\Phi$,
and the fact that our polishing operation $\pol{-}$ is length-preserving, it
cannot be the case that $\chi = \phi[z_{E}/x]$.
Hence, $\chi$ actually belongs to the set $\pol{\Psi}$.
That is, $\chi$ is of the form $\chi = \pol{\psi}$ with $\psi \in \Psi$.
This, however, implies that $\eta x.\phi \eqal \pol{\eta x. \phi} \eqal \chi = 
\pol{\psi} \eqal \psi$, which means that, contrary to our assumption, we are
in Case 1 after all.
\end{proof}

On the basis of the above we may (and will) define the subformula size of an
arbitrary $\mu$-calculus formula $\xi$ as the subformula size (i.e., number of
subformulas) of the formula $\pol{\xi}$, see Definition~\ref{d:szal} below.

\subsection*{Closure}

In the case of closure size, one approach would be to define the closure size 
of an arbitrary formula as the size of its closure set \emph{modulo 
$\alpha$-equivalence}.
That this definition makes sense follows from the Proposition below, which
states that if we identify $\alpha$-equivalent formulas, the closure sets of 
$\alpha$-equivalent formulas have the same size.

\begin{proposition}
\label{p:aleq-cl}
Let $\xi_{0}$ and $\xi_{1}$ be tidy $\mu$-calculus formulas such that $\xi_{0} 
\eqal \xi_{1}$.
Then 

1) for every $\phi_{0} \in \Clos(\xi_{0})$ there is a $\phi_{1} \in 
\Clos(\xi_{1})$ such that $\phi_{0} \eqal \phi_{1}$, and vice versa;

2) as a corollary, $\size{\Clos(\xi_{0})/{\eqal}} = 
\size{\Clos(\xi_{1})/{\eqal}}$.
\end{proposition}

\begin{proof}
We prove part 1) of this proposition by induction on the derivation of the 
membership of $\phi_{0}$ in $\Clos(\xi_{0})$.
In the base case we have $\phi_{0} = \xi_{0}$, so that we may take $\phi_{1} 
\isdef \xi_{1}$.

In the inductive case we assume some formula $\psi_{0} \in 
\Clos(\xi_{0})$ for which
membership can be proved by a shorter derivation, and which is such that 
$\phi_{0}$ is either (1/2) a direct modal or boolean subformula of $\psi_{0}$
or else (3) $\psi_{0}$ is  a fixpoint formula $\eta x_{0}. \chi_{0}$ of which 
$\phi_{0}$ is the unfolding.
An instance of the first case is where $\psi_{0}$ is of the form $\dia\phi_{0}$.
By the induction hypothesis this formula has an alphabetical variant $\psi_{1}$ 
in the closure set of $\xi_{1}$; it is then easy to see that $\psi_{1}$ must 
be of the form $\dia\phi_{1}$ for some formula $\phi_{1}$.
But then it is immediate that $\phi_{1} \in \Clos(\xi_{1})$ and that $\phi_{1}
\eqal \phi_{0}$, as required.
The case where $\phi_{0}$ is a boolean subformula of $\psi_{0}$ is dealt with 
in a similar way, and in the third case we use Proposition~\ref{p:aleq1}(\ref{aleq:7}).

For part 2) of the proposition, observe that as an immediate consequence of 
part 1), we find a bijection between the sets of $\Clos(\xi_{0})/{\eqal}$ and 
$\Clos(\xi_{1})/{\eqal}$.
\end{proof}

Note however, that Proposition~\ref{p:aleq-cl} on its own is not enough to 
consider the proposed definition (viz., of closure-size as the size of the 
closure set modulo $\al$-equivalence) as a proper \emph{size measure}.
It is not a priori clear that the definition meets our
requirement (\dag) that any reasonable size measure should be based on some 
transformation of a $\mu$-calculus formula into an equivalent parity formula.
The following theorem shows that, similar to the case of 
subformula-size, we can define a \emph{renaming} map $\spol{\cdot}$ on 
$\mu$-calculus formulas, such that our proposed closure-size of a formula $\xi$ 
corresponds to the size of the parity graph that we may associate with the 
renaming $\spol{\xi}$ of the original formula $\xi$.
Intuitively, $\spol{\xi}$ can be seen as an alphabetical variant of $\xi$ that 
is \emph{minimal} in terms of closure-size.

\begin{theorem}
\label{t:aleqcl}
There is a map $\spol{\cdot}: \muML \to \muML$ such that, for every 
$\mu$-calculus formula $\xi$:

1) $\spol{\xi}$ is tidy and $\xi \eqal \spol{\xi}$;

2) $\al$-equivalence is the identity relation on the closure set 
  $\Clos(\spol{\xi})$;
  
3) $\csz{\spol{\xi}} = \size{\Clos(\xi)/{\eqal}}$.
\end{theorem}
Since the proof of this theorem requires some work, we defer the details to 
a separate section (viz., section~\ref{sec:skel}).

\subsection*{Size measures, invariant under $\alpha$-equivalence}

As one of the main contributions of this paper, we can now provide the 
definitions of two size measures, corresponding to, respectively, 
subformula-size and closure-size, that are both invariant under alphabetical 
equivalence.

\begin{definition}
\label{d:szal}
Let $\xi$ be an arbitrary $\mu$-calculus formula. 
By putting
\begin{eqnarray*}
   \sszal{\xi} & \isdef & \ssz{\pol{\xi}}
\\ \cszal{\xi} & \isdef & \csz{\spol{\xi}}
\end{eqnarray*}
we define, respectively the \emph{subformula-size} $\sszal{\xi}$ and the 
\emph{closure-size} $\cszal{\xi}$ of $\xi$.
\end{definition}

We claim that both of these definitions provide $\al$-invariant size measures 
indeed.

\begin{theorem}
The maps $\sszal{\cdot}$ and $\cszal{\cdot}$ provide size measures 
for $\mu$-calculus formulas that satisfy the conditions (\dag), (\ddag a) and
(\ddag b) from the introduction.
\end{theorem}

\begin{proof}
For subformula-size, fix a formula $\xi \in \muML$, let $\xi' \isdef\pol{\xi}$
be its polished renaming, and let $\bbH_{\xi}$ be the (untwisted) parity formula
associated with $\xi'$, as in Theorem~\ref{t:1}.
Then we find $\xi \equiv \bbH_{\xi}$ and $\sszal{\xi} = \size{\bbH_{\xi}}$ so 
that $\sszal{\cdot}$ qualifies as a size measure, in the sense that it 
satisfies condition (\dag) given in the introduction.

We now show that $\sszal{\cdot}$ is invariant under $\al$-equivalence.
Condition (\ddag b) holds rather trivially: by Proposition~\ref{p:polpol} $\xi'$
is polished, and so by definition, $\al$-equivalence is the identity relation 
on the set $\Sfor(\xi')$.
For (\ddag a), assume that $\xi_{0} \eqal \xi_{1}$.
It follows that $\xi_{0}' \eqal \xi_{1}'$ by Proposition~\ref{p:rens2} (and the 
fact that $\eqal$ is an equivalence relation).
But then we obtain $\ssz{\xi_{0}'} = \ssz{\xi_{1}'}$ by the 
Propositions~\ref{p:polpol} and~\ref{p:alsz}, and so we find 
$\sszal{\xi_{0}} = \sszal{\xi_{1}}$ by definition of $\sszal{\cdot}$.
\smallskip

Similarly, for closure-size, take a formula $\xi \in \muML$, and let $\wh{\xi}
\isdef \spol{\xi}$ be any formula satisfying the conditions listed in 
Theorem~\ref{t:aleqcl} (a concrete definition of such a function is given in 
Definition~\ref{d:skren}).
Furthermore, let $\bbG_{\xi}$ be the
parity formula associated with $\wh{\xi}$ as in Theorem~\ref{t:clur}.
Then we obtain $\xi \equiv \bbG_{\xi}$ and $\cszal{\xi} = \size{\bbG_{\xi}}$, 
showing that $\cszal{\cdot}$ satisfies condition (\dag).
Turning to the question whether $\cszal{\cdot}$ is invariant under $\eqal$, we
see immediately that condition (\ddag b) follows from Theorem~\ref{t:aleqcl}.
For (\ddag a), assume that $\xi_{0} \eqal \xi_{1}$, then we find 
$\csz{\wh{\xi_{0}}} = \csz{\wh{\xi_{1}}}$ by Proposition~\ref{p:aleq-cl}(2) and
Theorem~\ref{t:aleqcl}(3).
Clearly then we have $\cszal{\xi_{0}} = \cszal{\xi_{1}}$ by definition of 
$\cszal{\cdot}$
\end{proof}

Our earlier observations in Proposition~\ref{p:szbas} on the connection between
various size measures hold in this setting as well.

\begin{theorem}
1) Every formula $\xi \in \muML$ satisfies $\cszal{\xi} \leq \sszal{\xi} \leq 
\len{\xi}$.

2) 
There is a family of formulas $(\xi_{n})_{n\in\om}$ such that 
$\sszal{\xi_{n}} \leq n+1$ while $\len{\xi_{n}} \geq 2^{n}$, for each $n$.

3) 
There is a family of formulas $(\xi_{n})_{n\in\om}$ such that $\cszal{\xi_{n}} 
\leq 2 \cdot n$ while $\sszal{\xi_{n}} \geq 2^{n}$, for each $n$.
\end{theorem}

\begin{proof}
For part 1), take an arbitrary formula $\xi \in \muML$.
By definition then we have $\sszal{\xi} = \ssz{\xi'}$ for some clean (and 
polished) formula $\xi' \eqal \xi$.
It follows that $\cszal{\xi} = \cszal{\xi'}$ by Proposition~\ref{p:aleq-cl}, 
while $\xi'$ obviously, like any formula, satisfies $\cszal{\xi'} 
\leq \csz{\xi'}$.
Finally, we saw already in Proposition~\ref{p:szbas} that $\csz{\xi'} \leq
\ssz{\xi'}$.
Putting all these comparisons together, we find
\[
\cszal{\xi} = \cszal{\xi'} \leq \csz{\xi'} \leq \ssz{\xi'} = \sszal{\xi},
\]
which takes care of the first inequality of part 1).

For the second inequality, it suffices  to observe that 
\begin{align*}
\sszal{\xi} 
   & = \ssz{\pol{\xi}} 
   & \text{(Definition)}
\\ & \leq \len{\pol{\xi}}
   & \text{(Proposition~\ref{p:szbas})}
\\ & = \len{\xi}.
   & \text{(Proposition~\ref{p:aleq0})}
\end{align*}

For part 2) we can take the same formulas as in the proof of
Proposition~\ref{p:szbas}.

Finally, let $\xi_{n}$ be the translation of the parity formula $\bbF$ with 
$2 \cdot n$ vertices in Example~\ref{ex:bfl}.
We saw that the \emph{fixpoint depth} of these formulas grows exponentially:
$\fdep{\xi_{n}} \geq 2^{n}$.
It is not hard to see that fixpoint depth is invariant under $\al$-equivalence, 
so that we obtain $\sszal{\xi_{n}} = \ssz{\pol{\xi_{n}}} \geq 
\fdep{\pol{\xi_{n}}} = \fdep{\xi_{n}} \geq 2^{n}$.
Finally, it follows from Theorem~\ref{t:cyc-fix} that $\cszal{\xi_{n}} \leq 
\csz{\xi_{n}} \leq 2 \cdot n$.
\end{proof}

\subsection*{Substitution revisited}

As promised in section~\ref{sec:bas}, we will now provide a proper definition 
of the substitution operation $[\psi/x]$, which is applicable to formulas $\phi$
where $\psi$ is \emph{not} free for $x$, in a way that avoids variable capture.
Our approach here is completely standard.

\begin{definition}
\label{d:sbs}
Given two $\mu$-calculus formulas $\phi$ and $\psi$, we define
\[
\phi[\psi/x] \isdef 
\left\{ \begin{array}{ll}
   \phi[\psi/x] & \text{if $\psi$ is free for $x$ in $\phi$}
\\ \subren{\phi}{\psi}[\psi/x] & \text{otherwise}.
\end{array} \right.
\]
where we let $\subren{\phi}{\psi}$ be a canonically chosen alphabetical variant
of $\phi$ such that $\psi$ is free for $x$ in $\subren{\phi}{\psi}$.
\end{definition}

\begin{remark}
\label{r:sbs}
Here is a concrete proposal for a renaming map as mentioned in 
Definition~\ref{d:sbs}.
Fix an enumeration $(x_{n})_{n\in\om}$ of the set of variables, then for a
natural number $n$ we let $r_{n}: \muML \to \muML$ be the operation that 
replaces every binding and every bound occurrence of a variable $x_{k}$
in a formula $\xi$ by the variable $x_{k+n}$. 
As an example, we obtain 
$r_{3}\Big(\mu x_{1}\, \Box (x_{1} \lor x_{5}) \land 
   \mu x_{5}\, \Box (x_{1} \lor x_{5})\Big) =
r_{3}\Big(\mu x_{4}\, \Box (x_{4} \lor x_{5}) \land 
   \mu x_{8}\, \Box (x_{1} \lor x_{8})\Big)$.
Furthermore, where $\xi$ is a formula in $\muML$, we let $n_{\xi}$ be the largest 
$k$ such that $x_{k}$ occurs in $\xi$ (free or bound).
Now we define, for two $\mu$-calculus formulas $\phi$ and $\psi$:
\[
\subren{\phi}{\psi} \isdef r_{\max(n_{\phi},n_{\phi})}(\phi).
\]
It is easy to check that this definition meets the requirements mentioned in 
Definition~\ref{d:sbs}.
\end{remark}

We leave it for the reader to verify that the notion of substitution given in 
Definition~\ref{d:sbs} is semantically correct (that is, we can prove equation
\eqref{eq:subs} in Proposition~\ref{p:subs1} for \emph{arbitrary} formulas
$\phi$ and $\psi$).
The following observation shows that it interacts well with both of our size
measures.

\begin{proposition}
\label{p:sbsz}
Let $\xi$ and $\psi$ be $\mu$-calculus formulas.

1) $\sszal{\xi[\psi/x]} \leq \sszal{\xi} + \sszal{\psi}$;

2) $\cszal{\xi[\psi/x]} \leq \cszal{\xi} + \cszal{\psi}$.
\end{proposition}

\begin{proof}
First of all, observe that by Definition~\ref{d:sbs} and the invariance under 
$\al$-equivalence of our size measures, we may without loss of generality assume
that $\psi$ is free for $x$ in $\phi$.

For part 1), we may in addition assume that $\xi$ and $\psi$ are both clean and
polished, and are such that no bound variable of $\xi$ occurs in $\psi$, and 
vice versa.
(If not, we may take alphabetical variants $\xi'$ and $\psi'$
of $\xi$ and $\psi$ for which this is the case, and use the fact that,
by Proposition~\ref{p:aleq1}(\ref{aleq:6}), $\xi'[\psi'/x] \eqal \xi[\psi/x]$, so that 
$\sszal{\xi'[\psi'/x]} = \sszal{\xi[\psi/x]}$, $\sszal{\xi'} = \sszal{\xi} $ and  
$\sszal{\psi'} = \sszal{\psi}$.)
But if $\BV{\xi} \cap \BV{\psi} = \nada$, the formula $\xi[\psi/x]$ must be 
clean as well, so that by Proposition~\ref{p:alsz}, polishing it will 
\emph{reduce} the number of subformulas.
But then we have 
\[
\sszal{\xi[\psi/x]} = \size{\Sfor(\pol{\xi[\psi/x]})}
\leq \size{\Sfor(\xi[\psi/x])}
= \ssz{\xi[\psi/x]} 
\leq \ssz{\xi} + \ssz{\psi}
= \sszal{\xi} + \sszal{\psi}
\]
as required.

For part 2), we first consider the case where $x \not\in \FV{\xi}$; here we find 
$\xi[\psi/x] = \xi$, so that the statement holds trivially.
In the case where $x \in \FV{\xi}$, recall that 
\[
\Clos(\xi[\psi/x]) = 
   \{ \chi[\psi/x] \mid \chi \in \Clos(\xi) \} \cup \Clos(\psi)
\]
by Proposition~\ref{p:clos3}.
Clearly then we have
\[
\Clos(\xi[\psi/x])/{\eqal}  \;=\; 
   \Big( \chi[\psi/x] \mid \chi \in \Clos(\xi) \}/{\eqal}\Big) 
   \cup \Big(\Clos(\psi)/{\eqal}\Big),
\]
from which it is immediate that
\[
\cszal{\xi[\psi/x]} \;\leq\;
   \size{\{ \chi[\psi/x] \mid \chi \in \Clos(\xi) \}/{\eqal}}  
   + \cszal{\psi},
\]
It thus remains to check that 
\begin{equation}
\label{eq:eqalcl1}
\size{\{ \chi[\psi/x] \mid \chi \in \Clos(\xi) \}/{\eqal}} 
\;\leq\; \cszal{\xi},
\end{equation}
and since (by definition) $\cszal{\xi} = \size{\Clos(\xi)/{\eqal}}$, it suffices
to prove that there is a surjection from the set $\Clos(\xi)/{\eqal}$ to 
the set $\{ \chi[\psi/x] \mid \chi \in \Clos(\xi) \}/{\eqal}$.
For this purpose, observe that we may consider the substitution $[\psi/x]$ as 
a candidate for such a map: 
\[
[\psi/x]: \Clos(\xi) \to \{ \chi[\psi/x] \mid \chi \in \Clos(\xi) \},
\]
and that by Proposition~\ref{p:aleq1}(\ref{aleq:6}) this map satisfies the condition that 
\[
\chi_{0} \eqal \chi_{1} \emph{ implies } \chi_{0}[\psi/x] \eqal \chi_{1}[\psi/x].
\]
From this observation it follows that the map $f: \Clos(\xi)/{\eqal} 
\to \{ \chi[\psi/x] \mid \chi \in \Clos(\xi) \}/{\eqal}$ given by
$f: [\chi] \mapsto [\chi[\psi/x]]$
is well-defined; it is then obviously the required surjection from
$\Clos(\xi)/{\eqal}$ to $\{ \chi[\psi/x] \mid \chi \in \Clos(\xi) \}/{\eqal}$.
This means that we have proved \eqref{eq:eqalcl1} indeed, and therewith, the 
statement 2) of the proposition.
\end{proof}

We can now fasten a loose thread concerning the subformula size of the formulas
that we may associate with a given parity formula, cf.~Remark~\ref{r:cyc-fix}.
Let $\tr$ be the translation given in Definition~\ref{d:cyc-tr}.
The following proposition states that, for any parity formula $\bbG$, the
subformula size of $\tr_{\bbG}$ is exponentially bounded by the number of states 
of $\bbG$.
We omit the proof, which closely follows the lines of the proof of 
Proposition~\ref{p:cyclin3}, using Proposition~\ref{p:sbsz}(1).

\begin{proposition}
\label{p:cfal}
Let $\bbG = (V,E,L,\Om,v_{I})$ be a parity formula, 
Then 
\begin{equation}
\sszal{\tr_{\bbG}(v)} \leq (\size{\bbG}+2)^{\size{\Dom(\Om)}+2}
\end{equation}
for all vertices $v$ of $\bbG$.
\end{proposition}

\section{Closure size: $\al$-invariance via renaming}
\label{sec:skel}

The aim of this section is to prove Theorem~\ref{t:aleqcl}; that is, we will
provide a renaming function which maps an arbitrary $\mu$-calculus formula $\xi$ 
to an alphabetical variant $\spol{\xi}$ such that $\cszal{\spol{\xi}} = 
\size{\Clos(\xi)/{\eqal}}$.
The key concept involved in the definition of $\spol{\xi}$ is that of a 
\emph{skeletal} formula:
As we will see below, if a formula is skeletal, then $\al$-equivalence is the
identity relation on its closure set.

\subsection*{Skeletal formulas} 

In this section we fix a placeholder variable $s$, which we assume to be `fresh'
in the sense that it does not occur in any formula in $\muML$.\footnote{%
   To do this in a precise way we could introduce the set $\muML_s$ of formulas 
   that is allowed to contained the placeholder variable $s$.}

\begin{definition}
Given a set $U$ of variables, we define the \emph{skeleton} $\sk_U(\phi)$ of a 
formula $\phi$ relative to a set of variables $U$ by induction on the complexity
of $\phi$.
Throughout this induction we will define
\[
\begin{array}{llll}
   \sk_U(\phi) & \isdef & s & 
   \text{if } U \cap  \FV{\phi} = \nada,
\end{array}
\]
so that in the inductive definition itself we may focus on the case where
$U \cap  \FV{\phi} \neq \nada$:
\[
\begin{array}{llll}
   \sk_U(x) & \isdef & x & 
   \text{for } x \in U
\\ \sk_U(\phi_{0} \odot \phi_{1}) & \isdef 
   & \sk_U(\phi_{0}) \odot  \sk_U(\phi_{1}) 
   & \text{where } \odot \in \{ \lor, \land \} \text{ and }
   U \cap  \FV{\phi_{0} \odot \phi_{1}} \neq \nada
\\ \sk_U(\hs\varphi) & \isdef & \hs  \sk_U(\phi) 
   & \text{where } \hs \in \{ \dia, \Box \} \text{ and }
       U \cap  \FV{\hs\phi} \neq \nada
\\ \sk_U(\eta z. \phi) & \isdef & \eta z. \sk_{U \cup \{z\}}(\phi) 
   & \text{where } \eta \in \{ \mu, \nu \} \text{ and } 
       U \cap  \FV{\eta z. \phi} \neq \nada
\end{array}
\]
For a single variable $x$ we write $\sk_x$ as abbreviation for $\sk_{\{x\}}$.
\end{definition}

The intuition is that we replace `closed' formulas - without any connection 
further up - by the place holder $s$. 
A couple of examples are in order.

\begin{example}
1)
Let $\phi = p \lor \Diamond x$. Then $\sk_x(\phi ) = s \lor \Diamond x$. 
 
\noindent 2) 
Let $\phi = 
\left((p \lor \mu z.( q \land \Box z)) \land \mu y.( (q \lor \Diamond y) \lor 
\Box x) \right)$, then  $\sk_x(\phi) = s \land \mu y. ((s \lor \Diamond y) \lor \Box x)$.
\end{example}

\begin{definition}
We call a set of formulas $\Phi$ \emph{skeletal} if for any pair of formulas 
$\phi_{0} = \eta_{0} x_{0} .\psi_{0}$ and $\phi_{1} = \eta_{1} x_{1}.\psi_{1}$
in $\bigcup_{\phi\in\Phi} \Sfor(\phi)$ we have 
\begin{eqnarray}
\label{eq:skelet} 
  x_{0} = x_{1} & \text{ iff } & 
\eta_{0} x_{0} .\sk_{x_{0}}(\psi_{0}) \eqal \eta_{1} x_{1}.\sk_{x_{1}}(\psi_{1}). 
\end{eqnarray}
We will call a single formula $\xi$ skeletal if the singleton $\{ \xi \}$ is
skeletal.
\end{definition}
To obtain some intuition about the skeleton and skeletal formulas note that
for a fixpoint formula $\eta x . \phi$, the formula
$\eta x. \sk_x(\phi)$ contains no proper free subformulas apart from the placeholder
$s$.

\begin{remark}
As we will see, the definition of skeletal will ensure (i) stability under taking closure, i.e., all elements of the closure of a skeletal formula will
be skeletal as well, and that (ii) alpha-equivalence
on the closure of a skeletal formula coincides with syntactic equality. For achieving (ii) one could be tempted 
to employ the polishing map from Definition~\ref{def:polishing} to rename variables appropriately. It is instructive to see what would go wrong and
why this would achieve neither (i) nor (ii). To this aim consider the formulas 
$\alpha = \mu x. \nu y . \Diamond x \land \Box y$ and $\beta = (\nu y. \Diamond \alpha \land \Box y) \vee \alpha$.

Let us first focus on $\alpha$: If we assume that $\alpha$ has already been renamed according to our polishing
map  from Def.~\ref{def:polishing}, it is easy to see that its closure will contain the formula $\nu y. \Diamond \alpha \land \Box y$ which would require another round of polishing
as the outer $\nu$-operator has a different body than the inner $\nu$-operator occurring in $\alpha$. Therefore the result of the polishing map is not ``stable'' under taking the
closure of a formula and thus does not achieve (i). On the other hand, it is not difficult to see that $\nu y. \Diamond \alpha \land \Box y$ is skeletal,
because the skeleton of the outer and inner $\nu$-formula is equal to  $\nu y. s \land \Box y$.

Let us now apply the polishing map to $\beta$. This would clearly yield a formula of the form
\[  \beta' = (\nu z. \Diamond \alpha \land \Box z) \vee \alpha  \mbox{ for some } z \not= y.\]
Again, this is because the body of the first $\nu$-formula is different from the body of  the formula $\nu y . \Diamond x \land \Box y$ within $\alpha$.
Now note that $(\nu z. \Diamond \alpha \land \Box z)$ and $\nu y . \Diamond \alpha \land \Box y$
will be both part of the closure of $\beta'$, they will be $\alpha$-equivalent but, violating our goal (ii), they will not be
syntactically equivalent. A skeletal renaming of $\beta'$, on the other hand, would
ensure that the variables $z$ and $y$ would be renamed into a single variable 
as both relevant $\nu$-formulas have a skeleton of the form $\nu y. s \land \Box y$.

We will now see in detail that skeletal formulas have indeed the desired properties. Finally, we will describe
a renaming that turns any given formula into an $\alpha$-equivalent skeletal one. 
\end{remark}

We start with some basic observations about the skeletal function.

\begin{proposition}\label{p:skalpha2}
Let $\phi$ be a formula and let $x \not\in \FV{\phi}$. Then
\[ 
\sk_{U \cup \{x\}} (\phi) = \sk_U(\phi).
\]
\end{proposition}

\begin{proof}
The proof of his Proposition is straightforward --- we omit the details.
\end{proof}

\begin{proposition}\label{p:skpreserve}
Let $\psi$ be a formula, let $U$ be a set of variables and let $x$ be a variable 
with $x \not\in U$.
Furthermore let $\beta$ be a formula which is free for $x$ in $\psi$, and such
that $U \cap \FV{\beta} = \nada$.
Then
\begin{equation}
\label{eq:zyx1}
\sk_U(\psi) = \sk_U(\psi[\beta/x]).
\end{equation}
In particular, if $x$ and $y$ are variables such that $x, y \not\in U$ and $y$
is free for $x$ in $\psi$, then $\sk_U(\psi) = \sk_U(\psi[y/x])$
\end{proposition}

\begin{proof}
Consider first the case where $U \cap \FV{\psi} = \nada$. 
We have $\FV{\psi[\beta/x]} \subseteq \FV{\psi} \cup \FV{\beta}$, which, together
with our assumption on $\FV{\beta}$, implies $U \cap (\FV{\psi[\beta/x]})
= \nada$.
Therefore we obtain $\sk_U(\psi) = s = \sk_U(\psi[\beta/x])$ as required.
 
In the case that $U \cap \FV{\psi} \neq \nada$ the claim is proven by
induction on $\psi$.
In the base step of the induction, we make a case distinction.
If $\psi \neq x$ then $\psi = \psi[\beta/x]$ so that \eqref{eq:zyx1} follows
immediately.
If, on the other hand, we have $\psi = x$, then $\sk_U(x) = s = \sk_U(\beta)
= \sk_{U}(x[\beta/x])$, 
where the second equality holds as $U \cap \FV{\beta} = \nada$.

The boolean and modal cases are easy. 
For instance, in the case of a Boolean operator, we have $\psi = \psi_{0} \odot 
\psi_{1}$, with $\odot \in \{ \land, \lor\}$.
By our assumption that $U \cap \FV{\psi} \neq \nada$, there is an $i$ with
$U \cap \FV{\psi_i} \neq \nada$.
Now for $j \in \{ 0,1 \}$ we may use the induction hypothesis in the case that 
$U \cap \FV{\psi_j} \neq \nada$, and the fact that the lemma is already 
proven for the case that $U \cap \FV{\phi_i} = \nada$.
Using these facts, we find
\begin{align*}
\sk_{U}(\psi_{0}\odot\psi_{1}) 
   & = sk_{U}(\psi_{0}) \odot \sk_{U}(\psi_{1}) 
   & \text{($U \cap \FV{\psi} \neq \nada$)}
\\ & = sk_{U}(\psi_{0}[\beta/x]) \odot \sk_{U}(\psi_{1}[\beta/x]) 
   & \text{(explained above)}
\\ & = sk_{U}(\psi_{0}[\beta/x] \odot \psi_{1}[\beta/x]) 
   & \text{($U \cap \FV{\psi[\beta/x]} \neq \nada$)}
\\ & = sk_{U}\big((\psi_{0}\odot \psi_{1})[\beta/x]) 
   & \text{(definition substitution)}
\end{align*}

Finally, in the case that $\psi = \eta z. \phi$, we recall that 
$U \cap \FV{\eta z.\phi} \neq \nada$, and calculate 
\begin{align*}
  \sk_U(\eta z .\phi) 
   & =  \eta z. \sk_{U \cup \{ z \}} (\phi) 
\\ & = \eta z. \sk_{U \cup \{z\}}(\phi[\beta/x])
   & \text{(induction hypothesis)}
\\ & = \sk_U(\eta z.\phi[\beta/x])
   & \text{(*)}
\end{align*}
Observe that the induction hypothesis is applicable, since by assumption $\beta$
is free for $x$ in $\psi$, which implies that $z \not\in \FV{\beta}$.
The final equality (*) uses the fact that $\nada \neq U \cap \FV{\eta z. \phi} 
\subseteq U \cap \FV{\eta z. \phi[\beta/x]}$, which holds since by assumption
$x \not\in U$.
\end{proof}

\begin{proposition}\label{p:skalpha1}
Let $\phi$ be a formula, let $U$ be a set of variables, and let $x$ and $z$ be
variables such that $x \in U$, $z \not\in U \cup \FV{\phi}$ and $z$ is free for
$x$ in $\phi$. 
Then 
\[ 
\sk_U(\phi)[z/x] = \sk_{U[z/x]}(\phi[z/x]) 
\]
where $U[z/x] \mathrel{:=} (U \setminus \{x\}) \cup \{z\}$.
\end{proposition}

\begin{proof}
In the case that $x \not\in \FV{\phi}$ we also have $x \not\in \FV{\sk_U(\phi)}$ 
and thus $(\sk_U(\phi))[z/x]  = \sk_U(\phi)$.
In addition, 
\[ 
\sk_{U[z/x]}(\phi[z/x]) = \sk_{U[z/x]}(\phi) =  \sk_{U}(\phi)
\]  
where the last equality follows from Proposition~\ref{p:skalpha2} as $z$ and $x$ do
not occur freely in $\phi$.
    
If, on the other hand, we have that $x \in \FV{\phi}$ we prove the claim by 
induction on $\phi$.
In the base case of this induction, where $\phi = x$, the claim is an easy 
calculation:
    $(\sk_U(x) [z/x] = z = \sk_{U[z/x]}(x[z/x])$.
    
If $\phi = \phi_1 \odot \phi_2$ with $\odot \in \{\lor,\land\}$, we have
\begin{align*}    
\sk_U(\phi_1 \odot \phi_2)[z/x] 
   & = \sk_U(\phi_1)[z/x] \odot \sk_U(\phi_2)[z/x]
\\ & = \sk_{U[z/x]}(\phi_1[z/x]) \odot   \sk_{U[z/x]}(\phi_2[z/x])
   & \text{(*)}
\\ & =  \sk_{U[z/x]}((\phi_1 \odot \phi_2)[z/x])    
\end{align*}
where (*) is either by the induction hypothesis, or by the previous case if $x$
does not occur in $\phi$.

The case where $\phi = \hs \psi$ for $\hs \in \{ \Box, \Diamond \}$ is similar
to the previous one.
      
Finally, we consider the case where $\phi = \eta y.\psi$. 
By our assumptions we have $y \not = x$ since $x \in \FV{\phi}$ and --- as $z$
is free for $x$ in $\phi$ --- we also have $z \neq y$.
We calculate:
\begin{align*}
   \sk_U(\eta y. \psi) [z/x] 
   & = \left( \eta y. \sk_{U \cup \{y\}}(\psi) \right) [z/x]
   & \text{(definition of $\sk$)}
\\ & = \eta y. \left(  \sk_{U \cup \{y\}}(\psi) [z/x] \right)  
   & \text{(induction hypothesis)}
\\ & = \eta y. \left( \sk_{(U \cup \{y\})[z/x]}(\psi [z/x]) \right) 
   & \text{(definition of $\sk$)}
\\ & = \sk_{U[z/x]}(\eta y. \psi [z/x]).
\end{align*}
\end{proof}

\begin{proposition}\label{p:skmoduloalpha}
Let $\phi_{0}$ and $\phi_{1}$ be formulas such that $\phi_{0} \eqal \phi_{1}$
and let $U$ be a set of variables. 
Then $\sk_U(\phi_{0}) \eqal  \sk_U(\phi_{1})$.
\end{proposition}

\begin{proof}
Assume that $\phi_{0} \eqal \phi_{1}$, then clearly $\FV{\phi_{0}} =
\FV{\phi_{1}}$ and so we find $U \cap \FV{\phi_{0}} = \nada$ iff $U \cap
\FV{\phi_{1}} = \nada$.
This means that in case $U \cap \FV{\phi_{0}} = \nada$ we have $\sk_U(\phi_{0}) =
\sk_U(\phi_{1}) = s$. 
 
In case that $U \cap \FV{\phi_{0}} \neq \nada$  we prove the claim by induction
on the structure of $\phi_{0}$.
We only treat the  fixpoint case, that is, where $\phi_{0}$ is of 
the form $\phi_{0} = \eta x_{0}. \psi_0$.
As $\phi_{0} \eqal \phi_{1}$ the formula $\phi_{1}$ must be of the form $\phi_{1}
= \eta x_{1}. \psi_{1}$.

Fix a fresh variable $z$, then we have $\psi_{i} = \psi_{i}[z/x_{i}][x_{i}/z]$.
We may now calculate, for $i = 0,1$:
\begin{align*}
    \sk_U(\phi_{i}) 
   & = \eta x_{i}. \sk_{U \cup \{x_{i}\}}(\psi_{i}) 
   & \text{(definition of $\sk$)}
\\ & = \eta x_{i}. \big( \sk_{U \cup \{z\}}(\psi_{i}[z/x_{i}])\big)[x_{i}/z] 
   & \text{(Proposition~\ref{p:skalpha1})}
\\ & \eqal \eta z. \sk_{U \cup \{z\}}(\psi_{i}[z/x_{i}]) 
   & \text{(Proposition~\ref{p:aleq1}(\ref{aleq:5}))}
\end{align*}

Now observe that by Proposition~\ref{p:aleq1}(\ref{aleq:2}) it follows from $\eta x_{0}. 
\psi_0 \eqal  \eta x_{1}. \psi_{1}$ that $\psi_{0}[z/x_{1}] \eqal 
\psi_{1}[z/x_{1}]$.
Hence by the induction hypothesis we obtain that 
$\sk_{U \cup \{z\}}(\psi_{0}[z/x_{0}]) \eqal 
\sk_{U \cup \{z\}}(\psi_{1}[z/x_{1}])$, so that by 
Proposition~\ref{p:aleq1}(\ref{aleq:8}),
we find that 
\[
\eta z.\sk_{U \cup \{z\}}(\psi_{0}[z/x_{0}]) \eqal 
\eta z. \sk_{U \cup \{z\}}(\psi_{1}[z/x_{1}]).
\]
But then from the above calculation of $\sk_U(\phi_{i})$ we may conclude that 
$\sk_U(\phi_{0}) \eqal \sk_U(\phi_{1})$, as required.
\end{proof}

\subsection*{Skeletal Formulas \& Closure}

The key property of skeletal formulas is that on the closure set of a tidy
skeletal formula, $\al$-equivalence coincides with syntactic identity.
As a first step we show that skeletal sets of formulas are polished.

\begin{proposition}\label{p:alpha}
Let $\Phi$ be a skeletal set of formulas. 
Then for any pair of formulas $\phi_{0}, \phi_{1} \in \bigcup_{\phi\in\Phi}
\Sfor(\phi)$ we have 
\begin{equation}\label{eq:skelclean}   
\phi_{0} \eqal \phi_{1} \quad \mbox{implies} \quad  \phi_{0} = \phi_{1}.
\end{equation}
\end{proposition}

\begin{proof}
Suppose $\Phi$ is skeletal, and let $\phi_{0}$ and $\phi_{1}$ be as stated.
We prove the claim by induction on the structure of $\phi_{0}$.
 
If $\phi_{0}$ is a literal, the claim is trivial. 
In case $\phi_{0}$ is a conjunction, disjunction or a modal formula of the form
$\hs \psi_1$, the claim easily follows by induction. 
  
Now suppose that $\phi_{0} = \eta_{0} x_{0} .\psi_{0}$, it is then easy to see
that $\phi_{1}$ must be of the form $\phi_{1} = \eta_{1} x_{1}.\psi_{1}$.
Furthermore, it follows from $\phi_{0} \eqal \phi_{1}$ that $\eta_{0} = \eta_{1}$ 
(so that we may write $\eta$ in the sequel), and that 
$\psi_0[z/x_0] \eqal \psi_1[z/x_1]$ for a fresh variable $z$.
It follows from  Proposition~\ref{p:skmoduloalpha} that $\sk_z(\psi_0[z/x_0] ) \eqal
\sk_z(\psi_1[z/x_1] )$. By Proposition~\ref{p:skalpha1}
we have $\sk_z(\psi_i[z/x_i] ) = \sk_{x_i}(\psi_i)[z/x_i]$ for $i=0,1$.
Therefore, as $z$ was fresh, we obtain
$\eta x_0.\sk_{x_0}(\psi_0) \eqal \eta x_1.\sk_{x_1}(\psi_1)$ by definition of
$\eqal$.
As $\Phi$ is skeletal this implies $x_0 = x_1 = x$ and thus
by Proposition~\ref{p:aleq1}(\ref{aleq:3}) that
$\psi_0 \eqal \psi_1$. The induction hypothesis yields
$\psi_0 = \psi_1$ which obviously implies
$\phi_0 = \phi_1$ as required. 
\end{proof}

The next proposition states the closure of a skeletal set is skeletal.

\begin{proposition}\label{p:closkel}
Let $\Psi$ be a skeletal set of tidy formulas.
Then $\Clos(\Psi)$ is skeletal as well.
\end{proposition}

\begin{proof}
Clearly it suffices to show that, if $\Phi'$ is obtained from a skeletal set
$\Phi$ of tidy formulas by applying one of the rules for deriving the closure,
then $\Phi'$ is also skeletal.

The only case where this is non-trivial is when $\Phi' = \Phi \cup 
\{\phi[\eta x \phi/x]\}$ for some formula $\eta x.\phi \in\Phi$.
Consider a pair of formulas $\phi_{0} = \eta_{0} x_{0}.\psi_{0}$ and 
$\phi_{1} = \eta_{1} x_{1}. \psi_{1}$ that are subformulas of some formulas in
$\Phi'$. 
In order to show that $\phi_{0}$ and $\phi_{1}$ satisfy \eqref{eq:skelet}, we
distinguish the following cases.

\begin{description}
\item[\it Case 1:] 
both $\phi_{0}$ and $\phi_{1}$ are subformulas of elements of $\Phi$. 
Then \eqref{eq:skelet} follows from the fact that $\Phi$ is 
skeletal.

\item[\it Case 2:] 
both $\phi_{0}$ and $\phi_{1}$ are subformulas of $\phi[\eta x \phi/x]$.
In this case, $\phi_{0}$ and $\phi_{1}$ are of the form $\phi_{0} = \eta x_{0}.
\psi_{0}'[\eta x. \phi/x]$ and $\phi_{1} = \eta x_{1}. \psi_{1}'[\eta x. \phi/x]$,
respectively, for  subformulas $\eta x_{0}. \psi_{0}'$ and $\eta x_{1}.
\psi_{1}'$ of $\phi$.
Then we have $\sk_{x_{i}}(\psi_{i}'[\eta x.\phi/x]) = \sk_{x_{i}}(\psi_{i}')$ 
for $i \in \{0,1\}$ by Proposition~\ref{p:skpreserve}.
Thus we find
\begin{align*}
x_{0} = x_{1} 
   & \text{ iff }  
  \eta_{0} x_{0}. \sk_{x_{0}}(\psi_{0}') \eqal \eta_{1} x_{1}. \sk_{x_{1}}(\psi_{1}') 
\\ & \text{ iff } 
  \eta_{0} x_{0}. \sk_{x_{0}}(\psi_{0}'[\eta x. \phi/x]) \eqal 
  \eta_{1} x_{1}. \sk_{x_{1}}(\psi_{1}'[\eta x.\phi/x])  
\end{align*}
where the first equivalence is a consequence of the fact that
property~\eqref{eq:skelet} holds for $\Phi$ by assumption.

\item[\it Case 3:] 
one of $\phi_{0}$ and $\phi_{1}$ is a subformula of a formula in $\Phi$, while 
the other is a subformula of $\phi[\eta x \phi/x]$.
Say, without loss of generality, that $\phi_{0}$ is a subformula of a formula in 
$\Phi$, while $\phi_{1} = \eta_{1} x_{1}.\psi'[\eta x . \phi/x]$ with 
$\eta_{1} x_{1} .\psi' \sforeq \phi$. 
As $\Phi$ is skeletal we have
\[ 
x_{0} = x_{1} \text{ iff } 
\eta_{0} x_{0}. \sk_{x_{0}}(\psi_{0}) = \eta_{1} x_{1} . \sk_{x_{1}}(\psi').
\]
By Proposition~\ref{p:skpreserve} we have $\sk_{x_{1}}(\psi') = 
\sk_{x_{1}}(\psi'[\eta_x.\phi/x]) = \sk_{x_{1}}(\psi_{1})$ and thus we obtain
\[  
x_{0} = x_{1} \text{ iff }
\eta_{0} x_{0}. \sk_{x_{0}}(\psi_{0}) = \eta_{1} x_{1}. \sk_{x_{1}}(\psi_{1})
\]
as required.
\end{description}
\end{proof}

As an immediate consequence of the Propositions~\ref{p:closkel} and~\ref{p:alpha},
we establish the key property of skeletal formulas.

\begin{proposition}
\label{p:skalid}
Let $\phi$ be a tidy skeletal formula. 
Then the relations of $\al$-equivalence and syntactic identity coincide on the 
set $\Clos(\phi)$.
\end{proposition}

\subsection*{The skeletal renaming}

We are now ready for the definition of the renaming map $\spol{\cdot}$.
For this purpose, recall from Definition~\ref{d:XZ} that, with $X$ and $Z$ 
disjoint sets of variables, we let $\muML_{X}$ denote the collection of 
$\mu$-calculus formulas taking their (free or bound) variables from $X$, while 
$\muML_{Z,X}$ denotes the formulas in $\muML_{X\cup Z}$ that take their 
\emph{bound} variables from $Z$.
In the definition below we assume that the set $Z$ contains a distinct variable
$\ul{z}_{E}$ for every $\al$-equivalence class $E$ of $\muML_{X}$-formulas.

\begin{definition}
\label{d:skren}
We define the renamed version $\spolform{}{\phi} \in \muML_{Z,X}$ of a formula 
$\phi \in \muML_{X}$  as follows: 
\[\begin{array}{llll}
   \spolform{}{\phi}        & \isdef 
  & \phi
  & \text{if $\phi$ is atomic} 
\\ \spolform{}{\hs\phi}     & \isdef
  & \hs \spol{\phi} 
  & \text{where } \hs \in \{ \dia, \Box \}
\\ \spolform{}{\phi_{0}\odot\phi_{1}} & \isdef 
  & \spolform{}{\phi_{0}}\odot\spolform{}{\phi_{1}} 
  & \text{where } \odot \in \{ \lor, \land \}
\\ \spolform{}{\eta x.\phi} & \isdef 
  & \eta \ul{z}_E. \spolform{}{\phi}[\ul{z}_E/x]
  & \text{where } \eta \in \{ \mu, \nu \} \text{ and } 
  E = \eqalc{\eta x.\sk_x(\phi)}
\end{array}\]
\end{definition}

\begin{remark}
The renamed version of $\phi \in \muML_{X}$ will only contain variables from 
the set $Z$ that are  bound. 
These bound variable can be replaced by fresh variables from $X$ in order to 
obtain a renamed version in $\muML_{X}$.
\end{remark}

\begin{example}
Consider the formula
\[
\phi =  \nu y. (\Diamond(\mu x. (\nu z. \Diamond (x \land z)) \land y)),
\]
which is $\alpha$-equivalent to the unfolding 
$( \nu y.\Diamond (x \land y) )[\psi/x]$ of 
$\psi = \mu x. \nu y.\Diamond (x \land y)$.
Furthermore let $E_1 = \eqalc{\nu y. \Diamond ( s \land y)}$ and $E_2 = 
\eqalc{\mu x. \nu y. \Diamond (x \land y)}$.
Then 
\[
\spol{\phi} =
\nu \ul{z}_{E_1}.(\dia(\mu  \ul{z}_{E_2}. (\nu  \ul{z}_{E_1}. 
\dia ( \ul{z}_{E_2} \land  \ul{z}_{E_1})) \land  \ul{z}_{E_1})),
\]
where we point out the re-use of the variable $\ul{z}_{E_{1}}$.
To obtain a standard formula (without underlined variables) we simply remove all
the underlining (as in this example there are no non-underlined variables 
present that could lead to clashes).
\end{example}

Our first goal is to show that the renamed version $\spolform{}{\phi}$ of a
formula $\phi$ is $\alpha$-equivalent to $\phi$.
To this aim we need the following rather technical lemma.

\begin{proposition}
\label{p:sksub}
Let $x$ and $y$ be variables, let $U$ be a set of variables with $y \in U$, and 
let $\phi$ and $\eta x. \psi$ be formulas such that $y \in \FV{\eta x. \psi}$ 
and $\eta x.\psi \sforeq \phi$,
while there is no formula of the form $\la y. \chi$ such that $\eta x.\psi
\sforeq \la y. \chi \sforeq \phi$.
Then $\sk_{x}(\psi) \not\eqal \sk_{U}(\phi)$.
\end{proposition}

\begin{proof}
By Proposition~\ref{p:aleq0} it suffices to show that 
\[
\len{\sk_{x}(\psi)} < \len{\sk_{U}(\phi)},
\]
and we will prove this by induction of the length of the shortest 
direct-subformula chain $\eta x.\psi \sfor_{0} \cdots \sfor_{0} \phi$ witnessing 
that $\eta x.\psi$ is a subformula of $\phi$.

In the base step of this induction we have $\eta x.\psi = \phi$, so that 
$\sk_{U}(\phi) = \eta x. \sk_{U\cup \{x\}}(\psi)$.
But it is easy to prove, using a straightforward inductive argument 
on the complexity of $\psi$, that $\len{\sk_{V}(\psi)} \leq \len{\sk_{V'}(\psi)}$
if $V \sse V'$. 
Thus we find $\len{\sk_{x}(\psi)} \leq \len{\sk_{U\cup \{x\}}(\psi)}
< \len{\eta x. \sk_{U\cup \{x\}}(\psi)} =  \len{\sk_{U}(\phi)}$, as required.

In the induction step of the proof, we make a case distinction as to the shape
of $\phi$.
Leaving the other cases as an exercise, we focus on the case where $\phi$ is of 
the form $\phi = \theta z. \phi'$.
It follows from the assumptions that $y \in \FV{\phi}$, so that $U \cap \FV{\phi}
\neq \nada$.
Thus we find $\sk_{U}(\phi) = \theta z. \sk_{U\cup \{ z \}}(\phi')$,
which immediately gives 
$\len{\sk_{U\cup \{ z \}}(\phi')} < \len{\sk_{U}(\phi)}$.
By the induction hypothesis we obtain
$\len{\sk_{x}(\psi)} < \len{\sk_{U\cup \{ z \}}(\phi')}$, so that we may infer
$\len{\sk_{x}(\psi)} < \len{\sk_{U}(\phi)}$ indeed.
\end{proof}

\begin{proposition}\label{p:spolalpha}
Let $\xi$ be a $\mu$-calculus formula. 
Then $\spol{\xi}$ is tidy and $\xi \eqal \spol{\xi}$.
\end{proposition}

\begin{proof}
The proof that $\spol{\xi}$ is tidy is easy and therefore left to the reader.
We prove the claim that $\xi \eqal \spol{\xi}$ by a formula induction on $\xi$. 
If $\xi$ is atomic, then $\xi$ and $\spol{\xi}$ are identical, and so, certainly
$\al$-equivalent.

For the induction step, distinguish cases.
If $\xi$ is of the form $\xi = \xi_{0} \odot \xi_{1}$ for $\odot \in \{\land,
\lor\}$, then the claim is an immediate consequence of the induction hypothesis 
and the fact that $\spol{\xi_{0} \odot \xi_{1}} = \spol{\xi_{0}} \odot 
\spol{\xi_{1}}$.
The case where $\xi$ is of the form $\xi = \hs \xi'$ for $\hs \in \{ \dia,
\Box\}$ is equally simple.

The interesting case is where $\xi$ is of the form $\xi = \la y.\phi$.
Then $\spol{\xi} = \eta \ul{z}_E. \spol{\phi}[\ul{z}_E/z]$, with $E = 
\eqalc{\eta y.\sk_y(\phi)}$.
We first claim that 
\begin{equation}
\label{eq:zfree}
\ul{z}_{E} \text{ is free for $y$ in } \spol{\phi}.
\end{equation}
To see this, suppose for contradiction that $y$ occurs freely in the scope of
a binder $\eta \ul{z}_E$ in $\spol{\xi}$. 
Then there must be a subformula $\eta x. \psi$ of $\phi$ with $\spol{\eta x.\psi}
= \eta \ul{z}_E.\spol{\psi}[\ul{z}_E/x]$ such that $y \in \FV{\psi}$.
By definition of $\spol{\cdot}$ we have $E = \eqalc{\eta x. \sk_{x}(\psi)}$ and
so $\eta x. \sk_{x}(\psi) \eqal \eta y. \sk_{y}(\phi)$ by our assumption that
$E = \eqalc{\eta y.\sk_y(\phi)}$.
It follows by Proposition~\ref{p:sksub} that there must be a formula $\la y.
\chi$ such that $\eta x.\psi \sforeq \la y. \chi \sforeq \phi$; without loss 
of generality we may take $\la y. \chi$ to be the smallest such formula (in
terms of the subformula ordering).
But from this we may infer that actually, when computing the formula 
$\spol{\xi}$, the variable $y \in \FV{\eta x.\psi}$ will be replaced by the 
variable $z_{E'}$, where $E' = \eqalc{\la y. \sk_{y}(\chi)}$.
In other words, the alleged free occurrence in $\spol{\xi}$ of the variable $y$,
within the scope of a binder $\eta \ul{z}_E$, is not actually possible.
Clearly this implies \eqref{eq:zfree}.

From this we reason as follows. 
By the induction hypothesis we obtain that $\spol{\phi} \eqal \phi$.
Now, because of \eqref{eq:zfree}, we may apply
Proposition~\ref{p:aleq1}\eqref{aleq:5} and obtain $\spol{\xi} = 
\eta z_E.\spol{\phi}[\ul{z}_E/z] \eqal \eta z.\phi = \xi$ as required.
\end{proof}

We now show that the renamed operation always produces skeletal formulas. 

\begin{proposition}
\label{p:rensk}
For each formula $\phi$ we have that $\spol{\phi}$ is skeletal.
\end{proposition}

\begin{proof}
As a preparatory step, consider any subformula $\eta \ul{z}_E. \psi$ of 
$\spol{\phi}$.
By definition of $\spol{\_}$ there is a subformula $\eta x.\xi$ of $\phi$ such
that $E= \eqalc{\eta x. \sk_x(\xi)}$ and 
$\psi = \spol{\xi}[\ul{z}_E/x] [\ul{z}_1/x_1]\dots [\ul{z}_n/x_n]$.
Then we have
\begin{align*}
  \eta \ul{z}_E. \sk_{z_{E}}(\psi)
   & = \eta \ul{z}_E. 
    \sk_{\ul{z}_E}( \spol{\xi}[\ul{z}_E/x] [\ul{z}_1/x_1]\dots [\ul{z}_n/x_n])
\\ & = \eta \ul{z}_E. \sk_{\ul{z}_E}( \spol{\xi}[\ul{z}_E/x])
   & \text{(Proposition~\ref{p:skpreserve})} 
\\ & = \eta \ul{z}_E. \sk_{x}(\spol{\xi})[\ul{z}_E/x] 
   & \text{(Proposition~\ref{p:skalpha1})} 
\\ & \eqal \eta x. \sk_x( \spol{\xi}
\\ & \eqal \eta x. \sk_x( \xi) 
   & \text{(Proposition~\ref{p:spolalpha})}
\end{align*}
where the last equality uses the fact that $\phi_1 \eqal \phi_2$ implies $\sk_x (\phi_1) \eqal \sk_x(\phi_2)$ - 
something that is easy to verify.
  
We now turn to the argument for why $\spol{\phi}$ is skeletal.
Suppose that we have two subformulas  $\eta_{0} \ul{z}_{E_{0}}. \psi_{0}$ and 
$\eta_{1} \ul{z}_{E_{1}}. \psi_{1}$ of $\spol{\phi}$.
We need to prove that
\begin{eqnarray}
\label{eq:skel} 
z_{E_{0}} = z_{E_{1}} & \text{ iff } & 
\eta_{0} z_{E_{0}}. \sk_{z_{E_{0}}}(\psi_{0}) \eqal 
\eta_{1} z_{E_{1}}. \sk_{z_{E_{1}}}(\psi_{1}). 
\end{eqnarray}
By the earlier observation there must be formulas $\eta_{i}x_{i}. \xi_{i} 
\sforeq \phi$ such that
$\eta_{i} \ul{z}_{E_{i}}. \sk_{\ul{z}_{E_{i}}}(\psi_{i}) \eqal
\eta_1 x_1. \sk_{x_1}( \xi_1)$ 
with $E_{i} = \eqalc{\eta_{i} x_{i}. \sk_{x_{i}}( \xi_{i})}$. 
  
In order to prove \eqref{eq:skel}, first assume that $z_{E_{0}} = z_{E_{1}}$.
Then $E_{0} = E_{1}$, so that 
$\eta_{0}x_{0}. \sk_{x_{0}}(\xi_{0}) \eqal 
\eta_{1}x_{1}. \sk_{x_{1}}(\xi_{1})$.
It follows that $\eta_{0} = \eta_{1}$ and so we find
\begin{align*}
\eta_{0} z_{E_{0}}. \sk_{z_{E_{0}}}(\psi_{0}) 
   & = \eta_{0} x_{0}. \sk_{x_{0}}(\xi_{0}) 
\\ & = \eta_{1} x_{1}. \sk_{x_{1}}(\xi_{1}) 
\\ & = \eta_{1} z_{E_{1}}. \sk_{z_{E_{1}}}(\psi_{1}) 
\end{align*}
as required.
  
Conversely, if $\eta \ul{z}_{E_1}. \sk_{ \ul{z}_{E_1}}(\psi_1) \eqal 
\eta \ul{z}_{E_2}. \sk_{ \ul{z}_{E_2}}(\psi_2)$, then we have 
$\eta x_1. \sk_{x_1}( \xi_1) \eqal  \eta x_2. \sk_{x_2}( \xi_2)$ which implies
$E_1 = E_2$ and thus $\ul{z}_{E_1}=\ul{z}_{E_2}$.
\end{proof}

\begin{proofof}{Theorem~\ref{t:aleqcl}}
Part 1) of the Theorem was proved as Proposition~\ref{p:spolalpha}, while part 2) 
follows from Proposition~\ref{p:rensk} and Proposition~\ref{p:skalid}.

As an immediate consequence of part 2), we see that $\csz{\spol{\xi}} = 
\size{\Clos(\spol{\xi})} = \size{\Clos(\spol{\xi})/{\eqal}}$, while we 
may infer from part 1) that $\size{\Clos(\xi)/{\eqal}}
= \size{\Clos(\spol{\xi})/{\eqal}}$.
From these observations, part 3) of the Theorem is immediate.
\end{proofof}

\section{Guarded transformation}
\label{sec:gua}

Finally, as an example of an important construction on $\mu$-calculus formulas,
we consider the operation of guarded transformation.
Recall that a $\mu$-calculus formula is \emph{guarded} if every occurrence of
a bound variable is in the scope of a modal operator which resides inside 
the variable's defining fixpoint formula.
Intuitively, the effect of this condition is that, when evaluating a guarded 
formula in some model, between any two iterations of the same fixpoint variable, 
one has make a transition in the model.
Many constructions and algorithms operating on $\mu$-calculus formulas 
presuppose that the input formula is in guarded form, which explains the need
for low-cost \emph{guarded transformations}, that is, efficient procedures for 
bringing a $\mu$-calculus formula into an equivalent guarded form.

In fact, one of the main contributions of Bruse, Friedmann \& Lange
in~\cite{brus:guar15} is to discuss size issues related to guarded 
transformations, and in the process to correct some mistaken claims in the
literature.
In this section we show how to perform guarded transformations on parity 
formulas, in the hope that this may help to further clarify some size issues 
related to this construction.
To start with, it is easy to translate the notion of guardedness to parity 
formulas; 

\begin{definition}
\label{d:guar}
A path $\pi = v_{0}v_{1}\cdots v_{n}$ is \emph{unguarded} if $n\geq 1$, $v_{0},
v_{n} \in \Dom(\Om)$ while there is no $i$, with
$0 < i \leq n$, such that 
$v_{i}$ is a modal node.
A parity formula is \emph{guarded} if it has no unguarded cycles, and 
\emph{strongly guarded} if it has no unguarded paths.
\end{definition}

In words, a parity formula is strongly guarded if every path, leading from one
state (node in $\Dom(\Om)$) to another contains at least one modal node
(different from the path's starting state).
We leave it as an exercise for the reader to verify that this notion is indeed
the parity-formula version of guardedness, in the sense that the constructions 
defined in the sections~\ref{s:fixpar} and~\ref{s:parfix} preserve guardedness.

The following theorem states that on arbitrary parity formulas, we can always 
give an exponential-size guarded transformation; note that the index of the 
formula does not change.

\begin{fewtheorem}
\label{t:guard1}
There is an algorithm that transforms a parity formula $\bbG = (V,E,L,\Om,v_{I})$
into a strongly guarded parity formula $\bbG^{g}$ such that 

\begin{enumerate}[topsep=0pt,itemsep=-1ex,partopsep=1ex,parsep=1ex,%
    label={\arabic*)}]

\item \label{eq:tg1:1} 
$\bbG^{g} \equiv \bbG$;

\item \label{eq:tg1:2}
$\size{\bbG^{g}} \leq 2^{1+\size{\Dom(\Om)}} \cdot \size{\bbG}$;

\item \label{eq:tg1:3}
$\idx(\bbG^{g}) \leq \idx(\bbG)$.

\item \label{eq:tg1:4}
$(E^{g})^{-1}[\Dom(\Om] \sse V^{g}_{m}$; that 
is, in $\bbG^{g}$ every predecessor of a state is a modal node.
\end{enumerate}
\end{fewtheorem}

For the proof of this theorem we need the following definition.

\begin{definition}
A parity formula $\bbG = (V,E,L,\Om,v_{I})$ is \emph{strongly $k$-guarded} if
it every unguarded path $\pi = v_{0}v_{1}\cdots v_{n}$ satisfies $\Om(v_{n}) > 
k$.
\end{definition}

Clearly, a parity formula is (strongly) guarded iff it is (strongly) $m$-guarded,
where $m$ is the maximum priority value of the formula.
Hence, we may prove Theorem~\ref{t:guard1} by successively applying the 
following proposition.
Recall that a parity formula is called \emph{linear} if its priority map is
injective.
We say that a parity formula has \emph{silent states only} if each of its states 
is labelled $\epsilon$.

\begin{proposition}
\label{p:guar2}
Let $\bbG$ be a linear, strongly $k$-guarded parity formula with silent states
only.
Then we can effectively obtain a linear, $k+1$-guarded parity formula $\bbG'$ 
with silent states only, and such that 
$\bbG' \equiv \bbG$, $\size{\bbG'} \leq 2 \cdot \size{\bbG}$ and 
$\idx(\bbG') \leq \idx(\bbG)$.
\end{proposition}

\begin{proof}
Let $\bbG = (V,E,L,\Om,v_{I})$ be an arbitrary linear, strongly $k$-guarded 
parity formula with silent states, that is, that is, $\Dom(\Om) \sse 
L^{-1}(\epsilon)$.
If $\bbG$ happens to be already $k+1$-guarded, then there is nothing to do:
we may simply define $\bbG' \isdef \bbG$.

On the other hand, if $\bbG$ is $k+1$-unguarded, then in particular there must
be a state $z \in V$ such that $\Om(z) = k+1$.
By injectivity of $\Om$, $z$ is unique with this property.
In this case we will build the parity formula $\bbG'$, roughly, on the disjoint
union of $\bbG$, a copy of a part of $\bbG$ that is in some sense generated from
$z$, and an additional copy of $z$ itself.

For the definition of $\bbG'$, let $W^{z}$ be the smallest set $W \sse V$ 
containing $z$, which is such that $E[w] \sse W$ whenever $w\in W$ is boolean 
or satisfies $w \in L^{-1}(\epsilon) \setminus \Dom(\Om)$.
Now define
\[
V' \isdef V \times \{ 0 \} \cup W^{z} \times \{ 1 \} \cup \{ (z,2) \}.
\]
In the sequel we may write $u_{0}$ instead of $(u,0)$, for brevity.
The edge relation $E'$ is now given as follows:
\[\begin{array}{lllll}
E' \isdef 
  & & \big\{ (u_{0},v_{0}) \mid (u,v) \in E \text{ and } v \neq z \big\}
    & \cup  
    & \big\{ (u_{1},v_{1}) \mid (u,v) \in E \text{ and } v \neq z \big\}
\\[1mm]
  & \cup & \big\{ (u_{0},z_{1}) \mid (u,z) \in E  \big\}
\\[1mm]
  & \cup & \big\{ (u_{1},v_{0}) \mid (u,v) \in E \text{ and }  u \in V_{m} \big\}
  & \cup 
  & \big\{ (u_{1},u_{0}) \mid u \in \Dom(\Om)\text{ and } \Om(u) > k + 1 \big\}
\\[1mm]
  & \cup 
  & 
    \big\{ (u_{1},z_{2}) \mid (u,z) \in E \text{ and }  u \not\in V_{m} \big\}
\end{array}\]
To understand the graph $(V',E')$, it helps, first of all, to realise that the
set $W^{z}$ provides a subgraph of $(V,E)$, which forms a dag with root $z$ and
such that every `leaf' is either a modal or propositional node, or else a state
$v \in \Dom(\Om)$ with $\Om(v) > k$.
(It cannot be the case that $\Om(v) \leq k$ due to the assumed $k$-guardedness of 
$\bbG$.)
Second, it is important to realise that the only way to move from the $V$-part
of $V'$ to the $W^{z}$-part is via the root $z_{1}$ of the $W^{z}$-part, while 
the only way to move in the converse direction is either directly following 
a modal node, or else by making a dummy transition from $u_{1}$  to its 
counterpart $u_{0}$ for any $u \in W^{z}$ with $\Om(u) > k$.
Finally, we add a single vertex $z_{2}$ to $V'$.

Furthermore, we define the labelling $L'$ and the priority map $\Om'$ of $\bbG'$ 
by putting
\[
L'(u_{i}) \isdef \left\{ \begin{array}{ll}
      L(u)   & \text{if } i = 0,1
   \\ \wh{z} & \text{if } u_{i} = z_{2} 
\end{array}\right. \]
where we recall that $\wh{z} = \bot$ if $\Om(z)$ is odd and $\wh{z} = \top$ if
$\Om(z)$ is even, and 
\[
\Om'(u_{i}) \isdef \left\{ \begin{array}{ll}
      \Om(u)     & \text{if } i = 0 \text{ and } u \in \Dom(\Om)
   \\ {\uparrow} & \text{otherwise}.
\end{array}\right. \]

In words, the label of a node $(v,i)$ in $\bbG'$ is identical to the one of $v$ 
in $\bbG$, with the sole exception of the vertex $(z,2)$.
To explain the label of the latter node, note that
by construction, any unguarded $E'$-path from $z_{1}$ to $z_{2}$ projects to an
unguarded $k+1$-cycle from $z$ to $z$ in $\bbG$.
If $\Om(z) = k+1$ is odd, any such cycle represents (tails of) infinite matches 
that are lost by $\eloi$; for this reason we may label the `second' appearance 
of $z$ in the $E'$-path, i.e., as the node $z_{2}$, with $\bot$.
\medskip

We now turn to the proof of the proposition.
It is not hard to show that $\bbG'$ is linear and that $\size{\bbG'} \leq 2 
\cdot \size{\bbG}$.
\medskip

To show that $\idx(\bbG') \leq \idx(\bbG)$, note that obviously, the projection
map $u_{i} \mapsto u$ preserves the cluster equivalence relation, i.e., 
$u_{i} \equiv_{E'} v_{j}$ implies $u \equiv_{E} v$.
Hence, the image of any cluster $C'$ of $\bbG'$ under this projection is part of 
some cluster $C$ of $\bbG$.
But then by definition of $\Om'$ it is easy to see that $\idx(C') \leq \idx(C)$.
From this it is immediate that $\idx(\bbG') \leq \idx(\bbG)$.

To see why $\bbG'$ is $k+1$-guarded, suppose for contradiction that it has 
a $k+1$-unguarded path $\pi = (v_{0},i_{0})(v_{1},i_{1})\cdots (v_{n},i_{n})$.
It is easy to see that this implies that the \emph{projection} $v_{0}v_{1}
\cdots v_{n}$ of $\pi$ is an unguarded path in $\bbG$  (here we ignore possible
dummy transitions of the form $(u_{1},u_{0})$), and so by assumption on
$\bbG$ it must be the case that $\Om'(v_{n},i_{n}) = \Om(v_{n}) = k+1$.
This means that $(v_{n},i_{n}) = (z,0)$; but the only way to arrive at the
node $(z,0)$ in $(V',E')$ is directly following a modal node (in $W^{z} \times
\{ 1 \}$), which contradicts the unguardedness of the path $\pi$.

In order to finish the proof of the Proposition, we need to prove the 
equivalence of $\bbG'$ and $\bbG$; but this can be established by a relatively
routine argument of which we skip the details.
\end{proof}

\begin{proofof}{Theorem~\ref{t:guard1}}
Let $\bbG$ be an arbitrary parity formula; without loss of generality we may 
assume that $\bbG$ is linear, i.e., $\Om$ is injective.
Let $\Ran(\Om) = \{ k_{1}, \ldots, k_{n} \}$; then $\size{\Dom(\Om)} = n$.
To ensure that all states are silent, we may have to duplicate some vertices;
that is, we continue with a version $\bbH$ of $\bbG$ that has at most twice as
many vertices, but the same index, the same number of states, and silent state 
only.

By a straightforward induction we apply Proposition~\ref{p:guar2} to construct,
for every $i \in \rng{1}{n}$, a linear, $k_{i}$-guarded parity automaton 
$\bbH^{i}$ with silent states only, and such that $\bbH^{i} \equiv \bbG$, 
$\size{\bbH^{i}} \leq 2^{i+1} \cdot \size{\bbG}$, and $\idx(\bbH^{k}) = 
\idx(\bbG)$.
Clearly then we find that $\bbH^{n}$ is the desired strongly guarded equivalent
of $\bbG$; and since $n = \size{\Dom(\Om)}$ we find that 
$\size{\bbH^{n}} \leq 2^{1+n} \cdot \size{\bbG}$ as required.

Finally, a closer inspection of the proof of Proposition~\ref{p:guar2} reveals
that inductively, we may assume that for every $i$, every predecessor of a state
in $\bbH^{i}$ with priority at most $k_{i}$ is in fact a modal node.
From this, the last item of Theorem~\ref{t:guard1} follows.
\end{proofof}


In the literature on the modal $\mu$-calculus there has been some confusion
concerning the existence of a \emph{polynomial} guarded transformation 
procedure; some authors erroneously claimed to have obtained a construction
producing a guarded equivalent of quadratic or even linear size in terms of 
the size of the input formula.
One of the main goals of Bruse, Friedmann \& Lange~\cite{brus:guar15} was to
clarify this situation; the results in section~4 of~\cite{brus:guar15} show that 
certain guarded transformation procedures are as hard\footnote{%
   It is an open question whether parity games can be solved in polynomial time. 
   Despite considerable efforts no polynomial algorithm has been found so far.
   In the recent literature, however, various quasi-polynomial algorithms have 
   been given, following the breakthrough work of Calude et 
   alii~\cite{calu:deci17}.
   }
as solving parity games. Theorem~\ref{t:gtlow} below can be seen as our
parity-formula version of this result. Our proof is in fact simpler
because we can exploit the close connection between parity games and
parity formulas and thus do not need the product construction from
\cite{kupfer:linbran05} that is used for the results from
\cite{brus:guar15}.

\newcommand{\ptime}{\textsc{ptime}}

\begin{fewtheorem}
\label{t:gtlow}
If there is a procedure that runs in polynomial time and transforms a parity
formula $\bbG$ to a guarded parity formula $\bbG^\gamma$ with $\bbG^\gamma 
\equiv \bbG$ then solving parity games is in \ptime.
\end{fewtheorem}

\begin{proof}
\newcommand{\game}{\mathcal{G}}
We describe a polynomial algorithm that given a parity game $\game$
and a position $p$ in $\game$ as input uses the polynomial guarded
transformation that exists by assumption to decide whether $\eloi$ has a
winning strategy starting from $p$ in $\game$.

On a high level the algorithm proceeds as follows.
We first transform $\game$ into a parity formula $\bbG_\game$.
Containing modalities nor literals, the truth of $\bbG_{\game}$ will in fact
be \emph{independent} of the Kripke model where it is evaluated.
$\bbG_{\game}$ is constructed such that $\bbG_\game$ is valid (true in any
pointed model) if $\eloi$ has a winning strategy starting from $p$, and 
$\bbG_\game$ is a contradiction otherwise. 
We then apply the assumed transformation to $\bbG_\game$ and obtain a guarded 
parity formula $\bbG_\game^\gamma$ that is equivalent to $\bbG_\game$. 
For all we know
the guarded transformation might have introduced modalities or literals
into $\bbG_\game^\gamma$. Using the equivalence $\bbG_\game^\gamma
\equiv \bbG_\game$ we show that these modalities and literals do not
play a crucial role and can be removed to obtain a new formula
$\bbF_\game$ without modalities and literals, but which still
expresses that $\eloi$ has a winning strategy starting from $p$ in $\game$.
Because $\bbF_\game$ is guarded and does not contain modalities it can not have 
cycles. 
Thus, being a dag, it can be evaluated in polynomial time, thereby establishing
whether $\eloi$ has a winning strategy at $p$ in polynomial time as well.

We continue by describing the different steps of this algorithm in
 detail.

The parity formula $\bbG_\game$ is defined such that it essentially uses the
same graph structure as the game $\game$. 
Additionally, we label $\eloi$'s nodes with $\lor$, and $\abel$'s nodes with
$\land$. 
The priority of a node in $\bbG_\game$ is the same as its priority in $\game$.
We initialise
$\bbG$ with the node corresponding to the position $p$. 
A minor difficulty in this construction is that positions in parity games can
have an arbitrary number of successors, whereas nodes in parity formula that
are labelled with $\lor$ or $\land$ have maximally two successors. 
We fix this problem by
introducing auxiliary intermediate nodes for positions in which there are more
than two available moves. 
It is clear that this requires no more than quadratically many new nodes.

Note that $\bbG_\game$ contains modalities nor literals.
Hence, its truth in a pointed Kripke model does not depend on the model at all.
For this reason we may focus on a very simple model $\bbB$ consisting of one
single state $b$, which has no successors and at which all propositional letters 
are true.
Because $\bbG_\game$ reproduces the structure from $\game$ it is not
hard to prove the following claim, using transformations between winning
strategies for $\eloi$ in the evaluation game $\EG(\bbG_\game,\bbB)$ and
winning strategies for $\eloi$ in $\game$:
\begin{equation}
\label{eq:guard1}
\bbB,b \Vdash \bbG_\game \text{ iff } 
p \text{ is a winning position in } \game.
\end{equation}

The hypothesized algorithm then computes, by assumption in polynomial time,
the guarded parity formula $\bbG_\game^\gamma$ with 
\begin{equation}
\label{eq:guard2}
\bbG_\game^\gamma \equiv \bbG_\game.
\end{equation}

In the next step we remove the modalities and literals of $\bbG_\game^\gamma$,
obtaining a parity formula $\bbF_\game$. 
More precisely, we assign $\bot$ to the label of each nodes in $\bbG$ that is
labelled in $\bbG_\game^\gamma$ with either $\Diamond$ or a negative literal, 
and we we assign $\top$ to each node that is labelled with either $\Box$ or a 
positive literal. 
To make sure that the resulting structure is a well-defined parity formula, we
also need to remove all outgoing edges from nodes that were labelled with a 
modality in $\bbG_\game^\gamma$. 
It is clear that $\bbF_\game$ does not contain any modalities or literals.
The following claim clarifies the relation between $\bbG_\game^\gamma$ and 
$\bbF_\game$:
\begin{equation}
\label{eq:guard3}
\bbB,b \Vdash \bbG_\game \text{ iff } \bbB,b \Vdash \bbF_\game.
\end{equation}

To prove this, note that since $\bbF_\game$ only contains $\lor$, $\land$,
$\top$ and $\bot$, the model $\bbB$ does not really play a role in the
evaluation game $\EG(\bbF_\game,\bbB)$. 
The evaluation game $\EG(\bbF_\game,\bbB)$ stops whenever we hit a vertex in 
$\bbF_{\game}$ that originated as a modal or literal node in the parity 
formula $\bbG_\game^\gamma$. 
But given the shape of the model $\bbB$, any match of the evaluation game 
$\EG(\bbG_\game^\gamma,\bbB)$ stops at modalities and literals, and it is
assigned the same winner as in $\EG(\bbF_\game,\bbB)$.
This proves \eqref{eq:guard3}.

Combining the equations \eqref{eq:guard1}, \eqref{eq:guard2} and
\eqref{eq:guard3}, we obtain the following:
\[
\bbB,b \Vdash \bbF_\game \mbox{ iff }
p \text{ is a winning position in } \game. 
\]
Thus, the algorithm just needs to decide whether $\bbB,b \Vdash \bbF_\game$,
while the model $(\bbB,b)$ is actually irrelevant because $\bbF_\game$ contains
no modalities and no literals. 
Moreover, we can evaluate $\bbF_\game$ in polynomial time because it is a dag,
labelled with just $\lor$, $\land$, $\top$ and $\bot$. 
To see that $\bbF_\game$ is a dag assume for a contradiction that there was a cycle in $\bbF_\game$. In this case this
cycle would also exist in $\bbG_\game^\gamma$ because the construction
of $\bbF_\game$ from $\bbG_\game^\gamma$ did not add new edges. 
Because $\bbG_\game^\gamma$ is guarded there must be a modality on this cycle.
But then the cycle can no longer exist in $\bbF_\game$ because we cut all the
edges going out of a modal node.
\end{proof}

\begin{remark}
Theorem~\ref{t:gtlow} stands in tension with Theorem~3.2 in~\cite{brus:guar15},
where the authors claim that for the ``classic'' guarded transformation
procedure $\tau_0$ described in the paper the number of the elements in the 
closure of the guarded output formula $\tau_0(\phi)$ is at most quadratic 
in the number of elements in the closure of the input formula $\phi$.
We doubt that the given proof actually shows this.
It relies on the assumption that there exists a hierarchical equation 
system (HES) that is equivalent to the input formula $\phi$, of size
equal to the number of elements in the closure of $\phi$, and which satisfies
the additional property that every variable occurs only once on the right hand 
side of an equation with higher priority. 
The existence of such a hierarchical equation system is unclear, however.

The authors seem to be aware of this issue when they write, on top of page 213,
that ``the usual guarded transformation procedures expect an HES of a special
form, namely one stemming from an $\mathcal{L}_{\mu}$-formula, and produce an
HES which does not fall into this class''.
It appears that this is exactly the problem with the construction they describe
in their proof of Theorem~3.2.
\end{remark}


Our third and last result about guarded transformation states that if we start
from an \emph{untwisted} parity formula, then we may obtain an equivalent
(strongly) guarded parity formula of \emph{quadratic} size.
Note, however, that this guarded equivalent will itself not necessarily be 
untwisted.
Our construction here is somewhat similar to normalisation procedures given
in for instance~\cite{vene:auto06,vard:auto08}, the nontrivial difference being
that here the input structures are not tree-based but dag-based.

\begin{fewtheorem}
\label{t:guard2}
There is an algorithm that transforms an untwisted parity formula $\bbG$ into
an equivalent, strongly guarded, parity formula $\bbG^{g}$ such that
$\size{\bbG^{g}} \leq \size{\bbG}^{2}$ and $\idx{(\bbG^{g})} \leq 
\idx(\bbG) + 1$.
\end{fewtheorem}

The guarded transformation referred to in Theorem~\ref{t:guard2} is provided
in Definition~\ref{d:gt} below.
Immediately after the formalities we provide some intuitions in 
Remark~\ref{r:gt1}.

\begin{definition}
\label{d:gt}
Let $\bbG = (V,E,L,\Om,v_{I})$ be an untwisted parity formula, witnessed by the
relations $D$ and $B$.
We define the structure $\bbG^{g} \isdef (V',E',L',\Om',v'_{I})$ as follows.
To start with, we take a fresh state $z$, and define $Y \isdef \Dom(\Om)$,
$Y_{*} \isdef Y \cup \{ z \}$. 
We define the carrier of $\bbG^{g}$ to be the set 
\[
V' \isdef 
\big\{ (v,x) \in V \times Y_{*} \mid v \in D^{*}(x) \text{ or } x = z
\big\}
\cup 
\big\{ (x, \wh{x}) \mid x \in \Dom(\Om) \big\},
\]
where we use the following auxiliary definition, for $x \in \Dom(\Om)$:
\[
\wh{x} \isdef \left\{\begin{array}{ll}
   \top & \text{ if } \Om(x) \text{ is even}
\\ \bot & \text{ if } \Om(x) \text{ is odd}.
\end{array}\right.
\]
Then we set
\[
E'(v,x) \isdef \left\{\begin{array}{ll}
   \nada & \text{ if } x \in \{ \top, \bot \}
\\[1mm] \nada & \text{ if } L(v) \in \Lit_{c}(\Prop) \text{ and } x \in Y_{*}
\\[1mm] \{ (u,u) \mid u \in E(v) \cap \Dom(\Om) \} 
\\[1mm] \multicolumn{1}{r}{\cup \{ (u, z) \mid u \in E(v) \setminus 
     \Dom(\Om) \}}
   & \text{ if } L(v) \in \{ \dia,\Box \} \text{ or } x = z
\\[1mm] \{ (u,x) \mid u \in D(v) \} 
    \cup \{ (u, \wh{u}) \mid u \in B(v) \cap D^{*}(x) \}
\\[1mm]  \multicolumn{1}{r}{\cup\; \{ (u,u) \mid u \in B(v) \setminus D^{*}(x)\}}
   & \text{ if } L(v) \in \{ \land,\lor,\epsilon \} \text{ and } x \in Y
\end{array}\right.
\]
The labelling map $L'$ and the priority map $\Om'$ are defined as follows:
\[
L'(v,x) \isdef \left\{\begin{array}{ll}
        x     & \text{ if } x \in \{ \top, \bot \}
\\[1mm] L(v)  & \text{ if } x \in Y_{*},
\end{array}\right.
\]
\[
\Om'(v,x) \isdef \left\{\begin{array}{ll}
   \Om(x) & \text{ if } L(v) \in \{ \Box, \dia \} \text{ and } x \in Y
\\[1mm] 0 & \text{ if } L(v) \in \{ \Box, \dia \} \text{ and } x = z
\\[1mm] \uparrow & \text{ otherwise }.
\end{array}\right.
\]
Finally, we let the starting point $v'_{I}$ of $\bbG^{g}$ be the pair
$(v_{I},z)$.
\end{definition}

\begin{remark}
\label{r:gt1}
For some first intuitions behind this definition, think of a node $(v,x)$ as the
vertex $v$, with $x$ representing a modality-free path $\pi$ through $\bbG$
which leads up to $v$.
In fact we need very little information about this path:
In case $x$ is a state, that is, $x \in Y = \Dom(\Om)$, it represents the
highest $D^{*}$-predecessor of $v$ encountered on the path $\pi$;
we have $x = z$ if there is no such predecessor state on $\pi$.
Similarly, a node of the form $(v,\wh{v})$, with $v \in \Dom(\Om)$, represents
the state $v$, together with a modality-free path $\pi$ which leads up to $v$ 
and happens to be a \emph{cycle} of which $v$ is the highest node; we may 
then consider such a node to be winning for either $\eloi$ or $\abel$, depending
on the parity of $\Om(v)$.

The above intuitions should explain the definitions of $V'$ and $L'$, and the
first two clauses of the definition of $E'$.
The third clause of the latter definition represents the situation where we
either start a new path, just after passing the modal vertex $v$, or else
continue a modality-free path that has not yet passed through a state.
The fourth clause represents the case where we continue a modality-free path
that did pass through a state already.

Here we encounter a key feature of the construction.
Consider a vertex $(v,x) \in V'$ such that $L(v) \in \{ \land, \lor, \epsilon
\}$ and $x \in Y$ is an ancestor of $v$; in addition consider a successor
vertex $u$ of $v$ that is reached by a back edge, i.e., $u \in B(v)$.
Now by untwistedness of $\bbG$ there are only two options for the relative 
position of $x$ and $u$: $u$ is either a proper ancestor, or a descendant of
$x$.
In the first case $u$ has become the new highest state on the path, so that 
we put the vertex $(u,u)$ in $E'(v,x)$; in the second case we have completed 
a loop from $x$ to $x$, and we put $(x,\wh{x})$ in $E'(v,x)$.

For the definition of $\Om'$, first observe that its domain consists exactly 
of the modal nodes of $\bbG^{g}$.
The idea behind the definition of $\Om'(v,x)$ with $x \in Y$ is that 
$\Om'(v,x)$ records the highest priority encountered on the path to $v$
represented by the pair $(v,x)$; but given that $x$ is the highest $\bbG$-state
on this path, this simply corresponds to the value $\Om(x)$.
A state of the form $(v,z)$ represents a path to $v$ that does not encounter
any $\bbG$-state, and so it is given priority $0$.
\end{remark}

More intuitions about the construction are given by the observations below.

\begin{proposition}
\label{p:gt1}
Let $\bbG = (V,E,L,\Om,v_{I})$ be an untwisted parity formula, witnessed by the
relations $D$ and $B$, let $\bbG^{g} = (V',E',L',\Om',v'_{I})$ be as defined in 
Definition~\ref{d:gt}, and let $\pi: (v,x) \mapsto v$ be the projection map 
$\pi: V' \to V$.
Then

\begin{enumerate}[topsep=0pt,itemsep=-1ex,partopsep=1ex,parsep=1ex,%
    label={\arabic*)}]
\item \label{i:10-9-1}
$\pi$ is a homomorphism from $(V',E')$ to $(V,E)$; that is, $E'(v_{0},x_{0})
(v_{1},x_{1})$ implies $Ev_{0}v_{1}$.

\item \label{i:10-9-2}
$\pi$ restricts to a bijection $\pi_{v,x}: E'(v,x) \to E(v)$, for every 
vertex $(v,x) \in V \times Y_{*}$.

\item \label{i:10-9-3}
With $k \geq 1$, let $\rho = (v_{i},x_{i})_{i \leq k}$ be an $E'$-path such
that,
either (i) $(v_{0},x_{0})$ is the only modal node on $\rho$, or (ii) $x_{0} = 
z$.
Then either $x_{k} \in \{ \top, \bot\}$ or there is some $n \in \rng{1}{k}$ such 
that $(v_{i})_{n\leq i \leq k}$ is a downward path and  $x_{j} = v_{n}$, for all
$j \in \rng{n}{k}$.

\item \label{i:10-9-4}
With $k \geq 1$, let $\rho = (v_{i},x_{i})_{i \leq k}$ be an $E'$-path such
that no $(v_{i},x_{i})$ with $0< i < k$ is a modal node, and 
$E'(v_{k},x_{k}) \neq \nada$.
Then either
\begin{equation}
\label{eq:g5a}
\Om'(v_{k},x_{k}) = 0 \text{ and } 
\Om(v_{i}) {\uparrow} \text{ for all } i \in \rng{1}{k}
\end{equation}
or 
\begin{equation}
\label{eq:g5b}
\Om'(v_{k},x_{k}) = \max \Big(\{ \Om(v_{i}) \mid i \in \rng{1}{k} \} \Big).
\end{equation}
\end{enumerate}
\end{proposition}

\begin{proof}
The proof of item~\ref{i:10-9-1} is straightforward.

For item~\ref{i:10-9-2} let $(v,x)$ be a state in $V \times Y_{*}$, then we
have $L'(v,x) = L(v)$ by definition of $L'$.
We make a case distinction.

If $L(v) \in \Lit_{c}(\Prop)$ it is easy to check that $E'(v,x) = E(v) = \nada$.
In case $v$ is a modal node or $x = z$, let $u$ be the (unique) $E$-successor
of $v$.
Then we have $E'(v,x) = \{ (u,u)\}$ if $u$ is a state, and  $E'(v,x) 
= \{ (u,z)\}$ otherwise.
In both cases the restriction of $\pi$ to $E'(v,x)$ is a bijection indeed.
Finally, in case $L(v) \in \{ \land, \lor, \epsilon \}$ and $x \in \Dom(\Om)$, 
since $E(v) = D(v) \uplus B(v)$ it is easy to see that 
\[
E(v) = D(v) \uplus \big(B(v) \cap D^{*}(x)\big) \uplus 
\big(B(v) \setminus D^{*}(x)\big),
\]
where both the second and third set have cardinality at most 1.
Given the definition of $E'(v,x)$ in this case it is easy to derive from this.
\medskip

Item~\ref{i:10-9-3} is proved by an induction on $k$, of which we omit the easy proof for
the base case, where $k = 1$.
In the inductive case, where $\rho = (v_{i},x_{i})_{i \leq k+1}$, it cannot be
the case that $x_{k} \in \{ \top, \bot\}$, since $(v_{k},x_{k})$ has a 
successor. 
Hence we may inductively assume the existence of an $n \in \rng{1}{k}$ such that
$v_{n} = x_{k}$ and $(v_{i})_{n\leq i \leq k}$ is a downward path.
Now distinguish cases.

If $v_{k+1} \in D(v_{k})$ then it is easy to verify that $n_{\rho} \isdef n$
meets the requirements.
If $v_{k+1} \in B(v_{k}) \cap D^{*}(x_{k})$, we find $x_{k+1} = \wh{v_{k+1}}
\in \{ \top, \bot\}$.
Finally, if $v_{k+1} \in B(v_{k}) \setminus D^{*}(x_{k})$, we have $x_{k+1} = 
v_{k+1}$, so that we may take $n_{\rho} \isdef k+1$.
\medskip

For the proof of item~\ref{i:10-9-4}, let $\rho$ be as described, then
we find $x_{k} \in Y_{*}$ since $E'(v_{k},x_{k}) \neq \nada$.
By induction on $k \geq 1$ we will prove the existence of an index $n \in 
\rng{1}{k}$ such that $D^{*}v_{n}v_{k}$ and either
\begin{equation}
\label{eq:g4a}
x_{k} = z, n = 1, \text{ and } \Om(x_{i}){\uparrow} \text{ for all }
   i \in \rng{1}{k}
\end{equation}
or
\begin{align}
\label{eq:g4b1}
   & x_{k} = v_{n} \in Y,
\\ \label{eq:g4b2}
   & (v_{i})_{n\leq i \leq k} \text{ is a downward path}
\\ \label{eq:g4b3}
   \text{ and } 
   & \Om(x_{k}) = \max \Big(\{ \Om(v_{i}) \mid 1 \leq i \leq k \} 
     \Big).
\end{align}

It is easy to see that this suffices to prove the statement.
For, observe that if $(v_{k},x_{k})$ is a modal node with $x_{k} = z$, then
we have \eqref{eq:g5a} because of \eqref{eq:g4a}.
On the other hand, if $x_{k} \in Y = \Dom(\Om)$, then by definition of $\Om'$
we find $\Om'(v_{k},x_{k}) = \Om(x_{k})$, so that we have \eqref{eq:g5b}
because of \eqref{eq:g4b3}.

In the base case of the induction, where $k = 1$, we distinguish cases, and 
recall that since $v_{0}$ is a modal node, we have
\[
E'(v_{0},x_{0}) = 
\{ (u,u) \mid u \in E(v_{0}) \cap \Dom(\Om) \} \cup
     \{ (u, z) \mid u \in E(v_{0}) \setminus \Dom(\Om) \} 
\]
Thus, if $\Om(v_{1}){\downarrow}$ we have $x_{1} = v_{1} \in \Dom(\Om)$.
Taking $n \isdef 1$, it is straightforward to verify 
\eqref{eq:g4b1} -- \eqref{eq:g4b3}.
On the other hand, if $\Om(v_{1}){\uparrow}$ we obtain $x_{1} = z$
and we may easily check \eqref{eq:g4a}, again with $n \isdef 1$.

In the inductive step for $k+1$, the inductive hypothesis yields an index 
$n \in \rng{1}{k}$ such that $D^{*}v_{n}v_{k}$ and either \eqref{eq:g4a} or
\eqref{eq:g4b1} -- \eqref{eq:g4b3}.
We need to find an $n' \in \rng{1}{k+1}$ satisfying the same conditions for
$k+1$.

Our first observation is that since $v_{k}$ is by assumption not modal, while
$E(v_{k}) \neq \nada$, we find $L(v_{k}) \in \{ \land, \lor, \epsilon \}$.
Thus by definition of $E'$ it follows  from $(v_{k+1},x_{k+1}) \in E'(v_{k},
x_{k})$ and $x_{k} \in Y_{*}$, that we can make the following case distinction.

\begin{ourlist}
\item[1)] 
   If $(v_{k},v_{k+1})$ is a downward edge, then we have $x_{k+1} = x_{k}$,
   and so we can take $n' \isdef n$ and easily obtain that 
   $x_{k+1} = v_{n} \in Y$ \eqref{eq:g4b1},    
   $D^{*}v_{n}v_{k+1}$,
   and $(v_{i})_{n\leq i \leq k+1}$ is a downward path \eqref{eq:g4b2}.
   Furthermore, note that $\Om(x_{k+1})$, if defined, is not bigger than 
   $\Om(v_{n})$ since $\bbG$ is dag-based and $D^{*}v_{n}v_{k+1}$.
   From this we derive that 
   $\Om(v_{n}) = \max \Big( \{ \Om(v_{i}) \mid m < i \leq k+1 \}\Big)$
   \eqref{eq:g4b3}.

\item[2)] 
   If $(v_{k},v_{k+1})$ is a back edge, then by definition of $E'$ we have 
   $x_{k+1} = v_{k+1}$, while $v_{k+1} \in Y = \Dom(\Om)$ since $\bbG$ is 
   dag-based; this proves \eqref{eq:g4b1} if we take $n' \isdef k+1$.
   By this choice \eqref{eq:g4b2} holds trivially.
   
   Furthermore, observe that by Proposition~\ref{p:dagcyc} we either have 
   $D^{*}v_{k+1}v_{n}$ or $v_{k+1} = v_{j}$ for some $j \in \rng{n}{k}$.
   (Here we crucially use the untwistedness of $\bbG$). 
   However, in the latter case we would find $D^{*}v_{n}v_{k+1}$, so that by
   definition of $E'$ we would obtain $x_{k+1} = \wh{v_{k+1}}$, contradicting
   our assumption that $x_{k+1} \in Y_{*}$.
   We may conclude that $D^{*}v_{k+1}v_{n}$, so that we find $\Om(v_{k+1}) \geq 
   \Om(v_{n})$ because $\bbG$ is dag-based.
   From this it is clear that 
   \[
   \Om(v_{k+1}) = \max \Big( \{ \Om(v_{i}) \mid 1 \leq i \leq k+1 \} 
     \Big)
   \]
   which proves \eqref{eq:g4b3}. 
\end{ourlist}

\noindent
This finishes the inductive proof, and hence the proof of item~\ref{i:10-9-4}
of the Proposition.
\end{proof}

\begin{proposition}
Let $\bbG = (V,E,L,\Om,v_{I})$ be an untwisted parity formula.
Then $\bbG^{g}$ is a strongly guarded parity formula such that 
$\size{\bbG^{g}} \leq \size{\bbG}^{2}$ and $\idx(\bbG^{g}) \leq 
\idx(\bbG) + 1$.
\end{proposition}

\begin{proof}
The statement about the size of $\bbG^{g}$ is immediate by the definitions.
Since $\Dom(\Om') = V'_{m}$, it is immediate that $\bbG^{g}$ has no unguarded 
paths.
It thus suffices to show that $\bbG^{g}$ is a parity formula, but, again using
$\Dom(\Om') = V_{m}$, we can do so by proving that every cycle through the graph
$(V',E')$ passes a modal node.
Let $D$ and $B$ be the relations witnessing the untwistedness of $\bbG$.

So let $(v_{i},x_{i})_{i\leq k+1}$, with $k \geq 1$ and $(v_{0},x_{0}) = 
(v_{k+1},x_{k+1})$, be a cycle in $(V',E')$,
and suppose for contradiction that the cycle does not contain a modal node.
It is not hard to see that this can only be the case if $L(v_{i}) \in 
\{\land, \lor, \eps \}$ and $L'(v_{i},x_{i}) = L(v_{i})$ for all $i$.
Furthermore, it follows from item (\ref{i:10-9-1} of 
Proposition~\ref{p:gt1} that $(v_{i})_{i\leq k+1}$ is a cycle in $(V,E)$, and
since the latter graph is untwisted, by Proposition~\ref{p:dagcyc}(2) 
this cycle must have a unique highest node with respect to the relation $D$.
Without loss of generality we may assume this node to be $v_{0}$; that is, we 
have $D^{*}v_{0}v_{i}$ for all $i \in \rng{0}{k+1}$.
In particular, it follows from $D^{*}v_{0}v_{k}$ that the edge from $v_{k}$ to
$v_{0}$ must be a back edge: $(v_{k},v_{0}) \in B$.
But since $L(v_{k}) \in \{\land, \lor, \eps \}$, by definition of $E'$ and the
fact that $(v_{0},x_{0} \in E'(v_{k},x_{k})$ we find $x_{0} = v_{0}$.

Now let $n \in \rng{0}{k}$ be the first index such that $(v_{n},v_{n 
\oplus 1})$ is a back edge (where we write $\oplus$ to denote addition modulo
$k+1$).
Then we have $Dv_{i}v_{i+1}$ for all $i \in \rng{0}{n-1}$, and from this it is
easy to derive that $x_{i} = x_{0}$ for all $i \in \rng{0}{n}$.
In particular we have that $x_{n} = x_{0}$, while also $D^{*}x_{0}v_{n+1}$.
But since $L'(v_{n},x_{n}) \in \{ \land, \lor, \epsilon \}$ and  
$L'(v_{n},x_{n}) = L(v_{n})$, by definition of $E'$ it must be the case that 
$x_{n\oplus 1} = \wh{v_{n \oplus 1}} \in \{\top,\bot\}$, and thus 
$E'(v_{n \oplus 1},x_{n \oplus 1}) = \nada$.
This provides the desired contradiction with the fact that 
$(v_{n \oplus 1},x_{n \oplus 1})$ lies on an $E'$-cycle.

To prove $\idx(\bbG^g) \leq \idx(\bbG)$ we show that for every
alternating chain of length $n + 1$ in some cluster of $\bbG^g$ there is
an alternating chain of length $n$ in some cluster of $\bbG$. So let
$(v_0,x_0),\dots,(v_n,x_n)$ be an alternating chain in some cluster of
$\bbG^g$. 
Because the nodes in this chain all are in the domain of the
parity function they must all be modal nodes. 
That these nodes are in the same cluster means that there is, for each $i \in
\rng{0}{n - 1}$, a path $\delta_{i+1}$ from $(v_i,x_i)$ to $(v_{i+1},x_{i+1})$,
and a path $\delta_0$ from $(v_{i+1},x_{i+1})$ to $(v_0,x_0)$. 
In each of these paths the starting node is modal, and distinct from the end
node; hence, for each $i \in \rng{0}{n}$ there is a node $(u_i,y_i)$ on
$\de_{i}$ which is the last occurrence of a modal node before the end
node $(v_i,x_i)$.
Then let $\rho_i$ be the final segment of the path $\delta_i$ from $(u_i,y_i)$ 
to node $(v_i,x_i)$. 
By item~\ref{i:10-9-4} in Proposition~\ref{p:gt1} it follows that for each 
$i \in \rng{0}{n}$ either there is some $(w_i,x_i)$ on $\delta_i$ with
\begin{equation} \label{eq:condition on priority}
 \Om'(v_i,x_i) = \Om(w_i),
\end{equation}
or $\Om'(v_i,x_i) = 0$. 
But the latter can only happen to $(v_0,x_0)$ since we are considering an 
alternating chain in which the priorities need to be strictly increasing. 
Thus, \eqref{eq:condition on priority} holds for all $i \in \rng{1}{n}$.
It follows that $w_1,\dots,w_n$ is an alternating chain of length $n$, and it
only remains to check that these nodes are from the same cluster of $\bbG$. 
To this aim observe that by concatenating all the paths $\delta_i$ for $i \in 
\rng{0}{n}$ we obtain a, not necessarily simple, cycle in $\bbG^g$ on which all 
the $(w_i,x_i)$ lie. 
Using the homomorphism $\pi$ from Proposition~\ref{p:gt1}(\ref{i:10-9-1},
we get a cycle in $\bbG$ that connects all the $w_i$ for $i \in \rng{1}{n}$.
Hence, all the elements in the chain are in the same cluster of $\bbG$.
\end{proof}

\begin{proposition}
Let $\bbG = (V,E,L,\Om,v_{I})$ be an untwisted parity formula, witnessed by the
relations $D$ and $B$.
Then
\begin{equation*}
\bbG \equiv \bbG^{g}.
\end{equation*}
\end{proposition}

\begin{proof}
Let $(\bbS,s_{I})$ be an arbitrary pointed Kripke model, with $\bbS = (S,R,V)$.
We need to show that 
\begin{equation}
\label{eq:g1}
\bbS,s_{I} \sat \bbG \text{ iff } \bbS,s_{I} \sat \bbG^{g},
\end{equation}
and in order to do so, by the determinacy of evaluation games it suffices to 
prove, for $\Pi \in \{ \eloi, \abel\}$, that 
\begin{equation}
\label{eq:g2}
(v_{I},s_{I}) \in \Win_{\Pi}(\EG(\bbG,\bbS))
\text{ implies } 
(v'_{I},s_{I}) \in \Win_{\Pi}(\EG(\bbG^{g},\bbS)).
\end{equation}
For reasons of symmetry between the roles of $\eloi$ and $\abel$ we will only 
prove \eqref{eq:g2} for the case where $\Pi = \eloi$.

Let $f$ be a positional winning strategy of $\eloi$ in the evaluation game 
$\EG \isdef \EG(\bbG,\bbS)@(v_{I},s_{I})$; we may think of $f$ as a function
mapping positions of the form $(v,s) \in L^{-1}(\lor) \times S$ to `disjuncts of
$v$', that is, elements of the set $E(v)$, and positions of the form $(v,s) 
\in L^{-1}(\dia) \times S$ to successors of $s$, i.e., elements of the set 
$R(s)$.

We will provide $\eloi$ with a positional winning strategy $f'$ in the game 
$\EG^{g} \isdef \EG(\bbG^{g},\bbS)@(v'_{I},s_{I})$.
For its definition, 
consider an arbitrary position $((v,x),s)$ for her in the game $\EG^{g}$, 
and distinguish cases.

\begin{ourlist}
\item[$\bullet$]
In case $(v,s) \in \Win_{\eloi}(\EG)$ and $L(v) = \lor$, $\eloi$ plays 
\[
f'((v,x),s) \isdef \pi_{v,x}^{-1}(f(v,s)) \in E'(v,x),
\]
where $f(v,s) \in E(v)$ is $\eloi$'s move at position $(v,s)$ in $\bbG$ as
prescribed by her strategy $f$, and $\pi_{v,x}$ is the bijection between
$E'(v,x)$ and $E(v)$ which is the restriction of the projection map $\pi$ to the 
set $E'(v,x)$, cf.~Proposition~\ref{p:gt1}(2).
\item[$\bullet$]
In case $(v,s) \in \Win_{\eloi}(\EG)$ and $L(v) = \dia$, $\eloi$ plays 
\[
f'((v,x),s) \isdef f(v,s) \in R(s),
\]
where $f(v,s) \in R(s)$ is $\eloi$'s move in $\bbG$ prescribed by her strategy 
$f$.
\item[$\bullet$] 
In case $(v,s) \not\in \Win_{\eloi}(\EG)$, $\eloi$ plays randomly.
\end{ourlist}

Given a (partial) $\EG^{g}$-match $\Si = ((v_{i},x_{i}),s_{i})_{i<\ka}$, define 
its \emph{projection} as the sequence $\Si^{\pi} \isdef (v_{i},s_{i})_{i<\ka}$.


\begin{claimfirst}
\label{cl:g1}
For any $f'$-guided (partial) $\EG^{g}$-match $\Si$, its projection $\Si^{\pi}$ 
is an $f$-guided (partial) $\EG$-match.
\end{claimfirst}

\begin{pfclaim}
We prove the claim by induction on the length $k$ of $\Si = ((v_{i},x_{i}),
s_{i})_{i\leq k}$.

The base case of the induction is immediate by the definitions, the two matches
being $\Si_{0} \isdef ((v_{I},z),s_{I})$ and $\Si_{0}^{\pi} = (v_{I},s_{I})$.

For the inductive case, let $\Si = ((v_{i},x_{i}),s_{i})_{i \leq k+1}$ be an
$f'$-guided $\EG^{g}$-match of length $k+1$, and let $\Si^{\pi}$ be its 
projection.
Write $\De = ((v_{i},x_{i}),s_{i})_{i \leq k}$, then $\Si = \De \cdot 
((v_{k+1},x_{k+1}),s_{k+1})$, $\Si^{\pi} = \De^{\pi} \cdot (v_{k+1},s_{k+1})$,
and inductively we know that $\De^{\pi}$ is an $f$-guided $\EG$-match.
Make the following case distinction.

If the position $((v_{k},n_{k}),s_{k})$ belongs to $\eloi$, then the assumption
that $\Si$ is $f'$-guided means that $((v_{k+1},n_{k+1}),s_{k+1})$ is picked by 
$f'$.
But because of how we defined $f'$ on the basis of the $\EG$-strategy $f$, it is
immediately clear that $(v_{k+1},s_{k+1})$ is the move picked by $f$ at position
$(v_{k},s_{k})$.
Thus $\Si^{\pi}$ is $f$-guided indeed.

If, on the other hand, the position $((v_{k},n_{k}),s_{k})$ belongs to $\abel$,
then $L'(v_{k},x_{k}) \in \{ \land, \Box \}$.
We only cover the first case, the other one is simpler.
Since $\Si$ is a partial $\EG^{g}$-match, its final position $((v_{k+1},x_{k+1}),
s_{k+1})$ must be such that $s_{k+1} = s_{k}$ and $(v_{k+1},x_{k}) \in 
E'(v_{k},x_{k})$.
But then it follows from the definition of $E'$ that $v_{k+1} \in E(v_{k})$, 
so that $(v_{k+1},s_{k+1}) = \pi(v_{k+1},x_{k+1})$ is a legitimate choice for 
$\abel$ as the next position in $\EG$ after $(v_{k},s_{k})$.
In particular, we have $\De \cdot (v_{k+1},s_{k+1}) = \Si^{\pi}$ is an
$f$-guided (partial) $\EG$-match, as required.
\end{pfclaim}

Note that it follows from Claim~\ref{cl:g1} that for every guided (partial)
$\EG^{g}$-match $\Si = ((v_{i},x_{i}),s_{i})_{i<\ka}$, every pair $(v_{i},
s_{i})$ is a winning position for $\eloi$ in $\EG$.

\begin{claim}
\label{cl:g3}
Let $\Si = ((v_{i},x_{i}),s_{i})_{i \leq k}$ be an $f'$-guided $\EG^{g}$-match.
Then $x_{k} \neq \bot$.
\end{claim}

\begin{pfclaim}
Suppose for contradiction that $x_{k} = \bot$.
First observe that since $(v_{0},x_{0}) = (v_{I},z)$, we find $x_{0} \neq \bot$,
so that $k$ cannot be zero.

Let $m < k$ be the last stage where $L'(v_{m},x_{m}) \in \{ \dia,\Box\}$, or
let $m \isdef 0$ if there is no such stage. 
It then follows by definition of $E'$ that $x_{m+1} \not\in \{ \bot, \top \}$, 
so that we must have $k \geq m+2$.
Furthermore, by our assumption on $m$ there must be an $s \in S$ such that 
$s_{i} = s$, for all $i$ with $m<i\leq k$.
That is, since stage $m$, the match has proceeded by Boolean moves played by 
$\eloi$ and $\abel$ (and by automatic $\epsilon$-transitions).

By definition of $E'$, it can only happen that $x_{k} = \bot$ if 
$Bv_{k-1}v_{k}$, $D^{*}x_{k-1}v_{k}$ and $\Om(v_{k})$ is odd.
By Proposition~\ref{p:gt1}(3), applied to the path $(v_{i},x_{i})_{m \leq i 
\leq k-1}$, we may find an $n \in \rng{m+1}{k-1}$ such that $x_{k-1} = v_{n}$ 
and
\begin{equation}
\label{eq:g11}
(v_{i})_{n\leq i \leq k-1} \text{ is a downward path in } \bbG.
\end{equation}

We now claim that 
\begin{equation}
\label{eq:g10}
v_{k} = v_{j} \text{ for some $j$ such that } n \leq j < k.
\end{equation}

To prove \eqref{eq:g10}, consider an arbitrary path $\si$ from $v_{I}$ to 
$v_{n}$, and compose this with the path $\rho = (v_{i})_{n\leq i \leq k-1}$.
Then by untwistedness, this path must pass through $v_{k}$, and so 
either $\si$ or $\rho$ must do the same (i.e., pass through $v_{k}$).
Clearly, if $\rho$ passes through $v_{k}$, \eqref{eq:g10} follows immediately, 
so assume that instead, $\si$ passes through $v_{k}$.
This means that we find $D^{*}v_{k}v_{n}$, while earlier on we already saw that 
$D^{*}v_{n}v_{k}$ (recall that $v_{n} = x_{k-1}$).
It follows that $v_{n} = v_{k}$ and so \eqref{eq:g10} holds in this case as 
well.

Formulating \eqref{eq:g10} in other words, we find that the partial $\EG$-match 
$(v_{i},s)_{j \leq i \leq k}$ induces an $E$-cycle.
This means that, against $\eloi$'s positional winning strategy $f$, $\abel$
has a local strategy, from position $(v_{j},s)$, that ensures a return to this 
same position $(v_{j},s) = (v_{k},s)$.

But it follows from \eqref{eq:g11} that $(v_{i})_{j\leq i \leq k-1}$ is a 
downward path in $\bbG$, and because $\bbG$ is dag-based, we obtain that 
$\Om(v_{j}) = \Om(v_{k})$ is the maximal priority reached on this cycle, and we
already saw that this is priority is odd.
As a consequence, if $\abel$ continues to play this local strategy against
$\eloi$'s winning strategy, the resulting match will be infinite, with its 
projection on $\bbG$ consisting of the cycle $v_{j}v_{j+1}\cdots v_{k}$ repeated
infinitely often.
Such a match would then constitute a loss for $\eloi$, contradicting the fact 
that, by Claim~\ref{cl:g1}, the position $(v_{j},s)$ is winning for $\eloi$ in
$\EG$.
\end{pfclaim}

To see why the Proposition follows from these claims, take an arbitrary full 
$f'$-guided $\EG^{g}$-match $\Si$.
First consider the case where $\Si = ((v_{i},x_{i}),s_{i})_{i \leq k}$ is 
finite.
We consider the projection match $\Si^{\pi}$, and make a case distinction as to 
whether $\Si^{\pi}$ is full or partial as an $\EG$-match.
In the first case it is not difficult to show that $\Si$ and $\Si^{\pi}$ will 
have the same winner, and since $\Si^{\pi}$ is guided by $\eloi$'s winning 
strategy $f$, this winner is $\eloi$.
This leaves the case where $\Si^{\pi}$ is not full but partial.
This can only occur if the last position of $\Si$ is of the form $((u,\wh{u}),s)$
for some state $u$.
It follows from Claim~\ref{cl:g3} that in this case we have $\wh{u} \neq \bot$,
so that $\wh{u} = \top$.
But then $\Si$ is won by $\eloi$, as required.

Finally, consider the case where $\Si = ((v_{i},x_{i}),s_{i})_{i < \om}$ is 
infinite.
Let $(m_{j})_{j\in\om}$ be the sequence of indices corresponding to the modal 
nodes in $\Si$; it easily follows from the fact that $\Dom(\Om') = 
(L')^{-1}(\{\dia,\Box\})$ and the definition of $\Om'$ that
\[
q \isdef \max \Big( \Inf \{ \Om'(v_{m_{j}}, x_{m_{j}}) \mid j \in \om \} \Big)
\]  
is the maximal priority occurring infinitely often in $\Si$, i.e.,
\[
q = \max \Big( \Om'[\Inf \{ (v_{i},x_{i}) \mid i \in \om \}] \Big).
\]
Thus, in order to prove that $\Si$ is won by $\eloi$, we have to show that 
$q$ is even.
To prove this, we consider the projection $\Si^{\pi} \isdef 
(v_{i},s_{i})_{i\in\om}$.
It easily follows by Claim~\ref{cl:g1} that $\Si^{\pi}$ is an $f$-guided 
$\EG$-match. 
By our assumptions then, $\Si^{\pi}$ is a win for $\eloi$; that is, we have 
\[
q^{\pi} \isdef \max \Big( \Om [ \Inf \{ v_{i} \mid i \in \om \} ] \Big) 
\text{ is even }.
\]
It is left to relate $q$ and $q^{\pi}$. 
This we do on the basis of Proposition~\ref{p:gt1}(\ref{i:10-9-1}, from which 
it readily follows that
\[
\Om'(v_{m_{j+1}},x_{m_{j+1}}) = \max \Big(
  \{ 0 \} \cup \{ \Om(v_{i}) \mid m_{j} < i \leq m_{j+1} \}
\Big)
\]
for all $j \in \om$.
But then it is almost immediate that 
\[
q = q^{\pi},
\]
so that $q$ is even, as required.
\end{proof}

Finally, as a corollary of the results in this section, we can say the following 
about the existence of a \emph{polynomial} guarded transformation.
There is a \emph{quadratic} guarded  transformation on $\mu$-calculus formulas
indeed, but only if one measures the input formula of the construction in
subformula-size, and the output formula in closure-size.
If we measure the input and output formula in the same way, then all known 
constructions are exponential, and if this measure is closure-size, then any
effective guarded transformation must be as hard as solving parity games.

\section{Conclusions}
\label{s:conc}

\subsection*{Summary}

In this paper we gave an in depth analysis of two complexity measures in the
modal $\mu$-calculus, viz., \emph{size} and \emph{alternation depth}.
Using alternating tree automata or parity formulas as yardstick, we studied and 
compared two size measures: subformula-size (number of subformulas) and 
closure-size (size of closure, or number of derived formulas).
The reason to take parity formulas for this purpose is that the size and index
of these devices are natural and obviously defined complexity measures that 
feature in most complexity theoretic studies, sometimes implicitly (for instance,
in a game-theoretic setting).

Our main findings are that
\begin{itemize}
\item 
If we measure the size of a $\mu$-calculus formula as closure-size, there are 
linear-size transformations between $\mu$-calculus formulas and arbitrary parity 
formulas (Theorem~\ref{t:clur} and Theorem~\ref{t:cyc-fix}).
\item
If we measure the size of a $\mu$-calculus formula as subformula-size, there are 
linear-size transformations between $\mu$-calculus formulas and so-called 
untwisted parity formulas (Theorem~\ref{t:1} and Theorem~\ref{t:cyc-fix-utw}).
\item 
In both cases we could obtain an (almost) exact match between the alternation 
depth of the regular formula and the index of its parity variant.
\end{itemize}

Probably the most important novelty of our approach is that we are explicit and 
precise about the notion of alphabetical equivalence and the impact of the 
identification of $\al$-equivalent formulas on the above correspondences.

First we work in a setting where alphabetical variants are \emph{not} identified.
For starters, this means that it only makes sense to define the subformula-size
$\ssz{\xi}$ for a \emph{clean} formula $\xi$ (cf.~Remark~\ref{r:sfsm}), and the
closure-size $\csz{\xi}$ for a \emph{tidy} formula $\xi$ 
(cf.~Remark~\ref{r:mottidy}).
In addition, in this setting we restricted attention to substitutions that are 
\emph{safe} in the sense that variable capture (which generally requires to 
consider alphabetical variants) is avoided all together.
Despite these constraints we worked out constructions taking care of the 
above-mentioned correspondences.

Then we moved to the arguably more natural setting where we do take alphabetical
equivalence into account --- note that in order to extend our definition of size 
to arbitrary formulas, we have to consider alphabetical variants.
We call a size measure \emph{$\al$-invariant} if it assigns the same value to 
alphabetical variants and counts $\al$-equivalent formulas only once, and our
first observation is that the standard size measures $\ssz{{\cdot}}$ and 
$\csz{{\cdot}}$ lack this property.
We also show that renaming the bound variables of a formula, in order to make it 
clean, may cause an exponential blow-up in size, measured in terms of 
closure-size (Proposition~\ref{p:closexp}).

On a more positive note, we introduced $\al$-invariant size measures 
$\sszal{{\cdot}}$ and $\cszal{{\cdot}}$, related to subformula-size and 
closure-size, respectively, for \emph{arbitrary} $\mu$-calculus formulas.
In either case, this measure can be defined via a suitable \emph{cautious}
renaming of bound variables; that is, in Definition~\ref{d:szal} we define 
\begin{eqnarray*}
   \sszal{\xi} & \isdef & \ssz{\pol{\xi}}
\\ \cszal{\xi} & \isdef & \csz{\spol{\xi}},
\end{eqnarray*}
where $\pol{\xi}$ is an alphabetical variant of $\xi$ such that $\eqal$ is the
identity relation on its set of subformulas, and similarly, $\spol{\xi}$ is an 
alphabetical variant of $\xi$ such that $\eqal$ is the identity relation on its 
closure set.
Finally, we show that both of these size measure interact properly with the 
standard definition of substitution (i.e., where we take alphabetical variants
in order to avoid variable capture).

As a working example, we consider the concept of a guarded transformation (that 
is, the construction of a guarded equivalent for an arbitrary $\mu$-calculus 
formula), from the perspective of parity formulas.
We prove the existence of an exponential construction for arbitrary parity 
formulas, and of a quadratic construction from untwisted to arbitrary parity 
formulas.
We also showed that any guarded transformation of parity formulas is as hard
as solving parity games.
For regular formulas this means that there is a quadratic construction if we
measure the input formula in subformula-size and the output formula in 
closure-size,
but for the time being there are only exponential 
constructions if we measure the input formula in the same way as the output
formula.

We tried to be fairly complete in stating the links between regular formulas
and their graph-based relatives, and proved all size-related claims about 
these links; as a consequence, this report is a blend of (well-)known and new
results.
We believe the following to be our main original contributions:
\begin{itemize}   
\item
the identification of $\al$-invariance as a desideratum for natural complexity 
measures on $\mu$-calculus formulas;
\item 
the construction of an equivalent parity formula on the closure graph of a given
tidy $\mu$-calculus formula, in such a way that the index matches the alternation
depth of the formula (Theorem~\ref{t:clur});
\item
the concept of an untwisted parity formula (Definition~\ref{d:dagcyc}) and its 
connection with the subformula dag of a formula (Theorem~\ref{t:1} and
Theorem~\ref{t:cyc-fix-utw});
\item
the realisation that a naive renaming of variables, turning a formula into
an equivalent clean one, comes at an exponential prize
(Proposition~\ref{p:closexp});
\item
the definition of size measures that are invariant under $\al$-equivalence
(Definition~\ref{d:szal}), and the cautious renamings on which these definitions
are based;
\item
the quadratic size guarded transformation for untwisted parity formulas 
(Theorem~\ref{t:guard2}).
\end{itemize}

\subsection*{What is the size of a $\mu$-calculus formula?}

After almost 100 pages of discussing size matters in the modal $\mu$-calculus, 
we still have not explicitly answered the question what `the' size of a 
$\mu$-calculus formula should be.
Implicitly, our response has been that the question as such is perhaps not that 
interesting, since the answer is very much context-dependent, so that the 
definition of size will be derived from the most natural graph representation 
for a $\mu$-calculus formula in the context.
That is, one should take closure-size as the definition if the closure graph is 
the most natural representation, and subformula-size if the subformula dag is 
more natural to work with.
Nevertheless, there is a bit more to say, and we will do so by pointing out 
some advantages and disadvantages of the two main definitions discussed here.
\medskip

\noindent
Advantages of working with the closure-size/closure graph include the following:

\begin{itemize}
\item
excellent correspondence with the parity formulas representation: 
the two most important measures that determine complexity-theoretic results 
about the modal $\mu$-calculus are hard-wired into the closure graph and the
global priority map $\gOm$;
\item 
sharper complexity results, due to the more succinct representation of 
$\mu$-calculus formulas using the closure graph;
\item
in case one wants to identify alphabetical variants: derived formulas seem
display smoother interaction with $\eqal$ than subformulas.
\item
in case one does not want to identify alphabetical variants, one needs to
restrict to tidy vs clean formulas, where tidyness is a far milder restriction 
than cleanness (recall that in the approach where propositional variables are 
separated from proposition letters, tidy formulas correspond to sentences);
\end{itemize}

\noindent
On the other hand, working with the subformula dag/subformula size has some
advantages as well:

\begin{itemize}
\item
the subformula relation is more natural and easier to work with than the trace
relation, since it corresponds to the inductive definition of formulas; the 
coinductive flavour of the closure graph can make it less transparent to work
with;
\item
in the context of logic, cleanness is a nice property of formulas, e.g. when
doing \emph{proof theory}; cleanness, however, is not preserved under taking 
unfoldings, and hence is less nature in the context of closure size.
\end{itemize}

\paragraph{Conclusion} 
Our \emph{conclusion} is that, when it comes to complexity results about model
checking or satisfiability, closure-size appears to be the most natural size 
measure.
Hence, if one wants to identify $\al$-equivalent formulas, the natural definition
of the size of a $\mu$-calculus formula $\xi$ would be to take the number 
$\size{\Clos(\xi)/{\eqal}}$ of derived formulas of $\xi$, modulo 
$\al$-equivalence.
This number coincides with the number $\cszal{\xi} \isdef \size{\Clos(\spol{\xi})}$
of derived formulas of the renaming $\spol{\xi}$ of $\xi$.
One has to realise, however, that in this approach, it does not make sense to
require formulas to be clean, since cleanness is not closed under taking 
unfoldings, and renaming variables to make a formula clean comes at an 
exponential cost.

This being said, there may be settings, where \emph{cleanness} is a very
desirable property of $\mu$-calculus formulas, or where the inductively defined
notion of a subformula is significantly easier to work with than that of a 
derived formula; these considerations then may make the number of subformulas 
a more natural size measure than closure-size.
This may be well be the case in a context where \emph{logical} considerations are as
important as complexity-theoretic ones, 

\subsection*{Questions for further research}

Below we mention some directions for future research.

\begin{enumerate}

\item 
To start with, one could take the identification of $\al$-equivalent 
$\mu$-calculus formulas one step further and take the `new-style' formulas to be
the $\al$-equivalence classes of `old-style' formulas.

In fact it is possible to give a \emph{direct} representation of a formula $\xi$
as a parity formula $\ol{\bbG}_{\xi}$ taking the quotient $\Clos(\xi)/{\eqal}$ 
as its carrier set.
We hope to provide some details in future work.

Note however, even thought this is an interesting road to take, following it we 
may lose the direct connection with standard \emph{logic}, since the notion of
a bound variable seems to vanish in this framework.
Perhaps the alternative, variable-free \emph{de Bruijn notation} would be useful
here.

\item
One may think of parity formulas as variable-free (or binding-free)
representations of ordinary $\mu$-calculus formulas.
Up til now they have only featured in algorithmic studies, but it may be 
interesting to take them seriously from the \emph{logical} perspective as well,
and develop, for instance, their model theory or proof theory.

\item
Related to the previous point, parity formulas seem interesting enough to
undertake a structural study of their theory.
In section~\ref{sec:par} we already made some first steps in this directions by
defining two natural notions of a \emph{morphism} between parity formulas, 
together with an associated relation of congruence/bisimilarity equivalence.

It would be interesting to further develop this theory, and to see whether it
helps us to understand some natural notions pertaining to formulas.
To mention two examples, we may show that the relation of $\al$-equivalence is
in fact a congruence, and so we may understand the definition of a parity formula 
on $\eqal$-cells (as mentioned above) from this perspective.

Or, to mention another question: is Kozen's map from $\Sfor(\xi)$ to $\Clos(\xi)$
(cf.~Definition~\ref{d:kozmap}) in fact a morphism of parity formulas? 
That is, is the parity formula $\bbG_{\xi}$ a quotient of $\bbH_{\xi}$ (defined 
in the proof of Theorem~\ref{t:1})?

\item
A key result in the theory of the modal $\mu$-calculus is the disjunctive normal
form theorem by Janin \& Walukiewicz~\cite{jani:auto95}, stating that every 
$\mu$-calculus formula is semantically equivalent to a so-called 
\emph{disjunctive} formula.
These disjunctive formulas have various nice properties, including a linear-time
solution of their satisfiability problem (in terms of subformula-size). 
It is therefore an interesting question what the best normalisation procedure 
is for rewriting a formula $\xi \in \muML$ into an equivalent disjunctive
formula $\xi^{d}$ of minimal \emph{size}.

The best normalisation constructions that are known from the literature are 
automata-theoretic in nature and consist of (in our terminology) a guarded 
transformation, constructing an equivalent guarded alternating automaton from 
a $\mu$-calculus formula, followed by a Simulation Theorem stating that any 
such alternating automaton can be transformed into an equivalent 
nondeterministic one.
Given the analysis of Bruse, Friedmann \& Lange~\cite{brus:guar15}, continued by
us in section~\ref{sec:gua}, both of these transformations are exponential
constructions, making the best normalisation procedure \emph{doubly exponential}.
We claim, however, that the two parts of the normalisation procedure can be
integrated, leading to a single-exponential procedure, and we hope to report on
the details of this construction in forthcoming work.

\item
Finally, we have defined the $\al$-invariant size measures for arbitrary
$\mu$-calculus formulas in terms of certain \emph{renaming} functions, 
the polishing map $\pol{{\cdot}}$ and the skeletal renaming $\spol{{\cdot}}$.
Our definitions of these functions are not very practical, however, and we leave
it as a matter of further research to come up with more useful 
algorithms for
defining these renamings (and hence, for defining the size of arbitrary 
$\mu$-calculus formulas).
\end{enumerate}

{\small
\bibliographystyle{plain}
\bibliography{references/mu,references/extra}

\begin{thebibliography}{10}

\bibitem{afsh:cutf17}
B.~Afshari and G.~Leigh.
\newblock Cut-free completeness for modal mu-calculus.
\newblock In {\em {P}roceedings of the 32nd {A}nnual {ACM\slash IEEE}
  {S}ymposium on {L}ogic {I}n {C}omputer {S}cience ({LICS}'17)}, pages 1--12.
  {IEEE} Computer Society, 2017.

\bibitem{arno:rudi01}
A.~Arnold and D.~Niwi{\'n}ski.
\newblock {\em Rudiments of {$\mu$}-calculus}, volume 146 of {\em Studies in
  Logic and the Foundations of Mathematics}.
\newblock North-Holland Publishing Co., Amsterdam, 2001.

\bibitem{brad:moda06}
J.~Bradfield and C.~Stirling.
\newblock Modal $\mu$-calculi.
\newblock In J.~{van B}enthem, P.~Blackburn, and F.~Wolter, editors, {\em
  Handbook of Modal Logic}, pages 721--756. Elsevier, 2006.

\bibitem{brus:guar15}
F.~Bruse, O.~Friedmann, and M.~Lange.
\newblock On guarded transformation in the modal $\mu$-calculus.
\newblock {\em Logic Journal of the IGPL}, 23(2):194--216, 2015.

\bibitem{calu:deci17}
C.S. Calude, S.~Jain, B.~Khoussainov, W.~Li, and F.~Stephan.
\newblock Deciding parity games in quasipolynomial time.
\newblock In H.~Hatami, P.~McKenzie, and V.~King, editors, {\em Proceedings of
  the 49th Annual {ACM} {SIGACT} Symposium on Theory of Computing, ({STOC}
  2017)}, pages 252--263. {ACM}, 2017.

\bibitem{dago:logi00}
G.~{D'A}gostino and M.~Hollenberg.
\newblock Logical questions concerning the $\mu$-calculus.
\newblock {\em Journal of Symbolic Logic}, 65:310--332, 2000.

\bibitem{demr:temp16}
S.~Demri, V.~Goranko, and M.~Lange.
\newblock {\em Temporal Logics in Computer Science: Finite-State Systems}.
\newblock Cambridge Tracts in Theoretical Computer Science. Cambridge
  University Press, 2016.

\bibitem{emer:tree91}
E.A. Emerson and C.S. Jutla.
\newblock Tree automata, mu-calculus and determinacy (extended abstract).
\newblock In {\em Proceedings~of the 32nd Symposium on the Foundations of
  Computer Science}, pages 368--377. IEEE Computer Society Press, 1991.

\bibitem{font:mode18}
G.~Fontaine and Y.~Venema.
\newblock Some model theory for the modal mu-calculus: syntactic
  characterizations of semantic properties.
\newblock {\em Logical Mewthods in Computer Science}, 14(1), 2018.

\bibitem{grae:auto02}
E.~Gr{\"a}del, W.~Thomas, and T.~Wilke, editors.
\newblock {\em Automata, Logic, and Infinite Games}, volume 2500 of {\em LNCS}.
\newblock Springer, 2002.

\bibitem{jani:auto95}
D.~Janin and I.~Walukiewicz.
\newblock Automata for the modal $\mu$-calculus and related results.
\newblock In {\em Proceedings of the Twentieth International Symposium on
  Mathematical Foundations of Computer Science, MFCS'95}, volume 969 of {\em
  LNCS}, pages 552--562. Springer, 1995.

\bibitem{jani:expr96}
D.~Janin and I.~Walukiewicz.
\newblock On the expressive completeness of the propositional $\mu$-calculus
  w.r.t.\ monadic second-order logic.
\newblock In {\em Proceedings~of the Seventh International Conference on
  Concurrency Theory, CONCUR '96}, volume 1119 of {\em LNCS}, pages 263--277,
  1996.

\bibitem{koze:resu83}
D.~Kozen.
\newblock Results on the propositional $\mu$-calculus.
\newblock {\em Theoretical Computer Science}, 27:333--354, 1983.

\bibitem{koze:fini88}
D.~Kozen.
\newblock A finite model theorem for the propositional $\mu$-calculus.
\newblock {\em Studia Logica}, 47:233--241, 1988.

\bibitem{koze:deci83}
D.~Kozen and R.~Parikh.
\newblock A decision procedure for the propositional $\mu$-calculus.
\newblock In {\em Proceedings~of the Workshop on Logics of Programs 1983},
  LNCS, pages 313--325, 1983.

\bibitem{kupfer:linbran05}
Orna Kupferman and Moshe~Y. Vardi.
\newblock From linear time to branching time.
\newblock {\em ACM Transactions on Computational Logic}, 6(2):273--294, 2005.

\bibitem{most:game91}
A.M. Mostowski.
\newblock Games with forbidden positions.
\newblock Technical Report~78, Instytut Matematyki, Uniwersytet Gda\'{n}ski,
  Poland, 1991.

\bibitem{niwi:fixp86}
D.~Niwi\'{n}ski.
\newblock On fixed point clones.
\newblock In L.~Kott, editor, {\em Proceedings of the 13th International
  Colloquium on Automata, Languages and Programming (ICALP 13)}, volume 226 of
  {\em LNCS}, pages 464--473, 1986.

\bibitem{stir:moda01}
C.~Stirling.
\newblock {\em Modal and Temporal Properties of Processes}.
\newblock Texts in Computer Science. Springer-Verlag, 2001.

\bibitem{vard:auto08}
M.Y. Vardi and T.~Wilke.
\newblock Automata: from logics to algorithms.
\newblock In J.~Flum, E.~Gr{\"{a}}del, and T.~Wilke, editors, {\em Logic and
  Automata: History and Perspectives [in Honor of Wolfgang Thomas]}, volume~2
  of {\em Texts in Logic and Games}, pages 629--736. Amsterdam University
  Press, 2008.

\bibitem{vene:auto06}
Y.~Venema.
\newblock Automata and fixed point logic: a coalgebraic perspective.
\newblock {\em Information and Computation}, 204:637--678, 2006.

\bibitem{vene:lect18}
Y.~Venema.
\newblock Lectures on the modal $\mu$-calculus.
\newblock {L}ecture Notes, ILLC, University of Amsterdam, 2018.

\bibitem{walu:comp00}
I.~Walukiewicz.
\newblock Completeness of {K}ozen's axiomatisation of the propositional
  {$\mu$}-calculus.
\newblock {\em Information and Computation}, 157:142--182, 2000.

\bibitem{wilk:alte01}
T.~Wilke.
\newblock Alternating tree automata, parity games, and modal $\mu$-calculus.
\newblock {\em Bulletin of the Belgian Mathematical Society}, 8:359--391, 2001.

\end{thebibliography}
}

\appendix
\section{Parity games}
\label{sec:games}

\begin{definition}
\label{d:game}
A {\em parity game} is a tuple $\bbG = (G_{\eloi},G_{\abel},E,\Om)$ where 
$G_{\eloi}$ and $G_{\abel}$ are disjoint sets, and, with $G \isdef G_{\eloi} 
\cup G_{\abel}$ denoting the \emph{board} or \emph{arena} of the game, the binary 
relation $E \subseteq G^2$ encodes the moves that are \emph{admissible} to the 
respective players, and the \emph{priority function} $\Om: G \to \om$, which is 
required to be of finite range, determines the \emph{winning condition}
of the game.
Elements of $G_{\eloi}$ and $G_{\abel}$ are called \emph{positions} for the 
players $\eloi$ and $\abel$, respectively; given a position $p$ for player 
$\Pi \in \{ \eloi, \abel\}$, the set $E[p]$ denotes the set of \emph{moves}
that are \emph{legitimate} or \emph{admissible to} $\Pi$ at $p$.
In case $E[p] = \nada$ we say that player $\Pi$ \emph{gets stuck} at $p$.

An \emph{initialised board game} is a pair consisting of a board game $\bbG$
and a \emph{initial} position $p$, usually denoted as $\bbG@p$.
\end{definition}

\begin{definition}
\label{d:match}
A {\em match} of a graph game $\bbG = (G_{\eloi},G_{\abel},E,\Om)$ is a (finite 
or infinite) path through the graph $(G,E)$.
Such a match $\Si$ is called \emph{partial} if it is finite and $E[\last\Si]
\neq\nada$, and \emph{full} otherwise.
We let $\PM{\Pi}$ denote the collection of partial matches $\Si$ ending in a 
position $\last(\Si) \in G_{\Pi}$, and define $\PM{\Pi}@p$ as the set of 
partial matches in $\PM{\Pi}$ starting at position $p$.

The \emph{winner} of a full match $\Si$ is determined as follows.
If $\Si$ is finite, it means that one of the two players got stuck at the 
position $\last(\Si)$, and so this player looses $\Si$, while the opponent
wins.
If $\Si = (p_{n})_{n\in\om}$ is infinite, we declare its winner to be $\eloi$ 
if the maximum value occurring infinitely often in the stream 
$(\Om p_{n})_{n\in\om}$ is even.
\end{definition}

\begin{definition}
A \emph{strategy} for a player $\Pi \in \{ \eloi,\abel \}$ is a map $f:
\PM{\Pi} \to G$.
A strategy is \emph{positional} if it only depends on the last position of a 
partial match, i.e., if $f(\Si) = f(\Si')$  whenever $\last(\Si) = 
\last(\Si')$; such a strategy can and will be presented as a map $f: 
G_{\Pi} \to G$.

A match $\Si = (p_{i})_{i<\kappa}$ is \emph{guided} by a $\Pi$-strategy 
$f$ if $f(p_{0}p_{1}\ldots p_{n-1}) = p_{n}$ for all $n<\kappa$ 
such that $p_{0}\ldots p_{n-1}\in \PM{\Pi}$.
A position is \emph{reachable} by a strategy $f$ is there is an $f$-guided
match $\Si$ of which $p$ is the last position.
A $\Pi$-strategy $f$ is \emph{legitimate} in $\bbG@p$ if the moves that it
prescribes to $f$-guided partial matches in $\PM{\Pi}@p$ are always
admissible to $\Pi$, and \emph{winning for $\Pi$} in $\bbG@p$ if in addition
all $f$-guided full matches starting at $p$ are won by $\Pi$.

A position $p$ is a \emph{winning position} for player $\Pi \in \{ \eloi, \abel 
\}$ if $\Pi$ has a winning strategy in the game $\bbG@p$; the set of these
positions is denoted as $\Win_{\Pi}$.
The game $\bbG = (G_{\eloi},G_{\abel},E,\Om)$ is \emph{determined} if every
position is winning for either $\eloi$ or $\abel$.
\end{definition}

When defining a strategy $f$ for one of the players in a board game, we can 
and in practice will confine ourselves to defining $f$ for partial matches 
that are themselves guided by $f$.

The following fact, independently due to Emerson \& Jutla~\cite{emer:tree91}
and Mostowski~\cite{most:game91}, will be quite useful to us.

\begin{fact}[Positional Determinacy]
\label{f:pdpg}
Let $\bbG = (G_{\eloi},G_{\abel},E,\Om)$ be a parity game.
Then $\bbG$ is determined, and both players have positional winning strategies.
\end{fact}

In the sequel we will often refer to a `positional winning strategy' for one
of the players in a parity game.
With this we mean any positional strategy which is winning for that player when
starting at any of his/her winning positions.

\end{document}